\pdfoutput=1
\RequirePackage{lhchiggs}
\documentclass{cernyrep}
\usepackage[T1]{fontenc}
\usepackage[utf8]{inputenc}
\usepackage{lmodern}
\usepackage{enumerate}
\usepackage{overpic}
\usepackage{graphicx}
\usepackage{epstopdf}
\usepackage{rotating}
\usepackage{placeins}
\usepackage{amssymb}
\usepackage{amsfonts}
\usepackage{subcaption}
\usepackage{booktabs}
\usepackage{lscape}
\usepackage{axodraw2}
\usepackage{graphpap}
\usepackage{cite}
\usepackage[nolist]{acronym}
\usepackage{hyperref}
\usepackage[capitalise]{cleveref}
\crefname{paragraph}{paragraph}{paragraphs}
\usepackage[export]{adjustbox}

\newcommand{\SMm}{\ensuremath{\text{SM}}} 
\newcommand{\kappaframework}{$\kappa$-framework\xspace}
\newcommand{\Lag}{\ensuremath{\mathcal{L}}} 

\newcommand{\Gf}{G_\text{F}}
\newcommand{\alphaS}{\ensuremath{\alpha_S} } 
\newcommand{\as}{\alphaS }
\newcommand{\mur}{\ensuremath{\mu_R} } 
\newcommand{\muf}{\ensuremath{\mu_F} } 

\newcommand{\eV}{\ensuremath{\text{e\kern-0.15ex{}V}}\xspace}
\newcommand{\MeV}{\ensuremath{\text{M\eV}}\xspace}
\newcommand{\GeV}{\ensuremath{\text{G\eV}}\xspace}
\newcommand{\TeV}{\ensuremath{\text{T\eV}}\xspace}
\newcommand{\fb}{\ensuremath{\text{fb}}\xspace}

\newcommand{\MGMC}{\texttt{MadGraph5\_aMC@NLO}}

\newcommand{\id}{\mathrm{d}}
\newcommand{\mrm}[1]{\mathrm{#1}}
\newcommand{\define}{\equiv}
\newcommand{\kur}[1]{\mathcal{#1}}
\newcommand{\Li}{\text{Li}}
\newcommand{\Litt}{\Li_{2,2}}

\newcommand{\mh}{\ensuremath{m_{\h}}\xspace} 
\newcommand{\mt}{\ensuremath{m_t}\xspace} 
\newcommand{\mb}{\ensuremath{m_b}\xspace} 
\newcommand{\mc}{\ensuremath{m_c}\xspace} 
\newcommand{\mz}{\ensuremath{M_Z}\xspace} 
\newcommand{\mw}{\ensuremath{M_W}\xspace} 
\newcommand{\MSUSY}{\ensuremath{M_{SUSY}}\xspace} 
\newcommand{\mtsq}{\ensuremath{\mt^2}\xspace} 
\newcommand{\ppf}{\ensuremath{\mh^2}\xspace} 

\newcommand{\phistar}{\ensuremath{\phi^{*}_{\eta}}\xspace}  
\newcommand{\thstar}{\ensuremath{\theta^{*}_{\eta}}\xspace}
\newcommand{\pth}{\ensuremath{p_{T\h}}\xspace}  
\newcommand{\Eh}{\ensuremath{E_{\h}}\xspace}  
\newcommand{\MTh}{\ensuremath{M_{T\h}}\xspace}  
\newcommand{\pthsq}{\ensuremath{\pth^2}\xspace} 
\newcommand{\Ehsq}{\ensuremath{\Eh^2}\xspace} 
\newcommand{\vecpth}{\ensuremath{\vec{p}_{T\h}}\xspace}   
\newcommand{\pthcut}{\ensuremath{p_{T\h}^{{\rm cut}}}\xspace}  
\newcommand{\phistarcut}{\ensuremath{\phi^{*, {\rm cut}}_{\eta}}\xspace}  

\newcommand{\pt}{\ensuremath{p_{T}}\xspace}
\newcommand{\ptz}{\ensuremath{p_{T Z}}\xspace}
\newcommand{\cq}{\ensuremath{\kappa_q}\xspace}
\newcommand{\ct}{\ensuremath{\kappa_{t}}\xspace}
\newcommand{\cqt}{\ensuremath{\kappa_{s}}\xspace}

\newcommand{\cg}{\ensuremath{c_g}\xspace}
\newcommand{\mq}{m_q\xspace}

\newcommand{\M}{{\cal M}}
\newcommand{\h}{\ensuremath{H}\xspace}
\newcommand{\AF}{\mathcal{M}^{gg}_{F}\xspace}
\newcommand{\AS}{\mathcal{M}^{gg}_{S}\xspace}
\newcommand{\AUV}{\mathcal{M}^{gg}_{UV}\xspace}
\newcommand{\MF}{\mathcal{M}_{F}\xspace}
\newcommand{\MS}{\mathcal{M}_{S}\xspace}
\newcommand{\MUV}{\mathcal{M}_{UV}\xspace}
\newcommand{\dOmega}{{\rm d}\Omega\xspace}
\newcommand{\dpth}{{\rm d}\pth\xspace}
\newcommand{\dsigma}{{\rm d}\sigma\xspace}
\renewcommand{\SM}{{\rm SM}\xspace}
\renewcommand{\LO}{{\rm LO}\xspace}
\renewcommand{\NLO}{{\rm NLO}\xspace}
\renewcommand{\NNLO}{{\rm NNLO}\xspace}
\newcommand{\BSM}{{\rm BSM}\xspace}
\newcommand{\MSSM}{{\rm MSSM}\xspace}
\newcommand{\HEFT}{{\rm HEFT}\xspace}
\newcommand{\EFTplus}{\ensuremath{\HEFT\!\oplus\!{\rm M}}\xspace}
\newcommand{\EFTtimes}{\ensuremath{\HEFT\!\otimes\!{\rm M}}\xspace}

\newcommand{\BR}{{\rm BR}}

\newcommand{\RunI}{Run\,1\xspace}
\newcommand{\RunII}{Run\,2\xspace}

\newcommand{\spc}{\,,}
\newcommand{\spp}{\,.}
\newenvironment{definition}[1][Definition]{\begin{trivlist}
\item[\hskip \labelsep {\bfseries #1}]}{\end{trivlist}}
\newcommand{\nl}{\nonumber\\}
\newcommand{\bqa}{\arraycolsep 0.14em\begin{eqnarray}}
\newcommand{\eqa}{\end{eqnarray}}

\makeatletter
\renewcommand{\thechapter}{\@Roman\c@chapter}
\renewcommand{\thechapter}{\Roman{chapter}}
\makeatother

\begin{acronym}
\acro{1PI}{One Particle Irreducible}
\acro{BFM}{Background Field Method}
\acro{BR}{Branching Ratio}
\acro{BSM}{Beyond the SM}
\acro{CDE}{Covariant Derivative Expansion}
\acro{EFT}{Effective Field Theory}
\acrodefplural{EFT}{Effective Field Theories}
\acro{EW}{Electroweak}
\acro{EWPD}{EW Precision Data}
\acro{IR}{infrared}
\acro{LEP}{Large Electron-Positron Collider}
\acro{LHC}{Large Hadron Collider}
\acro{LHE}{Les Houches Event}
\acro{LO}{Leading Order}
\acro{MC}{Monte Carlo}
\acro{NLO}{Next-to-Leading Order}
\acro{NNLO}{Next-to-Next-to-Leading Order}
\acro{N3LO}{Next-to-Next-to-Next-to-Leading Order}
\acro{NP}{New Physics}
\acro{NWA}{Narrow Width Approximation}
\acro{PDF}{Probability Density Function}
\acro{PO}{Pseudoobservable}
\acro{QCD}{Quantum Chromodynamic}
\acro{QED}{Quantum Electrodynamic}
\acro{QFT}{Quantum Field Theory}
\acro{RG}{Renormalisation Group}
\acro{SM}{Standard Model}
\acro{SMEFT}{Standard Model EFT}
\acro{STXS}{Simplified Template Cross Section}
\acro{UFO}{Universal FeynRules Output}
\acro{UV}{ultraviolet}
\acro{VEV}{Vacuum Expectation Value}
\acro{WST}{Ward-Slavnov-Taylor}
\acro{BW}{Breit-Wigner}
\acro{MPE}{multi-pole expansion}
\acro{PV}{Passarino-Veltman}
\acro{MHOU}{Missing Higher Order Uncertainty}
\end{acronym}

\begin{document}

\null\vskip-60pt \hfill
\begin{minipage}[t]{4cm}
IPPP/17/90 \\
\end{minipage}

{\centering{\LARGE{\bf{The HiggsTools Handbook:\\
Concepts and observables for deciphering \\the Nature of the Higgs Sector\par }}}}

\pagenumbering{roman}

\section*{Abstract}

\vspace{0.5cm}

This Report summarizes some of the activities of the HiggsTools Initial 
Training Network working group in the period 2015-2017.
The main goal of this working group was to produce a document discussing various 
aspects of state-of-the-art Higgs physics at the \ac{LHC} in a pedagogic manner.

The first part of the Report is devoted to a description of phenomenological searches for New Physics
at the \ac{LHC}.  All of the available studies of the couplings of the new resonance 
discovered in 2012 by the ATLAS and CMS experiments~\cite{Aad:2012tfa,Chatrchyan:2012xdj}
conclude that it is 
compatible with the Higgs boson of the \ac{SM} within present precision. So far 
the LHC experiments have given no direct evidence for any physical phenomena that cannot be described by the \ac{SM}.  As the 
experimental measurements become more and more precise, there is a pressing need 
for a consistent framework in which deviations from the \ac{SM} predictions can 
be computed precisely. Such a framework should be applicable to measurements in 
all sectors of particle physics, not only LHC Higgs measurements but also 
electroweak precision data, etc. We critically review the use of the \kappaframework, 
fiducial and simplified template cross sections,
effective field theories, pseudoobservables and phenomenological Lagrangians. 
Some of the concepts presented here are well known and were used already at the 
time of the Large Electron-Positron Collider (LEP) experiment.
However, after years of theoretical and experimental development, 
these techniques have been refined, and we describe new tools that have been introduced 
in order to improve the comparison between theory and experimental data.

In the second part of the Report, we propose \phistar as a new and complementary observable for 
studying Higgs boson production at large transverse momentum in the case where the Higgs boson 
decays to two photons. 
The \phistar variable depends on measurements of the angular directions and rapidities of 
the two Higgs decay products rather than the energies,  and exploits the information 
provided by the calorimeter in the detector. We show that, even without tracking information, the experimental 
resolution for \phistar is better than that of the transverse momentum of the photon pair, 
particularly at low transverse momentum. We make a detailed study of the phenomenology of the 
\phistar variable, contrasting the behaviour with the Higgs transverse momentum distribution 
using a variety of theoretical tools including  event generators and fixed order perturbative 
computations. We consider the theoretical uncertainties associated with both \pth and \phistar 
distributions. Unlike the transverse momentum distribution, the \phistar distribution is well 
predicted using the Higgs Effective Field Theory in which the top quark is integrated out -- 
even at large values of \phistar --  thereby making this a better observable for extracting 
the parameters of the Higgs interaction. In contrast, the potential of the \phistar distribution 
as a probe of new physics is rather limited, since although the overall rate is affected by 
the presence of additional heavy fields, the shape of the the \phistar distribution is 
relatively insensitive to heavy particle thresholds.

\newpage
{\centering{\bf{Authors\par}}}

\begin{flushleft}

M.~Boggia$^{1}$,
J.~M.~Cruz-Martinez$^{2}$,
H.~Frellesvig$^{3,4}$,
N.~Glover$^{2}$,
R.~Gomez-Ambrosio$^{2,5}$,
G.~Gonella$^{1}$,
Y.~Haddad$^{2,6}$, 
A.~Ilnicka$^{7,8}$,
S.~Jones$^{1,9}$,
Z.~Kassabov$^{5,10}$,
F.~Krauss$^{2}$,
T.~Megy$^{1}$,
D.~Melini$^{11,12}$,
D.~Napoletano$^{2}$,
G.~Passarino$^{5}$,
S.~Patel$^{13,14,15}$,
M.~Rodriguez-Vazquez$^{16}$,
T.~Wolf$^{17}$

\end{flushleft}

\begin{itemize}

\item[$^{1}$] Physikalisches Institut, Albert-Ludwigs-Universit{\"a}t Freiburg, 79104 Freiburg, Germany
\item[$^{2}$] Institute for Particle Physics Phenomenology, Department of Physics, University of Durham, Durham DH1 3LE, UK
\item[$^{3}$] Institute of Nuclear and Particle Physics, NCSR ``Demokritos'', Agia Paraskevi, 15310, Greece
\item[$^{4}$] Institute for Theoretical Particle Physics (TTP), Karlsruhe Institute of Technology, Engesserstra{\ss}e 7, D-76128 Karlsruhe, Germany
\item[$^{5}$] Dipartimento di Fisica Teorica, Universit\`a degli Studi di Torino and INFN, Sezione di Torino, Via Pietro Giuria 1, 10125 Turin, Italy
\item[$^{6}$] High Energy Physics Group, Blackett Lab., Imperial College, SW7 2AZ, London, UK
\item[$^{7}$] Institute for Theoretical Physics, University of Z\"urich, Winterthurerstrasse 190, 8057 Z\"urich, Switzerland
\item[$^{8}$] Institute for Particle Physics, Physics Department, ETH Z\"urich, 8093 Z\"urich, Switzerland
\item[$^{9}$] Max-Planck-Institut f\"ur Physik, Werner-Heisenberg-Institut, F\"ohringer Ring 6, 80805 M\"unchen, Germany
\item[$^{10}$] Dipartimento di Fisica, Universit\`a degli Studi di Milano and INFN, Sezione di Milano, Via Celoria 16, 20133 Milan, Italy
\item[$^{11}$] Departamento de F\'isica Te\'orica y del Cosmos, Avenida de la Fuente Nueva S/N C.P. 18071 Granada, Spain
\item[$^{12}$] IFIC, Universitat de Val\`encia and CSIC, Catedr\'atico Jose Beltr\'an 2, 46980 Paterna, Spain
\item[$^{13}$] DESY, Notkestrasse 85, 22607 Hamburg, Germany
\item[$^{14}$] Institute for Theoretical Physics (ITP),  Karlsruhe Institute of Technology, Wolfgang-Gaede-Stra{\ss}e 1, D-76131 Karlsruhe, Germany
\item[$^{15}$] Institute  for  Nuclear  Physics  (IKP),  Karlsruhe  Institute  of  Technology, 
Hermann-von-Helmholtz-Platz 1, D-76344 Karlsruhe, Germany
\item[$^{16}$] Laboratoire de Physique Th\'eorique, UMR 8627, CNRS, Universit\'e de Paris-Sud, Universit\'e
Paris-Saclay, 91405 Orsay, France
\item[$^{17}$] Nikhef, Science Park 105, 1098 XG Amsterdam, The Netherlands


\end{itemize}

\section*{Acknowledgements}

We thank Ramona Gr\"ober for useful comments, and Xuan Chen, Aude Gehrmann-De Ridder, Thomas Gehrmann, Alex Huss and Tom Morgan for their collaboration in developing the NNLOJET code.
We thank Marzieh Bahmani, Giulio Falcioni, Nicolas Gutierrez, Brian Le, Elisa Mariani, Joosep Pata and Cosimo Sanitate for their contributions to the HiggsTools Initial Training Network.
We gratefully acknowledge the Research Executive Agency (REA) of the European Union for funding through the Grant Agreement PITN-GA-2012-316704 (``HiggsTools''). M.B. and G.G. acknowledge the German Research Foundation~(DFG) and Research Training Group GRK~2044 for the funding and the support.




\newpage
\tableofcontents


\newpage
\pagenumbering{arabic}
\setcounter{footnote}{0}



\newpage

\chapter{Towards a theory of Standard Model deviations}
\renewcommand{\thesection}{\thechapter.\arabic{section}}
\label{cha:kappa}
\section{Introduction}
Before the discovery of the Higgs boson in $2012$, the hypothesis under test at the
\ac{LHC} was the Standard Model (SM)
with the Higgs boson mass, $\mh$, being the unknown parameter. Therefore, bounds 
on $\mh$ were derived through a comparison 
of theoretical predictions with high-precision data. Now, after the discovery, the  SM is fully specified and 
the unknowns are constrainable deviations from the SM. Of course, the definition of 
SM deviations requires a characterization of the underlying dynamics. So far, all of
the available studies of the couplings of the new resonance conclude that it is compatible with 
the Higgs boson of the SM with our current precision, and, as of yet, there is no direct evidence 
for new physics phenomena beyond the SM at the LHC.

This chapter is devoted to a description of phenomenological searches for \ac{NP} at the \ac{LHC}. 
Some of the concepts presented here are well known and were used already at the time of 
the \ac{LEP}. Now, after years of developments both on the experimental and the 
theoretical sides, these techniques have been refined, and new tools have been introduced 
in order to improve the comparison of theory to experiment.

In this work, different approaches are revisited, with special emphasis on the relations and connections between 
them, showing their limits, and how they complement each other.
In the following we will introduce their main definitions, necessary for a complete comprehension 
of the chapter. The technical details will be largely covered in the following sections.

\subsubsection*{The \kappaframework} 
The \kappaframework was proposed shortly after the 
discovery of the Higgs boson in order to try a systematic and model-independent search for 
\ac{SM} deviations. It is used at \ac{LO} and accommodates for factorisable \ac{QCD} corrections 
but not for \ac{EW} ones. The \kappaframework will be covered in~\cref{sec:kappa}, where its use during the \RunI  of the LHC
is discussed and its limitations highlighted.

\subsubsection*{Fiducial and Simplified Template Cross Sections}
 Fiducial cross sections 
are useful observables 
for particle physics, but they have always been a source of disagreement between theorists 
and experimentalists. The reason is that the phase space defined by the theory is not the same 
as the one of the detector and additionally, different detectors operate at their maximum efficiency in 
different regions of the phase space. The \acp{STXS} are one of the possible answers to this problem.
They provide a gain in experimental sensitivity by introducing a small theoretical model
dependence. \acp{STXS} are defined in greater detail in~\cref{sec:fiducial}.

\subsubsection*{\acp{EFT}} 
\acp{EFT} come into play when the solutions 
to a quantum field theory are provided as expansions in a small coupling constant and ratio of scales. Among 
these theories the \ac{SMEFT} is an example. In such theories an infinite number of higher 
dimensional operators must be included, which means that the theory is (strictly) 
non-renormalisable. Nevertheless, if $v$ is the Higgs vacuum expectation value and $E$ is the typical scale 
at which the process is measured, the \ac{EFT} amplitudes are expanded in powers of 
$v/\Lambda$ and of $E/\Lambda$, where $\Lambda$ is the scale of new physics. This expansion is computable 
to all orders and the introduction, order-by-order, of an increasing number of counter-terms 
can eliminate the UV divergences, which makes the theory renormalisable order-by-order in the $E/\Lambda$ perturbative expansion. An overview of \acp{EFT} is presented in \cref{sec:EFT} 
introducing definitions, descriptions and methodologies.

\subsubsection*{\acp{PO}} Pseudoobservables were introduced to allow a better 
interplay between theory and experiment. They answer the urgent need of extending the 
parameters of the \kappaframework to something with a more rigorous theoretical interpretation. Another main advantage of POs is that they deliver results which 
do not have to be totally unfolded by experiments to a model-dependent parameter space. \acp{PO} could also be the solution to 
the issue posed by the \kappaframework of not being able to describe theoretically any eventual \ac{SM} deviations (being only able to \emph{parametrise} but not to \emph{explain} such measurements), and 
at the same time, POs make it possible to get closer to quantities that are well defined QFT objects. \acp{PO} had been already used for \ac{LEP} analysis, and they 
have gained new relevance now given their close connection with \acp{EFT}. 
They are addressed in ~\cref{sec:POs} as a natural continuation of \cref{sec:EFT}.

\subsubsection*{Phenomenological Lagrangians} 
Several models have been built that aim at 
producing observable \ac{NP} effects based on phenomenological Lagrangians constructed by adding a 
limited number of additional interactions to the \ac{SM} Lagrangian. Theoretical tools and 
\ac{MC} generators with such implementations are already available. \cref{sec:tools}   
offers an insight into this topic, together with a discussion of the challenges faced when dealing 
with \ac{NLO} corrections.

\noindent
Finally, there are a couple of finer points in terminology that are worth keeping in mind: 
\begin{itemize}
\item[1)] technically speaking leading-order defines the order in perturbation theory where the process starts.
Sometimes, however, ``LO'' is used in the literature to denote tree level (as opposed to loop contributions).
\item[2)] The 
resonant (often called ``signal'') and the non-resonant (often called ``background'') are parts 
of a physical process (which may contain more than one resonant part).
Typically whenever we write $i \to X \to f$ we mean the $X\,$-resonant part of a process 
with initial state $i$ and final state $f$. However, we should keep in mind that separating 
and defining ``signals'' and ``backgrounds'' is not trivial. For example, $V h$ production  at \ac{LO} in QCD  with $V \to jj$ and Vector Boson Fusion (VBF) are not clearly separated. At \ac{NNLO} QCD everything becomes  much more complicated and even the definition of $V h, \; V \to jj$ and VBF is ambiguous.
\end{itemize}


\clearpage

\newpage
\section{The \kappaframework} \label{sec:kappa}
\subsection{Concept and description}
Theory and experiment aim at the same goal but, surprisingly, they often find it hard to communicate with each other. The search for a common language to be used 
by experimentalists to express their results and by theorists to interpret them has become more and more 
urgent after the observation of a massive neutral boson~\cite{Aad:2012tfa, Chatrchyan:2012xdj}, 
that is identified as being compatible with the quantum fluctuations of a field whose \ac{VEV} breaks 
the \ac{EW} symmetry. This is the well-known Brout-Englert-Higgs mechanism of Refs.~\cite{Higgs:1964ia,Higgs:1964pj,Englert:1964et,Guralnik:1964eu,Higgs:1966ev,Kibble:1967sv} and the LHC resonance has been interpreted so far in terms of the SM Higgs boson. 

The Higgs boson couplings to the other known particles play a central role in the investigation of the properties of this state, since they are predicted very accurately by the SM and its extensions. They also influence the Higgs boson production and decay rates: this is why an interim framework called 
the \kappaframework  was   proposed in Ref.~\cite{LHCHiggsCrossSectionWorkingGroup:2012nn} to parameterise
small deviations from the predicted \ac{SM} Higgs boson couplings and widths, 
for a recent review see Ref.~\cite{Mariotti:2016owy}. The basic assumptions are:
\begin{itemize}

\item The signals observed in the different search channels originate from a single narrow 
resonance with a mass near $125 \, \GeV$.

\item The zero-width approximation is used, meaning that the total width of the 
resonance is narrow enough to be negligible with respect to the mass. In this case, the signal cross section can be decomposed (for all $i \to \h \to f$ channels) as
\begin{equation}
\sigma_i \cdot \BR^f  = 
\frac{\sigma_{i} \cdot \Gamma_{f}}{\Gamma_{\h}} \spc
\end{equation}
where $\sigma_i$ is the production 
cross section  for $i \to \h$ with $i = (gg\textrm{F}, \, \textrm{VBF}, \, W\h, \, Z\h, \, t\bar{t}\h)$ and 
\BR$^f$ is the decay \ac{BR} for $\h \to f$ with $f = (ZZ, \, WW, \, \gamma\gamma, \, \tau\tau, \, bb, \, \mu\mu)$. $\Gamma_{f}$ and $\Gamma_{\h}$ are the partial and total decay 
widths of the Higgs boson respectively.

\end{itemize}
The couplings cannot be directly measured by the experiments and
events collected in the detector undergo an unfolding procedure, in order for us to be able to extract the information from some measurable observable(s). Typically, the measured observable is $\sigma \times \BR$, which is
defined within a certain acceptance and with some specific experimental cuts, and this leads to some model dependence. 


Obviously, it is not possible to fit the experimental data within the context of the SM while treating Higgs couplings as free parameters. Once the value of the Higgs boson  mass is specified, the couplings are specified as well. For this reason, it is only possible to test the  overall compatibility of the SM with the data. This kind of study can be used to extract or constrain deviations 
in the measured couplings with respect to the the SM ones. 

Some of the available approaches introduce specific modifications in the structure of the  couplings: besides including all the available higher order corrections in the model, they add additional terms to the Lagrangian, the so-called ``anomalous couplings''. This gives rise to modifications of the tensor structure of the amplitude leading not only to a modification of the coupling strengths, but also of the kinematic  distributions. This approach has turned out to be a difficult one when it comes to reinterpreting the results. An additional assumption is then made to simplify the framework:
\begin{itemize}

\item The tensor structure of the couplings is assumed to be the same as in the SM predictions, 
only modifications of coupling strengths are taken into account. This implies that the observed 
state is assumed to be a CP-even scalar.

\end{itemize}
The framework is built in such a way that the predicted SM Higgs cross section and partial decay 
widths are dressed with scaling factors $\kappa_{j}$, as shown in \cref{table:tableParametriz}. 
The cross section $\sigma_{i}$ and the partial decay width $\Gamma_{f}$ scale with 
$\kappa_{i}^2, \kappa_{f}^2$ when compared to SM predictions.
\begin{equation}
\label{eq:kappafac}
 \sigma_i \cdot \BR^{f} = 
(\sigma_{i} \cdot \BR^{f})_{\SMm} 
\cdot \frac{\kappa^{2}_{i}\cdot \kappa^{2}_{f}}{\kappa^{2}_{\h}} \spp
\end{equation}

In the SM, all the $\kappa_i$ are unity by definition and therefore, the best available (i.e. highest possible perturbative order) predictions are recovered.   However, when $\kappa_i \neq 1$ higher-order accuracy is in general lost due to the factorisation of Eq.\eqref{eq:kappafac} does not necessarily hold beyond LO. This will be illustrated later on with an example.   

The $\kappa$ are sometimes confused with the actual couplings in the Lagrangian.  They are the same at tree level,  but not the same once radiative corrections are taken into account. As an example we consider the process 
$gg \rightarrow t\bar{t}\h(bb)$. At tree level one can say that the squared matrix elements are 
proportional to the coupling of the interactions:
\begin{equation}
\left| \mathcal{M}\right|^2 \propto  \left( g_{t\h}  g_{b\h}\right)^2    \spc
\end{equation}
and then, assuming that the total width of the Higgs boson stays unchanged, 
$\Gamma_{\h} \equiv \Gamma^{\SMm}_{\h}$, one would have $\kappa_{i}=g_{t\h}$ and 
$\kappa_{f}=g_{b\h}$. 

Another quantity that can be expressed in this framework and is a common experimental observable, is the \emph{signal strength} $\mu$. 
Consider a specific process $i \to \h \to f$. For the production ($i \to \h$) the signal strength $\mu$ is defined as 
\begin{equation}
\mu_i = \frac{\sigma_i}{ \sigma_{i{\SMm}}} \spc
\end{equation}
whereas, for the decay ($\h \to f$)  we have
\begin{equation}
\mu_f = \frac{\BR^f}{\BR^f_{{\SMm}}} \spp
\end{equation}
By definition, in the \ac{SM} $\mu_i =$ 1 and $\mu_f =$ 1.
The only thing that can be measured experimentally is the product of $\mu_i$ and $\mu_f $, since 
it is not possible to separate them without further assumptions. In terms of the 
\kappaframework we obtain the following expression:
\begin{equation}
\mu\equiv \mu_i \mu_f \equiv \frac{\kappa^2_{i} \kappa^2_{f}}{\kappa^2_{\h}}   \spp
\end{equation}

To summarize, the original \kappaframework is a simplified picture which shows its limitations when more precise data necessitate the inclusion of higher order QCD and EW corrections. 
In many cases, the QCD corrections do not completely factorise and their impact in the context of the SMEFT can be 
sizeable, as was shown in Ref.~\cite{Gauld:2016kuu}.

Keeping this in mind, let us first discuss the LO strategy, which was applied during the analysis of the LHC \RunI data. Different production processes and decay modes probe different coupling modifiers. Together with the individual modifiers related to the coupling of the Higgs boson with different 
particles, two \textit{effective} modifiers need to be introduced to describe loop induced 
processes: the modifier $\kappa_g$ to describe the $gg$F production process and 
$\kappa_{\gamma}$ for the $\h \to \gamma \gamma$ decay. 

This is possible since it is expected 
that other \ac{BSM} particles which might be present in the loop do not change the kinematics 
of the process. The study of these processes can therefore be approached by either using effective 
coupling modifiers, which provide sensitivity to the presence of \ac{BSM} particles in the 
loops, or using \textit{resolved} coupling modifiers corresponding to the \ac{SM} particles.

\begin{table}
\[
\scalebox{0.9}{$
\begin{array}{lcccc}
\toprule
                       &            &              & \text{Effective}       & \text{Resolved} \\
\text{Production}             & \text{Loops}      & \text{Interference} & \text{scaling factors} & \text{scaling factors} \\ 
\midrule
\sigma(gg{\rm F})      & \checkmark & t\!-b          & \kappa_g^2      & 1.06\cdot \kappa^2_t + 0.01 \cdot \kappa_b^2 + 0.07 \cdot \kappa_t \kappa_b \\    
\sigma({\rm VBF})      & -          & -            & -               & 0.74\cdot \kappa_W^2 + 0.26 \cdot \kappa_Z^2 \\                       
\sigma(W\h)             & -          & -            & -               & \kappa^2_W   \\                   
\sigma(qq/qg \to Z\h)   & -          & -            & -               & \kappa_Z^2   \\                       
\sigma(gg \to Z\h)      & \checkmark & t\!-\!Z          & -               & 2.27 \cdot \kappa_Z^2 + 0.37 \cdot \kappa_t^2 - 1.64 \cdot \kappa_Z \kappa_t \\
\sigma(t\bar{t}\h)            & -          & -            & -               & \kappa_t^2   \\ 
\sigma(gb \to t\h W)     & -          & t\!-\!W          & -               & 1.84 \cdot \kappa_t^2 + 1.57 \cdot \kappa_W^2 - 2.41 \cdot \kappa_t \kappa_W \\                                
\sigma(qq/qb \to t\h q)  & -          & t\!-\!W          & -               & 3.40 \cdot \kappa_t^2 + 3.56 \cdot \kappa_W^2 - 5.96 \cdot \kappa_t \kappa_W  \\                              
\sigma(b\bar{b}\h)            & -          & -            & -               & \kappa_b^2 \\
\midrule
\text{Partial decay width}    &            &              &                 & \\
\midrule
\Gamma_{ZZ}             &  -         &   -          & -               &   \kappa^2_Z      \\                        
\Gamma_{WW}             &  -         &   -          & -               &   \kappa^2_W      \\                
\Gamma_{\gamma \gamma}    & \checkmark &   t\!-\!W        & \kappa_{\gamma}^2 &   1.59 \cdot \kappa_W^2 + 0.07 \cdot \kappa_t^2 - 0.66 \cdot \kappa_W \kappa_t            \\ 
\Gamma_{\tau \tau}       & -           & -           &-                 & \kappa^2_{\tau} \\    
\Gamma_{bb}             &  -         &   -          & -               &   \kappa^2_b      \\     
\Gamma_{\mu \mu}         & -          &  -           &-                &  \kappa^2_{\mu}  \\
\midrule
\text{Total width (\BR$_{\rm BSM}$ =0)} &  &  &  & \\
\midrule
             &             &              &               & 0.57 \cdot \kappa_b^2 + 0.22 \cdot \kappa_W^2 + 0.09 \cdot \kappa_g^2 +        \\
\Gamma_{\h}     &  \checkmark &  -           &  \kappa_{\h}^2   & 0.06 \cdot \kappa_{\tau}^2 +0.03 \cdot \kappa_Z^2 + 0.03 \kappa_c^2 +                \\
&             &              &             & 0.0023 \cdot \kappa_{\gamma}^2 + 0.0016 \cdot \kappa_{(Z\gamma)}^2 +                  \\
&            &             &              & 0.0001 \cdot \kappa_s^2 + 0.00022 \cdot \kappa_{\mu} \\
\bottomrule
\end{array}
$}
\]
\caption{Higgs boson production cross sections $\sigma_i$, partial decay widths $\Gamma_f$, 
and total decay width (in the absence of \ac{BSM} decays) parametrised as a function of the 
$\kappa$ coupling modifiers, including higher-order \ac{QCD} and \ac{EW} corrections to the 
inclusive cross sections and decay partial widths. The coefficients in the expression
for $\Gamma_{\h}$ do not sum exactly to unity because some negligible contributions  
are not shown. Table from Ref.~\cite{Khachatryan:2016vau}.}
\label{table:tableParametriz}
\end{table}
\subsection{The \kappaframework in the experiments} \label{subsec:kexperiment}

The measurement of the properties of the Higgs boson is one of the main goals of the two \ac{LHC} general purpose experiments: ATLAS, described in Ref.~\cite{Aad:2008zzm}, and CMS, described in Ref~\cite{Chatrchyan:2008aa}. 

For the interpretation of results in the light of a combination it is mandatory to have a global overview of the current situation and to understand how the \kappaframework has been used by the experiments so far. In the following we review the combination of measurements performed 
by the ATLAS and CMS experiments, which was presented in Ref.~\cite{Khachatryan:2016vau}, with a focus on the 
constraints on the couplings. 

The analysis used the data collected by the detectors from $pp$ collisions at the LHC in $2011$ and $2012$, corresponding to an integrated luminosity of 
approximately $5$ fb$^{-1}$ at $\sqrt{s} = 7 \, \TeV$ and $20 \, \fb^{-1}$ at 
$\sqrt{s} = 8 \, \TeV$. They considered multiple production processes: gluon fusion, vector boson fusion, and associated production with a $W$ or a $Z$ boson or a pair of top quarks,  and the  $\h \to ZZ$, $WW$, $\gamma \gamma$, $\tau \tau$, $bb$ and 
$\mu \mu$ decay modes.\footnote{Note that the definition of the $\h \to ZZ (WW)$ decay requires 
some additional information.}  
Two formalisms were used to interpret the results: the signal strength $\mu$, 
related to the yields, and the \kappaframework for the couplings. Usually, Higgs cross section measurements are given in two ways:
\begin{itemize}

\item Fiducial cross-section: this has basically no input from the signal Monte Carlo (MC), 
so it can be compared to any theory calculation, 
provided that it can reproduce the cross-section within these specific cuts

\item Signal strength ($\mu$). In this case, 

\begin{itemize}

 \item The event yield after cuts is computed.

 \item This is extrapolated to the full phase space using signal Monte Carlo (typically  POWHEG \cite{Alioli:2010xd} + JHUGen~\cite{Gao:2010qx,Bolognesi:2012mm,Anderson:2013afp,Gritsan:2016hjl} + Pythia8~\cite{Sjostrand:2014zea}+ Geant4~\cite{Agostinelli:2002hh}) in the full detector simulation to compute acceptance and efficiency.

 \item The result is compared with the best theory predictions, e.g. \ac{N3LO} QCD for production in the gluon fusion channel, and Prophecy4f, from Refs.~\cite{Bredenstein:2006rh,Bredenstein:2006nk,Bredenstein:2006ha}, for decays with four fermions in the final state.

\end{itemize}

\end{itemize}

To directly measure the individual coupling modifiers, an assumption for the Higgs boson width 
is necessary. It is predicted to be approximately $4 \, \MeV$ in the SM, which is assumed to be 
small enough for the \ac{NWA} to be valid and for the Higgs boson production and decay 
mechanisms to be factorised.

The relations among the coupling modifiers, the production cross-sections and the partial decay widths of \cref{table:tableParametriz} are used as a parametrisation to extract the coupling modifiers from the measurements.
Changes in the values of the couplings will lead to a change of the Higgs boson width. To characterise this variation a new modifier $\kappa_{\h}$ is introduced, defined as 
$\kappa_{\h}^2 = \sum_j \BR^j_{\SMm}\kappa_j^2$. If we only allow for \ac{SM} decays of the Higgs boson, the relation $\kappa_{\h}^2 = \Gamma_{\h} / \Gamma_{\h;\SMm}$ holds, 
otherwise $\Gamma_{\h}$ can be expressed as:
\begin{equation}
\Gamma_{\h} = \frac{\kappa^2_{\h} \cdot \Gamma_{\h;\SMm}}{1-\BR_{\rm BSM}} \spc
\end{equation}
where $\BR_{\rm BSM}$ indicates the total branching fraction into \ac{BSM} decays.  
Since $\Gamma_{\h}$ is not experimentally constrained in a model-independent manner with sufficient 
precision, only ratios of coupling strengths can be measured in the most generic parametrisation 
considered in the \kappaframework.
The individual ATLAS and CMS analysis of the Higgs boson production and decay rates are combined 
using the profile likelihood method.
\begin{figure}[]
  \begin{center}
    \includegraphics[scale=0.3]{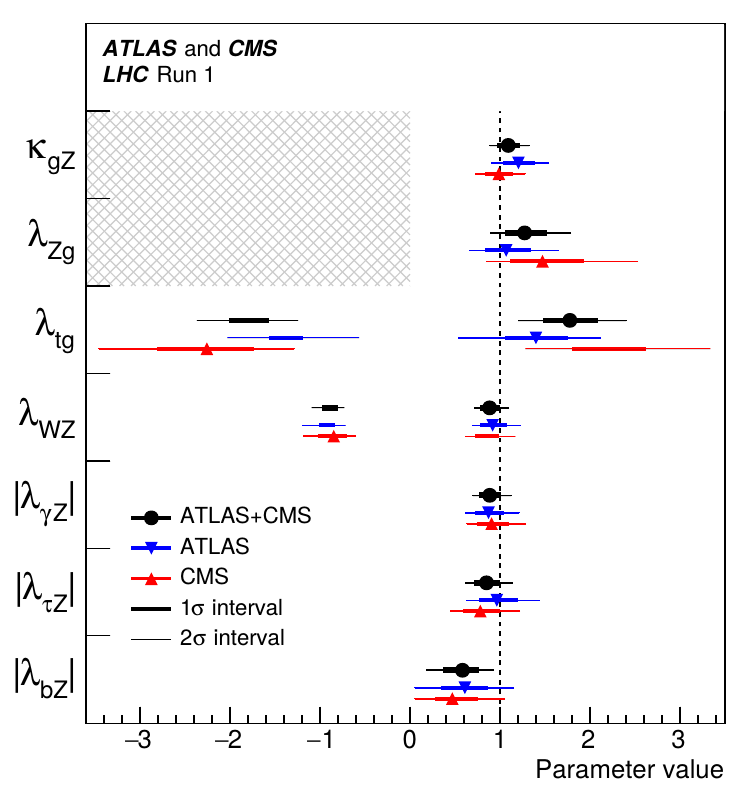}
  \end{center}
  \caption{Best fit values of ratios of Higgs boson coupling modifiers, as obtained from 
the generic parametrisation. The single results from each experiment are also shown. The thick 
error bars indicate the $1\,\sigma$ interval and the thin lines the $2\,\sigma$ one.
The hatched areas indicate the non-allowed regions for the parameters that are assumed to be 
positive without loss of generality. For those parameters with no sensitivity to the sign, 
only the absolute values are shown. Figure from Ref.~\cite{Khachatryan:2016vau}.}
  \label{fig:fitResults}
\end{figure}
\subsubsection*{Intermezzo: profile likelihood method} \label{par:likelihood}
The statistical data treatment used in the combination is the same as that used by the single analysis 
and described in Ref.~\cite{Cowan:2010js}. 
Let us consider a kinematic variable $x$ and the corresponding histogram of values 
$\mathbf{n} = (n_1,\dots n_n)$. The expectation value in a bin $i$ can be written as
\begin{equation}
E[n_i]= \mu s_i + b_i, \quad \mbox{where} \quad
s_i = s_{tot} \int_{bin\,i} f_s(x, \mathbf{\theta}_s)dx \spc \quad
b_i = b_{tot} \int_{bin\, i} f_b(x, \mathbf{\theta}_b)dx \spc
\end{equation}
and $\mu$ determines the strength of the signal process, with $\mu=0$ being the background-only 
hypothesis and $\mu=1$ the nominal signal hypothesis. 

The functions $f_s(x, \mathbf{\theta}_s)$ 
and $f_b(x, \mathbf{\theta}_b)$ are the probability density functions
of the variable $x$ for signal and 
background characterised by the set of parameters $\vec{\theta_s}$ and $\vec{\theta_b}$. 
The quantities $s_{tot}$ and $b_{tot}$ are the total mean numbers of events for signal and background: 
$s_{tot}$ is not considered to be an adjustable parameter but rather fixed to the value 
predicted by the nominal signal model. 

In addition to the histogram of interest $\mathbf{n}$, supporting measurements are 
usually made to help constrain the nuisance parameters. Therefore for a control 
sample where mainly background is expected, we may find a histogram $\mathbf{m}= (m_1, ..., m_M)$ for 
a particular kinematic variable, where the expectation value of $m_i$ is
\begin{equation}
E[m_i] = u_i(\mathbf{\theta}) \spp
\end{equation}
Here the $u_i$ are calculable quantities that depend on $\mathbf{\theta} = (\mathbf{\theta}_s, 
\mathbf{\theta}_b, b_{tot})$. This is often done to obtain information on the parameter 
$b_{tot}$. 

In each bin, the number of events follows the Poisson distribution and the 
likelihood function is the product of those probabilities for every bin,
\begin{equation}
\mathrm{L}(\mu, \mathbf{\theta}) = 
\prod_{j=1}^N \frac{(\mu s_i + b_i)^{n_j}}{n_j!} e^{-(\mu s_i + b_i)} 
\prod_{k=1}^M \frac{u_k^{m_k}}{m_k!}e^{-u_k} \spp
\end{equation}
The likelihood function is then used to build a so-caled ``test statistic''
\begin{equation}
\Lambda(\mu) = \frac{\mathrm{L}(\mu, 
\mathbf{\hat{\hat{\theta}}})}{\mathrm{L}(\hat{\mu}, \mathbf{\hat{\theta}})} \spc
\end{equation}
where $\mathbf{\hat{\hat{\theta}}}$ indicates the value of $\mathbf{\theta}$ that maximizes 
$L$ for a specific value of $\mu$ (conditional maximum-likelihood estimator for $\theta$), 
while $\hat{\mu}$ and $\mathbf{\hat{\theta}}$ are the unconditional maximum likelihood estimates 
of the parameter values.

Likelihood fits are then performed to obtain the values for the parameters of interest. The actual data is used for the observed values and Asimov datasets, see for instance Ref.~\cite{Cowan:2010js}, are used for the expected values. 
Asimov datasets are pseudo-data distributions equal to the signal plus background predictions 
for a given value of the parameters. By definition, when the Asimov dataset is used to  estimate the parameters, the ``true'' values are obtained. 

Often, not all of the parameters need to be estimated and in these cases a \textit{profile likelihood} analysis
is performed, which means that the parameters one is not interested in are written as a function 
of the parameters of interest. 

The combination is based on simultaneous fits to the data from both experiments taking into 
account the correlations between systematic uncertainties within each experiment and between 
the two experiments.
Almost all input analysis are based on the concept of event categorisation.
This consists of classifying the events in different categories, based on their kinematic properties.

This categorisation increases the sensitivity of the analysis and also allows separation of 
the different production processes on the basis of exclusive selections that identify the 
decay products of the particles produced in association with the Higgs boson: $W$ or $Z$ boson 
decays, VBF jets, etc. A total of approximately $600$ exclusive categories addressing 
the five production processes explicitly considered are defined for the five main decay channels.

The signal yield in a category $k$, $n_{signal}(k)$, can be expressed as a sum over all 
possible Higgs boson production processes $i$, with cross section $\sigma_{i}$  and decay 
modes $f$, with branching fraction \BR$^{f}$:
\begin{equation}
\begin{split}
n_{\rm{signal}}(k)& = \mathcal{L}(k) \cdot 
\sum_i \sum_f \left\{ \sigma_i 
\cdot A^f_{i; \SMm}(k) \cdot \epsilon_i^f(k) \cdot \BR^f \right\}
  \\
  & = \mathcal{L}(k) \cdot \sum_i \sum_f \mu_i \mu_f 
\left\{ \sigma_{i;\SMm} 
\cdot A^f_{i;\SMm}(k) \cdot \epsilon_i^f(k) \cdot \BR_{\SMm}^f \right\} \spc
\end{split}
\end{equation}
where $\mathcal{L}(k)$ represents the integrated luminosity, $A^f_{i;\SMm}(k)$ the detector 
acceptance assuming SM Higgs boson production and decay, and $\epsilon_i^{f}(k)$ the overall 
selection efficiency for the signal category $k$. Finally $\mu_i$ and $\mu_f$ are the production 
and decay signal strengths, respectively. 

It can be seen that the measurements are only 
sensitive to the products of the cross sections and branching fractions, 
$\sigma_i \cdot \BR^{f}$. The overall statistical methodology is the same as used by the single experiments. 

The parameters 
$\alpha$ are estimated via the profile likelihood ratio test statistics $\Lambda(\alpha)$, which depend on one or more parameters as well as on the nuisance parameters, $\theta$, which reflect various experimental and theoretical uncertainties.
The likelihood functions are constructed using products of signal and background \acp{PDF} 
of the discriminating variables. The probability density functions
are obtained from simulations in the case of the signal and 
from both data and simulation for the background case.

Three parametrisations of experimental results have been performed in the combination: two are 
based on cross-sections and branching fractions, one is based on ratios of couplings modifiers. 
The $\sigma \cdot \BR$ for $gg \to \h \to ZZ$ channel is parametrised as a function of 
$\kappa_{gZ} = \kappa_g \cdot \kappa_Z / \kappa_{\h}$ (where $\kappa_g$ is the effective coupling 
modifier). The measurement of VBF and $Z\h$ production probes $\lambda = \kappa_Z / \kappa_g$, 
the measurements of $t\bar{t}\h$ production processes are sensitive to 
$\lambda_{tg} = \kappa_t / \kappa_g$. The three decay modes $\h \to WW$, $\h \to \tau \tau$ and 
$\h \to bb$ probe the three ratios
$\lambda_{WZ} = \kappa_W / \kappa_Z$, $\lambda_{\tau Z} = \kappa_{\tau} / \kappa_Z$ and 
$\lambda_{bZ} = \kappa_b / \kappa_Z$, relative to the $\h \to ZZ$ branching fraction.

Other parametrisations were performed in the combination analysis with more specific and more restrictive assumptions.

\subsection{Limitations of the \kappaframework} \label{limits}
One problem that arises at higher orders is the violation of unitarity. 
Unitarity is connected to the renormalisability of a theory, and the introduction of coupling modifiers for the Higgs spoils unitarity and renormalisability.
\subsubsection*{Example $\h \to b \bar b$ decay width} \label{par:kframeNLO}
To show why the \kappaframework cannot be considered a fully consistent theoretical framework, 
we give an example of the problems arising when we try to compute an observable at \ac{NLO} 
precision in this framework. We consider a 
``simplified \kappaframework'' phenomenological Lagrangian, similar to the Lagrangian that is discussed in~\cref{sec:phenoLagSubsection}. 

We define this simplified theory as follows: 
we start from the \ac{SM} (following here the conventions of Ref.~\cite{book:Bohm2001}), and we 
modify all the fermionic couplings to scalar particles, multiplying them by a common 
factor $\kappa_{ffS}$. In this simplified framework, the generic scalar-fermion vertex is given by
\begin{equation}
\parbox[c]{62pt}{
\begin{picture}(62,50)(0,0)
\DashLine(0,25)(30,25){3}
\Text(15,30)[]{$\scriptstyle{h}$}
\Vertex(30,25){1}
\ArrowLine(60,5)(30,25)
\ArrowLine(30,25)(60,45)
\Text(62,48)[lt]{$\scriptstyle{\bar F_1}$}
\Text(62,2)[lb]{$\scriptstyle{F_2}$}
\end{picture}} \qquad  = i e \, \kappa_{ffS} \biggl( C_L \frac{1- \gamma_5}{2} + 
C_R \frac{1 + \gamma_5}{2} \biggr) \spc
\end{equation}
where the values of the coefficients $C_L$ and $C_R$ are specified in \cref{tab:ffsCouplings}.
\begin{table}[]
\[
\begin{array}{rcccc}
\toprule
S \bar F_1 F_2 	& h \bar f_i f_j				 & \chi \bar f_i f_j & \phi^+ \bar u_i d_j & \phi^- \bar d_j u_i \\
\midrule
C_L 		& -\frac{1}{2 s} \frac{m_{f,i}}{M_W} \delta_{ij} & -\frac{i}{2 s} 2 I^3_{W,f} \frac{m_{f,i}}{M_W} \delta_{ij} & \frac{1}{\sqrt{2} s} \frac{m_{u,i}}{M_W} V_{ij} &-\frac{1}{\sqrt{2} s} \frac{m_{d,j}}{M_W} V_{ji}^\dagger \\
\midrule
C_R 		& -\frac{1}{2 s} \frac{m_{f,i}}{M_W} \delta_{ij} & \frac{i}{2 s} 2 I^3_{W,f} \frac{m_{f,i}}{M_W} \delta_{ij}  &-\frac{1}{\sqrt{2} s} \frac{m_{d,j}}{M_W} V_{ij} & \frac{1}{\sqrt{2} s} \frac{m_{u,i}}{M_W} V_{ji}^\dagger \\
\bottomrule
\end{array}
\]
\caption{Scalar-fermion couplings in the \ac{SM}, from Ref.~\cite{book:Bohm2001}.}
\label{tab:ffsCouplings}
\end{table}
As for the observable that we want to compute, we consider the Higgs decay width into a pair 
of bottom quarks, i.e. $\Gamma(\h \to b \bar b)$.
The Born matrix element for the decay process reads as follows:
\begin{equation}
\label{eq:hbbLOME}
\mathcal{M}_0 = \frac{e \, m_b \, \kappa_{ffS}}{2 s M_W} \, \bar u(p) v(k) \spc
\end{equation}
and we can see how the \ac{LO} \ac{SM} decay width is modified using this simple 
phenomenological model: it is rescaled by a factor $\kappa_{ffS}^2$ coming from the square 
of \cref{eq:hbbLOME}. This rescaling reflects the effects of the \kappaframework on the 
fermion-fermion-Higgs coupling.
To achieve a more accurate theoretical prediction, we must include contributions containing 
higher powers of the coupling constant $\alpha$. The \ac{NLO} matrix element can be cast in the 
following form, see for example Ref.~\cite{Kniehl:1991ze},
\begin{equation}
\label{eq:hbbNLOME}
\mathcal{M} = \mathcal{M}_0 \biggl[ 1 + \frac{\alpha}{4 \pi} \biggl( \delta_\text{loop} + 
\delta_\text{CT} \biggr) \biggr] \spc
\end{equation}
where $\delta_\text{loop}$ contains the contributions due to loops with internal gauge and 
Higgs bosons, and $\delta_\text{CT}$ the counterterm contributions coming from the renormalisation procedure.

In this example, the purpose is not to show the complete derivation of the \ac{NLO} decay width, 
but we want to highlight the NLO inconsistency of the \kappaframework. To this end, we take 
advantage of the following simplifications: firstly, we drop \ac{QCD} contributions, and we 
focus on the \ac{EW} corrections. 

Then, by means of dimensional regularisation, we identify 
the UV divergent contributions in the \ac{NLO} matrix element, and we drop the UV finite part. This allows us to overlook the problems related to the infrared (IR) and collinear structure of the loop amplitude (involving the inclusion of the real emission process $\h \to b \bar b (\gamma, g)$). 
Finally, we neglect effects induced by the quark-mixing matrix $V$ in the $W$-boson couplings of \cref{tab:ffsCouplings}, setting $V_{ij} = \delta_{ij}$.

For the renormalisation procedure, we adopt the on-shell scheme, as presented 
in Ref.~\cite{Denner1991}. In this scheme, each renormalised mass is related to the corresponding physical mass, defined as the real part of the particle propagator pole.

\begin{figure}[]
\begin{subfigure}{1.\textwidth}
\centering
{
%
\begin{picture}(60,50)(0,0)
\DashLine(0,25)(20,25){3}
\Text(10,30)[]{$\scriptstyle{h}$}
\Vertex(20,25){1}
\ArrowLine(60,5)(40,5)
\ArrowLine(40,5)(20,25)
\ArrowLine(20,25)(40,45)
\ArrowLine(40,45)(60,45)
\Text(28,37)[rb]{$\scriptstyle{b}$}
\Text(28,13)[rt]{$\scriptstyle{b}$}
\Text(50,50)[b]{$\scriptstyle{b}$}
\Text(50,0)[t]{$\scriptstyle{b}$}
\DashLine(40,45)(40,5){3}
\Text(42,25)[l]{$\scriptstyle{h}$}
\Vertex(40,45){1}
\Vertex(40,5){1}
\end{picture}
\hspace{3mm}
%
\begin{picture}(60,50)(0,0)
\DashLine(0,25)(20,25){3}
\Text(10,30)[]{$\scriptstyle{h}$}
\Vertex(20,25){1}
\ArrowLine(60,5)(40,5)
\ArrowLine(40,5)(20,25)
\ArrowLine(20,25)(40,45)
\ArrowLine(40,45)(60,45)
\Text(28,37)[rb]{$\scriptstyle{b}$}
\Text(28,13)[rt]{$\scriptstyle{b}$}
\Text(50,50)[b]{$\scriptstyle{b}$}
\Text(50,0)[t]{$\scriptstyle{b}$}
\DashLine(40,45)(40,5){3}
\Text(42,25)[l]{$\scriptstyle{\chi}$}
\Vertex(40,45){1}
\Vertex(40,5){1}
\end{picture}
\hspace{3mm}
%
\begin{picture}(60,50)(0,0)
\DashLine(0,25)(20,25){3}
\Text(10,30)[]{$\scriptstyle{h}$}
\Vertex(20,25){1}
\ArrowLine(60,5)(40,5)
\ArrowLine(40,5)(20,25)
\ArrowLine(20,25)(40,45)
\ArrowLine(40,45)(60,45)
\Text(28,37)[rb]{$\scriptstyle{t}$}
\Text(28,13)[rt]{$\scriptstyle{t}$}
\Text(50,50)[b]{$\scriptstyle{b}$}
\Text(50,0)[t]{$\scriptstyle{b}$}
\DashLine(40,45)(40,5){3}
\Text(42,25)[l]{$\scriptstyle{\phi}$}
\Vertex(40,45){1}
\Vertex(40,5){1}
\end{picture}
\hspace{3mm}
%
\begin{picture}(60,50)(0,0)
\DashLine(0,25)(20,25){3}
\Text(10,30)[]{$\scriptstyle{h}$}
\Vertex(20,25){1}
\ArrowLine(60,5)(40,5)
\DashLine(40,5)(20,25){3}
\DashLine(20,25)(40,45){3}
\ArrowLine(40,45)(60,45)
\Text(28,37)[rb]{$\scriptstyle{h}$}
\Text(28,13)[rt]{$\scriptstyle{h}$}
\Text(50,50)[b]{$\scriptstyle{b}$}
\Text(50,0)[t]{$\scriptstyle{b}$}
\ArrowLine(40,5)(40,45)
\Text(43,25)[l]{$\scriptstyle{b}$}
\Vertex(40,45){1}
\Vertex(40,5){1}
\end{picture}
\hspace{3mm}
%
\begin{picture}(60,50)(0,0)
\DashLine(0,25)(20,25){3}
\Text(10,30)[]{$\scriptstyle{h}$}
\Vertex(20,25){1}
\ArrowLine(60,5)(40,5)
\DashLine(40,5)(20,25){3}
\DashLine(20,25)(40,45){3}
\ArrowLine(40,45)(60,45)
\Text(28,37)[rb]{$\scriptstyle{\chi}$}
\Text(28,13)[rt]{$\scriptstyle{\chi}$}
\Text(50,50)[b]{$\scriptstyle{b}$}
\Text(50,0)[t]{$\scriptstyle{b}$}
\ArrowLine(40,5)(40,45)
\Text(43,25)[l]{$\scriptstyle{b}$}
\Vertex(40,45){1}
\Vertex(40,5){1}
\end{picture}
\\
\vspace{8mm}
%
\begin{picture}(60,50)(0,0)
\DashLine(0,25)(20,25){3}
\Text(10,30)[]{$\scriptstyle{h}$}
\Vertex(20,25){1}
\ArrowLine(60,5)(40,5)
\DashLine(40,5)(20,25){3}
\DashLine(20,25)(40,45){3}
\ArrowLine(40,45)(60,45)
\Text(28,37)[rb]{$\scriptstyle{\phi}$}
\Text(28,13)[rt]{$\scriptstyle{\phi}$}
\Text(50,50)[b]{$\scriptstyle{b}$}
\Text(50,0)[t]{$\scriptstyle{b}$}
\ArrowLine(40,5)(40,45)
\Text(43,25)[l]{$\scriptstyle{t}$}
\Vertex(40,45){1}
\Vertex(40,5){1}
\end{picture}
\hspace{3mm}
%
\begin{picture}(60,50)(0,0)
\DashLine(0,25)(20,25){3}
\Text(10,30)[]{$\scriptstyle{h}$}
\Vertex(20,25){1}
\ArrowLine(60,5)(40,5)
\ArrowLine(40,5)(20,25)
\ArrowLine(20,25)(40,45)
\ArrowLine(40,45)(60,45)
\Text(28,37)[rb]{$\scriptstyle{b}$}
\Text(28,13)[rt]{$\scriptstyle{b}$}
\Text(50,50)[b]{$\scriptstyle{b}$}
\Text(50,0)[t]{$\scriptstyle{b}$}
\Photon(40,45)(40,5){2}{6}
\Text(42,25)[l]{$\scriptstyle{\gamma}$}
\Vertex(40,45){1}
\Vertex(40,5){1}
\end{picture}
\hspace{3mm}
%
\begin{picture}(60,50)(0,0)
\DashLine(0,25)(20,25){3}
\Text(10,30)[]{$\scriptstyle{h}$}
\Vertex(20,25){1}
\ArrowLine(60,5)(40,5)
\ArrowLine(40,5)(20,25)
\ArrowLine(20,25)(40,45)
\ArrowLine(40,45)(60,45)
\Text(28,37)[rb]{$\scriptstyle{b}$}
\Text(28,13)[rt]{$\scriptstyle{b}$}
\Text(50,50)[b]{$\scriptstyle{b}$}
\Text(50,0)[t]{$\scriptstyle{b}$}
\Photon(40,45)(40,5){2}{6}
\Text(42,25)[l]{$\scriptstyle{Z}$}
\Vertex(40,45){1}
\Vertex(40,5){1}
\end{picture}
\hspace{3mm}
%
\begin{picture}(60,50)(0,0)
\DashLine(0,25)(20,25){3}
\Text(10,30)[]{$\scriptstyle{h}$}
\Vertex(20,25){1}
\ArrowLine(60,5)(40,5)
\ArrowLine(40,5)(20,25)
\ArrowLine(20,25)(40,45)
\ArrowLine(40,45)(60,45)
\Text(28,37)[rb]{$\scriptstyle{t}$}
\Text(28,13)[rt]{$\scriptstyle{t}$}
\Text(50,50)[b]{$\scriptstyle{b}$}
\Text(50,0)[t]{$\scriptstyle{b}$}
\Photon(40,45)(40,5){2}{6}
\Text(42,25)[l]{$\scriptstyle{W}$}
\Vertex(40,45){1}
\Vertex(40,5){1}
\end{picture}
\hspace{3mm}
%
\begin{picture}(60,50)(0,0)
\DashLine(0,25)(20,25){3}
\Text(10,30)[]{$\scriptstyle{h}$}
\Vertex(20,25){1}
\ArrowLine(60,5)(40,5)
\Photon(40,5)(20,25){2}{6}
\DashLine(20,25)(40,45){3}
\ArrowLine(40,45)(60,45)
\Text(28,37)[rb]{$\scriptstyle{\chi}$}
\Text(28,13)[rt]{$\scriptstyle{Z}$}
\Text(50,50)[b]{$\scriptstyle{b}$}
\Text(50,0)[t]{$\scriptstyle{b}$}
\ArrowLine(40,5)(40,45)
\Text(43,25)[l]{$\scriptstyle{b}$}
\Vertex(40,45){1}
\Vertex(40,5){1}
\end{picture}
\\
\vspace{8mm}
%
\begin{picture}(60,50)(0,0)
\DashLine(0,25)(20,25){3}
\Text(10,30)[]{$\scriptstyle{h}$}
\Vertex(20,25){1}
\ArrowLine(60,5)(40,5)
\Photon(40,5)(20,25){2}{6}
\DashLine(20,25)(40,45){3}
\ArrowLine(40,45)(60,45)
\Text(28,37)[rb]{$\scriptstyle{\phi}$}
\Text(28,13)[rt]{$\scriptstyle{W}$}
\Text(50,50)[b]{$\scriptstyle{b}$}
\Text(50,0)[t]{$\scriptstyle{b}$}
\ArrowLine(40,5)(40,45)
\Text(43,25)[l]{$\scriptstyle{t}$}
\Vertex(40,45){1}
\Vertex(40,5){1}
\end{picture}
\hspace{3mm}
%
\begin{picture}(60,50)(0,0)
\DashLine(0,25)(20,25){3}
\Text(10,30)[]{$\scriptstyle{h}$}
\Vertex(20,25){1}
\ArrowLine(60,5)(40,5)
\DashLine(40,5)(20,25){3}
\Photon(20,25)(40,45){2}{6}
\ArrowLine(40,45)(60,45)
\Text(28,37)[rb]{$\scriptstyle{Z}$}
\Text(28,13)[rt]{$\scriptstyle{\chi}$}
\Text(50,50)[b]{$\scriptstyle{b}$}
\Text(50,0)[t]{$\scriptstyle{b}$}
\ArrowLine(40,5)(40,45)
\Text(43,25)[l]{$\scriptstyle{b}$}
\Vertex(40,45){1}
\Vertex(40,5){1}
\end{picture}
\hspace{3mm}
%
\begin{picture}(60,50)(0,0)
\DashLine(0,25)(20,25){3}
\Text(10,30)[]{$\scriptstyle{h}$}
\Vertex(20,25){1}
\ArrowLine(60,5)(40,5)
\DashLine(40,5)(20,25){3}
\Photon(20,25)(40,45){2}{6}
\ArrowLine(40,45)(60,45)
\Text(28,37)[rb]{$\scriptstyle{W}$}
\Text(28,13)[rt]{$\scriptstyle{\phi}$}
\Text(50,50)[b]{$\scriptstyle{b}$}
\Text(50,0)[t]{$\scriptstyle{b}$}
\ArrowLine(40,5)(40,45)
\Text(43,25)[l]{$\scriptstyle{t}$}
\Vertex(40,45){1}
\Vertex(40,5){1}
\end{picture}
\hspace{3mm}
%
\begin{picture}(60,50)(0,0)
\DashLine(0,25)(20,25){3}
\Text(10,30)[]{$\scriptstyle{h}$}
\Vertex(20,25){1}
\ArrowLine(60,5)(40,5)
\Photon(40,5)(20,25){2}{6}
\Photon(20,25)(40,45){2}{6}
\ArrowLine(40,45)(60,45)
\Text(28,37)[rb]{$\scriptstyle{Z}$}
\Text(28,13)[rt]{$\scriptstyle{Z}$}
\Text(50,50)[b]{$\scriptstyle{b}$}
\Text(50,0)[t]{$\scriptstyle{b}$}
\ArrowLine(40,5)(40,45)
\Text(43,25)[l]{$\scriptstyle{b}$}
\Vertex(40,45){1}
\Vertex(40,5){1}
\end{picture}
\hspace{3mm}
%
\begin{picture}(60,50)(0,0)
\DashLine(0,25)(20,25){3}
\Text(10,30)[]{$\scriptstyle{h}$}
\Vertex(20,25){1}
\ArrowLine(60,5)(40,5)
\Photon(40,5)(20,25){2}{6}
\Photon(20,25)(40,45){2}{6}
\ArrowLine(40,45)(60,45)
\Text(28,37)[rb]{$\scriptstyle{W}$}
\Text(28,13)[rt]{$\scriptstyle{W}$}
\Text(50,50)[b]{$\scriptstyle{b}$}
\Text(50,0)[t]{$\scriptstyle{b}$}
\ArrowLine(40,5)(40,45)
\Text(43,25)[l]{$\scriptstyle{t}$}
\Vertex(40,45){1}
\Vertex(40,5){1}
\end{picture}
}
\caption{}
\label{fig:hbbLoopDiag}
\end{subfigure}
\begin{subfigure}{1.\textwidth}
\centering
{
\begin{picture}(62,50)(0,0)
\DashLine(0,25)(30,25){3}
\Text(15,30)[]{$\scriptstyle{h}$}
\Vertex(30,25){1}
\Line(25,20)(35,30)
\Line(35,20)(25,30)
\ArrowLine(60,5)(30,25)
\ArrowLine(30,25)(60,45)
\Text(62,48)[lt]{$\scriptstyle{b}$}
\Text(62,2)[lb]{$\scriptstyle{b}$}
\end{picture}
}
\caption{} 
\label{fig:hbbCTDiag}
\end{subfigure}
\caption{(a) Loop contributions to the Higgs decay into a bottom quark pair. (b) Counterterm contribution to the Higgs decay into a bottom quark pair.}
\end{figure}
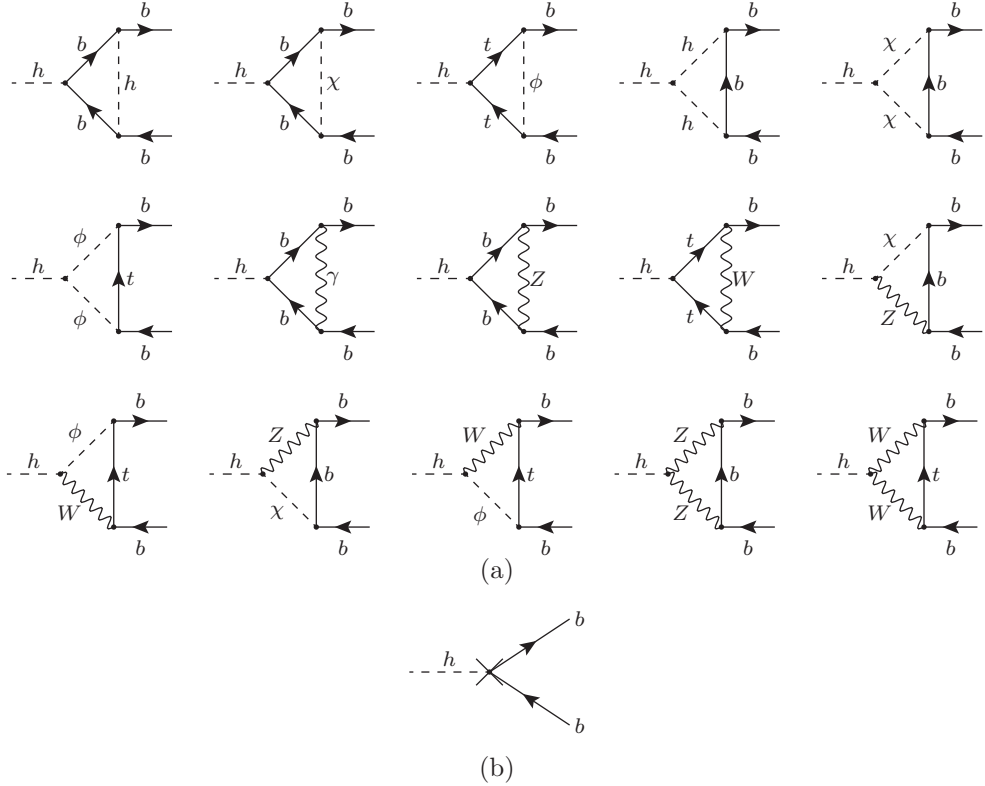

Considering \ac{EW} corrections, $\delta_\text{loop}$ of \cref{eq:hbbNLOME} contains contributions coming from the diagrams reported in \cref{fig:hbbLoopDiag} and 
$\delta_\text{CT}$, the counterterm contribution reported in \cref{fig:hbbCTDiag}.

The UV-divergent part of $\delta_\text{loop}$ turns out to be (in the MS scheme)
\begin{equation}
\label{eq:hbbDeltaLoopUV}
\bigl. \delta_\text{loop} \bigr|_\text{UV} = 
\frac{1}{36 \, s^2 M_W^2} \frac{2}{4 - D} \bigl( 26 M_W^2 + M_Z^2 + 
18 \, \kappa_{ffS}^2 \, \mt^2 \bigr) \spc
\end{equation}
where $D$ is the number of space-time dimensions considered to regularise the loop integrals, and the factor $2/(4 - D)$ shows the divergent contribution. 
To motivate this result, we can separate the loop contributions into four different classes, ordered as they appear in \cref{fig:hbbLoopDiag}:
\begin{itemize}

\item[I.] scalar-exchange diagrams, where a scalar particle is exchanged between the two 
outgoing $b$ quarks,

\item[II.] diagrams with two scalars running in the loop,

\item[III.] diagrams with one internal gauge boson line (exchanged by outgoing particles, 
or emitted by the incoming Higgs),

\item[IV.] diagrams with two gauge boson lines running in the loop.

\end{itemize}
As can be seen by naive power counting, type-II and type-IV diagrams are UV finite and do not contribute to $\delta_\text{loop}$. Conversely, type-I and III render UV contributions\footnote{Except for the $W$-exchange diagram, which is UV-finite because of the $\gamma^\mu (1-\gamma_5)$ structure of the $W$ coupling to two fermions.} that are respectively proportional to $\kappa_{ffS}^3$ and $\kappa_{ffS}$, leading 
to $\kappa_{ffS}^2$ and $\kappa_{ffS}^0$ terms in \cref{eq:hbbDeltaLoopUV}.

The counterterm contribution, written in terms of renormalisation constants, is given by
\begin{equation}
\label{eq:hbbDeltaCT}
\begin{split}
\frac{\alpha}{4 \pi} \delta_\text{CT} &=
	- \frac{\delta Z_{AA}}{2}
	+ \frac{\delta Z_{\h}}{2}
	- \frac{s}{2 c} \delta Z_{ZA}
	- \frac{1}{2 s^2 M_W^2} \delta M_W^2
	\\& \quad
	- \frac{c^2}{2 s^2} \delta M_Z^2
	+ \frac{1}{s^2 M_Z^2} \delta M_W^2
	+ \frac{\delta m_b}{m_b}
	+ \frac{\delta Z_{33}^{d,L} + \delta Z_{33}^{d,R}}{2} \spc
\end{split}
\end{equation}
where the renormalisation constants are fixed by the scheme chosen in the renormalisation procedure. In the on-shell scheme, the field renormalisation constants $\delta Z_{AA}$, 
$\delta Z_{\h}$, $\delta Z_{33}^{d,L(R)}$ are related respectively to the one-loop self-energies 
of the photon, the Higgs boson, and the bottom quark. Consequently, in the considered 
framework, we expect $\kappa_{ffS}^2$ contributions in the renormalisation constants 
$\delta Z_{\h}$ and $\delta Z_{33}^{d,L(R)}$, due to the presence of scalar-fermion couplings 
in the one-loop self-energies of the Higgs boson and the bottom quark. 

The expression for the constant $\delta Z_{AA}$, related to the photon self-energy, in which the modified couplings do not appear, will reproduce the \ac{SM} result. Repeating the same argument, we can argue 
that among the other constants, only $\delta m_b$ will present a $\kappa_{ffS}^2$ term. This is reflected in the expressions of the renormalisation constants, and their UV divergent parts:
\begin{equation}
\bigl. \delta Z_i \bigr|_\text{UV} = \frac{2}{4 - D} \frac{\alpha}{4 \pi} d Z_i \spc
\end{equation}
with,
\begin{equation}
\label{eq:hbbRenConstUV}
\begin{split}
d M_Z^2 &=
\frac{1}{6 M_W^2 s^2} \biggl[-3 M_Z^2 
\biggl( \sum_l m_l^2 + 3 \sum_q m_q^2 \biggr) -70 M_W^2 M_Z^2+22 M_W^4+47 M_Z^4 \biggr] \spc
\\
d Z_{AA} &= - \frac{23}{3} \spc
\quad
d Z_{ZA} = \frac{4 M_W}{M_Z s} \spc
\\
d M_W^2 &= 
	- \frac{1}{6 s^2} \biggl[ 7 M_W^2 - 6 M_Z^2+3 
\biggl( \sum_l m_l^2 + 3 \sum_q m_q^2 \biggr) \biggr] \spc
\\
d Z_{\h} &= 
\frac{1}{2 M_W^2 s^2} \biggl[ - \kappa _{ffS}^2 
\biggl( \sum_l m_l^2 + 3 \sum_q m_q^2 \biggr) +2 M_W^2+M_Z^2 \biggr] \spc
\\
d m_b &= 
	\frac{m_b}{24 M_W^2 s^2}  
\biggl[ 9 \kappa _{ffS}^2 \left(m_b^2-m_t^2 \right)+2 M_W^2+7 M_Z^2\biggr] \spc
\\
d Z_{33}^{d,L} &= 
- \frac{1}{36 M_W^2 s^2} 
\biggl[ 9 \kappa _{ffS}^2 \left(m_b^2+m_t^2\right)+26 M_W^2+M_Z^2 \biggr] \spc
\\
d Z_{33}^{d,R} &= 
	- \frac{1}{18 M_W^2 s^2} \Bigl( 9 m_b^2 \kappa _{ffS}^2-2 M_W^2+2 M_Z^2 \Bigr) \spp
\end{split}
\end{equation}
The compensation of divergent terms between loop diagrams and counterterms, when performing the renormalisation procedure is a delicate mechanism. We will see how the presence of $\kappa$-dependent terms spoils the cancellation of the divergent part.

Indeed, the UV component of the counterterm contribution, obtained inserting 
\cref{eq:hbbRenConstUV} into \cref{eq:hbbDeltaCT}, is the only source of divergences that might compensate the UV component of $\delta_\text{loop}$ given in \cref{eq:hbbDeltaLoopUV}, rendering the one-loop matrix element UV finite and thus valuable for the evaluation of the decay width. 

The substitution leads to
\begin{equation}
\label{eq:hbbproblem}
\begin{split}
\bigl. \delta_\text{CT} 
\bigr|_\text{UV} &= -\frac{1}{36 \, s^2 M_W^2} \frac{2}{4 - D} 
\biggl[ 26 M_W^2 + M_Z^2 - 9 \biggl( \sum_l m_l^2 + 3 \sum_q m_q^2 \biggr)
    \\& \qquad + 9 \, \kappa_{ffS}^2 \, 
\biggl( 2 m_t^2 + \sum_l m_l^2 + 3 \sum_q m_q^2 \biggr) \biggr] \spc
\end{split}
\end{equation}
and we see that the one-loop corrected transition matrix element of \cref{eq:hbbNLOME} gets 
a UV divergent contribution
\begin{equation}
\bigl. \mathcal{M} 
\bigr|_\text{UV} = \frac{\alpha}{4 \pi} \frac{\mathcal{M}_0}{4 s^2 M_W^2} \frac{2}{4 - D} 
\bigl( 1 - \kappa_{ffS}^2 \bigr) \biggl( \sum_l m_l^2 + 3 \sum_q m_q^2 \biggr) \spc
\end{equation}
which in general does not vanish for $\kappa_{ffS} \neq 1$.

Another unsatisfactory aspect of using the \kappaframework is that it cannot describe 
deviations in differential distributions.
This happens because global factors can predict how many times the Higgs decays in a specific 
channel, but not how the kinematics of the decay products are affected. 
Of course new physics could modify these differential distributions, but it would not be captured by 
the overall factors.

The \kappaframework is well motivated if one is looking for large deviations from the SM, as 
was the case in \RunI. The kappas parametrise these deviations and, if something new would be found, 
they would point which direction one should look into.  
ATLAS and CMS did not find any large differences in what was predicted by the 
SM, therefore the next step would be looking for small deviations.
\subsection{Extrapolation on achievable experimental accuracy in future Runs}
It starts to become clear that defining a theoretical framework able to describe possible 
deviations in the Higgs sector is not straightforward. If one wants to be as model independent as possible, an \ac{EFT} approach may be the answer. \acp{EFT} will be extensively 
discussed in~\cref{sec:EFT}, however, here we demonstrate how the testable hierarchies of scales are 
limited by the experimental accuracy.

The maximum testable hierarchy  of scales is determined by two ``sources'': the assumption of a maximum size of underlying  couplings and the experimental precision\footnote{The problem for the interpretation of results in terms of an \ac{EFT} during \RunI was indeed 
the limited experimental accuracy, see Ref.~\cite{Biekotter:2016ecg}.}
In the \ac{EFT} approach and for observables close to the on-shell Higgs region,
the higher dimensional operators are ordered by factors $g^2 \mh^2 / \Lambda^2$ (where 
$\Lambda$ is the scale of the momentum cut-off of the theory), i.e. the only relevant scale 
for ``on-shell'' Higgs production is given by $\mh$.

This scale should be well separated from the experimentally accessible scale, in our case 
the \ac{EW} scale $\Lambda \gg v$ .
The applicability of an \ac{EFT} approach is, however, limited when the hierarchy of scales is 
not guaranteed. Because hadron colliders do not have a well-defined partonic energy, strategies 
relying on boosted objects and large recoils are the most critical. 

While it is not clear 
that a marginal separation of scales invalidates the \ac{EFT} approach, such observables 
clearly pose a challenge. 

For this reasons, interpreting LHC physics in terms of an effective 
theory involves a delicate balance between energy scales. It is possible to roughly estimate  the physics scales which can be probed, and see if the deviation from the \ac{SM} Higgs production and decay rates lies within the experimental accuracy. For instance in the energy region around the Higgs peak,  
\begin{equation}
\label{eq:tilman}
\left\lvert \frac{\sigma \times \textrm{BR}}{\left(\sigma \times \textrm{BR} \right)_{\textrm{SM}}} -1  \right\lvert = \frac{g^2 \mh^2}{\Lambda^2} \spp
\end{equation}
The accuracy on a rate measurement can then be translated into a threshold for new physics as
\begin{equation}
\left\lvert \frac{\sigma \times 
\textrm{BR}}{\left(\sigma \times \textrm{BR} \right)_{\textrm{SM}}} -1  
\right\lvert = \frac{g^2 \mh^2}{\Lambda^2} \, > \, \Delta \spp
\end{equation}
This translates into the following relation:
\begin{equation}
\label{eq:approxLambda}
\Lambda \, < \frac{g \, \mh}{\sqrt{\Delta}}  \spp
\end{equation}
The precise value of the experimental accuracy $\Delta$ depends on the process under consideration.
If we assume a value of $\Delta = 10\%$ (which roughly covers the current experimental accuracy), 
then for a weakly interacting theory with $g^2 \sim 1/2$, one can probe scales up to $\Lambda \approx 280 \, \GeV$.
This means that for weakly coupled new physics and with current experimental accuracy, it is not possible to test very high energy scales.
However, for a more strongly interacting theory with 
$g \sim 1$ (but $g < 4\pi$, which is the theoretical limit for preserving unitarity), one can probe higher scales up to $\Lambda \approx 400 \, \GeV$.

The increased statistics and Higgs production cross sections at \RunII will enable us to add 
a wide range of distributions and off-shell processes to the Higgs observables, which can probe 
higher energy scales $\Lambda \gg \mh$.  It is important to note that, if we look at differential observables, and in particular in the tails of the distributions, Eq.~\eqref{eq:approxLambda} has to be modified by substituting $\mh \rightarrow \pT$. Depending on the $\pT$ regions that are reconstructed by the experiment, we will be able to access also different $\Lambda$ values.   
Of course, we have to keep in mind that when we move towards 
higher $Q^2$, the accuracy on the measurement will drastically decrease, and the value of $\Delta$ has to be reconsidered. 
%
\subsubsection*{Intermezzo: Higgs off-shellness}
In the \ac{NWA}, $\Gamma_{\h} << \mh$, the 
Higgs cross section factorises into on-shell production and on-shell decay, so the Higgs 
cross section can be written as:
\begin{equation}
\sigma_{i\to \h \to f} = 
\sigma_{i \to \h}\times \textrm{BR}_{\h \to f} 
\propto \frac{\sigma_{i \to \h}\Gamma_{\h \to f}}{\Gamma_{\h}} \spc
\end{equation}
and in terms of couplings
\begin{equation}
\sigma_{i\to \h \to f}  
\propto \frac{g_i^2 g_f^2}{\Gamma_{\h}} \sim \frac{g_i^2 g_f^2}{\sum_j g_j^2} \spc
\end{equation}
which means that the measurements in individual channels are complicated through dependence 
on the global Higgs properties due to the width.
The width of the $125 \, \GeV$ Higgs boson is predicted to be very small 
($4 \, \MeV$) compared to other heavy EW particles such 
as $Z$, $W$ or top ($\sim 2 \, \GeV$). This is because the width is calculated by 
summing over all the decays, and the Higgs mostly decays to $b$ quarks which are light 
compared to the EW scale
\begin{equation}
\Gamma_{\h} \sim \left( \frac{m_b^2}{m_{EW}^2} \Gamma_{EW} \right) \spp
\end{equation}
So what are the differences when looking at the resonance and at the tails?
In the resonance region, the propagator (ignoring 
differences between fixed width, running width and complex pole), 
\begin{equation}
\frac{1}{(s-M_X^2) + (i \Gamma_XM_X)}
\end{equation}
is dominated by the width in the denominator, so
\begin{equation}
\sigma^{on}_{i \to X \to f} \sim \frac{g_i^2g_f^2}{\Gamma_X} \spp
\end{equation}
In the off-shell region instead, the cross section does not depend on the width so that,
\begin{equation}
\sigma^{off}_{i \to X \to f} \sim g_i^2g_f^2 \spp
\end{equation}
For a detailed discussion of off-shell Higgs physics see Refs.~\cite{Kauer:2012hd,Caola:2013yja,Ghezzi:2014qpa,Englert:2015bwa}.

To summarise, the \kappaframework has been used in \RunI to interpret and deliver 
results, but a new era is approaching, where the limitations of this interim framework are 
becoming more and more evident. The lack of consistency when going to higher perturbative orders and the inability of 
the framework to describe \ac{BSM} effects that modify the kinematic distributions cannot   
be neglected.

 On one hand, the increasing experimental accuracy demands higher 
theoretical precision. On the other hand, the higher statistics and energies of the LHC 
\RunII will allow us to study off-shell processes and distributions that will require 
a robust theoretical description of differential quantities. 

The need for different frameworks is compelling, and in the following part of this chapter 
possible solutions will be offered, both on the experimental and the theoretical side. In general, the search for new physics at the LHC is performed via:
\begin{itemize}

\item A search for new states: resonances or more complicated structures searches.  
In this case, it is fundamental to have descriptive SM MC generators to achieve discoveries, i.e. the discovery in this case is data  driven. 
Further on we will need precision for characterisation.

\item Alternatively, a search for new interactions: deviations are expected to be small making this intrinsically a precision measurement.  There is also a need for predictive MC generators in this case, and additionally accurate theoretical predictions for both background (\ac{SM}) and signal (\ac{EFT}) hypotheses.

\end{itemize}
Having said that, we should keep in mind that
the matter content of the \ac{SM} has been experimentally verified and evidence 
for other light states has not yet been found. Therefore, 
\ac{SM}  measurements can always be interpreted as searches for deviations from the $\mathrm{dim} = 4$ \ac{SM} Lagrangian predictions. 

\newpage
\section{Fiducial and simplified template cross sections}
\label{sec:fiducial}
\subsection{Introduction}
In this chapter we describe the set of approaches used by the \ac{LHC} experiments to present their future results on the Higgs boson.
The former, attempt to optimize the results based on the following requirements:
\begin{itemize}
\item The measured quantities should be directly comparable to theory.
\item The measurements should be independent of the underlying theoretical assumptions.
\item The experimental and theoretical errors should be factorisable.
\end{itemize}
The $\kappa$-framework discussed in~\cref{sec:kappa} was the first proposed method to address these requirements.
Still, some limitations to its validity motivate going beyond this useful, but over-simplified, approach. For this purpose, a new framework is presented in this section.
\subsection{Fiducial cross sections}\label{subsec:fiducial}
As mentioned above, experiments should report quantities that are independent of theory, such that the result is comparable with theories other than the one that is more favoured at the time of the experiment. To have to redo an analysis for each new theoretical improvement is simply not feasible. 

Let us, as an example, consider a measurement of a signal strength (see \cref{sec:kappa}). If an experiment was to report its measurements in terms of
\begin{equation}
\mu_i^f = \frac{\left(\sigma_i \cdot 
\BR^f\right)^\text{exp}}{\left(\sigma_i \cdot \BR^f\right)^\text{theo}} \spc
\end{equation}
the results, and their errors, would be based on the theoretical prediction available at the time of the experiment. And thus the measurement would have to be repeated each time an improved formulation of the theory becomes available. Reporting only $\mu_i^f$, does not allow for a proper disentanglement of theoretical and experimental contributions to the reported errors.

One solution to this problem is to present experimental measurements of  
$\sigma_i \cdot \BR^f$, 
which do not contain a strong assumption on the theoretically expected value.
With such clean measurements, is is possible to deduce parameters in a way which is as theory-independent as possible.

Another area in which theoretical and experimental considerations become entangled, is the \emph{unfolding} procedure used to correct for detector effects and to extrapolate them outside the detector phase space.
Also here, it is desirable to report the results in a way that is independent on particular assumptions on the detector.

Experimentally, cuts are usually applied in order to increase 
the signal over background ratio, and to get a better sensitivity on the quantities of physical interest.
All these cuts are applied to objects constructed from signals coming directly from the detector, eventually caused by the interaction of the physical particles with the detector material.
\subsubsection*{Levels}
The phase space of signals in the detectors is called the \emph{detector level}.
Phase spaces defined for particles before their interaction with the detector are called \emph{truth levels}. The definition of truth levels is ambiguous and relies on a theoretical model.

One could define a level formed by particles with a relatively long lifetime ($\gtrsim 10^{-10}s$), which typically are the particles which interact with the detector. This is usually referred to as \emph{particle level}. The theoretical description of the particle level depends on shower and hadronisation modelling.

It is also possible to define a level based on partons, the \emph{parton level}, by considering particles before shower hadronisation. 
This level describes only elementary particles which exist in theoretical calculations, i.e. as defined in the Lagrangian.
As an example, gluons and quarks are valid objects at the parton level, while at the particle level the corresponding objects are jets.

If an experiment corrects a measurement to parton level, it must assume a theoretical model for showering and hadronisation, as well as a model for interaction of stable particles with the detector. 
At the particle level, only the  latter has to be assumed.  Thus parton level descriptions depend more on theoretical assumptions than the particle level.

Another thing to take into account is the detector efficiency.
Detectors do not allow the measurement of all the particles produced in collisions due to gaps and supporting structures.   Even when particles do interact 
with the detectors, they may not leave a clear enough trace to allow  
the object to be reconstructed correctly. Events whose reconstruction is not possible are usually discarded. 
When correcting a  measurement to truth level, only detector level events which passed detector level cuts are unfolded. 

To compare with a particular theoretical prediction, cuts on truth level objects 
have to be applied too. The set of cuts defines a volume in the truth level phase space, 
which is usually referred to as the \emph{fiducial volume}. In general, events which pass truth 
level cuts could also fail to be reconstructed at detector level, and vice versa.
To correct for this effect, a theoretical assumption has to be made on the correction of  
truth levels which were lost in detector level selections.
The larger the fraction of events from the fiducial volume that are reconstructed at detector level, 
the smaller the theoretical contribution to the correction. 
\begin{figure}[t]
  \centering
\includegraphics[trim= 0cm 15cm 0cm 0cm, clip, scale=0.4]{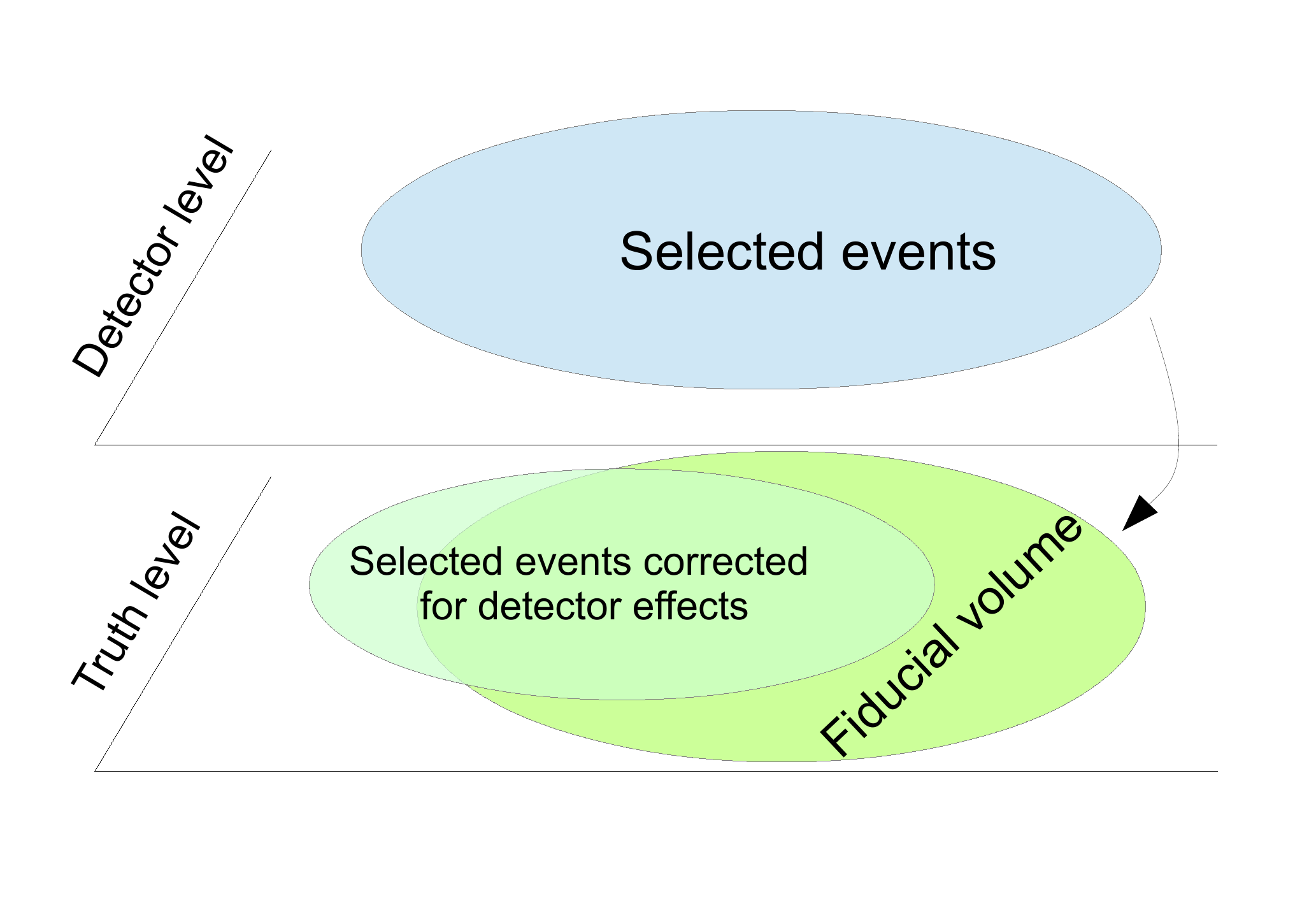}
  \caption{Schematic representation of the unfolding procedure. Events selected at detector level, with cuts enhancing the $S/B$ ratio (blue ellipse). 
  A correction is then applied to take into account particle interactions with the detector (light green ellipse). 
  A fiducial volume at truth level is then defined in order to compare the measurement with a theoretical predicition (green ellipse).
  }
  \label{fig:unfolding}
\end{figure}

In summary, if it is possible to define a truth level as close as possible to the detector 
level, the theoretical model dependence of such a level is minimised.
This is the idea which defines \emph{fiducial cross sections}, introduced in Ref.~\cite{Boudjema:2013qla}.
Fiducial cross sections are cross sections unfolded to particle level in a fiducial volume 
defined by some cuts that are as close as possible to the ones applied at detector level (but are applied on particle level objects instead).
Such cross sections are the measurements with the least theoretical assumptions. 

Experimentally, cross sections can be defined in terms of directly measurable quantities, such as:
\begin{itemize}

 \item Total cross sections 
\begin{equation*}  (\sigma_i )_{\text{exp}} = 
  \frac{N_{\text{ev} , if}}{\BR^{f} \mathcal{A}_{if} \epsilon_{if} \mathcal{L}}\spp \end{equation*}

 \item Total cross sections with fixed final state 
\begin{equation*}  (\sigma_i \times \BR^{f})_{\text{exp}} = 
   \frac{N_{\text{ev} , if}}{ \mathcal{A}_{if} \epsilon_{if} \mathcal{L}}\spp \end{equation*}

 \item Fiducial cross sections with fixed final state 
\begin{equation*}  (\sigma_i \times \BR^{f}\times \mathcal{A}_{if})_{\text{exp}} = 
\frac{N_{\text{ev} , if}}{ \epsilon_{if} \mathcal{L}}\spp \end{equation*}

 \end{itemize}
where $\mathcal{A}$ is the fiducial acceptance (i.e. a factor to extrapolate the 
measurement to the full particle level phase space), $\epsilon$ is the experimental efficiency 
(which includes all of the unfolding factors), $\mathcal{L}$ is the luminosity, 
$\BR^f$ is the branching ratio for final state $f$, $i$ determines a particular bin 
and/or initial state,  
and $N_{\text{ev}, if}$ is the number of events counted at detector level.
Apart from the luminosity and the collected number of events, all of these factors depend on a given theoretical model.

It is thus clear that fiducial cross sections, which have the fewest correction factors, depend least on the underlying theory assumptions.
It is also possible to extract differential fiducial cross sections: 
\begin{equation*} 
\frac{\text{d} \sigma^{\text{fid}}}{\sigma^{\text{fid}}} = 
\frac{\text{d} N_{\text{ev}}}{N_{\text{ev}}} \cdot 
\frac{\epsilon_{\text{d} N}}{\epsilon_N} \spc
\end{equation*}
where $\text{d} N_{\text{ev}}$ is the number of events in a particular bin, 
and $\epsilon_{\text{d} N}$ the experimental efficiency for this bin. 
In general, the measured and fiducial cross sections are related by,
\begin{equation} \label{eq:fiducial}
\sigma^{\text{fid}}_i=\sum_j \mathcal{A}_{ij} \sigma_j  \spc
\end{equation}
where $\sigma^{\text{fid}}_i$ are the fiducial cross sections, $\mathcal{A}_{ij}$ contains 
the factors which define the fiducial volumes and $\sigma_j$ are the measured cross sections.
The indices $i$ and $j$ specify the production mode, decay channel, or a bin of a 
differential cross section. 

From the theoretical point of view, fiducial cross sections are particle level 
predictions of a particular theoretical model, with the same fiducial 
cuts as those applied by the experiments for the observable of 
interest. Typically, fixed order theoretical calculations are implemented in a Monte Carlo event generator, in order to be able to apply the cuts on particle level objects and 
compare the result to the available measurements.

Fiducial measurements are, in principle, the cleanest possible way to compare data with theory and
generally all event selections which depend only on kinematic cuts allow 
the extraction of fiducial cross sections.
However, in order to increase sensitivity, experiments often use cuts which involve complex techniques, such as multivariate techniques (MVA) 
in the form of boosted decision trees (BDT) or neural networks (NN).

This type of cuts are generally needed to produce precise measurements, but are hard to reproduce in theoretical calculations. 
So, although minimally theory-dependent fiducial cross sections are fundamental to understand specific processes, which usually have a clean experimental signature, 
their application for studying the whole range of Higgs physics is limited.  
\begin{table}
\scriptsize
\centering
 \begin{tabular}{lcc}
 \toprule
  Cuts definition & ATLAS & CMS \\
   \toprule
   Lepton definition &  & \\
   \midrule
Electrons & $\pt>7 \, \GeV , |\eta|<2.47 $&  $\pt>7 \,\GeV , |\eta|<2.5$ \\ 
Muons & $\pt>6 \, \GeV , |\eta|<2.7 $&  $\pt>5 \, \GeV , |\eta|<2.4$ \\
\toprule
Event selection &  & \\
\midrule
Leptons $\pt$ cuts & $\pt>20, 15, 10, 10 \, \GeV $ & $\pt>20, 10, 7(5), 7(5) \, \GeV $\\
 & & \\
Invariant masses cuts & 
\begin{tabular}{c} $50 \, \GeV<m(l^+, l^-)<106 \, \GeV $ \\ $12 \, \GeV<m(l'^-, l'^+)< 115 \, \GeV$\\ $118 \, \GeV<m(l^+, l^-,l'^-, l'^+)< 129 \, \GeV$\\ $m(l^+, l^-)>5 \, \GeV$ \end{tabular} & 
\begin{tabular}{c}$40\, \GeV <m(l^+, l^-)<120 \, \GeV $\\ $ 12 \, \GeV<m(l'^-, l'^+)< 120 \, \GeV$ \\ $105 \, \GeV<m(l^+, l^-, l'^-, l'^+)< 140 \, \GeV$\\$m(l^+, l^-)>4 \, \GeV$\end{tabular} \\
 & & \\
 Lepton separation & \begin{tabular}{c}$\Delta R (l_i, l_j) >0.1(0.2)$ \\for same(opposite) sign\end{tabular}& \begin{tabular}{c}$\Delta R (l_i, l_j) >0.02$ \\ for every $i\neq j$\end{tabular}\\
\bottomrule
 \end{tabular}
\caption{Some of the differences in the definition of fiducial volumes for ATLAS, defined in Ref.~\cite{Aad:2014tca} and CMS, defined in Ref.~\cite{CMS:2015hja} for the
$\h\rightarrow 4l$ analysis.}
\label{table:fiducial_CMSvsATLAS}
\end{table}

Of course the definitions of the physical objects have to be shared between the various experiments and fiducial volumes have to be the same in order for all quantities to be comparable.
This is not obvious, since each experiment could be in principle most sensitive to an observable in a different region of the phase space than others. 

As an example, in the ATLAS and CMS analysis of $\h\rightarrow 4l$, published in Refs.~\cite{Aad:2014tca,CMS:2015hja}, the fiducial cuts 
are similar but not precisely the same - see \cref{table:fiducial_CMSvsATLAS}.
Global combinations can be done by correcting to a common fiducial volume. This  procedure introduces a small theoretical dependence, but the benefits of the combinations are  much larger, as discussed in~\cref{subsec:kexperiment}. 

In some cases, it is necessary to simply extrapolate the measurement to a larger phase space,  because the definition of minimally-theory dependent fiducial phase space is not possible. In that case, the formula for the fiducial cross section becomes:
\begin{equation}\label{eq:fiducial_accept}
  (\sigma_i)^{\text{fid}}_{\text{exp}} = 
\frac{N_{\text{ev}, if}}{ \alpha_{if} \epsilon_{if} \mathcal{L}} \spc
\end{equation}
where we introduced the factor $\alpha_{if}$ which extrapolates the measurement to the larger 
fiducial phase space.
A study of the extrapolation and unfolding factors can be found in Ref.~\cite{CMS:2015hja} 
from the CMS analysis of $\h\rightarrow 4l$ and is reported in \cref{table:theo_dep_fiducial} with the notation used in this section.

\begin{table}
\centering
\begin{tabular}{lcc}
  \toprule
  Signal process &  $\alpha_{if}$& $\epsilon_{if}$ \\
  \toprule
  &  Higgs production modes & \\
  \midrule
    $gg\h$ & $0.422 \pm 0.001$&$0.681 \pm 0.002$ \\
  $VBF$ & $0.476 \pm 0.003$& $0.678 \pm 0.005$\\
  $W\h$ & $0.342 \pm 0.002$& $0.672 \pm 0.003$\\
  $Z\h$ & $0.348 \pm 0.003$& $0.679 \pm 0.005$\\
  $tt\h$ & $0.250 \pm 0.003$& $0.685 \pm 0.010$\\
  \toprule
 & Non-SM models & \\  
  \midrule
    $q\bar q \rightarrow \h(J^{\text{CP}}=1^-)$ & $0.238 \pm 0.001$& $0.642 \pm 0.002$\\
    $q\bar q \rightarrow \h(J^{\text{CP}}=1^+)$ & $0.283 \pm 0.001$& $0.651 \pm 0.002$\\
    $gg \rightarrow \h \rightarrow Z\gamma^*$ & $0.156 \pm 0.001$ & $0.667 \pm 0.002$\\
    $gg \rightarrow \h \rightarrow \gamma^*\gamma^*$ & $0.238 \pm 0.001$& $0.671 \pm 0.002$\\
    \bottomrule
\end{tabular}
 \caption{Fiducial volume and detector/unfolding efficiencies and their 
 errors, for $\h\rightarrow4l$. Table from Ref.~\cite{CMS:2015hja}.}
 \label{table:theo_dep_fiducial}
\end{table}
\cref{table:theo_dep_fiducial} shows that the acceptance factors depend strongly on the  Higgs production modes, and that the unfolding factors are compatible within their errors, because the different production modes have very different kinematics and 
final states. 
Thus, it makes sense to report Higgs fiducial cross sections in large fiducial  volumes in terms of Higgs production modes.

A last thing to notice is that in \cref{eq:fiducial_accept}, the experimental and 
theoretical uncertainties factorise. 
In fact, a naive calculation of the relative error on the fiducial cross section gives:
\begin{equation}\label{eq:fiducial_err}
 \frac{\Delta \sigma^{\text{fid}}}{\sigma^{\text{fid}}} = 
\frac{\Delta N_{\text{ev}}}{N_{\text{ev}}}  \oplus 
\frac{\Delta \epsilon_{if}}{\epsilon^2_{if}} \oplus \frac{\Delta \alpha_{if} }{\alpha^2_{if} } \spp
\end{equation}
When a more precise measurement becomes available, there is no need to repeat the analysis 
from the beginning. Only the extrapolation factors $\alpha_{if}$ need to be re-computed, starting 
from the unfolded distribution.
To summarise, the new framework for Higgs physics should:
\begin{itemize}

\item Measure cross sections.

\item Unfold cross sections to fiducial volumes, taking advantage of positive aspects of 
      fiducial measurements while not renouncing to maximise experimental sensitivity.

\item Yield cross section measurements in terms of production modes, since acceptance 
      factors depend on them.

\end{itemize}
The simplified template cross section (STXS) framework, introduced in Ref.~\cite{deFlorian:2016spz} is designed to take these considerations into account.
It tries to combine the ease with which signal-strength-like fits are performed experimentally with the  theoretical needs of fitting to well defined and calculable predictions, aiming to find a good 
compromise between theory-independence of the measurements and their experimental sensitivity.
The STXS framework will be explained in detail in the next section.

\subsection{Simplified template cross section framework}\label{subsec:STXS}
\acp{STXS} are fiducial cross sections defined in simplified fiducial volumes.
The definition of the fiducial volumes allows the use of advanced experimental techniques, thereby
gaining in experimental sensitivity, at the cost of a small dependence on the theoretical model.
This is necessary, since at the moment all the measurements on Higgs boson cross sections are dominated by statistical uncertainty. Also the combination of different decay channels reduces the statistical uncertainty.
This has been shown, for example, in Ref.~\cite{ATLAS:2016hru}, where the combination reduced 
the statistical error on the production cross section by a factor of $1.4$. 
\begin{equation}
\begin{split}
& \frac{\Delta\sigma}{\sigma}(\h\rightarrow\gamma \gamma)=35\% \spc  \qquad
  \frac{\Delta\sigma}{\sigma}(\h\rightarrow Z Z)=23\% \spc\\ 
& \frac{\Delta\sigma}{\sigma}(\h\rightarrow\gamma \gamma \  \oplus \  \h\rightarrow Z Z)= 17\% \spp\\  
\end{split}
\end{equation}
Combining the different Higgs decay channels would further improve the precision of 
the measurements. This is the main difference between the usual fiducial differential cross 
sections and the STXS framework. 
While fiducial measurements maximise the independence on the theoretical modelling, simplified cross sections maximise the experimental sensitivity.
Instead of using fully differential distributions, cross sections are divided into sub-cross 
sections, called \emph{bins}. 
How to choose such \emph{bins} is an interesting topic in its own right and 
is discussed in~\cref{subsec:STXS_bin}.

The definition of the physics objects in the STXS fiducial volume, is aimed at taking advantage of combinations.
It is independent of the Higgs decay modes in such a way that combinations do not 
introduce further theory dependent factors.
In STXS, the particle level is defined with undecayed Higgs bosons while jets are reconstructed 
from the particles which are not associated with the Higgs decay.
\begin{figure}[t]
\centering 
\includegraphics[scale=0.25]{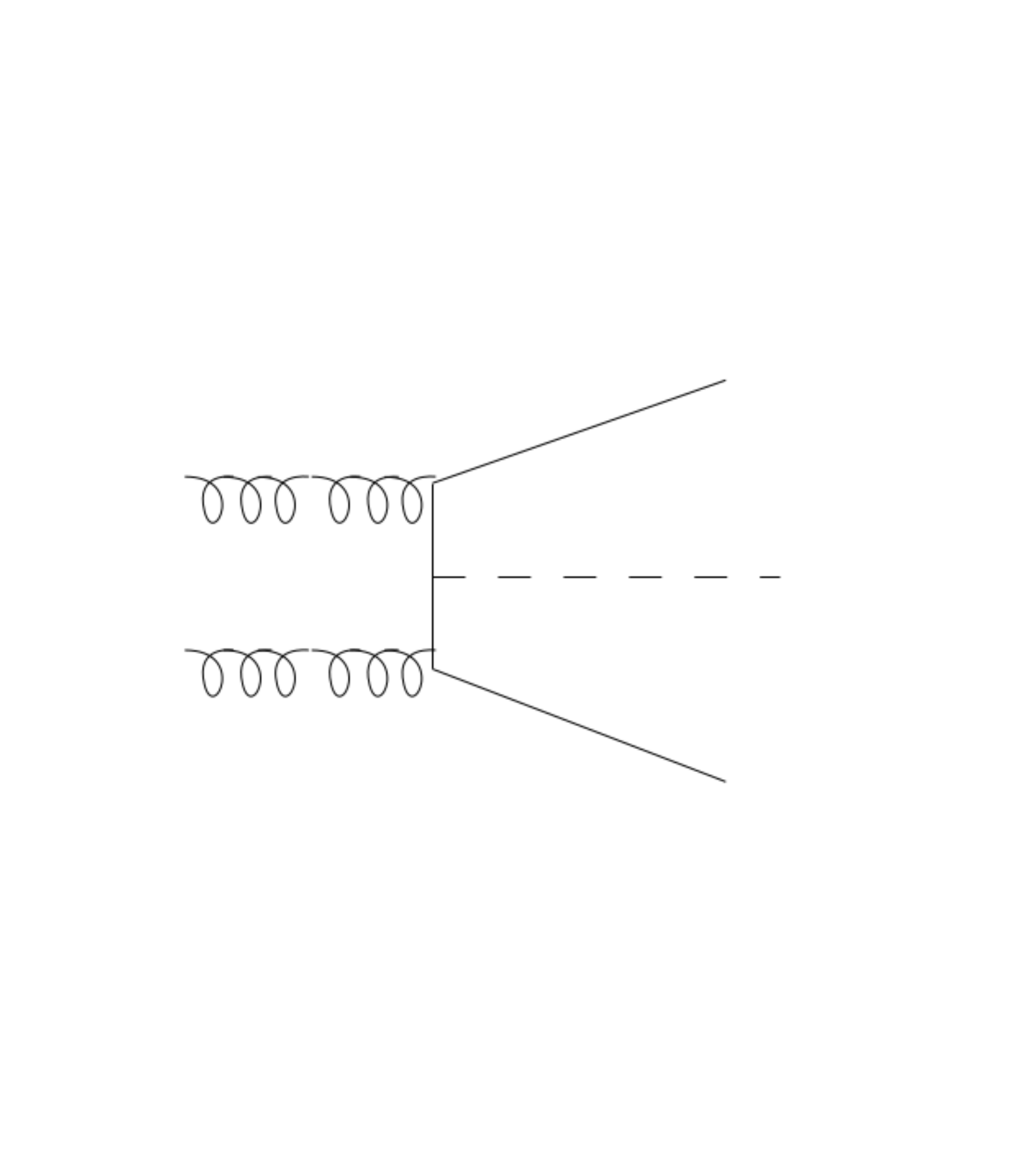} 
\includegraphics[scale=0.25]{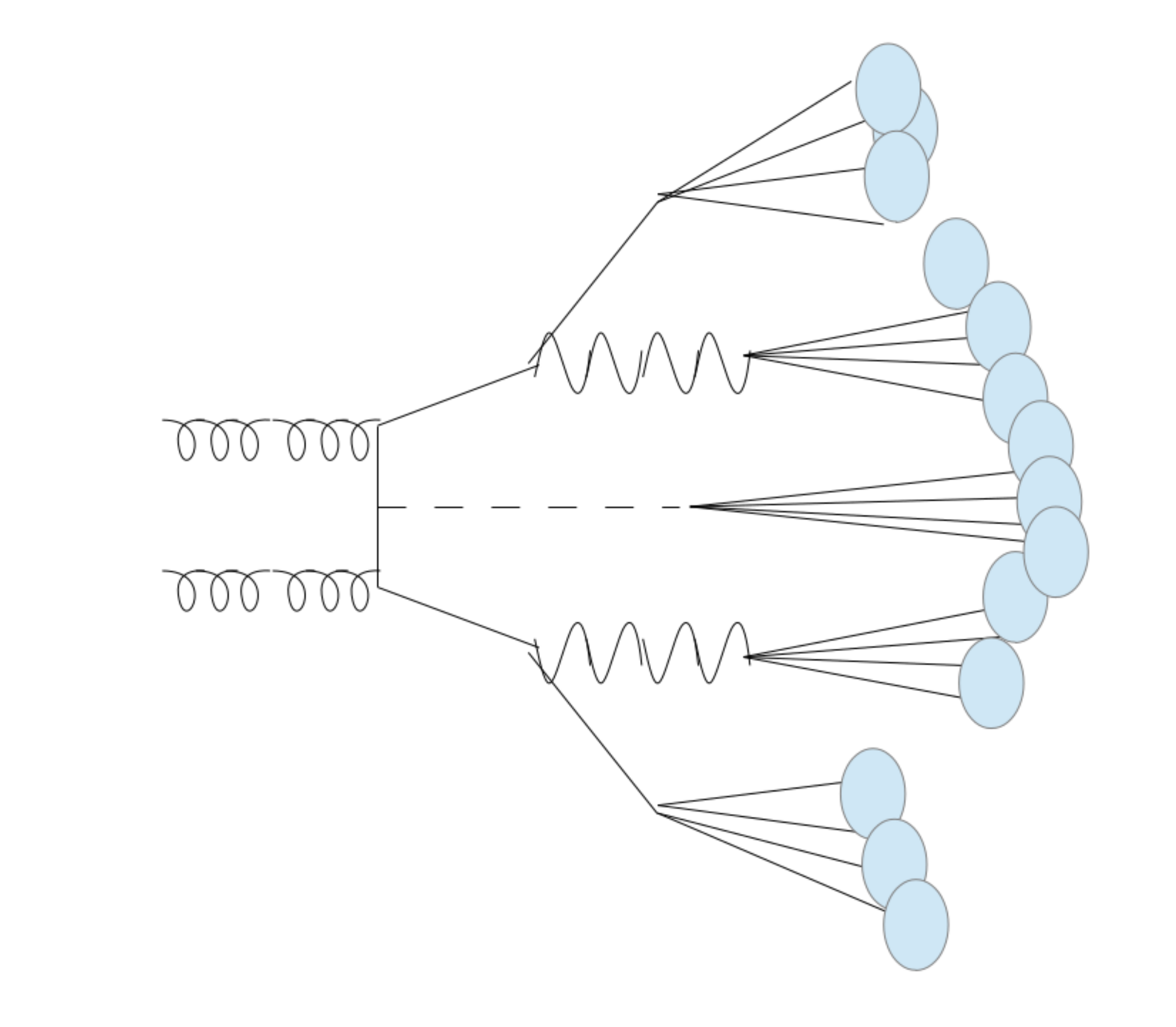} 
\includegraphics[scale=0.25]{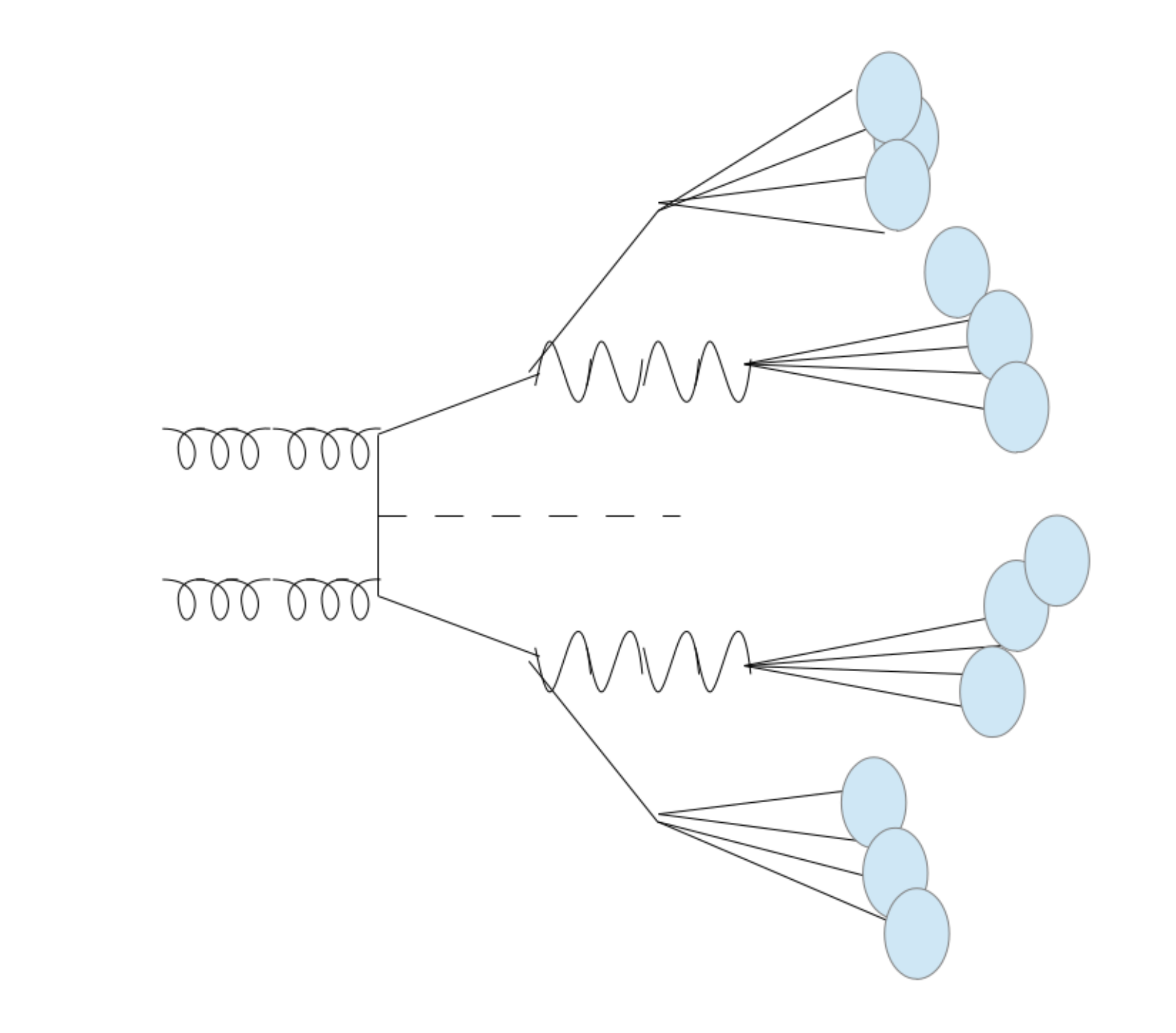}
\caption{Parton level (left), usual particle level definition (center) and corresponding particle level definition for \acp{STXS} (right), for the $t\bar t \h$ production mode. Blue blobs represent particles with long lifetime ($\gtrsim 10^{-10}s$).}
\label{fig:STXS_phasespace}
\end{figure}
The formula connecting single channel measurements to the \ac{STXS}  is:
\begin{equation} \label{eq:STXS_formula}
\begin{split}
& \sigma^{f}_{i}(\text{exp}) = A^{i}_{f , \alpha_1  }  \sigma^{\alpha_1 }_{i} \\
& \sigma^{f}_{i}(\text{exp}) = A^{i}_{f , \alpha_1 \alpha_2}  \sigma^{\alpha_1 \alpha_2 }_{i} \\
& \cdots \\
& \sigma^{f}_{i}(\text{exp}) = 
A^{i}_{f , \alpha_1 \alpha_2 \cdots \alpha_n }  
\sigma^{\alpha_1 \alpha_2 \cdots \alpha_n }_{i} \spc \\ 
\end{split}
\end{equation}
where $i$=\{$ggF$, $VBF$, $V\h$, $tt\h$, $bb\h$, $t\h$\} and 
$f$=\{$b\bar b$, $\gamma \gamma$, $ZZ$, $WW$, $\tau \tau$\} denote production and decay modes respectively,
$\sigma^{\alpha_1}_{i}$ , $\sigma^{\alpha_1 \alpha_2}_{i}$, 
$\sigma^{\alpha_1 \alpha_2 \cdots \alpha_n }_{i}$ are the STXSs 
and the $A^{f}_{i , \alpha_1 \alpha_2 \cdots \alpha_n }$ are efficiency/acceptance factors 
for each STXS. 

Each line of \cref{eq:STXS_formula} is called a \emph{stage}. The first \emph{stage} or 
\emph{stage}(0) is very close to the signal strength measurements performed during \RunI 
of the \ac{LHC}.
The indices $\alpha_{k}$ indicate the \emph{bin} divisions and the sum over repeated indices is 
understood.

Every bin of \emph{stage}($n$) is split into smaller bins in \emph{stage}($n+1$). 
The 
sub-bins are defined in independent regions of the full fiducial phase space.
This is repeated at every stage. The procedure could be repeated recursively, but given the prediction for the amount of data that \ac{LHC} experiments are going to collect during \RunII and the HL-\ac{LHC} phases, 
only stages up to stage(2) will have enough statistical significance.
Notice that the structure of Eq.~\eqref{eq:STXS_formula} is identical to \cref{eq:fiducial}, which 
allows the factorisation of the experimental and theoretical contributions from the total uncertainty.

\subsection{STXSs \emph{bins} and the $t \bar{t} \h$ binning proposal}\label{subsec:STXS_bin}

The \ac{STXS} framework is incomplete without a definition of a \emph{bin}.  The binning aims to 
reduce the theoretical dependence of the measurement within each bin so that
limits derived from a certain \emph{bin} do not strongly depend on a particular theoretical model.

A \emph{bin} should therefore be defined through cuts on truth level objects. 
If, for instance, the cut would be on the transverse momentum of the reconstructed object, the theoretical uncertainty would need to be convoluted with other effects and thereby be enlarged.

Another aspect to take into account is the identification of regions where \ac{BSM} physics 
has a higher probability of being observed. Separating such phase space regions from the SM dominated ones, 
would increase the chance of seeing potential deviations from the SM.
Since no big deviation from the SM has been observed, \ac{BSM} effects are expected to have the
largest impact in the tails of distributions or in extreme kinematic regions. 
For example, one could put a cut on Higgs transverse momentum, $\pth$, beyond which no SM events are expected (for a given integrated luminosity). All entries in such a bin would then be 
sign of \ac{BSM} physics. 

The binning depends on the integrated luminosity and on experimental systematic uncertainties, therefore it can only be fully understood  a posteriori. To allow the framework to be 
more flexible, the combination of bins is possible. In case a bin is found to be not significant, 
it can be combined with others to increase the global significance. Of course, not all bins can be  combined with all the others, but only adjacent ones. 

The best solution is then starting with a very general binning, similar to the one used during \ac{LHC} \RunI. This is called \emph{stage}(0).
Then, by taking into account more recent measurements and expected results, a second stage can 
be already defined for most processes.
The current proposal can be found in Ref.~\cite{deFlorian:2016spz} where the gluon 
fusion, vector boson fusion and associated vector boson production modes for the Higgs have 
been studied up to \emph{stage}(1), and hints for a future stage are given.

The $t\bar t \h$ production mode has not been studied in depth yet. It has a  more complicated topology than the other channels, and the cross section is relatively small.
It is difficult then, given a reasonable luminosity, to have enough events to fill a lot 
of \emph{bins}. 
This means that the statistical experimental uncertainty will be big, which allows for a relaxation of the theoretical independence criterion.
As an exercise, we are going to propose a possible \emph{binning} for $t \bar t \h$ production mode up to \emph{stage}(2).

Stage(0) has been already defined in Ref.~\cite{deFlorian:2016spz}, as inclusive 
$t\bar t \h$ production with Higgs pseudo-rapidity ($Y_\h$) less than $2.5$. 
The cut on $Y_\h$ avoids an extrapolation to the full phase space. Similar to all production modes, this bin is the most similar to a \RunI-like signal strength measurement.

Top quark pair associated production with a Higgs allows direct 
access to the top-Higgs interaction. One could naively expect that \ac{NP} would show up at high energies, hence a kinematic region which deserves consideration is the boosted regime. 
This is the phase space where the Higgs and one (or two) top(s) are produced at high 
transverse momentum, ideally much larger than their mass, but realistically starting from
around $200 \, \GeV$. 

Boosted analysis usually have a smaller background 
contamination, as explained in Ref.~\cite{Plehn:2009rk}.
On the other side, objects with high transverse momentum are less likely to be produced 
than objects with mild/low $p_T$. Furthermore, the region with $\pth <200 \, \GeV$ is sensitive 
to the CP properties of the top-Higgs interaction, see Ref.~\cite{Demartin:2014fia}.

Finally, a reasonable choice for the first \emph{stage} of $t\bar t \h$ production mode 
seems to be the definition of a boosted \emph{bin} more sensitive to potential \ac{NP} 
phenomena, and unboosted bins sensitive to CP properties. 
Since the former would probably contain few events, the latter topology could be 
divided in even smaller bins. 

Using the transverse momentum of the Higgs to define fiducial sub-volumes, we could 
define the \emph{bins}: 
\begin{equation}
 \begin{split}
                        & \text{bin(0) : } \qquad 0 <\pth < 100~\GeV \\ 
\text{stage(1)} \quad   & \text{bin(1) : } \qquad 100~\GeV <\pth < 200~\GeV \\
                        & \text{bin(2) : } \qquad 200~\GeV <\pth \\
 \end{split}
\end{equation}
where \emph{bin}(2) contains the boosted category where \ac{NP} effects are 
more probable. 
Although not a requirement for a high energy top-Higgs interaction, most of the boosted analysis require at least one top quark to be boosted to reduce background contamination.

It has been shown in Ref.~\cite{AguilarSaavedra:2009mx} that the general contribution arising from $\mathrm{dim} =6$ operators to the Yukawa-like 
interaction between fermions and the Higgs boson, is: 
\begin{equation}
 \mathcal{L}\propto \bar{f} \left( a_f + b_f \gamma^5 \right) f \h \spc
\end{equation}
where $f$ is the fermion (top quark in our case), $a_f$ and $b_f$ are two real factors.
In Ref.~\cite{Demartin:2014fia} the top quark Yukawa interaction has been studied for 
$t\bar t \h$ production mode.

Another distribution which is sensitive to the Yukawa 
coupling is the rapidity difference between the two tops, $\eta(t, \bar t)$. 
In particular, the region with $0<\eta(t, \bar t)<2$ is most sensitive, the region 
with $2<\eta(t, \bar t)<4$ has a smaller dependence, while the region $4<\eta(t, \bar t)<6$ 
has no sensitivity and low statistics.
A second stage seems possible (and useful when enough statistics will be available).

In principle, top quarks are not available among the objects we defined in the fiducial space,  but strategies exist to define ``pseudo-tops'' in a way which is as theory independent as possible. Furthermore, when removing Higgs decay products from the list of particles from  which jets are constructed, we are left with a 
$t\bar t$-only signature, for which such  algorithms are optimised.
The use of pseudo-tops is currently a standard procedure in many experimental analysis, e.g., Ref.~\cite{Aad:2015eia}. 

\begin{figure}[h]
\def\svgwidth{17.5cm}
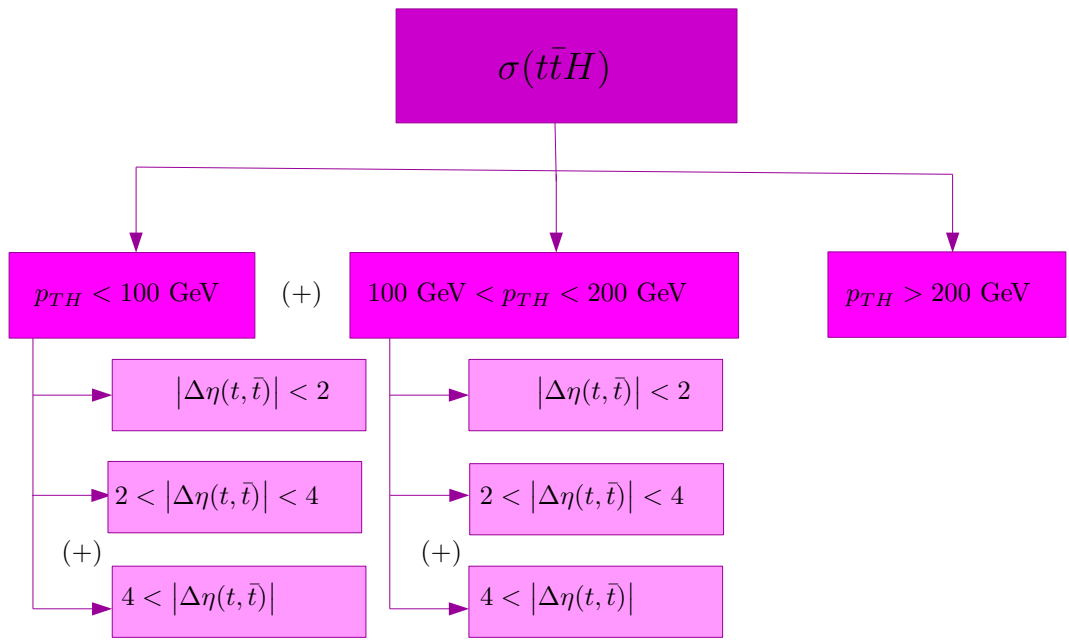
 \caption{Staging and binning for the $t\bar{t}\h$ production mode. $\pth$ is the transverse momentum 
of the Higgs boson, $\left| \Delta \eta (t, \bar t)\right|$ is 
the absolute rapidity difference between the two tops.  The ``$+$'' sign means that the 
bins can be added together.}
  \label{figure:ttH_bins}
\end{figure}

The second stage we propose is then:
\begin{equation}
 \begin{split}
                        & \text{bin(0) : } \qquad 0 <\left| \Delta \eta (t, \bar t)\right | < 2 \\ 
\text{stage(2) } \quad  & \text{bin(1) : } \qquad 2 <\left| \Delta \eta (t, \bar t)\right | < 4 \\
                        & \text{bin(2) : } \qquad 4 <\left| \Delta \eta (t, \bar t)\right | < 6 \\
 \end{split}
\end{equation} 
Compared to other production modes, $t\bar t \h$ has a clearer signature, therefore its staging and binning has been more focused on finding regions sensible to \ac{BSM} effects. 
And there is no need for defining bins with cuts which allow to distinguish it from other production channels, as it is the case for VBF, studied in Ref.~\cite{deFlorian:2016spz}. The final staging and binning we propose for Higgs boson associated production with top pairs is shown in~\cref{figure:ttH_bins}. Summarising the choice of bins, we have:
\begin{itemize}

 \item One bin where the Higgs is boosted and hypothetical \ac{NP} can be reached with 
       a higher probability because of the high energies involved.  

 \item Other bins sensitive to the CP characteristics of the Yukawa interaction.

\end{itemize}

After the definition of \emph{stages} and \emph{bins}, we end up with well defined measured quantities which have a small residual theoretical dependence. 

The next step is to interpret such measurements within a particular theory model through  the extraction of Wilson coefficients.
This is a difficult step, since different theories contain different coefficients, and there can be a large number of them, making  a simultaneous fit to all of them a priori impossible. Additionally, if an experiment would perform a fit to Wilson coefficients of a chosen theory,  the results would be valid only for this particular theory (and for all theories which can be matched to it).

Instead of fitting Wilson coefficients directly, one could fit objects, such as Pseudo-Observables (POs), which are well defined from both the theoretical and the experimental points of view.
These objects should be general enough (and fewer in number) to allow for variety of different theoretical models to be studied.
The \acp{PO} framework will be described in~\cref{sec:POs}.

In summary, \acp{STXS} and \ac{BR} measurements, together with physical \acp{PO}, allow  a wide variety of Higgs measurements to be studied in a well defined theoretical framework.
\subsection{Interplay with pseudo-observables}\label{subsec:STXS_vs_POs}
For Higgs production and decay modes, a list of Pseudo-Observables already exists, proposed in Refs.~\cite{Gonzalez-Alonso:2014eva, Greljo:2015sla}.
As discussed in~\cref{subsec:STXS}, STXSs are divided according to the various production modes. 

In order to maximally disentangle measurements from theory assumptions, a good option is to connect the STXSs with \acp{PO} 
of Higgs decays, which introduces fewer theoretical-experimental correlations than interpreting them  directly through \acp{PO} of Higgs production.

In Ref.~\cite{Greljo:2015sla} Higgs \ac{EW} production modes were written in terms of Higgs \acp{PO}. Most of these \acp{PO} also contribute to Higgs decays. 
This allows for a direct connection between VBF and VH production modes (and therefore their STXSs)  and the  Higgs \ac{EW} decays.
Note that \acp{PO} coming from the second order terms of the momentum expansion around physical poles cannot be directly connected to \acp{PO} of Higgs decays and only describe the production processes.

Such effects also appear in off-shell decay cross sections and distributions, where the kinematic regime (high transferred momentum) is similar.
\ac{QCD} \acp{PO} are not generally available yet, only a few have been introduced for production modes, 
which are  modifiers of the $gg\h$ and $t\bar t \h$ production rates.

Since \acp{PO} come from a momentum expansion around physical poles, such as vector boson masses, 
the validity of using \acp{PO} has to be checked carefully. 
Of course, transferred momentum is not directly accessible from an experimental point of view, but 
other variables can be used, which are related through kinematic constraints.
For instance, for the VH production mode $q \rightarrow \ptz$ for $\ptz\rightarrow \infty$, while for VBF production $q \rightarrow p_{T \text{forward-jet}}$, where $q$ is the 
transferred momentum, as defined in Ref.~\cite{Greljo:2015sla}.
Once the missing \acp{PO} are defined, an important task will be to find the variables which allow us to infer the value of the transferred momentum and thereby check the validity of the approximation.

In the case of the $t\bar t \h$ production mode,  the problem is more complex, since many \acp{PO} can be defined (taking into account the decays of the top quarks)
while there is limited information from the measurements (i.e. approximately 10 correlated STXS bins of~\cref{subsec:STXS_bin}).

From the definition of the \acp{PO} that will be given in~\cref{sec:POs}, the amplitudes depend linearly on \acp{PO}. 
Therefore cross section-like observables have a quadratic dependence, since integration over phase space does not change this behaviour.
For this reason, a theoretical prediction for the expected number of events can be written in the form:
\begin{equation}
 N^{\text{ev}} = X_{ij}k_i k_j \spc
\end{equation}
where $k_i$ is a vector of \acp{PO} and $X_{ij}$ is a matrix of coefficients that can be computed with \ac{MC} simulations.
This is in general true for every theory which introduces a linear dependence on a coefficient at scattering amplitude level.
For a \ac{STXS} bin, the procedure is the same, although the factors $X_{ij}$ are different.

Using the notation of section~\cref{subsec:STXS}, then
for \emph{stage}$(n)$ with a bin described by $a_1, \dots, a_n$:
\begin{equation}\label{eq:STXStheo}
 \sigma_{i}^{a_1 \dots a_n} = C_{i ,\ a_1 \dots a_n}^{lm}k_l k_m \spc
\end{equation}
where $C_{i ,\ a_1 \dots a_n}^{lm}$ are factors computed theoretically.
The question one should ask is \emph{which values of $\{k_i\}$ are more likely, given 
the experimentally observed $\sigma_i^{a_1 \dots a_n}$?}. 
To try to answer this question, a number of technical tools and results are needed:
\begin{itemize}

\item \ac{MC} generators producing theoretical predictions for various NP scenarios.

\item Detector level simulations of NP models, in order to verify their independence from the data correction.

\item All the relations between observables and theory parameters.

\item A software framework to perform the fit.

\end{itemize}
All these aspects introduce additional technical difficulties which will be discussed 
in~\cref{subsec:PO_k_Wilson}.

\newpage
\section{Effective field theories} \label{sec:EFT}
\subsection{Introduction}
At the heart of effective field theories lies the idea of adding a series of higher dimensional operators to a Lagrangian. In principle, adding operators of mass dimension bigger than four to a Lagrangian in four space-time dimensions renders the latter non-renormalizable in the classical sense. However, if this effective Lagrangian is interpreted as a series expansion in operator dimensions, it can be shown that the new Lagrangian is renormalizable order by order in such expansion.  

This result is closely related with the Wilsonian interpretation of ultraviolet divergences, where the higher dimensional operators are suppressed by powers of some energy scale (a \emph{cut-off} scale). This way, they can be understood as a collection of non-local operators, parametrising the effects of the local, renormalizable, operators of the theory in the UV regime.  

For this reason, such an effective Lagrangian can be used to parametrize SM deviations and ultimately lead to the development of an improved version of the Standard Model, valid in higher energy regimes. Any effective Lagrangian, describing SM deviations, can be written as
\begin{equation}
\mathcal{L}_{\rm{Eff}} = \mathcal{L}_{\rm{SM}} + \sum_{n > 4} 
\sum_i  \frac{a_i}{\Lambda^{n-4}} \mathcal{O}_i^{(n)} \spc
\end{equation}
where $\mathcal{O}_i^{(n)}$ are higher dimensional operators, built out of \ac{SM} fields,
$\Lambda$ is the cut-off scale for the effective theory, and the $a_i$ are the Wilson 
coefficients for the new operators, acting as effective couplings. In this section we will 
discuss how to build such a Lagrangian and make predictions with it. 


\subsection{Fermi Effective Theory} \label{subsec:FermiTheory}
Fermi Theory can be seen as the prototype of all EFTs, and as such, we will briefly discuss it in this section. The first Lagrangian for weak interactions was written by Fermi in $1934$, based on the 
electromagnetic one, $\mathcal{L}_{em}$. Its purpose was to explain the neutron decay, also 
known as $\beta$-decay, $n \rightarrow p + e^{-} + \overline{\nu}_{e} $.
After many theoretical efforts Fermi wrote an effective four-fermion Lagrangian
\begin{equation}
\mathcal{L}_{F} = \frac{G_{F}}{\sqrt{2}} J^{\mu}(x) J^{\dagger}_{\mu}(x) \spc
\label{eq:fermi}
\end{equation}
where the $J^{\mu}$ are point-like interaction currents. In particular,
$J^{\mu}(x) = L^{\mu}(x) + H^{\mu}(x)$, where $L^{\mu}(x)$ is the weak \emph{leptonic} 
current and $H^{\mu}(x)$ the weak \emph{hadronic} current. This theory considers the proton 
and the neutron fields to be fundamental and does not take parity violation into account.
Although it violates the unitarity of the scattering matrix and it is not 
renormalisable, Fermi theory proved to be a good effective theory with phenomenological success. Now we understand that the Fermi theory is a low-energy version of a Yang-Mills QFT.  However, the theory of \ac{EW} interactions did not appear as a UV completion of Fermi theory, and in fact it took much longer before the connection between the two was properly 
understood, for a discussion on this connection see ref.~\cite{Veltman:1968ki}. 

The decades after Fermi theory was proposed, namely the $40$'s and the $50$'s were times when 
particle physicists were extremely active. Many small experiments were taking place all around 
the world and experimental methods were developing very fast. This way parity violation was 
discovered, the V-A structure was detected, as well as $SU(2)$ symmetry and neutral currents, 
all these led to the postulation and experimental confirmation of the \ac{EW} theory. 

Without these experimental discoveries, Fermi and his theory colleagues would have been challenged to discover the \ac{SM} starting  only from the Fermi theory of $\beta$-decay and fundamental principles. 
In this case, a natural way forward could have been to enhance the theory by adding higher 
dimensional operators, which respect the known symmetries, as proposed in ref.\cite{David:2015waa}. Then, it would be possible to make predictions with this 
new theory and see in which manner the latter deviate from experiments. This way, information about  previously unknown symmetries such as $SU(2)$ could have been revealed. Theorists may then have written :
\begin{equation}
\mathcal{L}_{F} = G_F \bar{\psi} \psi \bar{\psi} \psi + 
\sum_{i,j} \underbrace{(\bar{\psi} 
\mathcal{O}_i \psi)(\bar{\psi} \mathcal{O}_j \psi)}_{\mathcal{O}_{i,j} =  
\substack{\text{any other fields}}} = G_F \bar{\psi} \psi \bar{\psi} \psi + 
a_{\Box} G_F^2 \bar{\psi} \psi  \Box 	\bar{\psi} \psi + \dots 
\end{equation} 

History did not happen like this, but this example illustrates how the theoretical effort done 
nowadays in the search for the \emph{underlying theory} complements what we know at the 
energies accessible for us. This procedure by which we add higher dimensional operators to our 
low-energy theory in a model-independent way is the so called ``bottom-up'' approach. 
\subsection{EFT: Top-down approach}
In practise, \ac{EFT} comes in two different flavours, the bottom-up approach
that we just outlined, and the top-down approach. In this section, we will focus on the latter.

The top-down approach is model dependent: it represents a scenario where we want to study a 
particular high-energy or UV-complete theory and we try to infer the behaviour and 
footprints of this theory in the experimentally accessible low energy regime. One can 
understand this technique by looking at the Fermi theory where the UV completion was 
the \ac{SM}, or more concretely, the theory of \ac{EW} interactions. Back then, the energy 
regime of the \ac{SM} was far beyond what could be probed experimentally, therefore, if one would 
have wanted to test the validity of the \ac{SM} at energies of say, the order of $10 \, \GeV$, 
what would be the best strategy to follow? 

The first observation is that particles heavier than the energy regime accessible by our 
experiments will not be created. Therefore we can remove them from our theory, 
since we do not expect them to appear. However, particles are not always on-shell, and, heavy particles that are not directly produced in the experiment can 
nevertheless contribute via loop corrections. This has the direct 
implication that we cannot simply remove them from the theory by setting the masses to infinity and/or the couplings to zero. We have to integrate out the relevant 
degrees of freedom using legitimate QFT methods. 

The technique of removing heavy particles but keeping the consistency of the QFT, goes back  to the early $80$'s~\cite{Weinberg:1980wa}, in the context of quantum gravity, and makes use  of a technique called \ac{BFM} to integrate out the heavy states in the path integral. 

\vspace{0.5cm}
\begin{figure}[h]
\begin{center}
\def\svgscale{0.8}
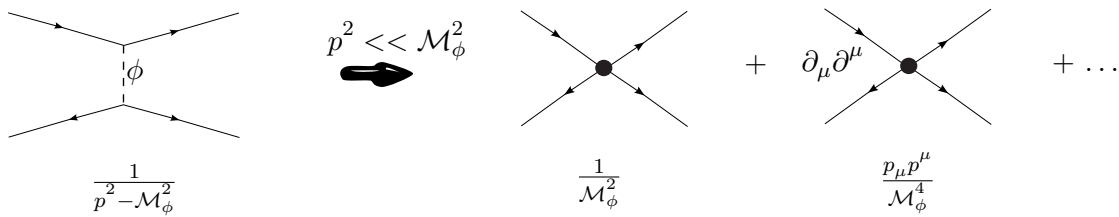 
\end{center}
 \caption{In the top-down approach the heavy particles (a scalar $\phi$ field in this case) are \emph{integrated out} by taking the limit $M^2_{\phi} >> \mid p^2 \mid$. This way 
the propagator contracts to become a vertex. }\label{fig:exampleEFT}

\end{figure}

After we identify the the cut-off scale of the UV theory, and integrate out 
the particles above this scale, the last step is to make predictions for the low energy 
regime using the new and simplified effective scenario. 

To understand how heavy particles are integrated out, we can do a simple exercise: imagine some 
heavy singlet under $SU(2)$, represented by a scalar field $S$ with a mass $M_S$ much bigger than the \ac{SM}  
masses and construct the singlet extension of the SM (SESM). This SM extension was born in the 80's as a dark matter candidate, in  Ref.\cite{Silveira:1985rk}, it is also often used in the context of Higgs physics and LHC phenomenology see for instance Refs.\cite{Schabinger:2005ei,Pruna:2013bma, Robens:2015gla, Robens:2016xkb, Gorbahn:2015gxa}, as well as more exotic scenarios: to guarantee the stability of the EW vacuum at very high energies, in Ref. \cite{EliasMiro:2012ay}, or to describe baryogenesis, in Ref.\cite{Anderson:1991zb,Espinosa:1993bs}, 


\begin{equation}
\begin{split}
\label{eq:Dittmaier1}
\mathcal{L}_{\rm{UV}}  &= \mathcal{L}_{\rm{SM}} + \mathcal{L}_{S}, \\ 
\mathcal{L}_{S}  & =   \frac{1}{2} ( \partial_{\mu} S ) (\partial^{\mu} S) 
- \frac{1}{2} M_S^2 S^2 - \lambda_1 S^4 
- \frac{\lambda_{12}}{2} S^2 (2 v H + H^2 + 2 \phi^+ \phi^- + \chi^2 ) \spc
\end{split}
\end{equation}
where $\varphi$ is the scalar doublet of the SM and $\phi$, $\chi$ the unphysical Goldstone fields.
To remove the heavy field $S$ from the Lagrangian, we have to \emph{integrate it out}.  This 
can be done either by diagrammatically or by using functional methods. Here we will outline the latter. 

To integrate out the field $S$, we write down a path integral over that field. This can be 
understood as keeping all \ac{SM} fields fixed (also called ``background fields'') with 
the heavy ones being dynamic. The effective action then becomes:
\begin{equation} \label{eq:seff}
S_{\rm{eff}} = 
\exp \Bigl\{i \int  {d}^4 x \mathcal{L}_{ {eff}}  \Bigr\} \sim \int \mathcal{D} S \exp \Bigl\{\frac{i}{2} \int  {d}^4 x [ \frac{1}{2} ( \partial_{\mu} S ) (\partial^{\mu} S) - 
M_S^2 S^2 - S^2 F(\varphi)] \Bigr\} \spp
\end{equation}
For the interested reader, this procedure is discussed in detail in 
Refs.~\cite{Dittmaier:1995cr,Dittmaier:1995ee} in the context of a heavy Higgs in the \ac{SM}. Here 
we just show the basic steps. 
\begin{equation}
\begin{split}
\exp \Bigl\{ i \int  {d}^4 x \mathcal{L}_{ {eff}}  \Bigr\} &=
\int \mathcal{D} S 
\exp \Bigl\{\frac{-i}{2} \int  {d}^4 x  \int  {d}^4 y  
S(x) ( \partial_{x}^2 + M_S^2 + F(x) ) \delta(x-y) S(y) ) \Bigr\}  \\ &= 
 \exp \Bigl\{\mathcal{D}et \left[ ( \partial_{x}^2 + M_S^2 + F(x) ) 
\delta(x-y) \right]^{-\frac{1}{2}} \Bigr\}   \\ &= 
 \exp \Bigl\{ -\frac{1}{2} \mathcal{T}r \lbrace {\rm ln} 
(\partial_{x}^2 + M_S^2 + F(x) ) \delta(x-y))	\rbrace \Bigr\} \spp
\end{split}
\end{equation}
Now we can take the \emph{decoupling limit} $M_S \rightarrow \infty$
\begin{equation}
\begin{split}
& \mathcal{T}r \biggl\{ {\rm ln} \left[ \left( -\partial_x^2 - M_S^2 - F(x) \right) \delta (x-y) \right] \biggr\} \\ & \qquad
= \int  {d}^4 x \int \frac{ {d}^4 p}{(2 \pi)^4} {\rm ln} \left( -(\partial_x + ip)^2 - M_S^2 - F(x) \right) \\ & \qquad
= \int  {d}^4 x \int \frac{ {d}^4 p}{(2 \pi)^4} {\rm ln} \left( p^2 - M_S^2 - 2i p_{\mu} \partial_x^{\mu} - \partial_x^2 - F(x) \right) \\ & \qquad
= - \sum_{n=1}^{\infty} \frac{1}{n} \int  {d}^4 x \frac{ {d}^4 p}{(2 \pi)^4}  \left( 
\frac{2i p_{\mu} \partial_x^{\mu} + \partial_x^2 + F(x)}{p^2 - M_S^2}
\right)^n + \rm{const.} \\ & \qquad
= \int  {d}^4 x \frac{ {d}^4 p}{(2 \pi)^4}  \left[ 
\frac{F(x)}{p^2 - M_S^2} + \frac{(F(x))^2}{2(p^2 - M_S^2)^2} +  
 \frac{F(x) \partial_x^2 F(x) + F(x)^3   }{3(p^2 - M_S^2)^3} + \mathcal{O}(M_S^{-4})
\right] + \rm{const.} \spc
\end{split}
\end{equation}
where we used the following identity,
\begin{align*}
{\rm ln} & (-\partial_{x}^2 - M_S^2 - F(x) ) \delta(x-y)) = {\rm ln} (-\partial_{x}^2 - M_S^2 - F(x) )  \int \frac{ {d}^4 p }{(2 \pi)^4} e^{i p (x-y) } =  \\
& = \int \frac{ {d}^4 p }{(2 \pi)^4} e^{i p (x-y) } {\rm ln}  ( - (\partial_x + ip)^2 - M_S^2 - F(x) ) \spp
\end{align*} 
Written in a more compact way: 
\begin{equation}
\begin{split}
& \mathcal{T}r \lbrace {\rm ln} \left[ \left( -\partial_x^2 - M_S^2 - F(x) \right) \delta (x-y) \right] \rbrace \\  
& \qquad =  \frac{-i}{16 \pi^2} \int  {d}^4 x \biggl[ I_{01} F(x) + \frac{I_{02}}{2} F(x)^2 + \frac{I_{03}}{3} \biggl( F(x) \partial_x^2 F(x) + F(x)^3 \biggr) \biggr] \spc
\end{split}
\end{equation} \label{eq:traceLog}
with the momentum integrals defined as, 
\begin{equation}
\begin{split}
I_{kl} g_{\mu_1 \dots \mu_2k} &= \frac{(2 \pi \mu)^{4-D}}{i \pi^2} \int  {d}^D p \frac{p_{\mu_1} \dots p_{\mu 2k}  }{(p^2 - M_S^2)^l} \\
& = g_{\mu_1 \dots \mu_{2k}} \frac{(-1)^{(k+l)}}{2^k} \frac{\Gamma (l-k- \frac{D}{2} )}{\Gamma(l)} (4 \pi \mu^2)^{\frac{4-D}{2}} M_S^{D+2k-2l} \spp
\end{split}
\end{equation} \label{eq:integralsI12}
Substituting the integrals of Eq.\eqref{eq:integralsI12} in the expression in Eq.\eqref{eq:traceLog}, and this in $\mathcal{L}_{ {eff}}$ we find:
\begin{equation}
\begin{split}
\label{eq:DittmaierLast}
\mathcal{L}_{eff} & = \frac{1}{32 \pi^2} \left[ c_1 (\varphi^\dagger \varphi) + c_2 (\varphi^\dagger \varphi)^2 + c_3 (\varphi^\dagger \varphi)\partial_x^2 (\varphi^\dagger \varphi) + c_4 (\varphi^\dagger \varphi)^3 \right] \\ 
& = \frac{1}{32 \pi^2} \left[ 
- \frac{\lambda_{12}^2}{3 M_S^2} (\varphi^\dagger \varphi) \partial_x^2 (\varphi^\dagger \varphi) -
\frac{4 \lambda_{12}^3}{3 M_S^2} (\varphi^\dagger \varphi)^3 \right]
+ \mathcal{O} (M_S^{-4}) \spp
\end{split}
\end{equation}

In the last step we have used the Appelquist-Carazzone theorem of Ref.~\cite{Appelquist:1974tg} 
to remove some of the operators whose Wilson coefficients are mass-suppressed.
Observe that in this case, the scale of new physics is: $\Lambda^2 = M_S^2$, however if we were 
working with the mass eigenbasis instead of with the weak eigenbasis, the scale would have 
to be extracted from: 
\begin{equation}
M_S^2 = \Lambda^2 \sum_{n=0} \xi_n \left( \frac{\mw^2}{\Lambda^2} \right)^n \spc
\end{equation} 
where $\xi_n$ is a parameter dependent on the model we are studying. We can see that the 
difference between both scales is sub-leading, however the question of the scale choice 
is a non-trivial one and should be carefully addressed before choosing the cut-off for the 
\ac{EFT} calculation. 
\subsubsection{The Background Field Method}

To quantize a gauge theory in the conventional approach it is necessary to choose a gauge
and the corresponding Faddeev-Popov ghost Lagrangian. 
Then, when the gauge has been fixed, the Feynman rules can be read from the Lagrangian which is, 
by construction, not invariant any more under a gauge transformation of the fields and requires
the Becchi-Rouet-Stora-Tyutin (BRST) global supersymmetry introduced in the 
mid-$1970$s in Ref.\cite{Becchi:1981jx}. 
Consequently, quantities with no direct physical meaning, like Green's functions, are not 
gauge invariant. On the other hand, the gauge dependence consistently cancels when computing 
physical quantities like the $S\,$-matrix.

The Background Field Method introduced in Refs.\cite{DeWitt:1967ub,tHooft:1973bhk,Abbott:1980hw,Abbott:1981ke,Denner:1994nn,Denner:1994xt} 
is a technique that was developed in the $80$'s in the context of quantum gravity, to preserve  some degree of gauge invariance in every step of the calculation; for instance, in any gauge  theory the gauge-fixing term can be chosen to involve only the quantum fields and not the 
classical ones. 

Another benefit of working with the \ac{BFM} at the one-loop level, 
comes from the fact that 
it is possible to take apart, for a generic field $\phi$ of the theory, a classical 
field $\hat \phi$ and its quantum fluctuation $\phi$. The so-called background field 
$\hat \phi$ can only appear on external lines, while the quantum field $\phi$ only as an 
internal line in loops. Then, considering the generating functional
\begin{equation}
 Z[J] = \int \mathcal{D} \phi \, \exp i \{ S[\phi] + J \cdot \phi \}
\end{equation}
of the Green's functions of the considered theory (for simplicity here, $\phi$ is a scalar 
field and we do not consider a gauge theory), it is possible to define the analogous functional
\begin{equation} \label{eq:genFuncBFM}
 \tilde Z[J, \hat \phi] = \int \mathcal{D} \phi \, \exp i \{ S[\hat \phi + \phi] + J \cdot \phi \} \spc
\end{equation}
obtained by the substitution $\phi \to \hat \phi + \phi$ in the action. Here, the background 
field $\hat \phi$ can be considered as an additional source. When defining the generating 
functional for the \ac{1PI} Green's functions of the theory
\begin{equation}
\begin{split}
 \Gamma[\bar \phi] &= W[J] - J \cdot \bar \phi \spc \\
 W[J] &= - i \ln Z[J], \quad \bar \phi = \frac{\delta W}{\delta J} \spc
\end{split}
\end{equation}
a similar functional can be defined in presence of the background field $\hat \phi$,
\begin{equation}
\begin{split}
 \tilde \Gamma[\tilde \phi, \hat \phi] &= \tilde W[J, \hat \phi] - J \cdot \tilde \phi \spc \\
 \tilde W[J, \hat \phi] &= 
- i \ln \tilde Z[J, \hat \phi], \quad \tilde \phi = \frac{\delta \tilde W}{\delta J} \spp
\end{split}
\end{equation}
Applying the previous definitions one recovers the main result of Ref.~\cite{Abbott:1980hw}, that
\begin{equation}
\label{eq:BFMequiv}
 \tilde \Gamma[0,\hat \phi] = \Gamma[\hat \phi] \spp
\end{equation}
This equation provides an alternative way to compute the generating functional of \ac{1PI} 
Green's functions $\Gamma$ by using the background field functional 
$\tilde \Gamma[\tilde \phi, \hat \phi]$, i.e. the conventional functional in presence of 
the background field $\hat \phi$. 

Functional derivatives of $\tilde \Gamma[\tilde \phi, \hat \phi]$ with respect to 
$\tilde \phi$ would generate \ac{1PI} Green's functions with external $\phi$ field, 
in presence of the background field $\hat \phi$. 
Since the latter does not appear in the functional integral, it 
cannot be present in the internal lines of the Green's functions. Moreover, since
the functional $\tilde \Gamma[0,\hat \phi]$ in~\cref{eq:BFMequiv} does not depend on 
$\tilde \phi$, it cannot generate Green's functions with external $\phi$ lines, but only 
with external $\hat \phi$ lines. These considerations explain why $\hat \phi$ and $\phi$ 
are called ``background'' and ``quantum'' fields respectively, and why they can only appear respectively on external and loop lines.

For this reason, the \ac{BFM} provides a good bookkeeping framework that can be used when 
integrating out heavy degrees of freedom from the path integral: when performing the path 
integral over the heavy fields (reported in~\cref{eq:PathIntEFT}), we have a solid argument 
to state that only the Lagrangian terms that are bilinear in the integration variable 
$\Phi$ matter for the computation of the one-loop effective action. Then,
\begin{itemize}

\item couplings with exactly one quantum field are not relevant for one-loop diagrams (however, terms with one \emph{heavy} quantum field are retained in the functional integration, in couplings with two quantum fields, of which only one is heavy),

\item couplings with more than two quantum fields are only needed beyond one loop (analogously to the previous remark, in this case terms containing more than two heavy fields are retained, given that no more than two quantum fields are present).

\end{itemize}
After the application of the \ac{BFM}, to work at the one-loop level, the action of 
Eq.\eqref{eq:genFuncBFM} contains only linear and bilinear terms in the quantum heavy field. 
Linear terms can be shifted away by a redefinition of the heavy quantum field, and the 
functional integral over the heavy quantum field can be performed by means of Gaussian 
integration. This has been used, for example, in Refs.~\cite{Dittmaier:1995cr,Dittmaier:1995ee} in the context of the SM.
\subsubsection{The Covariant Derivative Expansion}

A very similar, equivalent, approach was first proposed by 
Mary K.~Gaillard in Ref.\cite{Gaillard:1985uh} and  went relatively unadvertised until 2012 when an \ac{EFT} review, Ref.\cite{Henning:2014wua}, put it back on the scope. This approach is known as the \ac{CDE}.  
It relies on the same physical ideas as the \ac{BFM} but focusses on the steps needed
to derive $\Gamma[\hat \phi]$.

Given the action of a theory with a heavy field $\Phi$ and a lighter one $\phi$: $S[\phi,\Phi]$ 
the effective action after integrating out the heavy field at the cut-off scale $\mu$, 
can be written in terms of a path integral, just as we did in~\cref{eq:seff}
\begin{equation}
\label{eq:PathIntEFT}
e^{i S_{eff} [\phi](\mu)} = \int \mathcal{D} \Phi e^{i S[\phi,\Phi]} \spp
\end{equation} 
Up to this point,  everything is similar to the example shown in Eqs.~\eqref{eq:Dittmaier1}-\eqref{eq:DittmaierLast}. The \ac{CDE} approach introduces 
an elegant way of solving this path integral analytically in a gauge-invariant way. 

To solve \eqref{eq:PathIntEFT}, we use the saddle point approximation, namely, we 
do a Taylor expansion of the exponent on the r.h.s. such that we can solve the path integral 
term by term of the power expansion, in particular we expand around the minimum of the action:
\begin{equation} \begin{split}
& \frac{\delta S}{\delta \Phi} = 0 \quad \Rightarrow \quad \Phi_c[\phi]  \\
& S[\phi, \Phi_c + \eta] = S[\Phi_c] + \frac{1}{2} \frac{\delta^2 S}{\delta \Phi^2} \Bigg\vert_{\Phi_c} \eta^2 + \mathcal{O} (\eta^3) + \dots
\end{split} 
\end{equation}
this way, the path integral becomes easier to solve,
\begin{eqnarray}  
e^{i S_{ {eff}}} &=& \int \mathcal{D} \eta e^{i S[\phi, \Phi_c + \eta]}  \nonumber \\
& \approx& e^{i S[\Phi_c]} \left[  \rm{det} \left( - \frac{\delta^2 S}{\delta \Phi^2} \Bigg\vert_{\Phi_c} \right) \right]^{-1/2} \nonumber \\
 S_{\rm{eff}} &\approx& S[\Phi_c] + \frac{1}{2} \rm{Tr} \, \, \rm{log} \left( - \frac{\delta^2 S}{\delta \Phi^2} \Bigg\vert_{\Phi_c} \right) \spp  \label{eq:seff1loop}
\end{eqnarray}
Eq.~\eqref{eq:seff1loop} can be applied to any UV action that we want to solve. If we want to find 
a more explicit expression, we observe that almost any Lagrangian 
can be cast in the following form, 
\begin{equation} \label{eq:ellipticLag}
\mathcal{L} =  \underbrace{\Phi^{\dagger} B}_{B(\phi(x))} + B^{\dagger} \Phi + \Phi^{\dagger} (-D^2 - m^2 - \underbrace{U}_{U(\phi(x))}) \Phi  + \mathcal{O}(\Phi^3, \Phi^4, \dots) \spp
\end{equation}
Solving the equation of motion for $\Phi$ we find: 
\begin{equation}
\Phi_c = \frac{B}{D^2 + m^2 + U} + \mathcal{O}(\Phi^2) \spp
\end{equation}
In order to simplify this expression, we make an inverse mass expansion of $\Phi_c$:
\begin{equation}
\Phi_c \approx \frac{1}{m^2} B + \frac{1}{m^2}(-D^2 - U) \frac{1}{m^2} B + \frac{1}{m^2} (-D^2 - U) \frac{1}{m^2} (-D^2 -U) \frac{1}{m^2} B + \dots
\end{equation}
The name \ac{CDE} comes from the fact that we leave $D^2$ contracted rather than making it 
explicit and expanding in powers of\footnote{This seems counter-intuitive, but there are many 
examples where an expansion in powers of $\frac{\partial^2}{m^2}$ is more useful, due to 
the $x$ integration in $\int \rm{d} x \, \mathcal{L}$} $\left(\frac{\partial^2}{m^2} \right)$. 
Replacing $\Phi_C$ in \eqref{eq:ellipticLag}, we find the effective Lagrangian at tree level:
\begin{equation}
\mathcal{L}_{\rm{eff, tree}} = 
\underbrace{B^\dagger \frac{1}{m^2} B}_{\text{dim-6 operator}} +   
\underbrace{  B^\dagger\frac{1}{m^2}  (-D^2 - U) \frac{1}{m^2} B}_{\text{dim-8 operator}}  
+ \, \text{higher dim. ops} \spp
\end{equation}

To find the one-loop part of the effective Lagrangian, we have to solve the last term 
in \eqref{eq:seff1loop}. The details about the calculation of the functional trace and log can 
be found in Ref.~\cite{Henning:2014wua}, as well as tables of ``universal'' results and many 
detailed examples. Repeating these examples would be an useful exercise for the reader 
interested in functional methods. Here, however, we simply focus on the physical 
content rather than on the algebra, and we will not repeat the derivations. 

\begin{figure}
\centering
\def\svgscale{0.55}
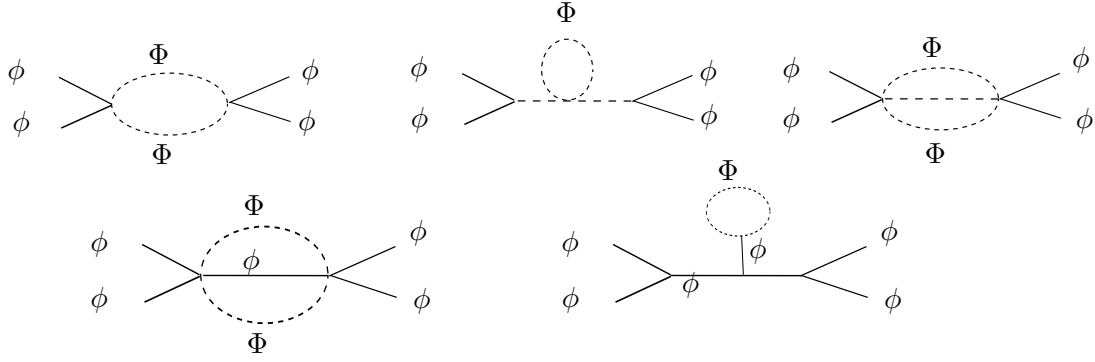
\caption{Diagrammatic interpretation of the \ac{BFM}. Diagrams with mixed heavy 
and light fields, like the ones in the bottom row, are not taken into account by 
the \emph{naive} \ac{CDE} method.}
\label{fig:BFM1}
\end{figure}

As discussed earlier, both the \ac{CDE} and \ac{BFM} methods rely on fixing the heavy fields 
while making the light ones dynamic. Looking at the diagrammatic interpretation of the method 
(see~\cref{fig:BFM1}) one can rapidly realize a flaw, namely the diagrams with heavy fields inside the loops are not being taken into account. It can be argued 
that the contribution of these diagrams to the Wilson coefficients of interest is generally 
small, however it is not advisable to use the previous equations without making sure of the 
size of the neglected contribution. 

This problem was first pointed out in Ref.~\cite{delAguila:2016zcb} and further discussed in Ref.~\cite{Boggia:2016asg}. Later, the authors of the original 
\ac{CDE} paper proposed the following solution in Ref.~\cite{Henning:2016lyp}. Eq.~\eqref{eq:seff1loop} for the one-loop effective action, 
\begin{equation}
S_{eff} \approx \underbrace{S \left[ \Phi_C \right]}_{\text{tree-level}} + 
\underbrace{\frac{i}{2} \text{Tr} 
\log \left( - \frac{\delta^2 S}{\delta \Phi^2} {\Bigg\vert_{\Phi_C}} \right) }_{\text{one-loop}}
\spc
\end{equation} 
has to be corrected by adding a non-local term,
\begin{equation}
\Gamma[\phi] = 
S_{eff}[\phi] + \underbrace{\frac{i}{2	} 
\rm{log } \, \, \rm{det} \left( \frac{- \delta^2 S}{\delta \phi^2}
\right)}_{\text{non-local term}} \spc
\end{equation}
which accounts for the neglected terms. Unfortunately it is no longer possible to find a 
universal result since the non-local term depends on the tadpoles and self energies of the particular theory that we are studying. In that case, the remaining non-universal loop diagrams can be solved diagrammatically, as it was done for example in Ref.\cite{Boggia:2016asg}.

The connection with Fermi theory is now clear: the UV theory in this case is the 
electroweak sector of the \ac{SM}, and the energy scale above which we will \emph{cut} is related 
to the mass of the heavy particles, $\mw$. We can thus apply either the \ac{BFM} or the \ac{CDE} and 
integrate out the $W$ bosons from our theory, shrinking the propagator to a point, and the full 
diagram becomes a vertex, as shown in~\cref{fig:exampleEFT}.
\subsubsection{Matching of the UV and SM scales}

After deriving the effective Lagrangian, there is one more step necessary before we can make 
predictions. As we know from a general QFT, the masses and couplings are not constant, 
and their precise value depends on the scale we are looking at. We commonly say that these 
parameters \emph{run} with the scale. 

Therefore the value of the effective couplings that we extracted using the previous methods has to be evolved down to the scale of interest to us. As an illustration, in the singlet example of Eq.~\eqref{eq:DittmaierLast}, the coefficients in front of 
the $\mathrm{dim}=6$ operators, 
\begin{equation}
(\varphi^{\dagger} \varphi) \partial_x^2 (\varphi^{\dagger} \varphi) \spc \quad 
(\varphi^{\dagger} \varphi)^3 \spc
\end{equation}
at the scale $\Lambda^2 = M_S^2$ are given by
\begin{eqnarray*}
c_1 (M_S) &=& \frac{1}{32 \pi^2} \frac{- \lambda_{12}^2}{3} \spc \\ 
c_2 (M_S) &=& \frac{1}{32 \pi^2} \frac{- 4 \lambda_{12}^3}{3} \spc
\end{eqnarray*}
respectively. In order to find these coefficients at the scale of our experiment (which 
might be $\mw$, $m_b$, $\mh$, \dots, depending on the process we are looking at) there are 
two possible approaches: 
\begin{itemize}

\item The fastest and most commonly used approach relies on the \ac{RG} 
flow equations, see for example Ref.~\cite{Alonso:2013hga}. At LO, one has 
\begin{equation}
c_i (\mw) \neq c_i (\Lambda), \quad \Rightarrow \quad 
c_i (\mw) =  c_i (\Lambda) - \sum_j \frac{1}{16 \pi^2} 
\gamma_{ij} c_j(\Lambda) \rm{log} \frac{\Lambda}{\mw} 
\end{equation} \label{eq:RGflowEq}
where $\gamma_{ij}$ is the anomalous dimension matrix.
The full equation is: 
$$ \frac{d c_i (\mu)}{d \rm{log} \mu} = \sum_j \frac{1}{16 \pi^2} \gamma_{ij} c_j \spp$$

\item The other possible approach is more rigorous: we can perform the full 
renormalisation of the newly found effective Lagrangian, in Eq.~\eqref{eq:DittmaierLast}, and 
the values of the coefficients at the new scale will be fixed at the point when we choose  the renormalisation scheme and the set of input parameters. For the example of a heavy singlet, the full procedure was carried out in Ref.~\cite{Boggia:2016asg}. Further details on 
the renormalisation procedure will be given in \cref{subsec:ren} in the context of 
the bottom-up approach. 

\end{itemize}
After we have computed the values of $c_i (\mw)$ we can start to make predictions for physical 
processes with the effective theory. However, as pointed out in Ref.~\cite{Freitas:2016iwx},
the effective operator expansion is only slowly converging at best.  
The agreement between the SMEFT and a range of UV-complete models depends
sensitively on the appropriate definition of the matching and can be systematically improved through 
an appropriate matching procedure at the one loop level;
matching schemes based solely on leading logarithms are not useful for any kind of precision physics.
\subsection{Fine points in EFT}
It is important to note that the EFT that we are discussing here is the ``continuum EFT'' introduced by 
Georgi in Refs.~\cite{Georgi:1992xg,Georgi:1994qn}. The main question is: how can an EFT be constructed 
without appealing to a cut-off that will prevent us from using dimensional regularisation? 
For top-down constructions, the initial momentum splitting of the fields in a Wilsonian EFT
and the integration over the heavy modes, is replaced in a continuum EFT by the following steps,
\begin{itemize}

\item First, start with a dimensionally-regularised theory with Lagrangian density
$\mathcal{L}(\phi\,,\,\Phi)$ where $\phi$ are the light fields and $\Phi$ the heavy ones. 

\item Then, evolve the theory to a lower scale using RG flow equations, Eq. \eqref{eq:RGflowEq}.

\item When below some scale, say $M$, the EFT is changed to a new one without the $\Phi\,$-fields
$\mathcal{L}(\phi) + \Delta\,\mathcal{L}(\phi)$ where $\Delta\,\mathcal{L}$
encodes a ``matching correction'' that includes any new nonrenormalisable interactions that 
may be required. The matching condition is defined so that the physics of the light fields is the
same in the two theories at the boundary $M$. At leading order this condition is trivial: it represents the continuity of the couplings at the matching scale. At NLO and beyond this is not the case any more. 

\item For explicit calculations, $\Delta\mathcal{L}$ is expanded in a complete set of local operators.

\end{itemize}
To summarise, in the construction of a Wilsonian EFT, the heavy fields
are first integrated out of the underlying high-energy theory and the
resulting Wilsonian effective action is then expanded in a series of local operator terms. 
The cutoff in a Wilsonian EFT plays a double role: first, it explicitly separates the 
low-energy physics from the high-energy physics;
and second, it regulates divergent integrals in the calculation of
observable quantities.

In the construction of a continuum EFT, the heavy fields initially
remain in the underlying high-energy theory, which is first evolved down to the
appropriate energy scale. The continuum EFT is then constructed by completely removing the
heavy fields from the high-energy theory, as opposed to integrating them out; and this
removal is compensated for by an appropriate matching calculation.
For a detailed description see Ref.~\cite{Bain:2013aca}.
\subsection{EFT Bottom-Up approach}
If we think about the phenomenological application of EFTs, we can say that the top-down option generates the following chain: UV complete theory $\to$
EFT (preferably at NLO) + Parton Shower $\to$ data analysis $\to$ experimental fit to Wilson coefficients. Whilst in the bottom-up approach instead the chain is: SM data analysis $\to$ observable $\to$ EFT at NLO fit to Wilson coefficients.

An \ac{EFT} can be used also in those cases in which we do not integrate out explicitly the 
heavy degrees of freedom. Once we have a theory which describes quite well the physics on 
the scale which we explore, we can build up new interactions which would account for the 
small deviations. This requires the assumption that the ``real'' \ac{NP} (i.e. new resonances), lie high above the currently available experimental scale. Note that this description  is model independent - we do not make any a priori assumptions about the underlying UV 
theory. The \ac{SMEFT} approach is therefore very different to studies of specific models such as supersymmetry or compositeness. Nevertheless, the two approaches are complementary and they can each benefit from the other. For reviews we refer to Refs.~\cite{Weinberg:1980wa,Callan:1969sn,Manohar:1983md,Georgi:1994qn,
Kaplan:1995uv,Manohar:1996cq,Cohen:1997rt,Luty:1997fk,Polchinski:1992ed,
Rothstein:2003mp,Skiba:2010xn,Burgess:2007pt,Jenkins:2013fya,
Jenkins:2013sda,Buchalla:2014eca,Buchalla:2013eza,Gavela:2016bzc,Brivio:2017vri}.

To consistently build the \ac{EFT} operators, we start by taking all fields and derivatives 
which are present in the renormalisable theory and combine them in the gauge and Lorentz 
invariant way building the higher-dimension operators. 

\begin{table}[]
\begin{center}
\begin{tabular}{c|cccc}
\toprule
Field & $\phi$ & $\Psi$ & $F_{\mu \nu}$ & $\mathcal{D}_{\mu}$ \\
\midrule
$[M]$ & 1 & $\frac{3}{2}$ & 2 & 1 \\
\bottomrule
\end{tabular}
\end{center}
\caption{The mass dimensions of the fields in Lagrangian.}
\label{tab:dim}
\end{table}

\begin{table}[]
\begin{center}
\begin{tabular}{cccr}
\toprule
Field & SU(3) & SU(2) & U(1) \\
\midrule
$\varphi$ & 1 & 2 & $\frac{1}{2}$ \\ 
$L_L$ & 1 & 2 & $-\frac{1}{2}$ \\
$e_R$ & 1 & 1 & -1 \\
$Q_L$ & 3 & 2 & $\frac{1}{6}$ \\
$u_R$ & 3 & 1 & $\frac{2}{3}$ \\
$d_R$ & 3 & 1 & $\frac{1}{3}$ \\
\bottomrule
\end{tabular}
\end{center}
\caption{Gauge representations for the \ac{SM}. A singlet is denoted by $1$, a doublet by $2$ etc.}
\label{tab:gauge}
\end{table}

If we do so with the \ac{SM} fields we would find that the only operator with dimension $5$ which can be built is the neutrino Majorana mass term,
\begin{equation}
(\tilde{\varphi}^{\dagger}l_i)^T C (\tilde{\varphi}^{\dagger} l_j) \spc
\end{equation}
which was first described by Weinberg in Ref.~\cite{Weinberg:1979sa}.
Since neutrinos cannot be directly observed at the LHC, this operator is mostly explored in neutrino experiments and is generally not included in the SMEFT description.

If we combine fields further we find out that there are many more operators arising at 
$\mathrm{dim}= 6$. As a first example, there are the the four fermion interactions which we saw in
the Fermi theory, see~\cref{tab2WB}. These four fermion interactions 
are nowadays used to study deviations from \ac{SM} predictions in B meson decays, but can appear as well in many other SM processes. 

Considering the whole particle content of the \ac{SM}, there are many other operators that can be built at $\mathrm{dim}= 6$. Let us first note that the square of the Higgs doublet, $\varphi^{\dagger} \varphi$, is both Lorentz and 
gauge invariant, as can be read from~\cref{tab:gauge}. 
From~\cref{tab:dim} we easily conclude that $\varphi^{\dagger} \varphi$ has mass dimension $2$, and therefore, multiplying any of the \ac{SM} terms by $\varphi^{\dagger} \varphi$ produces terms of dimension $6$. 
This way, we can obtain the following terms (see~\cref{tab1WB}):
\begin{itemize}

\item Scalar interactions, $(\varphi^{\dagger} \varphi)^3$ (class $\varphi^6$ of~\cref{tab1WB}).

\item Gauge boson kinetic terms multiplied by $\varphi^{\dagger} \varphi$ 
      (class $X^2\varphi^2$). 

\item Yukawa couplings multiplied by $\varphi^{\dagger} \varphi$  (class $\psi^2 \varphi^3$).

\end{itemize}

Since $\varphi$ can be expanded around the constant \ac{VEV} ($v$),  
these operators generally lead to modifications of the \ac{SM} couplings. For example, 
in the \ac{SM}, the cubic Higgs coupling is fully determined by the $\lambda$ - quartic coupling 
constant and \ac{VEV}, while the $(\varphi^{\dagger} \varphi)^3$ operator changes this relation. Similarly the \ac{SM} Yukawa couplings depend only on the 
particle mass and \ac{VEV}, while using operators of the form $v^2\bar{l}\varphi e$ modifies this dependence. 

There are other classes of operators which are not simply multiplications of the \ac{SM} ones. For instance: 
\begin{itemize}

\item The cubic field strength operators (class $X^3$ in \cref{tab1WB}), which need to be multiplied by the appropriate structure constants ($SU(3)$ or $SU(2)$) in order to be gauge invariant.

\item The magnetic dipole operators, which combine the Higgs doublet, fermions and gauge fields in a way not present in the \ac{SM}. This is the class $\psi^2 X \phi$ in \cref{tab1WB}. 

\end{itemize}
Looking at~\cref{tab:dim} and~\cref{tab:gauge}, we can think of other classes of 
$\mathrm{dim} = 6$ operators which we can write down, e.g. $D^4\varphi^2$ or $XD^2\varphi^2$. 
However, these can be reduced into the other classes of operators mentioned above using the Equations of Motion (EoM) and Fierz identities. Why and how the other classes of operators can be reduced to the presented minimal basis (i.e. a complete and non-redundant set of  higher-dimensional operators) is described in Ref.~\cite{Grzadkowski:2010es}. 
\begin{table}
\begin{center}
\includegraphics[scale=0.23]{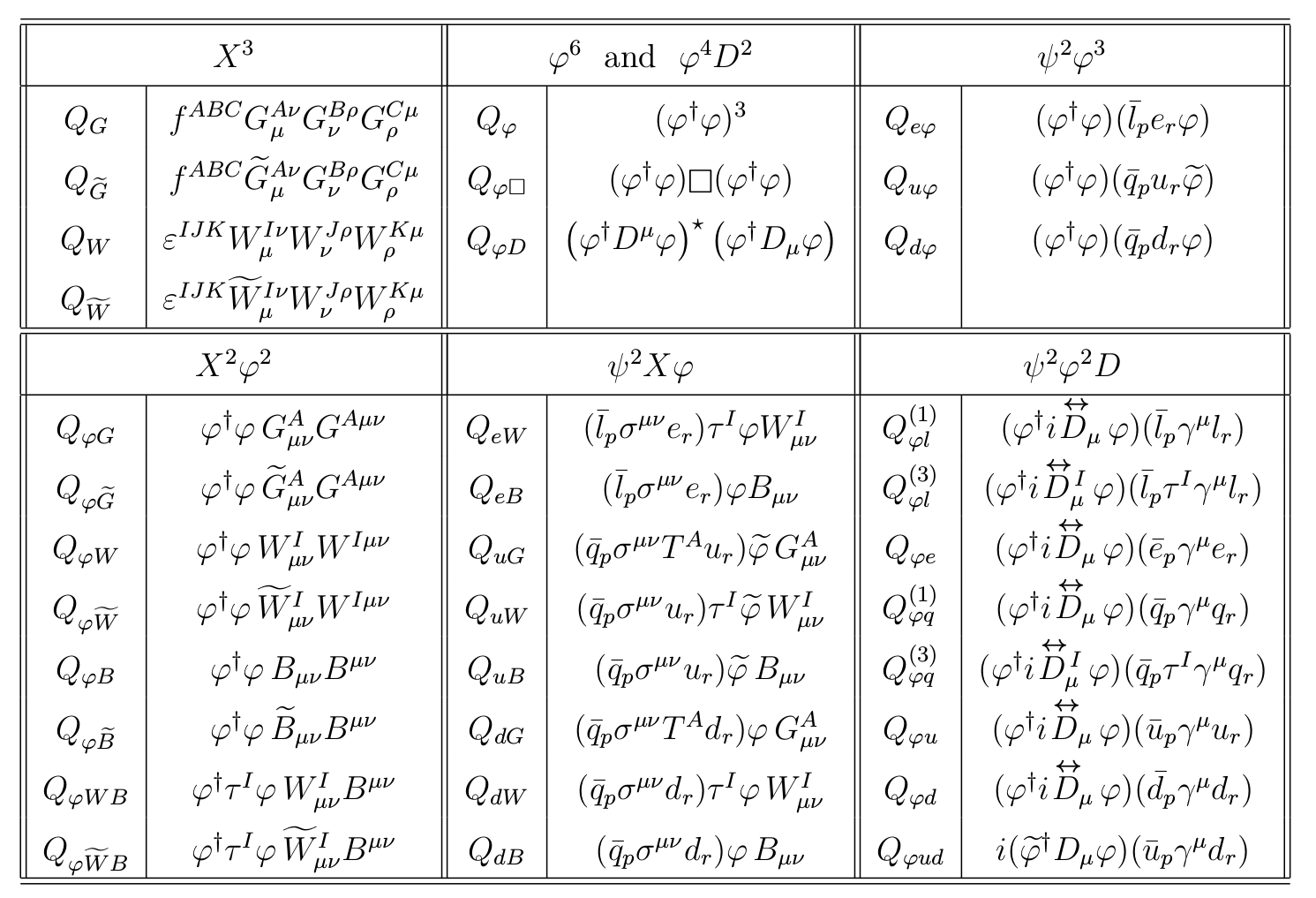}
\end{center}
\caption{Bosonic operators of the Warsaw basis. Table from Ref.~\cite{Grzadkowski:2010es}.}
\label{tab1WB}
\end{table}
\begin{table}
\begin{center}
\includegraphics[scale=0.2]{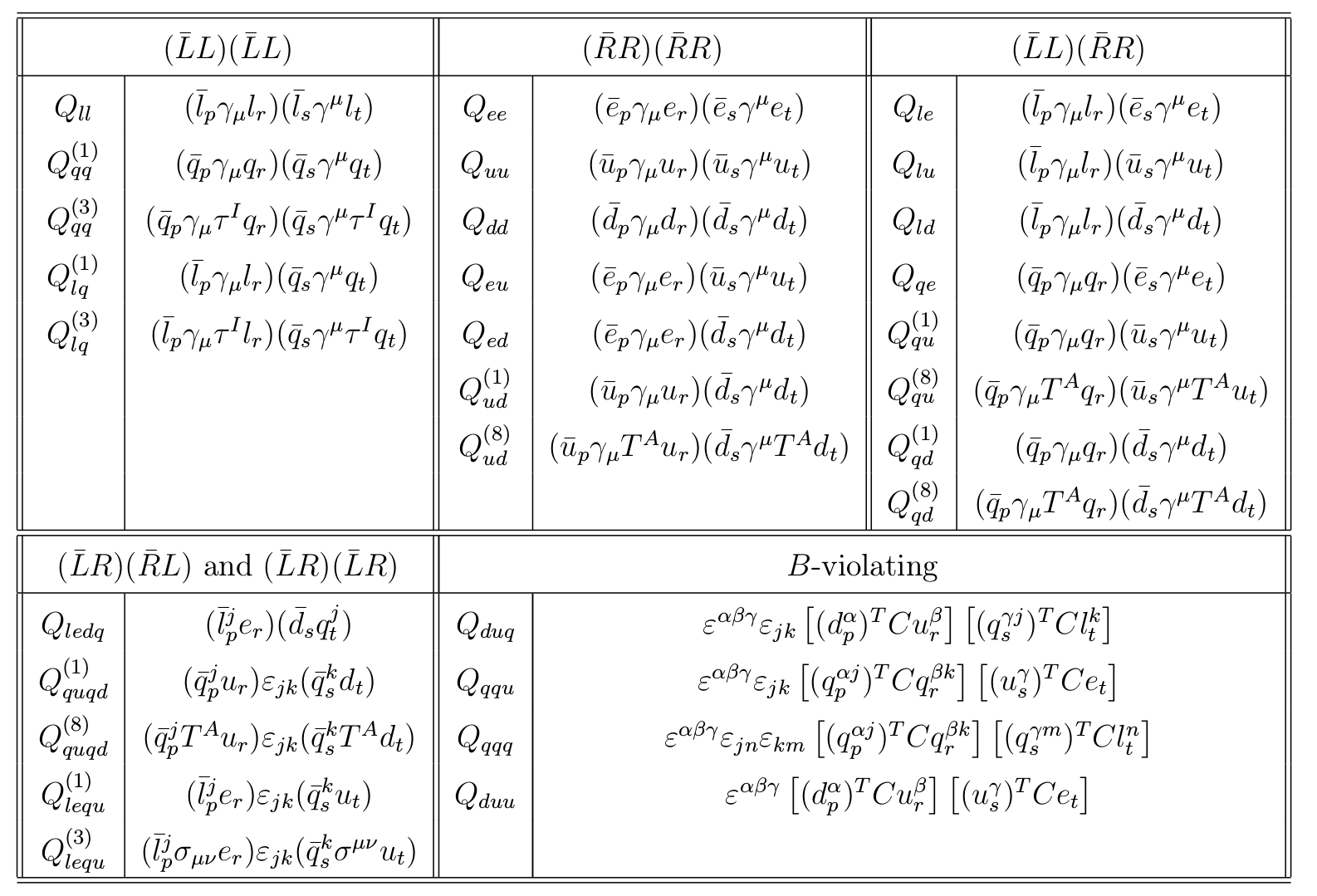}
\end{center}
\caption{Fermionic operators of the Warsaw basis. Table from Ref.~\cite{Grzadkowski:2010es}.}
\label{tab2WB}
\end{table}

The general form of the \ac{SMEFT} Lagrangian reads:
\begin{equation}\label{eq:SMEFTLag}
\mathcal{L}=\mathcal{L}_{SM} + \frac{a_5}{\Lambda} 
\mathcal{O}_5 + \Sigma_i  \frac{a_6^{(i)}}{\Lambda^2} \mathcal{O}_6^{(i)} + O(\Lambda^{-3})
\end{equation}
where the dimensionless coefficients $a_i$ are the Wilson coefficients of the higher 
dimensional operators, which are suppressed by powers of the mass scale, $\Lambda$, such that each term in the Lagrangian has a mass dimension equal to $4$. 
As we can recall from the example from the previous section, $\Lambda$ plays the analogous role 
as the $W$ boson mass in the Fermi theory. Indeed, the meaning of the $\Lambda$ is the common 
scale above which we expose the heavy degrees of freedom of the full theory 
Since we do not make any assumption about the \ac{NP} (other that it is well separable 
from \ac{SM}, and obeys Lorentz and gauge invariance), we do not automatically identify it with 
the onset of new resonances. 

It is also important to note that there is not a one-to-one correspondence between 
coefficients and measurements, since one operator can contribute to many observables and one observable may depend on more that one operator. 
\subsubsection*{Examples: SMEFT at LO}

Let us show this with the example of gluon fusion Higgs production at LHC. At \ac{LO} in 
the \ac{SM}, there is just one diagram that contributes to this process:
\[
\scalebox{1.2}{
\begin{picture}(60,50)(0,0)
\Gluon(20,5)(0,5){3}{3}
\Gluon(0,45)(20,45){3}{3}
\ArrowLine(20,5)(20,45)
\ArrowLine(20,45)(40,25)
\ArrowLine(40,25)(20,5)
\DashLine(40,25)(60,25){3}
\Vertex(20,5){1}
\Vertex(20,45){1}
\Vertex(40,25){1}
\end{picture}
}
\]
\noindent
In the \ac{SMEFT}, there are contributions from three different 
dimension $6$ operators:
\[
\scalebox{1.2}{
\begin{picture}(60,50)(0,0)
\Gluon(20,5)(0,5){3}{3}
\Gluon(0,45)(20,45){3}{3}
\ArrowLine(20,5)(20,45)
\ArrowLine(20,45)(40,25)
\ArrowLine(40,25)(20,5)
\DashLine(40,25)(60,25){3}
\Vertex(20,5){1}
\Vertex(20,45){1}
\BBox(37.5,22.5)(42.5,27.5)
\end{picture}
}
\quad
\scalebox{1.2}{
\begin{picture}(60,50)(0,0)
\Gluon(20,5)(0,5){3}{3}
\Gluon(0,45)(20,45){3}{3}
\ArrowLine(20,5)(20,45)
\ArrowLine(20,45)(40,25)
\ArrowLine(40,25)(20,5)
\DashLine(40,25)(60,25){3}
\Vertex(20,5){1}
\BBox(17.5,42.5)(22.5,47.5)
\Vertex(40,25){1}
\end{picture}
}
\quad
\scalebox{1.2}{
\begin{picture}(60,50)(0,0)
\Gluon(30,5)(0,5){3}{4}
\Gluon(0,45)(30,25){3}{5}
\ArrowArcn(30,15)(10,270,90)
\ArrowArcn(30,15)(10,90,270)
\DashLine(30,25)(60,25){3}
\BBox(27.5,22.5)(32.5,27.5)
\Vertex(30,5){1}
\end{picture}
}
\quad
\scalebox{1.2}{
\begin{picture}(60,50)(0,0)
\Gluon(30,25)(0,5){3}{5}
\Gluon(0,45)(30,25){3}{4}
\DashLine(30,25)(60,25){3}
\BBox(27.5,22.5)(32.5,27.5)
\end{picture}
}
\]
\begin{itemize}

\item $\frac{c_1}{\Lambda^2} \mathcal{O}_1 \sim  
      \frac{\alpha_s}{ \pi v } a_g \h G_{\mu \nu}^a G^{\mu \nu a}
  \quad \rightarrow  $  Higgs-gluon-gluon coupling

\item $\frac{c_2}{\Lambda^2} \mathcal{O}_2 \sim  \frac{m_t}{ v } a_t t \bar{t} \h  
\quad \rightarrow  $  Higgs-$\,t \bar{t}$  

\item $ \frac{c_4}{\Lambda^2} \mathcal{O}_4 \sim  
   \frac{g_s m_t}{ 2 v^3 } a_{tg} (v + \h) G_{\mu \nu}^a (\bar{t}_L 
    \sigma^{\mu \nu} \rm{T}^a t_R + h.c. )  \quad \rightarrow  $  chromomagnetic operator

\end{itemize}       
Another example is top pair production, where there is a contribution from 
the chromomagnetic operator $\bar{Q}_L \h \sigma^{\mu\nu}T^a u_R G_{\mu\nu}^a + h.c.$:
\[
\begin{picture}(60,50)(0,0)
\Gluon(0,50)(30,25){3}{5}
\Gluon(0,0)(30,25){3}{5}
\ArrowLine(60,0)(30,25)
\ArrowLine(30,25)(60,50)
\BBox(27.5,22.5)(32.5,27.5)
\end{picture}
\qquad
\begin{picture}(60,50)(0,0)
\Gluon(0,0)(30,0){3}{5}
\Gluon(0,50)(30,50){3}{5}
\ArrowLine(60,0)(30,0)
\ArrowLine(30,0)(30,50)
\ArrowLine(30,50)(60,50)
\Vertex(30,50){1}
\BBox(27.5,-2.5)(32.5,2.5)
\end{picture}
\qquad
\begin{picture}(60,50)(0,0)
\ArrowLine(0,0)(30,25)
\ArrowLine(30,25)(0,50)
\ArrowLine(60,0)(30,25)
\ArrowLine(30,25)(60,50)
\BBox(27.5,22.5)(32.5,27.5)
\end{picture}
\]

\noindent
or $t\bar{t}\h$ associated production:
\[
\begin{picture}(60,50)(0,0)
\Gluon(0,50)(30,25){3}{5}
\Gluon(0,0)(30,25){3}{5}
\ArrowLine(60,0)(30,25)
\ArrowLine(30,25)(60,50)
\DashLine(30,25)(60,25){3}
\BBox(27.5,22.5)(32.5,27.5)
\end{picture}
\qquad
\begin{picture}(60,50)(0,0)
\Gluon(0,0)(30,0){3}{5}
\Gluon(0,50)(30,50){3}{5}
\ArrowLine(60,0)(30,0)
\ArrowLine(30,0)(30,50)
\ArrowLine(30,50)(60,50)
\DashLine(30,0)(60,25){3}
\Vertex(30,50){1}
\BBox(27.5,-2.5)(32.5,2.5)
\end{picture}
\qquad
\begin{picture}(60,50)(0,0)
\Gluon(0,0)(30,0){3}{5}
\Gluon(0,50)(30,50){3}{5}
\ArrowLine(60,0)(30,0)
\ArrowLine(30,0)(30,25)
\ArrowLine(30,25)(30,50)
\ArrowLine(30,50)(60,50)
\DashLine(30,25)(60,25){3}
\Vertex(30,50){1}
\Vertex(30,25){1}
\BBox(27.5,-2.5)(32.5,2.5)
\end{picture}
\qquad
\begin{picture}(60,50)(0,0)
\Gluon(0,0)(30,0){3}{5}
\Gluon(0,50)(30,50){3}{5}
\ArrowLine(60,0)(30,0)
\ArrowLine(30,0)(30,25)
\ArrowLine(30,25)(30,50)
\ArrowLine(30,50)(60,50)
\DashLine(30,25)(60,25){3}
\Vertex(30,50){1}
\Vertex(30,0){1}
\BBox(27.5,22.5)(32.5,27.5)
\end{picture}
\qquad
\begin{picture}(60,50)(0,0)
\ArrowLine(0,0)(30,25)
\ArrowLine(30,25)(0,50)
\ArrowLine(60,0)(30,25)
\ArrowLine(30,25)(45,37.5)
\ArrowLine(45,37.5)(60,50)
\DashLine(45,37.5)(60,25){3}
\BBox(27.5,22.5)(32.5,27.5)
\Vertex(45,37.5){1}
\end{picture}
\]

Observe that we have always inserted precisely one \ac{EFT} operator per diagram. The motivation for this 
will be discussed later.

\subsection{SMEFT at NLO} \label{subsec:ren}

SMEFT is an active area of research that is moving towards NLO predictions. 
NLO in the SMEFT is important for two reasons: from a phenomenological point of view  NLO is important for capturing potentially 
large QCD K-factors in total rates, gaining greater sensitivity. In this way we can verify stability of differential information  
beyond leading order with consistent estimates of the scale uncertainty. 
This programme has already been started with Refs.~\cite{Maltoni:2016yxb,Bylund:2016phk}.

Thus, interpreting the data using theoretical results developed beyond LO can often be crucial 
for the SMEFT. NLO calculations (not only NLO QCD) should be used if they are 
available, as argued in Ref.~\cite{Passarino:2016pzb}.
NLO calculations help characterize (and reduce) theoretical uncertainties
and allow the consistent incorporation of precise measurements in the SMEFT. 
NLO interpretations of the data will be critical in the event that deviations 
from the SM emerge over the course of LHC operations.  

From a more formal point of view we can say that, although different in the UV, the SM and SMEFT are both examples of a QFT; therefore, NLO SMEFT
requires renormalisation. Details of renormalisation in the SMEFT can be found in Refs.~\cite{Ghezzi:2015vva,Passarino:2016pzb}.
However, we stress that consistent removal of UV poles in SMEFT, i.e.
closure of SMEFT under renormalisation, requires starting from a basis of higher dimensional operators, that respects the full gauge symmetries. Inevitably, the renormalisation scheme will be a 
mixed one, on-shell for the SM parameters and ${\overline{\mathrm{MS}}}$ for the Wilson
coefficients. 

One question often raised concerns the ``optimal'' parametrization of
the $\mathrm{dim} > 4$ basis; once again, all sets of gauge invariant, dimension
$d > 4$ operators, none of which is redundant, form a basis and all bases are
equivalent. For a formal definition of redundancy see Sect. 3 of Ref.~\cite{Einhorn:2013kja}.
\subsubsection{Renormalisation abridged}

It is worth discussing how the problems described in \cref{par:kframeNLO}. in particular in \cref{eq:hbbproblem},
are solved in the SMEFT. Consider again the process $\h \to \bar{b} b$ at NLO in SMEFT.
Ultraviolet poles in self-energies and transitions, e.g. the $\h$ self-energy etc, are
eliminated by extending the usual definition of counterterms for fields and SM parameters,
\begin{equation}
\delta Z_i = \frac{1}{{\bar{\epsilon}}}\,\frac{g^2}{16\,\pi^2}\,
\Bigl[ \delta Z^{(4)}_i + g_{_6}\,\delta Z^{(6)}_i \Bigr] \spc
\end{equation}
more details can be found in Ref.~\cite{Ghezzi:2015vva}. Here we use the ${\overline{\mathrm{MS}}}$ prescription
where ${\bar{\epsilon}}^{-1} = 2/(4 - D) - \gamma - \ln\pi - \ln\mu^2_R$, $\gamma$ is the
Euler-Mascheroni constant, $g$ is the $SU(2)$ coupling constant and $\mu_R$ is the renormalisation scale. 
Furthermore, $g_{4+2k} = 1/(\sqrt{2} G_F \Lambda^{2})^k$, where $G_F$ is the Fermi coupling constant.
Note that we have made no attempt to go beyond $\mathrm{dim} = 6$. Extension of the basis for $\mathrm{dim} > 6$
has been considered in Refs.~\cite{Lehman:2014jma,Kobach:2016ami,Lehman:2015via,Liao:2016qyd}.

\noindent
First we define combinations of Wilson coefficients as follows:
\begin{equation}
\begin{split}
& {\rm a_{Z\!Z}}= s^2_W {\rm a}_{{\rm \phi \!B}} + c^2_W {\rm a}_{{\rm \phi \!W}} - s_Wc_W {\rm a}_{\rm W\!B}\spc \\
& {\rm a_{A\!A}}= c^2_W {\rm a}_{{\rm \phi \!B}} + s^2_W {\rm a}_{{\rm \phi \!W}} + s_W c_W {\rm a}_{{\rm \phi \!B}} \spc\\
& {\rm a_{A\!Z}}= 2c_Ws_W \left( {\rm a}_{{\rm \phi \!W}} - {\rm a}_{{\rm \phi \!B}}\right) + 
  \left( 2c^2_W - 1 \right) {\rm a}_{{\rm \phi W\!B}} \spp \\ 
\end{split}
\label{rela}
\end{equation}
We present the result for the renormalisation of the Higgs 
mass, in the limit where only $m_t$ is kept finite (5-flavour scheme). The SM counterterm is simple,
\begin{equation}
\delta Z^{(4)}_{\mh} =
           \frac{3}{2}\,\frac{\mw^2}{\mh^2}\,\frac{1 + 2\,c^4_{\theta}}{c^4_{\theta}}
          - \frac{3}{2}\,\frac{4\,m^4_t - \mh^2\,m^2_t - \mh^4}{\mw^2\,\mh^2}
          - \frac{1}{2}\,\frac{1 + 2\,c^2_{\theta}}{c^2_{\theta}} \spc
\end{equation}
where $c_{\theta} = \mw/M_Z$. For $\mathrm{dim} = 6$ there are more terms and we write
\begin{equation}
\delta Z^{(6)}_{\mh} = \sum_{i\,\in\,A}\,C_i\,a_i \spc
\end{equation}
where $\{A\} = \{\,AA\,,\,ZZ\,,\,AZ\,,\,\phi D\,,\,\phi \Box\,,\,\phi\,,\,t \phi\,\}$.
The result is as follows:
\bqa
C_{AA} &=& - \frac{9 - 8\,s^2_{\theta}}{c^2_{\theta}}\,s^2_{\theta}
\nl
{}&-& 3\,\frac{4\,m^4_t - \mh^2\,m^2_t - \mh^4}{\mw^2\,\mh^2}\,s^2_{\theta}
          + 3\,\frac{\mw^2}{\mh^2}\,\frac{1 + 10\,c^4_{\theta}}{c^4_{\theta}}\,s^2_{\theta}
\spc \nl\nl
C_{ZZ} &=& - \frac{3 + c^2_{\theta} + 8\,c^4_{\theta}}{c^2_{\theta}}
\nl
{}&-& 3\,\frac{4\,m^4_t - \mh^2\,m^2_t - \mh^4}{\mw^2\,\mh^2}\,c^2_{\theta}
          + 3\,\frac{\mw^2}{\mh^2}\,\frac{4 + c^2_{\theta} + 10\,c^6_{\theta}}{c^4_{\theta}}
\spc \nl\nl
C_{AZ} &=& - 3\,\frac{4\,m^4_t - \mh^2\,m^2_t - \mh^4}{\mw^2\,\mh^2}\,c_{\theta}\,s_{\theta}
\nl
{}&+& 3\,\frac{s_{\theta}}{c_{\theta}}\,\frac{\mw^2}{\mh^2}\,
      \frac{1 + 10\,c^4_{\theta}}{c^2_{\theta}}
      - \frac{s_{\theta}}{c_{\theta}}\,(1 + 8\,c^2_{\theta})
\spc \nl\nl
C_{\phi\,D} &=& - \frac{1}{8}\,\frac{5 - 4\,c^2_{\theta}}{c^2_{\theta}}
\nl
{}&+& \frac{3}{8}\,\frac{8\,m^4_t - 2\,\mh^2\,m^2_t - 7\,\mh^4}{\mw^2\,\mh^2}
      + \frac{3}{4}\,\frac{\mw^2}{\mh^2}\,\frac{3 - 2\,c^4_{\theta}}{c^4_{\theta}}
\spc \nl\nl
C_{\phi\,\Box} &=& - \frac{1 + 2\,c^2_{\theta}}{c^2_{\theta}}
\nl
{}&-& \frac{12\,m^4_t - 3\,\mh^2\,m^2_t - 11\,\mh^4}{\mw^2\,\mh^2}
          + 3\,\frac{\mw^2}{\mh^2}\,\frac{1 + 2\,c^4_{\theta}}{c^4_{\theta}}
\spc \nl\nl
C_{\phi} &=& - 33
          - 3\,\frac{\mw^2}{\mh^2}\,\frac{1 + 2\,c^2_{\theta}}{c^2_{\theta}}
\spc \nl\nl
C_{t\,\phi} &=& - 3\,\frac{8\,m^2_t - \mh^2}{\mw^2\,\mh^2}\,m^2_t \spp
\eqa
This example is enough to show the complexity of renormalisation in the SMEFT.
In fact, this is only the first step of the renormalisation procedure since $\mw, \mh$ and $m_t$ are the renormalized masses but we still need a new set of lengthy equations (notreported here) of the form $\mh = \mh(\hbox{IPS})$ etc, where IPS stands for the input parameter set, in order to connect these quantities with experimentally measured ones and so, complete the finite renormalisation procedure.

Further, UV poles in Green's functions with three or more 
legs contain residual $\mathrm{dim}=6$ divergences that can only be removed by introducing a mixing of Wilson coefficients,
\begin{equation}
a_i = \sum_{j}\,Z_{ij}\,a^{\mathrm{ren}}_j \spc
\quad
Z_{ij} = \delta_{ij} + \frac{1}{{\bar{\epsilon}}}\,\frac{g^2}{16\,\pi^2}\,\delta Z_{ij} \spp
\end{equation}
For $\h \to \bar{b} b$ we will have non-zero entries when $a_{b\,\varphi}$ mixes with the
other $25$ Wilson coefficients (without neglecting light quark masses) introduced in \cref{tab1WB} and in \cref{tab2WB}.
For instance, the mixing of $a_{b\,\varphi}$ with $a_{t\,\varphi}$ is
given by
\begin{equation}
 \frac{3}{4}\,\frac{m^2_t}{\mw^2} - \frac{3}{2}\,\sum_{i=u,c}\,\frac{m^2_i}{\mw^2} \spp
\end{equation}
It is worth noting again that the renormalisation of Wilson coefficients is performed in
the ${\overline{\mathrm{MS}}}\,$-scheme, i.e. giving a residual dependence on the
renormalisation scale. 

Operator mixings are important: mixing means that in general the Wilson coefficients at low scale 
are related. One immediate consequence is that 
assumptions about some coefficients being zero at low scales are in general 
not valid. A global point of view is required and contributions from individual couplings 
may not make sense and only their sum is meaningful.
Note also that operator mixing is not symmetric.
A simplified example is as follows: consider the sub-set $\{\,a_{_t\,\phi}\,,\,a_{t\,g}\,\}$.
Starting with $a_{t\,g} = 1$ and $a_{t\,\phi} = 0$ at $1\,\TeV$ one obtains $a_{t\,g} = 0.98$
and $a_{t\,\phi} = 0.45$ at $m_t$, see Ref.~\cite{Maltoni:2016yxb}.

The cancellation of UV divergences from all the $\mathrm{dim} = 6$ operators in
the Warsaw basis gives a highly non-trivial check on the calculation. 
The logarithmic corrections could have been 
deduced from the RG analysis. However, the calculation of the full NLO calculation identifies
non-logarithmic terms which would be otherwise missed and which are not always negligible.

We can summarise the situation as follows: NLO is the first order where a non-trivial 
SMEFT structure becomes manifest. A nice example of SMEFT scale dependence can be found in
Ref.~\cite{Maltoni:2016yxb}.
Another example can be found in Ref.~\cite{Hartmann:2016pil} where it is shown that
SMEFT parameters contributing to LEP data are formally unbounded when the accuracy of loop corrections is reached. In other words, 
the number of SMEFT parameters contributing at one-loop is larger than the number of available measurements.

The size of these loop effects is generically 
a correction of the order of a few percent compared to leading effects in the SMEFT, but even so, multiple large numerical coefficients have been found at this order. 
Furthermore, in Ref.~\cite{Gauld:2016kuu} it has been shown that there are 
contributions from $\mathrm{dim} = 6$ operators, which alter the $gbb$ vertex and introduce sizeable corrections to the $hbb$ vertex which are unrelated to the SM corrections and cannot be anticipated through a renormalisation-group analysis. 

We will skip here all technical details related to the
treatment of IR/collinear divergences necessary for computing $\h \to \bar{b}\,b\,\gamma(g)$ in the SMEFT. Once a finite $S\,$-matrix has been obtained we can start making approximations,
LO SMEFT, NLO SMEFT in the PTG scenario~\footnote{For a definition of
potentially-tree-generated (PTG) operators see Ref.~\cite{Einhorn:2013kja}} and  the full NLO SMEFT.
Note that the LO SMEFT involves the following Wilson coefficients:
\begin{equation}
{a_{\varphi\,\scriptscriptstyle{W}}} \spc
\quad
{a_{\varphi\,\scriptscriptstyle{D}}} \spc
\quad
{a_{\varphi\,\scriptscriptstyle{\Box}}} \spc
\quad
{a_{b\,\varphi}} \spc
\end{equation}
but only the combination
\begin{equation}
{a_{\varphi\,\scriptscriptstyle{W}}} -
\frac{1}{4}\,{a_{\varphi\,\scriptscriptstyle{D}}} +
{a_{\varphi\,\scriptscriptstyle{\Box}}} -
{a_{b\,\varphi} }
\end{equation}
appears when computing observables. In the NLO SMEFT additional Wilson coefficients will be present, i.e. we will have more ``generalized'' kappas and more sub-amplitudes, and some of them will not factorize onto the SM LO/NLO amplitudes.

The renormalisation procedure continues until all UV poles have been eliminated. Therefore, we will
have the following scheme:
\begin{itemize}

\item Compute the (on-shell) decay $h(P) \to A_{\mu}(p_1)\,A_{\nu}(p_2)$; 
the amplitude, containing only one Lorentz structure, is made UV finite by mixing 
$a_{AA}$ with $a_{AA}\,,\,a_{ZZ}\,,\,a_{AZ}$ and $a_{QW}$.

\item Compute the (on-shell) decay $h(P) \to A_{\mu}(p_1)\,Z_{\nu}(p_2)$; 
the amplitude, containing only one Lorentz structure, is made UV finite by mixing 
$a_{AZ}$ with $a_{AA}\,,\,a_{ZZ}\,,\,a_{AZ}$ and $a_{QW}$.

\item Compute the (on-shell) decay $h(P) \to Z_{\mu}(p_1)\,Z_{\nu}(p_2)$.
The amplitude contains a part proportional to $g^{\mu\nu}$ ($\mathcal{D}$) and 
a part proportional to $p^{\mu}_2\,p^{\nu}_1$ ($\mathcal{P}$). 
Mixing of $a_{ZZ}$ with other Wilson coefficients makes $\mathcal{P}$ UV finite, while the 
mixing of $a_{\phi\,\Box}$ makes $\mathcal{D}$ UV finite.

\item Continue until there are no Wilson coefficients left free, so that UV finiteness follows 
from gauge cancellations, which follow from having selected a ``basis'. 

\end{itemize}
\subsection{Predictions using the Dimension 6 Lagrangian}
\subsubsection{Amplitudes in SMEFT}
Any amplitude for $1 \to 2$ processes, in the \ac{SMEFT}, can be written in the 
following way, as proposed in Ref.\cite{Ghezzi:2015vva}:
\begin{equation}\label{eq:AmpEFT}
\mathcal{A} = \sum_{n=N}^{\infty} g^n \mathcal{A}_n^{(4)} + 
\sum_{n=N_6}^{\infty} \sum_{\ell=0}^{n} 
\sum_{k = \ell}^{\infty} g^n g_{4+ 2k}^{\ell} \mathcal{A}_{n \ell k}^{(4+2k)} \spc 
\qquad  g_{4+2k} = \frac{1}{(\sqrt{2} G_F \Lambda^{2})^k} \spc
\end{equation}
where $g$ is the $SU(2)$ coupling constant, and $g_{4+2k} = g_{_6}^k $ is the new coupling constant. 
$\Lambda$ is the scale of new physics. $N$ is the number of vertices at leading order (i.e. $1$ 
for $\h\rightarrow VV$, $3$ for $\h\rightarrow \gamma \gamma$ etc.), and $N_6$ is $1$ for 
tree initiated processes and $N-2$ for loop initiated ones.

\begin{figure}
\begin{centering}  
 \includegraphics[scale=0.6]{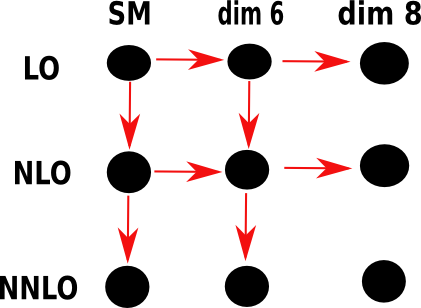}
 \caption{The power counting in perturbative \ac{SMEFT} grows in two directions: higher loops, 
          and higher dimensional operators.} \label{fig:powerCounting} 
\end{centering}
\end{figure}

As we can see from~\cref{eq:AmpEFT}, there is more than one expansion parameter 
for the amplitude, and this is a key fact to consider when doing \ac{EFT} predictions. 
Namely, the power counting for the perturbative expansion grows simultaneously in two 
directions: we can go to higher loops, like in the \ac{SM} perturbation theory, or we can 
go to higher orders in the $1/\Lambda$ expansion. Because both the former and the latter corrections 
depend on the size of $\Lambda$ it is a priori not possible to determine which are more 
relevant, as sketched in ~\cref{fig:powerCounting}. 

Even when we consider going to higher orders 
in perturbation theory, but restricting ourselves to dimension $6$, there is the question 
of how many operator insertions should we include at each level. For example: if a tree level diagram with one $\mathrm{dim} = 6$ operator is ``\ac{LO} \ac{EFT}'' and a one-loop  diagram with one $\mathrm{dim}= 6$ operator and one \ac{SM} operator is ``\ac{NLO} \ac{EFT}'', 
where do we put a diagram with tree topology but two $\mathrm{dim} = 6$ operators?. 

Traditionally, the community works with only one $\mathrm{dim} = 6$ operator per diagram, 
both at tree and loop level, since each of these vertices is suppressed by two powers of 
the cut-off scale ($1/\Lambda^2$), which we consider to be \emph{big enough}. However this 
raises the question of which terms shall we keep or discard when squaring the amplitude 
(terms with only one operator insertion squared will become sizeable to the interference 
term of those that we just neglected with the \ac{SM}, i.e. of order $1/\Lambda^4$. 
This last fact will be discussed thoroughly in the next section.

Moreover, in the \ac{SM}, when a particular process is calculated, a common practice is that 
a theoretical error is assigned. It can be subtle to assign such an error if we do not have an estimate for the missing higher order perturbative terms in the \ac{SM}, and this part of the 
calculation is often overlooked, since there are allegedly enough hints perturbative expansions are legit. However, the need to include theoretical errors when perturbatively expanding  the \ac{SMEFT} is tied to the fact that different truncations of such expansions can be 
constructed, and to the fact that no assumption is done on the size of the Wilson coefficients 
of the UV theory nor the scale $\Lambda$, and it becomes even more necessary than in 
the \ac{SM} case.  

In particular, suppose that at some point in the future, we observe some deviation from the \ac{SM}. Then, 
if the experimental precision allows, we could be able to test one-loop corrections and/or 
$\mathrm{dim} = 8$ effects. If, on the other hand, no deviation from the \ac{SM} is
observed and limits on the $\mathrm{dim} = 6$ coefficients are set through a \ac{LO}
procedure, we think that one-loop effects can be used to express a rough estimate of the
corresponding MHOU (\ac{SMEFT} truncation).

To be more precise, the higher order corrections in \ac{SMEFT} are normally classified 
as follows: 
\begin{itemize}

\item $g g_{_6} \mathcal{A}_{111}^{(6)}$ defines \ac{LO} \ac{SMEFT} 

\item $g^3 g_{_6} \mathcal{A}_{311}^{(6)}$ defines \ac{NLO} \ac{SMEFT}, 
or the Missing Higher Orders Uncertainty (MHOU) for \ac{LO} \ac{EFT}. 
$g_{_6}$ stands for a single $\mathrm{dim} = 6$ operator insertion, and it is therefore 
known as \emph{linear \ac{EFT} term.}

\item $g g_8 \mathcal{A}_{112}^{(8)}$, $g^3 g_{_6}^2 \mathcal{A}_{321}^{(6)} $ defines the 
MHOU for \ac{NLO} \ac{SMEFT}. $g_{_6}^2$ stands for 2 $\mathrm{dim} =6$ operator insertions, and 
it is called \emph{quadratic \ac{EFT} term}.

\end{itemize}
Written in a more schematic way we have the following situation:
\begin{equation}  \label{eq:eftAmp}
\begin{split}
\vert \mathcal{A} \vert^2= &  \vert 
\mathcal{A}_{\SMm} + \mathcal{A}_{\mrm{dim 6}} + \mathcal{A}_{\mrm{dim 8}} + ... \vert^2 =  \\
 = & \vert \mathcal{A}_{SM} \vert^2 + \underbrace{ \vert \mathcal{A}_{SM} \times  
\mathcal{A}^{(6)}\vert}_{\color{red} \text{``linear \ac{EFT}''}} 
+ \underbrace{ \vert \mathcal{A}^{(6)} \vert^2}_{\color{red} 
\text{``quadratic \ac{EFT}''}} 
+ \underbrace{ \vert \mathcal{A}_{SM} \times  
\mathcal{A}^{(8)}\vert}_{\color{red}\text{not available (th.uncertainty)}} + \dots
\end{split} 
\end{equation}
\subsubsection{To square or not to square?}
\begin{figure}
\begin{subfigure}{.5\textwidth}
  \centering
  \includegraphics[width=.9\linewidth]{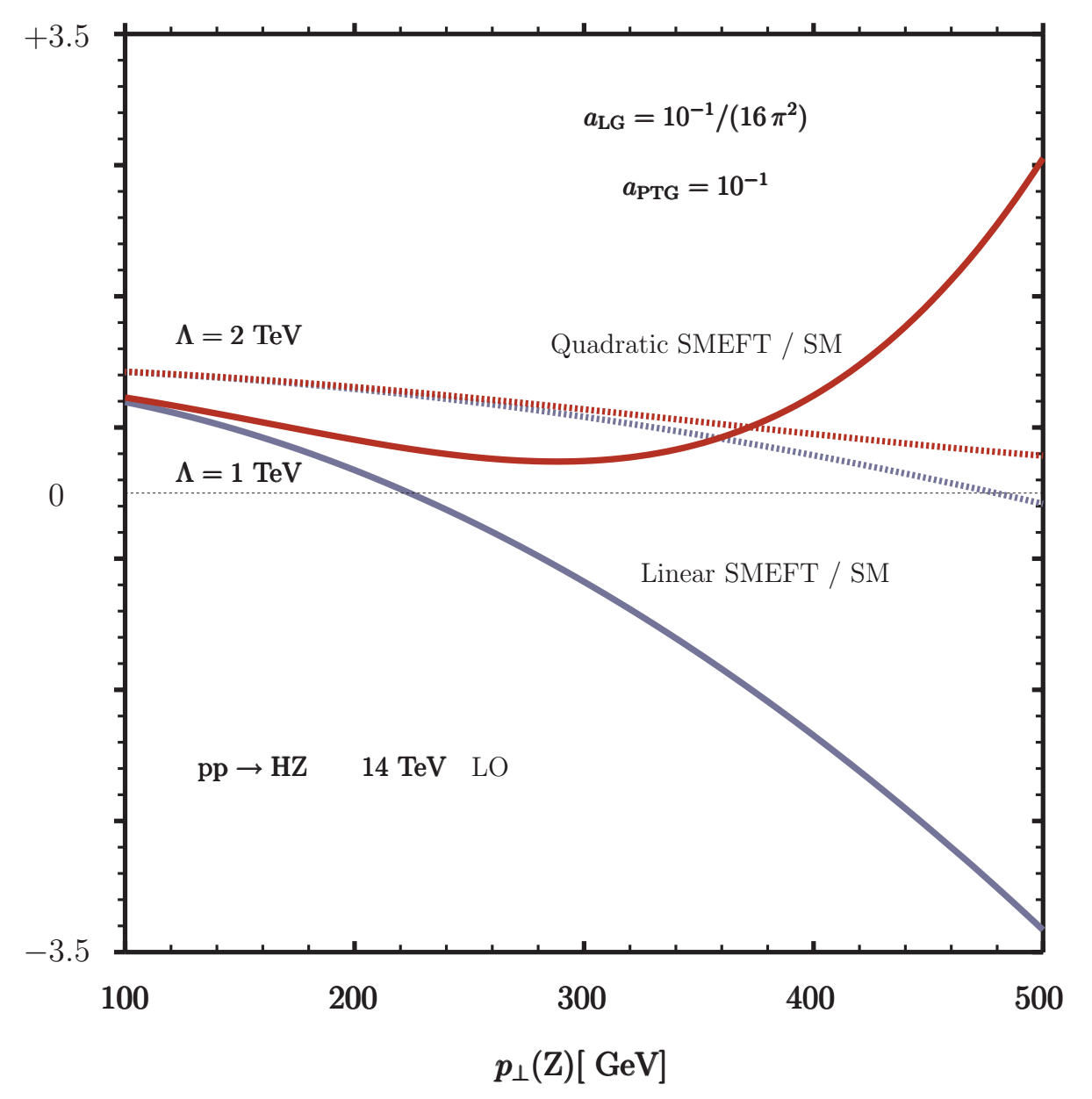}
  \caption{}
    \label{fig:linsq}
\end{subfigure}
\begin{subfigure}{.5\textwidth}
  \centering
  \includegraphics[width=.98\linewidth]{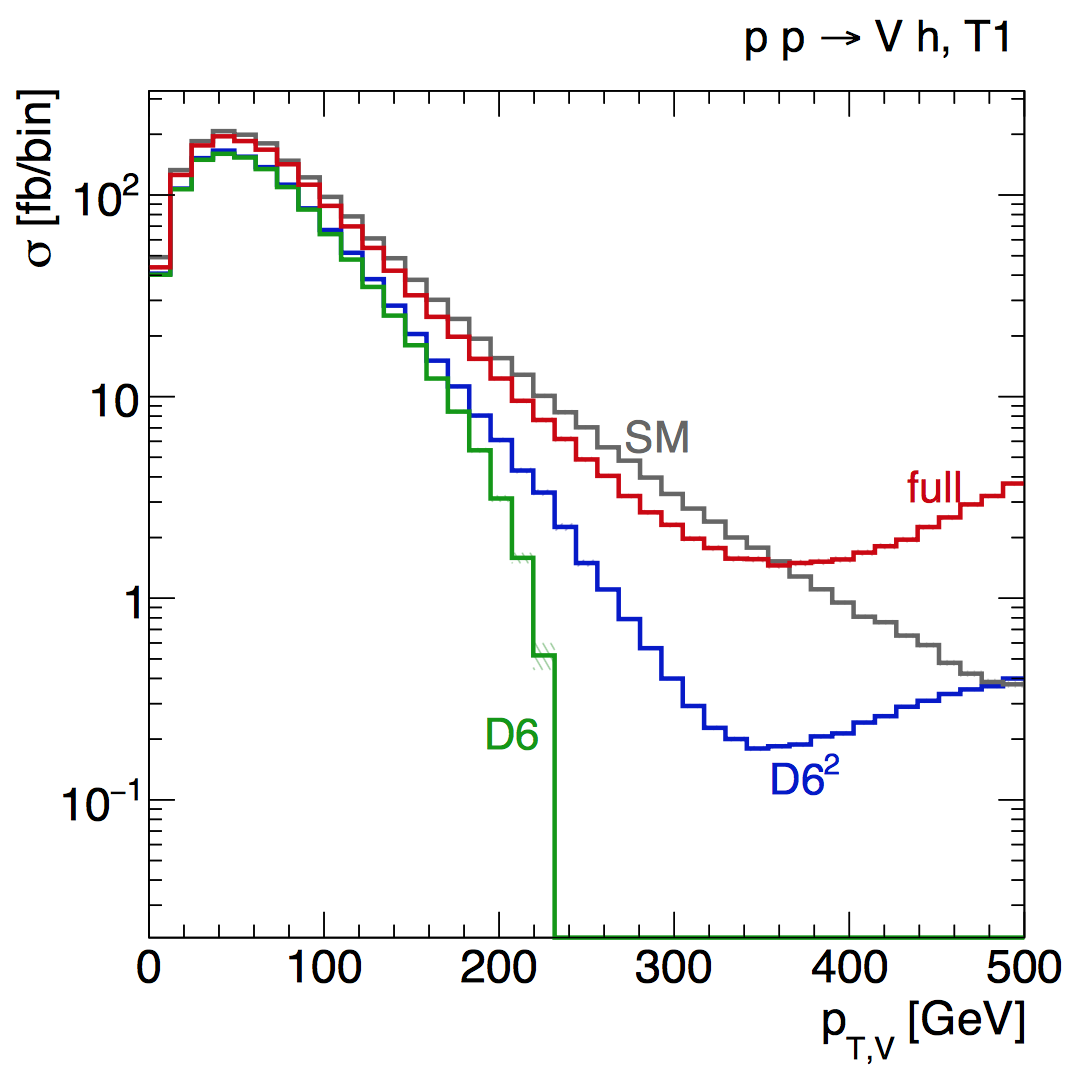}
  \caption{}
  \label{fig:Plehn}
\end{subfigure}
\caption{The transverse momentum distribution of the vector boson in $V h$ production. In \cref{fig:linsq}, different ratios of SMEFT and SM cross sections are depicted, the blue lines represent cross sections where only the linear EFT terms are considered, while the red lines include also quadratic contributions. Solid and dashed functions represent two different choices for the $\Lambda$ scale. In 
\cref{fig:Plehn}, we see a similar prediction, obtained by means of the top-down approach, where the effective Lagrangian for a UV vector triplet model is studied. This figure is taken from Ref.~\cite{Brehmer:2015rna} with permission from the authors.  
}
\label{fig:EFTsq}
\end{figure}

In order to make predictions with the effective Lagrangian, it is necessary to calculate amplitudes and cross sections. Amplitudes in SMEFT can generically be expressed as in Eq. \eqref{eq:AmpEFT}.
As we discussed previously, the square term $| \mathcal{A}_{\mrm{dim 6}} |^2$ is of the order of $1/\Lambda^4$, just as  the interference term of $\mathrm{dim} = 8$, which is usually unavailable. For this reason, 
it can be argued that it is inconsistent to include just part of the $1/\Lambda^4$ terms. 
However, it is interesting to analyse the impact of the inclusion of the squared term in the differential 
observables in~\cref{fig:EFTsq}.

In the following, we will refer as ``linear \ac{SMEFT}'' to the case where 
only the interference term is considered, and we will refer to ``quadratic \ac{SMEFT}'' in the case where $| \mathcal{A}_{\mrm{dim 6}} |^2$ is also included.
In~\cref{fig:linsq}, the transverse momentum spectra of the $Z$ boson from $Z\h$ production 
is showed, in scenarios with two scales $\Lambda = 1, 2 \, \GeV$ (solid and dashed lines 
respectively) and in linear and quadratic \ac{SMEFT} (blue and red lines respectively). 
The linear \ac{SMEFT} for $\Lambda = 1 \, \GeV$ starts to 
have negative cross sections above $\pt \approx 200 \, \GeV$, which is cured by the inclusion of squared terms, which make the distribution positive by definition. It is also clearly visible that both approaches 
deviate significantly at higher $\pt$. This behaviour signals the breakdown of the \ac{EFT} 
approximation. 

In~\cref{fig:Plehn}, where again the $\pt$ spectra of the associated vector boson 
are presented, four cases were plotted: \ac{SM}, full theory of a UV triplet vector boson of mass $591 \, \GeV$ and the low-energy EFT for the previous theory, comparing both linear and linear-plus-quadratic contributions. The linear and quadratic realisations have very similar behaviour at low $\pt$, but at higher $\pt$ the linear approximation 
breaks down. The quadratic EFT, although being far from the full theory, 
reproduces the behaviour better.

Since the effective operators should be used to describe small  deviations from the \ac{SM}, the situation in which the inclusion of 
$| \mathcal{A}_{\mrm{dim 6}} |^2$ which should be actually suppressed, becomes necessary, shows the end of the validity regime of the EFT. Once the \ac{SMEFT} amplitudes are known, 
both linear and quadratic \ac{SMEFT} observables can be generated, and in that case, the comparison of their 
behaviour is a good check for the validity of \ac{EFT} approximation and the difference of 
them should definitely be included into the theoretical uncertainty. Additionally there are other contributions of order $1/\Lambda^4$ that should also be studied and included in the theoretical uncertainty.   

The example of how the inclusion of the $| \mathcal{A}_{\mrm{dim 6}} |^2$ term can change 
the behaviour of \ac{SMEFT} prediction, calls for a more detailed look on the  validity of the approach. Although we roughly describe $\Lambda$ as the ``scale of new physics'', 
it is evident from the previous example, that it does not correspond to the breakdown 
of the approach, since in~\cref{fig:linsq} problems started at $p_t \approx 200 \, \GeV$ , 
well below $1 \, \TeV$. One of the usual checks of the validity of a perturbative expansion is a study of
perturbative unitarity. Perturbative unitarity is in general violated in \ac{EFT}, 
however it can be used to check energies lower than $\Lambda$, since above \ac{EFT} should be definitely replaced by the full theory.  

To summarise: experiments occur at finite energy and ``measure'' an effective action 
$S^{\mathrm{eff}}(\Lambda)$; whatever QFT should give low energy 
$S^{\mathrm{eff}}(\Lambda)\,,\;\forall\,\Lambda < \infty$.
One also assumes that there is no fundamental scale above which 
$S^{\mathrm{eff}}(\Lambda)$ is not defined and $S^{\mathrm{eff}}(\Lambda)$ loses its 
predictive power in a process where the scale $E$ approaches $\Lambda$; indeed, 
$E = \Lambda$ requires $\infty$ renormalized parameters. 
\subsubsection{SMEFT validity and unitarity}

The SMEFT behaviour is similar to that of renormalisable theories that satisfy unitarity in perturbation theory; a perturbative expansion in $E/\Lambda$ always becomes difficult at  high enough energy, and the applicability of the SMEFT will break down in the ``tails'' of kinematic distributions 
(or new physics will be seen before the breakdown).  Therefore, projecting data into 
the SMEFT will have 
a large intrinsic uncertainty, i.e we do not know what exactly is going on because the 
SMEFT interpretation becomes a series where the expansion parameter is close to $1$ and/or 
the perturbative unitarity bound is saturated.

The range of validity of the perturbative expansion in SMEFT is poorly known and unitarity conditions 
can be used to get additional information. Having said that, EFTs could enjoy 
some range of non-perurbative validity before new physics is manifest (which goes under the name of classicalization, asymptotic safety, etc $\dots$).

Finally, one should keep in mind that these are not statements that unitarity is violated in 
the SMEFT. Unitarity {\it{would be violated}}, if we could trust the perturbative expansion, which 
we cannnot. Perturbative unitarity bounds can be computed, but the bounds also imply that loops 
{\it{must}} be important.

To give a brief example, we consider the scattering of longitudinally polarized $W$ bosons. In the \ac{SM},
\begin{equation}
T_{\SMm}(W^+_L W^-_L \rightarrow W^+_L W^-_L) 
\sim - \frac{G_F \mh^2}{4 \sqrt{2} \pi} \ \ \ \ s \rightarrow \infty \spp
\end{equation}
Unitarity is not violated, since the amplitude is constant in limit of large $s$. In the \ac{SMEFT},
\begin{equation}
T_{\mrm{SMEFT}}(W^+_L W^-_L \rightarrow W^+_L W^-_L) 
\sim - \frac{1}{32 \pi} g_{_6} G_F (a_{\phi D} + 
2 a_{\phi \Box}) \frac{s}{\Lambda^2} + \mathcal{O}(1) \spc
\end{equation}
where $g_{_6}$ is the common coupling of $\mathrm{dim} = 6$ \ac{SMEFT} operators and $a_X$ 
are specific Wilson coefficients. The \ac{SM} contributes to the constant part, 
while the part growing with $s$ is driven by the new higher dimension interactions. 
From these equations, 
we can, by imposing $T_{\mrm{SMEFT}} = 1$, estimate the scale at which perturbative unitarity 
is broken, $s_C$:
\begin{equation} \begin{split}
|T_{\mrm{SMEFT}}| &\sim \frac{1}{32 \pi} g_{_6} G_F (a_{\phi D} + 
2 a_{\phi \Box}) \frac{s}{\Lambda^2} < 1 \\
s_C &< \frac{32 \pi}{g_{_6} G_F (a_{\phi D} + 2 a_{\phi \Box})} \Lambda^2 \spp
\end{split} 
\end{equation}
Note, that from this equation it is not obvious that the scale $s_C$ is smaller 
than $\Lambda^2$. However, it makes no sense to talk about \ac{EFT} for $s_C$ above $\Lambda^2$. 
Thus, in general, 
we can define number of scales in \ac{EFT}, at which we suspect the breakdown of the approximation will occur, and they should be studied case by case, to find the one which leads to the strongest constraints. 
Performing calculations above these scales, although still possible, should be treated with 
extreme caution. 

Finally, imposing a form-factor-like suppression to EFT coefficients in order to 
avoid unitarity violations at scales $E \sim \Lambda$ corresponds (at best)
to choosing some specific UV model; in particular form-factors introduce a second
cut-off scale which a priori has nothing to do with the scale
appearing in front of the Wilson coefficients.

In principle, one could introduce a cut-off to forbid unphysical events manually (a prescription also partially used by ATLAS and CMS, known as ``event clipping''). Such a cut-off could also be motivated theoretically by the
argument that these events could have never arisen in a UV-complete theory. However, this leads to
a sharp edge in the distribution which does not resemble any sensible approximation to a UV-complete theory, 
see Ref.~\cite{Sekulla:2016yku} for details and for a description of EFT, perturbative unitarity and unitarization.

Ref.~\cite{Aydemir:2012nz} indicates that the two phenomena of 
tree-unitarity violation and the onset of new physics are separate in practice in QCD-like theories with
different numbers of colors and flavours. 
Unitarity violation happens in elastic scattering well below the energy where
the QCD degrees of freedom are produced. In such a case,
there is no other option but that the apparent unitarity
violation must be solved within the effective field theory
without recourse to the new degrees of freedom.
\subsection{Amplitudes in SMEFT, asymptotic behaviour}
To fully understand the ``linear'' vs. ``quadratic'' problem we provide the asymptotic behaviour of the helicity amplitudes for the process $q \bar q \to \h Z$ as a function of the $\h Z$ invariant mass.
From~\cref{tab:EFThelicityAmpl} it is easy to understand when and why the linear approach 
starts giving unphysical results.
\begin{table}[h!]
\[
\begin{array}{ccccccc}
\toprule
\text{Helicity} & \qquad \,	& \text{SM}		& \qquad \,	& \text{one insertion}	&	& \text{two insertions} \\
\midrule
-+-		& & \frac{\Gf M_Z^3}{M_{\h Z}}	&  & \frac{M_Z M_{\h Z}}{\Lambda^2} & 	& \frac{M_Z M_{\h Z}}{\Gf \Lambda^4} \\
\text{Wilson}	 & & -				& & a_{ZZ}, a_{\phi q}^{(1)}, a_{\phi q}^{(3)}	&  \qquad \, & a_{AA}, a_{AZ}, a_{ZZ}, a_{\phi D}, a_{\phi \Box}, a_{\phi q}^{(1)}, a_{\phi q}^{(3)}	 \\
\midrule
-+0 &  & \Gf M_Z^2 & & \frac{M_{\h Z}^2}{\Lambda^2} & & \frac{M_{\h Z}^2}{\Gf \Lambda^4} \\
\text{Wilson} & & -  & & a_{\phi q}^{(3)}, a_{\phi q}^{(1)} & & a_{AA}, a_{AZ}, a_{ZZ}, a_{\phi D}, a_{\phi \Box}, a_{\phi q}^{(3)}, a_{\phi q}^{(1)} \\
\midrule
-++ & & \frac{\Gf M_Z^3}{M_{\h Z}} &  & \frac{M_Z M_{\h Z}}{\Lambda^2} & & \frac{M_Z M_{\h Z}}{\Gf \Lambda^4} \\
\text{Wilson} & &  - & & a_{ZZ}, a_{\phi q}^{(1)}, a_{\phi q}^{(3)} & & a_{AA}, a_{AZ}, a_{ZZ}, a_{\phi D}, a_{\phi \Box}, a_{\phi q}^{(1)}, a_{\phi q}^{(3)} \\
\bottomrule
\end{array}
\]
\caption{\ac{LO} \ac{EFT} helicity amplitudes for the process $q \bar q \to \h Z$. For each 
helicity amplitude, the perturbative order and the involved Wilson coefficients are reported.}
\label{tab:EFThelicityAmpl}
\end{table}


\begin{figure}
\begin{center}
\includegraphics[scale=0.3]{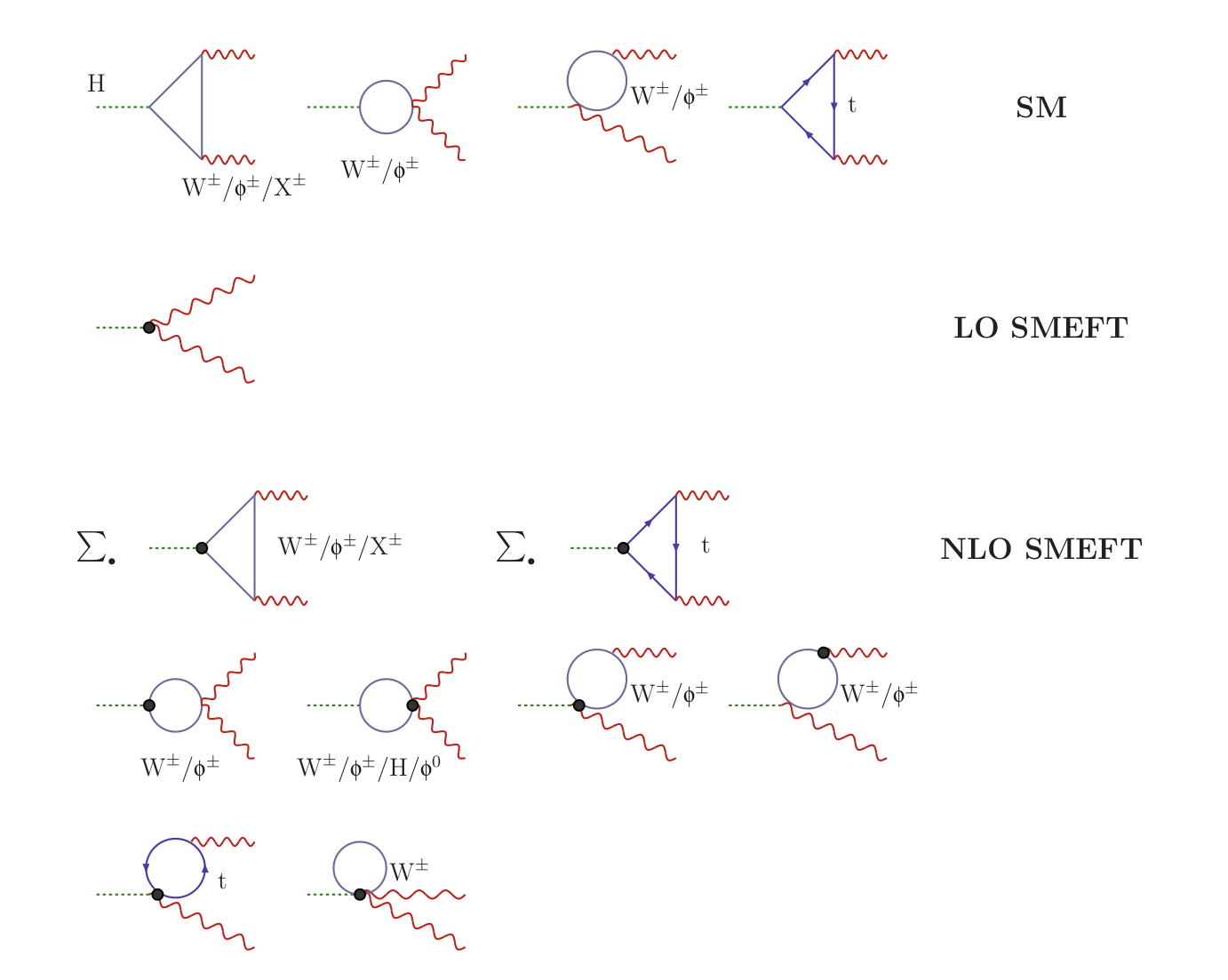}
\end{center}
\label{fig:nloSMEFTHgammagamma}
\caption{Original figure from Ref.~\cite{David:2015waa}. Diagrams contributing to the amplitude for $\h \rightarrow \gamma \gamma$  in the $R_{\xi}$-gauge: \ac{SM} (first
row), \ac{LO} \ac{SMEFT} (second row), and \ac{NLO} \ac{SMEFT}. Black circles denote the insertion of one dim-6 operator.
$\Sigma_{\bullet}$ implies summing over all insertions in the diagram (vertex by vertex). For
triangles with internal charge flow ($t$, $W^\pm$ , $\phi^\pm$ , $X^\pm$ ) only the clockwise orientation is shown.
Non-equivalent diagrams obtained by the exchange of the two photon lines are not shown. Higgs and photon wave-function factors are not included. The Fadeev-Popov ghost fields are denoted by $X$.}
\end{figure}

%
%
%
%
%

\clearpage

\newpage
\pagebreak
\section{Pseudoobservables for the LHC}
\label{sec:POs}
\subsection{Pseudoobservables from the LEP times}
The \ac{LEP} collider was operating between $1990$ and $2000$, in the same tunnel that is now used by the LHC; at the beginning it was operating at a c.m. energy close to the $Z$ boson  mass, more precisely at $| \sqrt{s} - \mz| < 3 \, \GeV$, and on its second run the energy was raised in order to produce also $W$ boson pairs, reaching $209 \, \GeV$. 

With  each successive energy upgrade of the LEP collider, some hope re-emerged that the discovery of the Higgs boson was about to happen. Just prior to the planned shutdown of LEP in 2000, few events that resembled a Higgs boson with a mass  of $\sim 115 \,\GeV$ were observed. This led to the extension of the final LEP run by a few months. But in the end the data was inconclusive and insufficient to justify another run and the final decision to shut down and dismantle LEP was taken. This way it was possible to make room for the new Large Hadron Collider.  The analysis of the direct search for the Higgs boson at LEP resulted in a final lower bound of the Higgs mass $114.4 \,\GeV$ at the $95\%$ confidence level.


The concept of \acp{PO} was born in the frame of the LEP analysis, see for instance Refs.~\cite{Bardin:1999ak,Z-Pole}, as a counterpart to the traditional Fiducial Observables (FOs), also called Realistic Observables (ROs). The POs were designed to have two main features: 
\begin{itemize}

\item to allow comparison between experiments (i.e. independent of the various detector cuts) 

\item to be as independent as possible of changes in the underlying theory.
 
\end{itemize}
For this purpose, it was needed to find a set of quantities that are well defined in QFT and as independent as possible of detector fiducial volumes. 
To start the discussion on the \acp{PO} used at LEP we define the matrix element for a $Z$ boson decaying into a fermion pair:
\begin{equation}\label{eq:Zff}
\mathcal{M}_{\rm{Z \rightarrow f \bar{f}}} = 
\bar{u}_f \slashed{\epsilon}_Z (\mathcal{G}^f_V + \gamma_5 \mathcal{G}^f_A) v_f \spc
\end{equation}
where $\mathcal{G}^f_{V/A}$ are complex valued effective coefficients, which absorb all 
correction factors and are evaluated at a scale $Q^2 = - \mz^2$. 
 
The strategy followed a LEP in defining the POs was, first of all, to de-convolute the initial state QED radiation and the final state QED/QCD radiation. This is possible  given our good understanding of the structure of radiation: therefore, we can define a  \emph{radiator}, e.g. for initial state QED radiation, such that the observed cross section 
becomes
\begin{equation}
\sigma (s) = \int^{1- x_{cut}}_{0} \rm{dx} \, \,  H(x,s) \, \sigma_0 ((1-x) s) \spc
\end{equation}
where $\sigma_0$ is the de-convoluted cross section and $H(x,s)$ is the radiator absorbing the corrections. One can think of this radiator as similar to the parton distribution functions, PDFs, so important nowadays for the LHC strategy. Furthermore, around the $Z$ peak, $\sigma_0$ only contains the $Z$ resonant part of the amplitude. 

Observe that this approximation will only work if we know the corrections that we are neglecting, or at least have a good estimate of their size. This was the case at LEP but opens a debate in the case of the SMEFT analysis, where we still define POs but we do not have such a good knowledge of the SMEFT-extended radiator. For instance Bremsstrahlung has a very important effect for a circular electron-positron collider. 
 
This way, after we subtract the real emission and the non-resonant part from the process, we are left with a kernel which looks more like the \ac{BW} function we expect, for instance for $f \bar{f} \rightarrow Z$ we have: 
\begin{equation}
\sigma_{\bar{f} f} (s) = 
\sigma_0^{\bar{f} f} \frac{s^2  \Gamma_Z^2}{ (s- \mz^2)^2 + 
s^2 \Gamma_Z^2/\mz^2} , \qquad \sigma_0^{\bar{f} f} = 
\frac{12 \pi}{\mz^2} \frac{\Gamma_e \Gamma_f}{\Gamma_Z^2}  \spp
\end{equation}
The partial and total $Z$ widths were defined as \acp{PO} (i.e. $\Gamma_Z,\Gamma_f, \Gamma_e$), with additional definitions of:
\begin{equation}
	\begin{split}
\Gamma_{\rm{hadr}} &= \Gamma_u + \Gamma_d + \Gamma_c + \Gamma_s + \Gamma_b \spc \qquad
\Gamma_{\rm{inv}} = \Gamma_Z - \Gamma_e + \Gamma_{\mu} + \Gamma_{\tau} + \Gamma_{\rm{hadr}}  \spc \\
R_l &= \frac{\Gamma_{\rm{hadr}}}{\Gamma_e} \spc \ \ \  
R_{b,c} = \frac{\Gamma_{b,c}}{\Gamma_{\rm{hadr}}} \spc   \qquad
\sigma_{\rm{hadr}} = \frac{12 \pi}{\mz^2} \frac{\Gamma_e \Gamma_{\rm{hadr}}}{\Gamma_Z^2}  \spp
	\end{split}
\end{equation}
These partial and total widths were calculated including the final state QCD and QED corrections. 
 
Another class of \acp{PO} used at LEP are the ones connected to distributions, such as forward-backward asymmetries and polarisations. The assumption for this de-convoluted \acp{PO} was that QED and QCD corrections were subtracted from the experimental data. From the theoretical point of view they can be calculated from the differential cross section,
\begin{equation}
\frac{\rm{d}\sigma_f}{\rm{d} \Omega} = 
\frac{\alpha}{4 \pi} \cos(\theta) \beta_f 
\Big{[} (1 + \cos(\theta)^2) F_1(s) + 4 \mu^2_f 
(1 + \cos(\theta)^2) F_2(s) + 2 \beta_f \cos(\theta) F_3(s) \Big{]} \spc
\end{equation}
where $\beta_f^2 = 1 - 4\mu_f^2$ with $\mu_f^2 = m_f^2/s$, while $F_i(s)$ are form 
factor functions which depend on reduced $\gamma/Z$ propagator ratio, effective $Z$ couplings 
and electric charges, both for electron and final state fermions. Then asymmetries 
and polarisations can be expressed by the form factors, e.g.
\begin{equation}
\mathbf{A}_{FB}^f(s) = \frac{3}{4} \frac{\beta_f F_3(s)}{F_1(s)+2\mu_f^2F_2(s)} \spp
\end{equation}
The definitions of POs were significantly simplified in the limit of massless fermions 
and vanishing $\Gamma_Z^2$ terms, yielding:
\begin{equation}
\mathbf{A}_{FB}^f = 
\frac{3}{4}\mathcal{A}_e \mathcal{A}_f \spc \ \ \ 
\mathbf{A}_{LR}^e=\mathcal{A}_e \spc \ \ \ 
\mathbf{P}^f= - \mathcal{A}_f \spc \ \ \ \mathbf{P}_{FB}(\tau) = -\frac{3}{4}\mathcal{A}_e \spc
\end{equation}
where $\mathcal{A}_f$ is defined as:
\begin{equation}
\mathcal{A}_f = 
\frac{2\Re[\mathcal{G}^f_V (\mathcal{G}^f_A)^*]}{|\mathcal{G}^f_V|^2 + |\mathcal{G}^f_V|^2} \spc
\end{equation}
and the couplings $\mathcal{G}$ were defined in~\cref{eq:Zff}. 
\begin{figure}
\begin{center}
\begin{subfigure}[b]{0.45\textwidth}
\includegraphics[width = 0.9\textwidth]{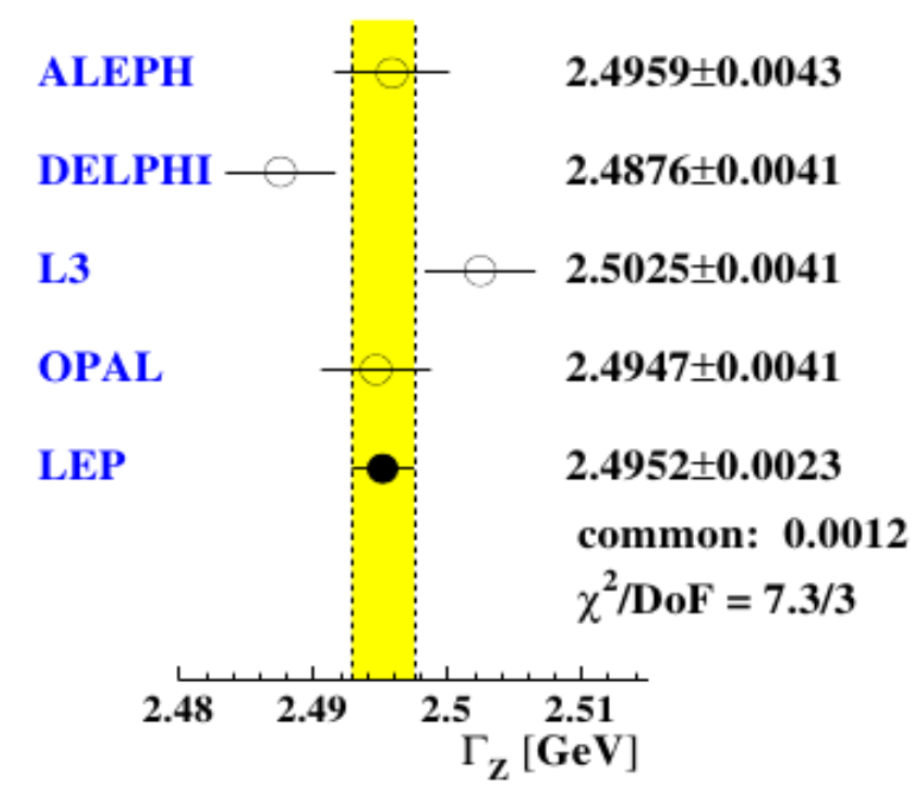}
\caption{}
\label{fig:combLEP}
\end{subfigure}
\begin{subfigure}[b]{0.45\textwidth}
\includegraphics[width = 0.9\textwidth]{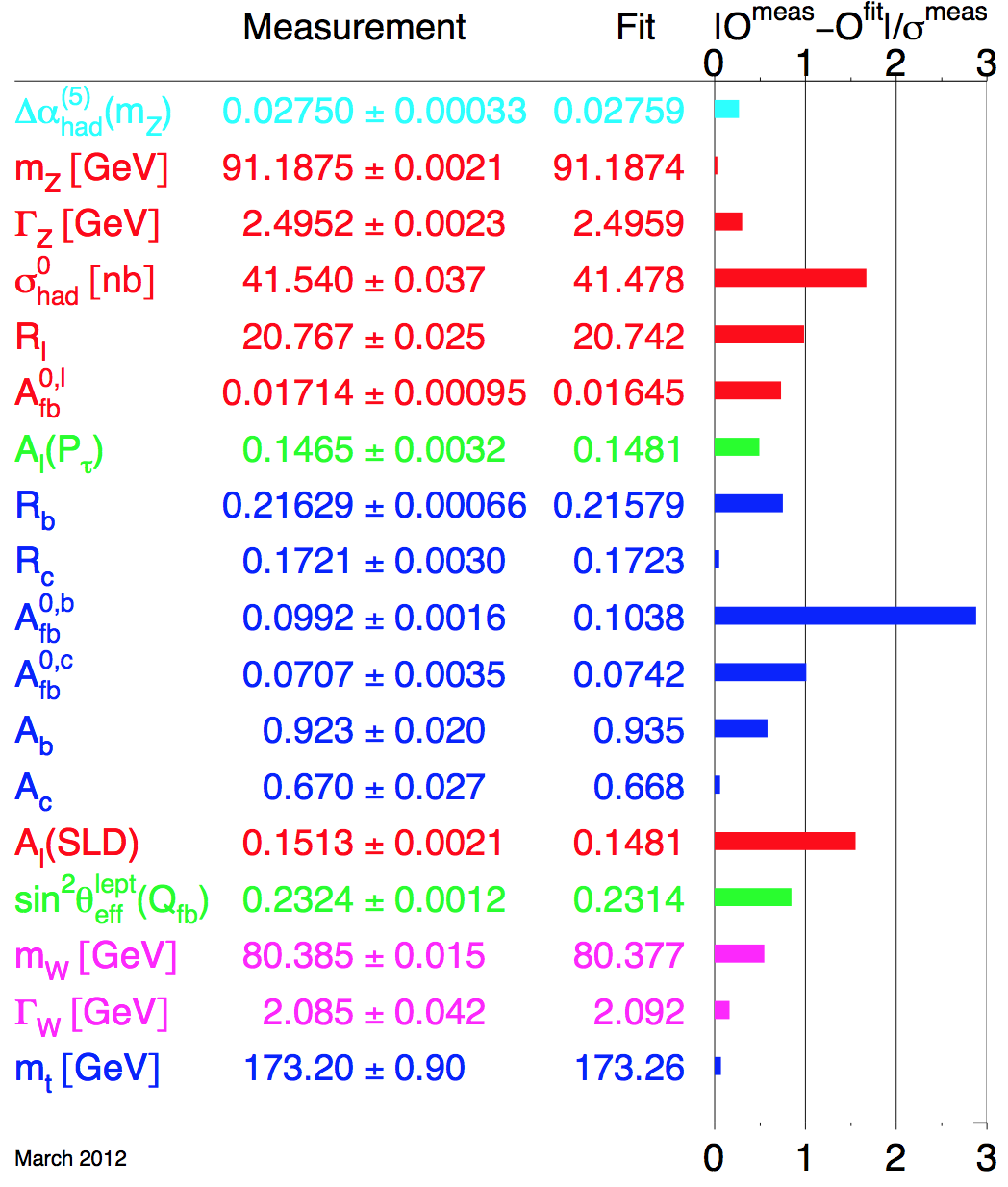}
\caption{}
\label{fig:EWPD}
\end{subfigure}
\end{center}
\caption{The example of combination of PO ($\Gamma_Z$) between experiments (a) and legacy list of 
POs combined between all experiments, representing the EWPD \cite{LEP-2} (b).}
\end{figure}

The experiments performed fits to the \acp{PO} described above, and then were able to combine them to present legacy LEP results. ~\cref{fig:combLEP} shows the  combined measurement for the total $Z$ width, from the analysis of the four LEP experiments. The table in~\cref{fig:EWPD}  shows the set of combined \acp{PO} measurements, altogether known as \ac{EWPD}. These were measured with an 
unprecedented precision, and up to now they still represent a strong constraint on all BSM models. Additionally, due to their theoretical connection with the unknown Higgs and top quark masses, they pointed at the correct mass ranges, which were then confirmed by direct observation of these particles. 
\subsubsection{LEP summary}
The rationale for the deconvolution is based on the fact that all experiments
used different kinematic cuts and selection criteria, while an objective
requirement was put forward by the scientific community for having universal
results expressed in terms of the on-shell $Z$ region.

Assuming a structure function representation for the initial fermions, in
turn, allows us to deconvolute the measurements and to access the
hard scattering at the nominal peak. The transition from FO's to PO's involves certain assumptions that  reflect our understanding of QED effects but, within those assumptions, there is a well-defined mathematical procedure. 

Insofar as this procedure is an explicitly specified and mathematically meaningful 
transformation, the PO's possess the status of observability.
The strategy is as follows: begin with the predicted amplitude, dressed by 
the weak loop corrections, and use the fact that in the SM there are several effects, 
such as 
\begin{itemize}

\item[a)] the imaginary parts or 
\item[b)] the $\gamma{-}Z$ interference or 
\item[c)] the pure QED background, 

\end{itemize}
that have a usually negligible influence on the $Z$ line shape.
Therefore, PO's are determined by fitting FO's but we will have some ingredients 
that are still taken from the SM, making the final results dependent upon the SM.
In this way the exact (de-convoluted) cross-section is successively reduced to 
a $Z$-resonance. It is a modification of a pure Breit-Wigner resonance because of the $s\,$-dependent width.
\subsection{Key differences between LEP and LHC in context of POs}
Before we start to consider a proposal for \acp{PO} at LHC let us discuss what are the differences, not only in the collider properites but also in the theory status between LEP and LHC times.  LEP was an electron-positron collider, well suited for precision measurements and, as we will discuss now, with much cleaner event productions. Already at LEP, the important quest was to separate the hard 
effects from the soft physics, which was then parametrised by the ``radiator'' function. 

At LHC there are more sources of model dependency. As we will see below, the POs will  be defined at the parton level, while in the detector we have stable particles and jets. Therefore, we need a bridge through parton distribution functions, parton showers, etc. 

\begin{figure}[h]
\begin{centering}
 \includegraphics[scale=0.5]{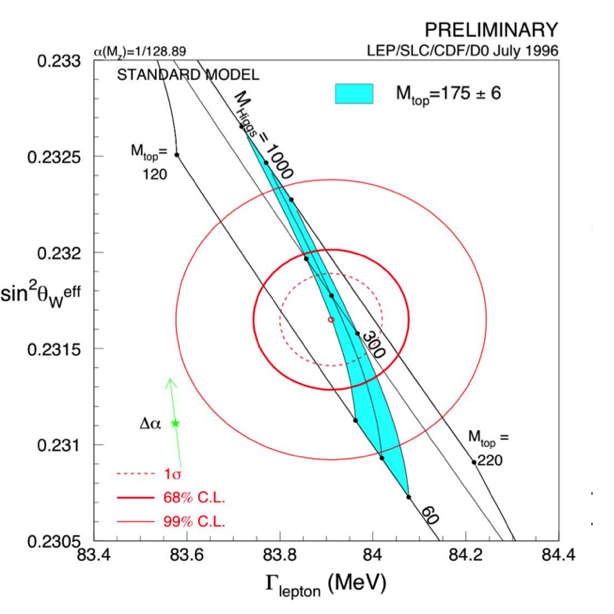}
 \caption{The plot presenting measured values of two LEP \acp{PO}. On the plot also 
predictions for different Higgs and top quark masses are plotted. See Ref.\cite{LEP-2}}
  \label{fig:POsMH}
\end{centering}
\end{figure}

There is another difference to be taken into account: the status of the SM was also different. At LEP, the SM was still missing a key piece, the Higgs boson; therefore, the hypothesis was the SM, the Higgs mass was the unknown and measurements were presented in a way to see bounds on $\mh$, as we can see in~\cref{fig:POsMH}. 

Now the paradigm has changed: all particles  of the SM have been observed and we do not really have a clear direction in which to look for NP. What is mostly missing, 
are precise measurements to be confronted with the SM or with a theory of SM deviations. 
\subsection{Pseudoobservables for the LHC}
\acp{PO} proved to be a very successful tool for storing the LEP results, and allowed for better communication between experiments and theory; therefore, there is 
an ongoing discussion on reintroducing \acp{PO} for LHC. 

Although the experimental situation is quite different, the advantages of POs should be clear: a high degree of model independence and the possibility of reinterpreting experimental results when more precise theoretical calculations become available.
In that regard, POs act as a bridge between the data collected in 
the experiment and the theory calculations. The POs can be obtained from the fiducial cross sections, by deconvoluting effects such as parton distribution functions and radiative corrections. 

It is important to keep in mind that this procedure will be the source of an extra uncertainty, not necessarily small. It is of the highest importance to keep the uncertainties under control, requiring a careful cross-check of theoretical tools used in the analysis. 

As it was described in detail in~\cref{sec:fiducial}, the fiducial or \acp{STXS} allow to store the experimental data, with a minimal dependence on the precision of the currently available calculations, and are the closest measurement to what is actually seen in the detectors. However, one of their drawbacks, is that they cannot be used for the combinations and comparisons between the two experiments. For this reason, the proposal is to perform combination in terms of POs, as it was done at LEP. 

Furthermore, theory upgrades can be applied at the level of fiducial quantities, rather than starting from raw data. POs, being well defined objects from the theoretical point of view, can be then interpreted in terms of more fundamental quantities such as Wilson coefficients or Lagrangian parameters of some UV complete theory. 

In that regard, it would be much more handy for model builders to interpret the nature of NP and parameters of the specific models from POs, rather than trying to extract them directly  from fiducial or template cross sections. 

Also, some of the limitations of the $\kappa$-framework are reduced thanks to the flexibility of the POs. However, this goes with a cost: more parameters to fit and simulate. The set of POs  used for parametrising Higgs properties was proposed both in the \ac{EW} decays in Ref.\cite{Gonzalez-Alonso:2014eva} and in the \ac{EW} production in Ref.\cite{Greljo:2015sla}. From the theoretical point of view, the POs are inspired by the $\mathrm{dim} = 6$ EFT Lagrangian (SMEFT), which we described in~\cref{sec:EFT}. 

In this case we end up with $20$ real parameters for the Higgs decay observables, 
and additional $32$ parameters in case of productions (corresponding to the interactions  with the light quarks). The amount of the parameters can be significantly reduced when 
we consider some of the SM symmetries, such as flavour universality or CP conservation. 
The POs flexibility manifests in fact that we allow the modification not only in the Higgs couplings but also in coupling of the \ac{EW} bosons with the final (initial) state fermions. Also the description of vertices is more detailed since it considers the different modifications for different tensor structures.

It is worth noting that we do not claim any theoretical supremacy of the POs with respect to the best SM or BSM calculations; the main reason of their use is that they represent an optimal way of keeping record of experimental data. 

Let us now present how the LHC POs should be defined. In this review we will discuss the \ac{MPE} of amplitudes and we will briefly mention in ~\cref{hicsunt}
the factorization of a process into sub-processes, the details can be found in Ref.~\cite{David:2015waa}.

We start from the easiest case of the Higgs coupling to two fermions. At LHC this is directly measured by $\h \rightarrow \tau^+ \tau^-$ and $\h \rightarrow b \bar{b}$. In the most general form we can write this (tree level) amplitude decomposed in the CP conserving and violating part, although the latter is not present in the SM:
\begin{equation}
\mathcal{A}(\h\rightarrow f \bar{f}) = 
- \frac{i}{\sqrt{2}} ( y_S^f \bar{f} f + i y_P^f \bar{f} \gamma_5 f) \spp
\end{equation}
The coefficients $y_S^f$ and $y_P^f$ are \acp{PO} and need to be measured experimentally. 
We can also express them in the well known $\kappa$ framework as:
\begin{equation}
\kappa_f = \frac{ y_S^f}{ y_{S,\SMm}^f} \spc \ \ \ 
\delta_f^{\mrm{CP}} =  \frac{y_P^f}{ y_{S,\SMm}^f} \spp
\end{equation}
We need to note here, that the measurement of the total rate of Higgs \ac{BR} to fermions, 
would not allow to differentiate between the two contributions, since it scales as:
\begin{equation}
\Gamma(\h\rightarrow f \bar{f}) = 
\left[ {\kappa_f}^2 + {\left(\delta_f^{\mrm{CP}} \right)}^2 \right] \Gamma(\h\rightarrow f \bar{f})^{\SMm} \spp
\end{equation}
To access separately the information about the CP violating part of the above amplitude, the spins of the fermions need to be determined. It can be accessed, e.g. by the measurements of angular distributions of the $\tau$s decay products. Note also that the CP violating part is not present in the EFT formalism we presented in the previous chapter. It is so, because we assumed all the symmetries of the \ac{SM}, including the CP one.
\begin{figure}[h]
\begin{centering}
 \includegraphics[scale=0.3]{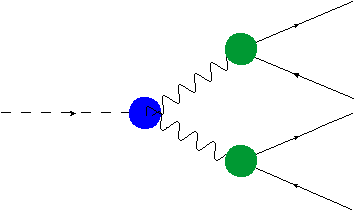}
 \caption{The Higgs decaying to 4 fermions via vector bosons.}
  \label{fig:h4f}
\end{centering}
\end{figure}

Obviously, POs must always match the best, available, calculations process by process. In the previous example we have considered $\h \rightarrow b \bar{b}$ at LO in QCD; however, consider the EFT approach and analyse the decay at NLO in QCD. At this order, new real contributions enter at $\mathrm{dim} = 6$ which either do not have a Born amplitude (chromomagnetic interaction) or are built starting from another amplitude. In these latter channels the same final state is reached through a completely different propagator structure.

For example, let us imagine a situation in which the Yukawa of the $b\,$-quark is SM-like and that other Wilson coefficients are greatly enhanced such that the resulting distribution in $m(b \bar{b}$) peaks towards lower masses in a way that cannot be described by simple QCD ``radiation''.
In this case, POs still represent the optimal way of separating $\h \to b \bar(b) (g)$, $\h \to gg$ (with one gluon converting into a $\bar{b} b$ pair) and $\h \to \bar{b} b g$ (with a hard gluon).

The situation is more complicated when the Higgs boson decays into $4$ fermions. 
In the \ac{SM} the doubly-resonant part of the process goes via Higgs decaying into two vector bosons decaying then into fermions (detected or giving measured through their missing transverse energy in case of neutrinos). 

A similar situation is also present in the SMEFT, although here a second decay channel is open: when the Higgs boson decays via the chromomagnetic operator into a pair of fermions and one vector boson. 

We will construct the corresponding POs in the end of this Section, but here we start with introducing the doubly-resonant part of the process,
$h \rightarrow VV \rightarrow f_1 f_2 \bar{f}_3 \bar{f}_4$: this is shown in~\cref{fig:h4f}. 

To ensure full flexibility we will proceed as follows:

\begin{enumerate}[a)]
\item define couplings for Higgs decaying to vector bosons (as it is done in the $\kappa$-framework),
\item define couplings for vector bosons decaying to fermions (blue and green blobs respectively).
\end{enumerate}

If we want to study distributions and not only total rate modifiers, 
(as done in the $\kappa$-framework), different tensor structures of the amplitude need 
to be considered. 
Here we will present the case for the process of the Higgs effectively decaying into two $Z$ bosons, which then decay into a pair of muons and a pair of electrons. 

This reasoning can be easily repeated for the different decays in the second stage, and also while including the $W$ bosons in the intermediate stage. For the general description we refer to the original paper, see Ref.~\cite{Gonzalez-Alonso:2014eva}. 

\begin{figure}[h]
\begin{centering}
\includegraphics[scale=0.3]{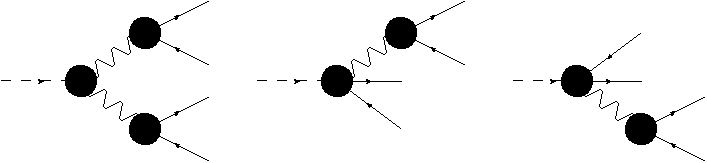}
 \caption{The pictorial representation of the expansion around physical poles 
for the longitudinal form factor for Higgs decaying in muon and electron pair.}
  \label{fig:momexp}
\end{centering}
\end{figure}

The full description can be started with the definition of POs corresponding to $Z \rightarrow l^+ l^-$, the green blob from~\cref{fig:h4f}. The $Z$-fermion interaction is governed by a term,

\begin{equation}
\Sigma_{f=f_L,f_R} Z_{\mu} c_Z^{\SMm} \bar{f} \gamma^{\mu} f, \qquad c_Z^{\SMm}(f) =\frac{g}{c_W} (T_3(f) -Q(f) s_W^2)
\end{equation}
from which we can read out the generalised amplitude,
\begin{equation}\label{zff}
\mathcal{A}(Z\rightarrow f \bar{f}) = 
\sum_{f=f_L,f_R} \epsilon_{\mu} g_Z^{ff'} \bar{f}' \gamma^{\mu} f \spc
\end{equation}
where $\epsilon_{\mu}$ is the polarisation of the $Z$ boson and $g_Z^{ff'}$ is a generalised coefficient which takes the SM value $g_Z^{ff}=c_{Z}^{\SMm}(ff)$ (being zero for different flavours). Now, having two additional couplings in mind (note, that we do not include flavour violating neutral currents), i.e. $g_Z^{\mu},g_Z^{e}$, we need to define the coupling between $\h$ and two $Z$ bosons. Let us decompose this amplitude as follows:
\begin{equation}\label{ffac}
\mathcal{A}(h \rightarrow \mu^+(p_1) \mu^-(p_2) e^+(p_3) e^-(p_4) ) = 
i \frac{2 m_Z^2}{v} \sum_{\mu=\mu_L,\mu_R} \sum_{e=e_L,e_R} 
(\bar{\mu} \gamma_{\mu} \mu) (\bar{e} \gamma_{\nu} e) \mathcal{T}_{nc}^{\mu \nu}(q_1,q_2) \spc
\end{equation}
where $q_1 = p_1 + p_2$ and $q_2 = p_3 + p_4$. Now we can decompose the tensor part of the
amplitude $\mathcal{T}_{nc}^{\mu \nu}(q_1,q_2)$ into Lorentz-allowed tensor structures and their form factors:
\begin{equation}
\mathcal{T}_{nc}^{\mu \nu}(q_1,q_2) = 
g^{\mu\nu} F^{\mu e}_L (q_1, q_2) + \frac{g^{\mu\nu} q_1 \cdot q_2 + 
q_1^{\nu} q_2^{\mu}}{m_Z^2} F^{\mu e}_T (q_1, q_2) + 
\frac{\epsilon^{\mu\nu\rho\sigma}q_{1,\sigma} q_{2,\rho}}{m_Z^2} F^{\mu e}_{\mrm{CP}} (q_1, q_2) 
\spp
\end{equation}
Now, the form factors can be be expanded around the $Z$ pole\footnote{We will not discuss here 
the fact that the $Z$ on-shell mass should be replaced by the corresponding complex pole}. 
Let us note here that the transverse and CP-violating tensor structures, second and third one 
in~\cref{ffac}, can also result from off-shell photons, mediated by the usual electromagnetic, 
charge dependent, coupling ($e Q$),
\begin{equation}
	\begin{split}
F^{\mu e}_L (q_1, q_2) =&  
\kappa_{ZZ} \frac{g_Z^{\mu}g_Z^{e}}{P_Z(q_1^2) P_Z(q_2^2)} +
\frac{\epsilon_ {Z\mu}}{m_Z^2} \frac{g_Z^{e}}{ P_Z(q_2^2)} +
\frac{\epsilon_ {Z e}}{m_Z^2} \frac{g_Z^{\mu}}{ P_Z(q_1^2)} + 
\Delta_L^{\SMm}(q_1^2, q_2^2)  \\
F^{\mu e}_T (q_1, q_2) =&  \epsilon_{ZZ} 
\frac{g_Z^{\mu}g_Z^{e}}{P_Z(q_1^2) P_Z(q_2^2)} +
\epsilon_ {Z\gamma} \left( \frac{e Q_{\mu} g_Z^{e}}{ q_1^2 P_Z(q_2^2)} + 
\frac{e Q_e g_Z^{\mu}}{q_2^2 P_Z(q_1^2)} \right) + 
\epsilon_ {\gamma\gamma} \frac{e^2 Q_{\mu} Q_{e}}{ q_1^2 q_2^2} + 
\Delta_T^{SM}(q_1^2, q_2^2)   \\
F^{\mu e}_{\mrm{CP}} (q_1, q_2) =&  
\epsilon^{\mrm{CP}}_{ZZ} \frac{g_Z^{\mu}g_Z^{e}}{P_Z(q_1^2) P_Z(q_2^2)} +
\epsilon^{\mrm{CP}}_ {Z\gamma} 
\left( \frac{e Q_{\mu} g_Z^{e}}{ q_1^2 P_Z(q_2^2)} + 
\frac{e Q_e g_Z^{\mu}}{q_2^2 P_Z(q_1^2)} \right) + 
\epsilon^{\mrm{CP}}_ {\gamma\gamma} \frac{e^2 Q_{\mu} Q_{e}}{ q_1^2 q_2^2}  \spp
	\end{split}
\end{equation}
The $\Delta$s correspond to the SM sub-leading non-local contributions, which remain 
unaffected by $\mathrm{dim} = 6$ operators. As we can see,  
on top of the $Z$ coupling constants, $g_Z^{f}$, also new constants appeared: 
$\kappa_{ZZ}$ and $\epsilon^{\mrm{(CP)}}_{XX}$. In the longitudinal form factor we can see the momentum expansion around the pole, and it is pictorially represented 
in~\cref{fig:momexp}. 

These are additional POs, which describe the $\h \rightarrow 2\mu 2e$  decays and would need to be determined in the experiments by measuring appropriate \acp{BR}. We can note here, that the $Z$ POs, i.e. $g_Z^{f}$, were already determined by LEP with a good precision, and thus may be used as input parameters when we concentrate on 
determining the Higgs couplings. 

We can also build the amplitudes for the Higgs decaying via charged vector bosons, 
e.g. $\h \rightarrow \mu^+ e^- \nu_{\mu} \bar{\nu}_{e}$. Keeping the lepton flavour 
conservation as a valid symmetry in the vector boson decays 
we can see that these final states can only be mediated by the $W^+W^-$ pair. However let 
us consider a slightly different final state: 
$\h \rightarrow \mu^+ \mu^- \nu_{\mu} \bar{\nu}_{\mu}$. Although the final state looks pretty 
similar, here the situation is more complicated, since the decay can be mediated both by 
$ZZ$ and $W^+W^-$ pairs, and we need to take all the possibilities into account. To handle 
this situation we can perform a ``top layer'' decomposition into neutral currents 
(described above), and charged currents:
\begin{equation}
	\begin{split}
\mathcal{A}(\h \rightarrow 
\mu^+ \mu^- \nu_{\mu} \bar{\nu}_{\mu}) &= 
\mathcal{A}_{nc}(\h \rightarrow (Z \rightarrow \mu^+ \mu^- ) 
(Z \rightarrow \nu_{\mu} \bar{\nu}_{\mu})) \\
&- \mathcal{A}_{cc}
(\h \rightarrow (W^+ \rightarrow \mu^+ \nu_{\mu}) (W^- \rightarrow \mu^- \bar{\nu}_{\mu}))  \spp
	\end{split}
\end{equation}

The \acp{PO}, as described by the different tensor structures, can be directly implemented into \ac{MC} tools, which then enable the comparison with experiment.
This condition is already satisfied with the POs that we have introduced and they 
have been implemented in the common \ac{UFO} format, described in Ref.~\cite{Degrande:2011ua}, allowing to use the model in many different \ac{MC} tools. Extensive comparisons have also been performed with Prophecy4f and \MGMC$\ $ generators. 

The POs we defined so far are thought of as a form of effective couplings;
in~\cref{tab:physPO} we can see the relations between the partial widths of 
the Higgs boson and the POs at the amplitude level. For some of these widths, a measurement would not be enough to distinguish between two CP even and CP odd \acp{PO}. It is important to keep in mind also that the quality of PO measurements will depend heavily on the uncertainty introduced by deconvoluting soft radiation and by the PDFs. 

The numerical factors were obtained by calculating the full partial decay width into a given final state with only one PO \emph{switched on} and by dividing by the \acp{BR} corresponding to vector boson decaying into two of final state particles. 
Therefore these coefficients depend on the accuracy of the theory of the presented calculations. 
\begin{table}[h]
\begin{center}
\begin{tabular}{ccc}
\toprule
Physical PO & Connection with effective coupling PO & effective coupling PO \\
\midrule
$\Gamma(\h \rightarrow f \bar{f})$ & $\Gamma(\h \rightarrow f \bar{f})^{\SMm} [ (\kappa_f)^2 + (\delta_f^{CP})^2 ]$ & $\kappa_f, \delta_f^{CP} $ \\[0.2cm]
$\Gamma(\h \rightarrow \gamma \gamma)$ & $\Gamma(\h \rightarrow \gamma \gamma)^{\SMm} [ (\kappa_{\gamma\gamma})^2 + (\delta_{\gamma\gamma}^{CP})^2]$ & $\kappa_{\gamma\gamma}, \delta_{\gamma\gamma}^{CP} $ \\[0.2cm]
$\Gamma(\h \rightarrow Z \gamma)$ & $\Gamma(\h \rightarrow Z \gamma)^{\SMm} [ (\kappa_{Z \gamma})^2 + (\delta_{Z \gamma}^{CP})^2]$ & $\kappa_{Z \gamma}, \delta_{Z \gamma}^{CP} $ \\[0.2cm]
$\Gamma(\h \rightarrow Z_L Z_L)$ & $ (0.209 \, \MeV) \times |\kappa_{ZZ}|^2$ & $\kappa_{ZZ}$ \\[0.2cm]
$\Gamma(\h \rightarrow Z_T Z_T)$ & $(1.9 \times 10^{-2} \, \MeV) \times |\epsilon_{ZZ}|^2$ & $\epsilon_{ZZ}$ \\[0.2cm]
$\Gamma^{\mrm{CPV}}(\h \rightarrow Z_T Z_T )$ & $(8.0 \times 10^{-3} \, \MeV) \times |\epsilon^{\mrm{CP}}_{ZZ}|^2$ & $\epsilon^{\mrm{CP}}_{ZZ}$ \\[0.2cm]
$\Gamma(\h \rightarrow Z f \bar{f})$ & $(3.7 \times 10^{-2} \, \MeV) \times N_c^f |\epsilon_{Zf}|^2$ & $\epsilon_{Zf}$ \\[0.2cm]
$\Gamma(\h \rightarrow W_L W_L)$ & $ (0.84  \, \MeV) \times |\kappa_{WW}|^2$ & $\kappa_{WW}$ \\[0.2cm]
$\Gamma(\h \rightarrow W_T W_T)$ & $(0.16 \, \MeV) \times |\epsilon_{WW}|^2$ & $\epsilon_{WW}$ \\[0.2cm]
$\Gamma^{\mrm{CPV}}(\h \rightarrow W_T W_T)$ & $(6.8 \times 10^{-2} \, \MeV) \times |\epsilon^{\mrm{CP}}_{WW}|^2$ & $\epsilon^{\mrm{CP}}_{WW}$ \\[0.2cm]
$\Gamma(\h \rightarrow W f' \bar{f})$ & $(0.14 \, \MeV) \times N_c^f |\epsilon_{Wf}|^2$ & $\epsilon_{Wf}$ \\[0.2cm]
\bottomrule
\end{tabular}
 \caption{\label{tab:physPO}The relation between the effective coupling POs and physical ones. $N_c^f$ corresponds 
to the number of colours. }
\end{center}
\end{table}%

Using crossing symmetry we find that the same diagrams describing $\h$ decay into four fermions also govern the \ac{EW} production modes of 
vector boson fusion and Higgsstrahlung. This means that, in principle, we can use the same set of POs. However, we are still missing dedicated POs for Higgs production via gluon fusion or associated with top quark pair. The authors of Ref.\cite{deFlorian:2016spz} advise to stick to $\kappa_g$ and $\kappa_t$ for these cases respectively. In the following we will consider the Higgs-gluon interactions in context of $\h$ decay into 
four quarks. 

In the context of POs in production, addressed in Ref.~\cite{Greljo:2015sla}, the momentum expansion around physical poles is valid only in limited kinematic regions. This is so because in the production it is not always the case that the Higgs boson is close to its threshold, and thus the standard definition of POs will not hold. In principle one should cut on the momentum transfer, something that is not directly available in the experiment. 

Indeed, the cut can only be based on the  $\pt$ of (VBF) jets or $Z$ boson in $Z \h$ production (which is correlated with the ``theoretical'' cut), however this correlation is becoming less and less clear for large values of the momentum transfer, and that could spoil the whole approach. 
\subsection{Extending the POs basis}
\subsubsection*{Four-quark final states resulting from the effective coupling of 
the Higgs to gluons}

\begin{figure}[h]
\begin{centering}
 \includegraphics[scale=0.3]{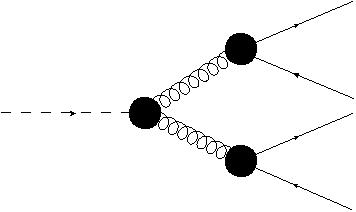}
 \caption{The Higgs decaying to $4$ quarks via gluons.}
  \label{fig:h4q}
\end{centering}
\end{figure}

The Higgs decaying into two quark-antiquark pairs via an effective gluon coupling is presented in~\cref{fig:h4q}

First of all we shall define the POs corresponding to the gluon-quark sector. Based on the convention used in~\cref{zff} we write:
\begin{equation}
\mathcal{A}(g \rightarrow q \bar{q}) = 
i \sum_q g^q_g \epsilon_{\mu} \bar{q} T^a \gamma^{\mu} q  \spp
\end{equation}
Now we can rewrite the expression from~\cref{ffac}, in this case
\begin{equation}
\mathcal{A}(h \rightarrow q(p_1) \bar{q}(p_2) q'(p_3) \bar{q}'(p_4)) = 
i \alpha_s^2 \sum_{q,q'} (\bar{q} 
\gamma^{\mu} q) (\bar{q}' \gamma^{\nu} q') \delta_{ab} \mathcal{T}^{\mu \nu}(q_1,q_2) \spp
\end{equation}
where $q_1 = p_1 + p_2$ and $q_2 = p_3 + p_4$ correspond to the momenta of the gluons. The tensor can be expressed through the form factors,
\begin{equation}
\mathcal{T}^{\mu \nu}(q_1,q_2) = 
(g^{\mu \nu} q_1 \cdot q_2 - q_1^{\nu} q_2^{\mu}) F^g(q_1,q_2) + 
\epsilon^{\mu \nu \rho \sigma} q_{1,\rho} q_{2,\sigma} F_{\mrm{CP}}^g(q_1,q_2) \spc
\end{equation}
which can be expressed in terms of the new POs,
\begin{equation}
\begin{split}
F^g(q_1,q_2) &= \kappa_{gg} \frac{g_g^q g_g^{q'}}{q_1^2q_2^2} \\
F^g_{\mrm{CP}}(q_1,q_2) &= \kappa_{gg}^{\mrm{CP}} \frac{g_g^q g_g^{q'}}{q_1^2q_2^2} \spp
\end{split}
\end{equation}
It is important to note here that the above POs have been introduced for illustrative purposes and are as of today of no relevance in the framework of the LHC. 

The QCD background makes it impossible to look for the $4$ jets final state. 
Additionally, the huge uncertainties are connected with jet reconstructions and their assignment to quarks and gluons. It would probably make it also impossible to measure these POs even in an electron-positron collider, which would be able to see the $4$ jet final state. 
\subsubsection*{Helicity-violating amplitudes resulting from effective dipole 
interactions of the Higgs field to (light) fermions and weak gauge bosons}

\begin{figure}[h]
\begin{centering}
 \includegraphics[scale=0.3]{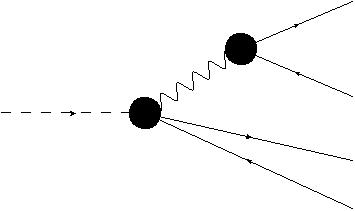}
 \caption{The Higgs dipole interaction, the vector can correspond to EW bosons or gluons.}
  \label{fig:hdip}
\end{centering}
\end{figure}

Let us now define the most general form of the amplitude for the $4$ fermion final state of the Higgs decay. 
\begin{equation}
\mathcal{A}(h \rightarrow f(p_1) 
\bar{f}(p_2) f'(p_3) \bar{f}'(p_4)) = 
i \sum_{f,f'} (\bar{f'} \Gamma^{(1)}_{\{\mu\}} f') 
(\bar{f} \Gamma^{(2)}_{\{ \nu\}} f) \mathcal{T}^{(1,2)\{\mu \nu\}} \spp
\end{equation}
In the general case, each of the $\Gamma^{(i)}_{\{ \mu\}}$ can be of the following form:
\begin{equation}
\Gamma = \Big{\{} \mathbf{1}, \gamma_5 \spc 
\gamma_{\mu}, \gamma_{\mu}\gamma_{5}, \sigma_{\mu\nu} = 
[\gamma_{\mu}, \gamma_{\nu}] \Big{\}} \spp
\end{equation}
So far we have adopted $\Gamma^{(i)}_{\{ \mu\}} = \gamma_{\mu}$; now we will investigate the case with the magnetic operator (one of the operators of $\rm{dim} = 6$ SMEFT). In that scenario, one of the $\Gamma^{(i)}_{\{ \nu\}}$ will have a tensor structure $\sigma_{\mu\nu}$.
The diagram for this interaction is presented in~\cref{fig:hdip}.

We will use the usual POs for the vector bosons decaying into two 
fermions (see~\cref{zff}), whereas the amplitude for the magnetic dipole  
interaction reads:
\begin{equation}
\mathcal{A}(h \rightarrow f(p_1) \bar{f}(p_2) f'(p_3) \bar{f}'(p_4)) = 
i \Sigma_{f,f'} (\bar{f}' 
\gamma^{\mu} f) (\bar{f} \sigma_{\mu \nu} f) \mathcal{T}^{\nu}(q) \spc
\end{equation}
where $q=p_3+p_4$ is the vector boson momentum transfer. Again we expand into form factors, however here we have just one tensor term:
\begin{equation}
\begin{split}
\mathcal{T}^{\nu}(q) &= q^{\nu} F^{\mrm{(dip)}}(q) \\
F^{\mrm{(dip)}}(q) &= 
\kappa^{\mrm{(dip)}}_{Zf} 
\frac{g_Z^f}{P_Z(q)} + \frac{\epsilon^{\mrm{(dip)}}_{Zf} }{m_Z^2} \spp \\
	\end{split}
\end{equation}
Note that the lack of a CP odd tensor structure comes from the fact that the term 
$\sigma_{\mu\nu} \gamma_5$ can always be reduced into lower rank terms of basis $\Gamma$.
\subsection{An alternative layer of POs} \label{hicsunt}
As we have seen, residues of resonant poles, $\kappa\,$-parameters and Wilson coefficients are different layers of POs. At LEP we had $g^{e}_{V \backslash\,A}$ but also $\Gamma(Z \to \bar{f}f)$ and $\sigma^{\mathrm{had}}_{\mathrm{peak}}$, therefore the question is: can we define $\Gamma(\h \to ZZ)$, $\sigma(\bar{t} t \h)$ etc. at the PO level?
Furthermore, can ``physical'' POs bypass ad-hoc constructions like ``diagram removal'' (not gauge invariant) or ``diagram subtraction'' (ad hoc prefactor and Breit-Wigner profile)?

It is clear that POs are conventionally defined but their definition should be consistent with first principles; here we concentrate on physical POs other than the ones defined in~\cref{tab:physPO} or defined with a different set of conventions. In particular, we are interested in going beyond fully extrapolated POs.

We have to consider several steps in constructing physical POs:
multi-pole expansion, principal-value cuts, phase space factorization and
helicity factorization. One should keep in mind the LEP example, the hadronic peak 
cross section extracted from $\sigma(e^+ e^- \to \,\mbox{hadrons})$, even if LEP 1 was a ``one resonance'' problem while LHC is always a ``multi-resonance'' problem.
\begin{itemize}

\item \textbf{Multi Pole Expansion:} poles and their residues are intimately related to the gauge 
invariant splitting of the amplitude (Nielsen identities); residues of poles (after squaring 
the amplitude and after integration over residual variables) can be interpreted as physical POs, 
which requires factorization into sub-processes.

\item \textbf{Phase Space Factorization:} gauge invariant splitting is not the same as factorization of the process into sub-processes, indeed phase space factorization requires the pole to be inside the physical region. It is the amplitude \emph{squared} that matters. 

\end{itemize}
Given an unstable particle we decompose the square of its propagator:
\begin{equation}
\Delta = \frac{1}{( s - M^2)^2 + \Gamma^2\,M^2} =
\frac{\pi}{M\,\Gamma}\,{\delta( s - M^2)} +
\mathrm{PV}\,\left[ \frac{1}{( s - M^2)^2}\right] 
\end{equation}
and use the $n$-body decay phase space
\begin{equation}
d\Phi_n( P, p_1 \dots p_n) =
\frac{1}{2\,\pi}\,dQ^2\,d\Phi_{n-j+1}( P, Q, p_{j+1} \dots p_n)\,
d\Phi_j( Q, p_1 \dots p_j )  \spp
\end{equation}
To complete the decay ($d\Phi_j$) we need the $\delta\,$-function.
We can say that the $\delta\,$-part of the resonant (squared) propagator opens the corresponding 
line allowing us to define physical POs ($t\,$-channel propagators cannot be cut). 

Principal-value cuts are better illustrated in terms of diagrams. First, we need the definition
of $\mathrm{K}\,$-diagrams.

\begin{definition}[$\mathrm{K}\,$-diagrams]
Consider a diagram, and then consider the diagram obtained by joining the final states of the 
original diagram and its complex conjugate. We next consider a diagram in which the initial
state lines are also connected together. The resulting diagram looks like a vacuum to vacuum
Feynman diagram except that the initial legs connected together are really ``cut'' lines,
i.e. we do not integrate over the momenta of the initial state lines but only average
over their spins, as usual. Such a diagram is called a $\mathrm{K}$ 
diagram, as introduced in Ref.\cite{Kinoshita:1962ur}. From a $\mathrm{K}$ diagram we can read the corresponding 
$S\,$-matrix.
\end{definition}

\begin{figure}
\begin{centering}  
 \includegraphics[scale=0.6]{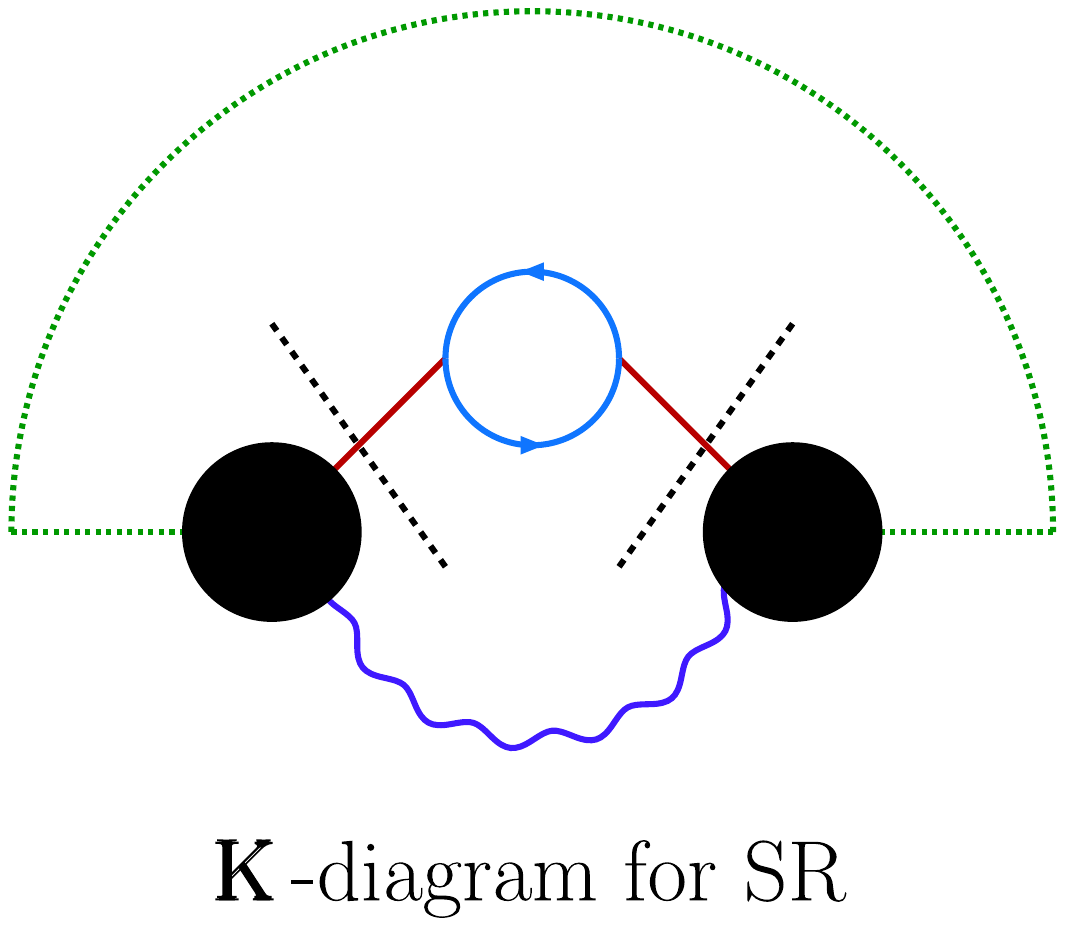}
 \vspace{-7.cm}
 \caption{$\mathrm{K}$ diagram for the $Z$ resonant part of the process
$\h \;\hbox{(dashed green line)} \to \bar{f} f \gamma \;\hbox{(wavy blue line)}$. } \label{fig:Kdia} 
\end{centering}
\end{figure}

\begin{figure}
\begin{centering}  
 \includegraphics[scale=0.6]{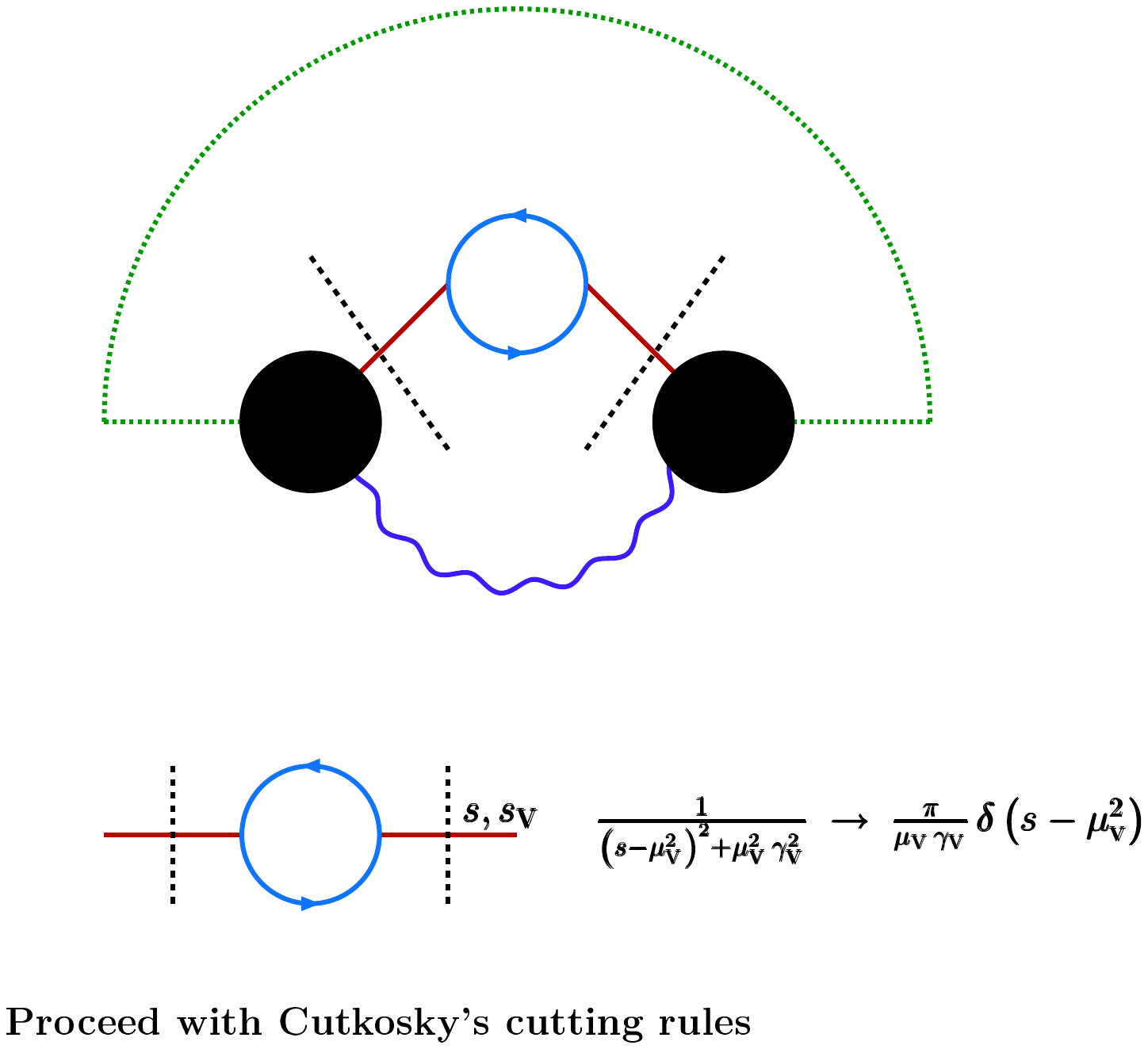}
 \vspace{-5.cm}
 \caption{PV cuts defining the sub-processes $\h \to Z \gamma$ and
$Z \to \bar{f} f$; $\mu_V$ and $\gamma_V$ parametrise the $V = Z$ complex pole.} \label{fig:PVcut} 
\end{centering}
\end{figure}

The $\mathrm{K}$ diagram for the single ($Z$) resonant part of $\h \to \bar{f} f \gamma$ is
shown in~\cref{fig:Kdia}.
The PO definition of $\h \to Z \gamma$ is based on~\cref{fig:PVcut}, illustrating the PV cut
(not to be confused with Cutkosky's cutting rules).

Consider the process $q q \to \bar{f}_1 f_1 \bar{f}_2 f_2 j j$, according to the structure of 
the resonant poles we have different options in extracting physical POs, e.g.
\begin{equation}
\sigma(q q \to \bar{f}_1 f_1 \bar{f}_2 f_2 jj) \stackrel{PO}{\longmapsto}
\sigma(q q \to h jj)\,\mathrm{Br}(\h \to Z \bar{f}_1 f_1)\,
\mathrm{Br}(Z \to \bar{f}_2 f_2) \spc
\end{equation}
\begin{equation}
\sigma(q q \to \bar{f}_1 f_1 \bar{f}_2 f_2 jj) \stackrel{PO}{\longmapsto}
\sigma(q q \to  Z Z jj)\,\mathrm{Br}(Z \to \bar{f}_1 f_1)\,
\mathrm{Br}(Z \to \bar{f}_2 f_2)  \spp
\end{equation}
There are fine points when factorising a process into physical sub-processes 
(POs): extracting the $\delta$ from the (squared) propagator is not enough. 
Consider an amplitude that can be factorised as follows:
\begin{equation}
\mathrm{A} = \mathcal{A}^{(1)}_{\mu}\,\Delta_{\mu\nu}(p)\,\mathcal{A}^{(2)}_{\nu} \spc
\end{equation}
where $\Delta_{\mu\nu}$ enters the propagator for a resonant, spin-$\,1$ particle.
We would like to replace, using conserved currents,
\begin{equation}
\Delta_{\mu\nu} \to \frac{1}{s - s_{c}}\,\sum_{\lambda}\,
\epsilon_{\mu}(p,\lambda)\,\epsilon^*_{\nu}(p,\lambda) \spc
\end{equation}
where $s_{c}$ is the complex pole and $\epsilon_{\mu}$ are spin-$\,1$ polarization vectors.
What we obtain is
\begin{equation}
\mid \mathrm{A} \mid^2 = \frac{1}{\mid s - s_{c} \mid^2}\,\mid \sum_{\lambda}\,
\Bigl[ \mathrm{A}^{(1)}\,\cdot\,\epsilon(p,\lambda) \Bigr]\,
\Bigl[ \mathrm{A}^{(2)}\,\cdot\,\epsilon^*(p,\lambda) \Bigr]\,\mid^2 \spc
\end{equation}
which means that we do not have what we need,
\begin{equation}
\sum_{\lambda} \mid \mathrm{A}^{(1)}\,\cdot\,\epsilon(p,\lambda) \,\mid^2\,
\sum_{\sigma}  \mid \mathrm{A}^{(2)}\,\cdot\,\epsilon(p,\sigma)  \,\mid^2 \spp
\end{equation}
Is there a solution? If cuts are not introduced, the interference terms among different
helicities oscillate over the phase space and drop out, i.e. we achieve factorisation,
see Ref.~\cite{Uhlemann:2008pm}. In any case, the effects of cuts can be computed.

\begin{figure}[t]
\begin{centering}  
 \includegraphics[scale=0.8]{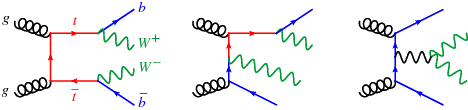}
 \caption{Sample of diagrams contributing to $gg \to WWbb$.}
\label{fig:twb} 
\end{centering}
\end{figure}

Another example will be the following:
\[
\left.
\begin{array}{ll}
\bar{t}\,t                          & \mbox{DR part of} \\
t (\bar{t})\,W^- (W^+)\,\bar{b} (b) & \mbox{SR part of} \\
\end{array}
\right\}
W\,b\,W\,b \quad \mbox{production}
\]
where DR stands for doubly-resonant and SR for single-resonant.
Our proposal is: do not ``kill'' diagrams,
write the $\mathrm{K}\,$-diagrams corresponding to~\cref{fig:twb}
for $W\,b\,W\,b$ production and find the appropriate factorisation (i.e. the corresponding $\delta\,$-parts). Of course, having few accessible discriminating kinematic variables, low statistics and high background is always a problem.

Summarising: the MPE should be understood as an ``asymptotic expansion'', not as a \ac{NWA}. The phase space decomposition is obtained by using the two parts in the propagator expansion: the $\delta\,$-term is what we need to reconstruct POs, the PV-term (understood as a distribution) gives the remainder and POs are extracted without making any approximation. It is worth noting that, in extracting POs, analytic continuation (on-shell masses into complex poles)  is performed only after integrating over residual variables, as done in Ref.\cite{Goria:2011wa}.

Finally, let us stress again that the extraction of POs comes with a remnant of the underlying theory, e.g. at LEP there was a SM-remnant which, however, was shown to be negligible. At LHC this should be examined with great care, process-by-process and PO-by-PO.

\clearpage
\section{Tools and phenomenological Lagrangian}
\label{sec:tools}
In~\cref{sec:kappa} we discussed the need for an upgrade of the \kappaframework to other frameworks with a more solid theoretical basis. However, for frameworks such as those discussed in~\cref{sec:fiducial,sec:EFT,sec:POs} to be usable in practice, it is fundamental that they are implemented in the software tools that are capable of interfacing theory and experiment. 

Such tools  should provide predictions for observables as a function of all relevant input parameters as well as some way of estimating the uncertainty from theory and from phase space effects, such as different acceptances/efficiencies compared to the SM-only predictions used in experimental analyses.

The goal of this section is to provide an overview of some of the available Monte Carlo event generators, and the theoretical basis that underpins automatic computations, highlighting some aspects related to the choice of phenomenological models and \acp{EFT}.
\subsection{Theory: The quest for higher accuracy}
With the growing precision of experimental analysis, there is a need to increase the accuracy of the corresponding theoretical predictions. So far, there has been a concerted effort to automate the \ac{NLO} \ac{QCD} corrections and this is the current standard that has been implemented in a number of
\ac{MC} generators, some of which are discussed below.

In addition, the \ac{NLO} \ac{EW} corrections are known for many processes, for example in Refs.\cite{Biedermann:2016lvg,Dittmaier:2017bnh}, while the \ac{QCD}
corrections for a range of $2 \to 2$ scattering processes are now known at \ac{NNLO} and  two Higgs production channels have reached  
N3LO precision, see Refs.\cite{Anastasiou:2015ema,Anastasiou:2016cez,Dreyer:2016oyx}.  Given the considerable
effort that is required to reach this accuracy, and \ac{NNLO} (even \ac{NLO} \ac{EW}) computations are far from being automatized.   

\begin{figure}[b]
\[
\parbox{80pt}{
\begin{picture}(80,80)(0,0)
\DashLine(0,40)(40,40){3}
\Vertex(40,40){1}
\ArrowLine(80,0)(60,20)
\ArrowLine(60,20)(40,40)
\ArrowLine(40,40)(60,60)
\ArrowLine(60,60)(80,80)
\Gluon(60,60)(60,20){3}{3}
\Vertex(60,20){1}
\Vertex(60,60){1}
\end{picture}
}
\quad
\raisebox{-1.5pt}{\scalebox{1.5}{$\times$}}
\quad
\parbox{80pt}{
\begin{picture}(80,80)(0,0)
\DashLine(0,40)(40,40){3}
\Vertex(40,40){1}
\ArrowLine(80,0)(40,40)
\ArrowLine(40,40)(80,80)
\end{picture}
}
\quad
\raisebox{-1.5pt}{\scalebox{1.5}{+}}
\quad
\parbox{80pt}{
\begin{picture}(80,80)(0,0)
\DashLine(0,40)(40,40){3}
\Vertex(40,40){1}
\ArrowLine(80,0)(40,40)
\ArrowLine(40,40)(60,60)
\ArrowLine(60,60)(80,80)
\Gluon(60,60)(80,40){3}{3}
\Vertex(60,60){1}
\end{picture}
}
\quad
\raisebox{-1.5pt}{\scalebox{1.5}{$\times$}}
\quad
\parbox{80pt}{
\begin{picture}(80,80)(0,0)
\DashLine(0,40)(40,40){3}
\Vertex(40,40){1}
\ArrowLine(80,0)(40,40)
\ArrowLine(40,40)(60,60)
\ArrowLine(60,60)(80,80)
\Gluon(60,60)(80,40){3}{3}
\Vertex(60,60){1}
\end{picture}
}
\]
  \caption{Virtual loop contribution and real emission diagrams included in a 
           typical \ac{NLO} \ac{QCD} computation.}
  \label{fig:QCDNLO}
\end{figure}
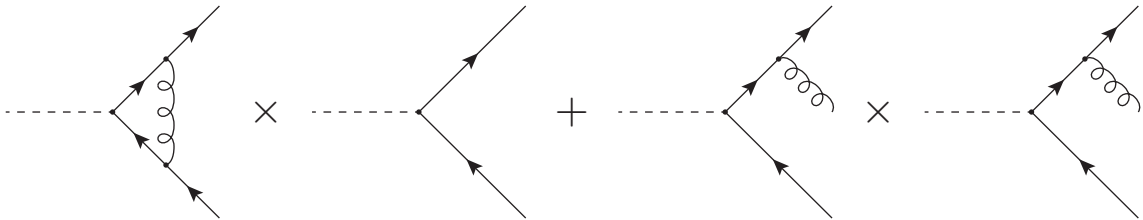

When considering higher-order contributions, \ac{UV} and \ac{IR} divergences, arising from virtual and real emission diagrams (see~\cref{fig:QCDNLO}) must be isolated and either removed through the renormalisation procedure (\ac{UV}) or explicitly cancelled (\ac{IR}). 
The \ac{UV} divergences are cured by using divergent renormalisation counterterms so that the singularities are absorbed into the bare parameters.  The counterterms are model-specific, and the
renormalisation (especially when considering \ac{EW} corrections) has to be implemented on
a case-by-case basis.

The \ac{IR} and collinear divergences typically cancel after performing the phase-space
integration. 
In  \ac{QED} and \ac{QCD}, the Kinoshita-Lee-Nauenberg theorem \cite{Kinoshita:1962ur,Lee:1964is} ensures \ac{IR}-finiteness of physical quantities. However,  it has not yet been understood how this would work in full generality for an \ac{EFT} (for example the computation of  exclusive \ac{QCD} jet observables at higher orders requires a method for the subtraction  of \ac{IR} singular configurations arising from multiple radiation of real partons).

\subsubsection*{SMEFT in the IR}
As an example, we derive the \ac{NLO} partial width for the decay
$Z \to \bar l l (\gamma)$, considering the \ac{QED}-like operators of the Warsaw basis 
(i.e. involving photon fields), after the proper \ac{UV} 
renormalisation, done for the SMEFT in Refs.~\cite{Hartmann:2015oia,Hartmann:2015aia,Ghezzi:2015vva,Passarino:2016pzb,
Passarino:2016owu}, the LO amplitude for $Z \to ll$ reads as follows:
\begin{equation}
\mathrm{A}_{\mu} = g\,\mathrm{A}^{(4)}_{\mu} + g g_{_6}\,\mathrm{A}^{(6)}_{\mu} \spc
\end{equation}
where the $\mathrm{dim} = 6$ part can be written as
\begin{equation}
\mathrm{A}^{(6)}_{\mu} = \frac{1}{4}\,\gamma_{\mu}\,\bigl(V_l + A_l\,\gamma^5 \bigr) \spc
\end{equation}
\begin{equation}
\begin{split}
 V_l &=
  \frac{s_\theta^2}{c_\theta} \left( 4 s_\theta^2 - 7 \right) a_{AA}
  + c_\theta \left( 1 + 4 s_\theta^2 \right) a_{ZZ}
  + s_\theta \left( 4 s_\theta^2 -3 \right) a_{AZ}
  \\ & \qquad
  + \frac{1}{4 c_\theta} \left( 7 - s_\theta^2 \right) a_{\phi D}
  + \frac{2}{c_\theta} a_{\phi l V} \spc
\end{split}
\end{equation}
\begin{equation}
A_l = \frac{s^2_{\theta}}{c_{\theta}}\,a_{AA} + c_{\theta}\,a_{ZZ} + s_{\theta}\,a_{AZ} -
\frac{1}{4\,c_{\theta}}\,a_{\phi\,D} + \frac{2}{c_{\theta}}\,a_{\phi l\,A} \spc
\end{equation}
with $a_{\phi l\,A} - a_{\phi l\,V} = 2\,a_{\phi l}$ and $a_{\phi l\,A} + a_{\phi l\,V} = 2\,(a^{(3)}_{\phi l} - a^{(1)}_{\phi l})$.
The \ac{IR}/collinear part of the one-loop virtual and real emission corrections 
take respectively the forms
\begin{equation}
\begin{split}
\left. \Gamma(Z \to \bar l l) \right|_\text{div} &= 
- \frac{g^4}{384 \pi^3} M_Z s_\theta^2 \mathcal{F}^\text{virt} 
\biggl[ \Gamma^{(4)}_0 \bigl( 1 + g_{_6} \Delta\!\Gamma \bigr) + g_{_6} \Gamma^{(6)}_0 \biggr],
\\
\left. \Gamma(Z \to \bar l l \gamma) \right|_\text{div} 
&= \frac{g^4}{384 \pi^3} M_Z s_\theta^2 \mathcal{F}^\text{real} 
\biggl[ \Gamma^{(4)}_0 \bigl( 1 + g_{_6} \Delta\!\Gamma \bigr) + g_{_6} \Gamma^{(6)}_0 \biggr] \spc
\end{split}
\end{equation}
where \ac{IR}/collinear divergences are contained in $\Gamma^{(4)}_0$ and $\Gamma^{(6)}_0$, 
$\mathcal{F}^\text{virt}$ and $\mathcal{F}^\text{real}$ are squared matrix elements and 
$\Delta\!\Gamma$ contains contributions from $\mathrm{dim} = 6$ operators. Adding up these 
contributions and defining the linear combination of Wilson coefficients,
the final result for the decay width is an \ac{IR}/collinear-free quantity, given by
\begin{equation}
 \Gamma^1_\text{QED} = 
\frac{3 \alpha}{4 \pi} \frac{\Gf M_Z^3}{24 \sqrt{2} \pi} \Bigl[ (v_l^2 + 1)\, 
\biggl( 1 + g_{_6} \delta_\text{QED}^{(6)} \biggr) + g_{_6} \Delta_\text{QED}^{(6)} \Bigr] \spc
\label{eq:deconv}
\end{equation}
where $v_l = 1 - 4\,s^2_{\theta}$, $\delta_\text{QED}^{(6)}$, $\Delta_\text{QED}^{(6)}$ are driven by the presence 
of $\mathrm{dim} = 6$ operators and can be cast in the following form:
\begin{equation}
 \begin{split}
  \delta_\text{QED}^{(6)} &= 2 \left( 2 - s_\theta^2 \right) a_{AA} + 2 s_\theta^2 a_{ZZ} + 
   2\,\frac{c_\theta^3}{s_\theta}\,a_{AZ} - \frac{1}{2} \frac{c_\theta^2}{s_\theta^2} a_{\phi D} \spc 
  \\
  \Delta_\text{QED}^{(6)} &= \left( 1 - 6 v_l - v_l^2 \right) \frac{1}{c_\theta^2} 
   \biggl( s_\theta a_{AA} - \frac{1}{4} a_{\phi D}  \biggr)
  + \left( 1 + 2 v_l - v_l^2 \right) \biggl( a_{ZZ} + \frac{s_\theta}{c_\theta} a_{AZ} \biggr)
  \\ & \qquad  
  + \frac{512}{13} v_l a_{AZ} 
  + \frac{2}{c_\theta^2} \left( a_{\phi l A} + v_l a_{\phi l V} \right) \spp
\label{eq:smeftqed}
 \end{split}
\end{equation}
This result is important, not only for extending IR/collinear finiteness to the SMEFT but also 
because it shows that higher dimensional operators enter everywhere: signal, background and radiation. The latter is particularly relevant when one wants to include (SM-deconvoluted) EW precision observable constraints in a fit to Higgs data. Since LEP POs are (mostly) SM-deconvoluted\footnote{There are several levels of deconvolution and the most important are:  single-deconvolution (SD), giving the kernel cross-sections without initial state QED radiation, but including all final-state correction factors; double-deconvolution (DD), giving the
kernel cross-sections without initial- and final-state QED radiation and without any final-state QCD radiation}, the effect of $\mathrm{dim} = 6$ operators on the deconvolution procedure should be checked carefully.

In~\cref{eq:deconv} $\Gamma_0 = \Gf M_Z^3/(24 \sqrt{2} \pi) = 82.945(7)\,\MeV$ is the standard
partial width and the LEP definition is
\begin{equation}
\Gamma^0_l = \frac{\Gamma_l}{1 + \frac{3}{4}\,\frac{\alpha(M^2_Z)}{\pi}} \spp
\end{equation}
Once again, fitting $\Gamma^0_l$ as reported at LEP with the SMEFT is not consistent; note that
in~\cref{eq:smeftqed} there are three PTG coefficients (i.e. not suppressed), $a_{\phi\,D}$ and
$a_{\phi l\,V}, a_{\phi l\,A}$.

\subsection{Phenomenological Lagrangians}
\label{sec:phenoLagSubsection}
Since we do not know the structure of the next \ac{SM}, it is necessary to have a systematic way to look for deviations in the experimental data. Also, the moment we observe a deviation, it will be important to have the technology to interpret it properly. 

There are are two distinguished steps in this procedure:
(a) the observation and (b) the interpretation of the deviation. In order to address (a), 
an approximate framework can be sufficient, but to move forward and interpret successfully the outcome of the experiments, a more solid and rigorous framework (for instance an \ac{EFT}) 
is necessary. 

We consider here \textit{phenomenological models}, intended as models that are not completely rigorous from the theoretical point of view, but that can be used to guide searches.  Indeed, fully consistent models like \acp{EFT} often require a large number of parameters, that makes them impractical.  With \textit{phenomenological Lagrangian} we refer here to the Lagrangian related to
a \textit{phenomenological model}. Because of their approximate nature, these Lagrangians are not adequate to provide \ac{NLO} results.

In the literature it is possible to find many phenomenological models, but it is not always  clear how to distinguish between these models (conceived to address point (a)), from 
theoretically motivated models (which are essential to address (b)). In order to  look for deviations in a particular observable, phenomenological Lagrangians usually consist  of the \ac{SM} Lagrangian extended by some effective interactions selected from a basis of higher-dimensional operators. Depending on the procedure used to select such additional  interactions, the 
obtained model will be also usable to address (b). 

It is important to note that applying a certain field transformations to an \ac{EFT} basis some operators might drop out, and the basis obtained in this way should give the same physics as before. It was shown  in Refs.~\cite{Chisholm:1961tha,Kallosh:1972ap} that this is in principle true, but only at the level of the (renormalised) $S$-matrix, and not at the Lagrangian level. Hence, a field can be transformed using  linear or non-linear transformations (even non-local, at the price of introducing ghosts) and the $S$-matrix (hence the predictions for the physical quantities) will not change. 

Nevertheless, as the physics is encoded in the $S$-matrix and not in the Lagrangian, it is not straightforward to get the same physics from the transformed Lagrangian. This problem was raised in Ref.~\cite{Passarino:2016saj} (see also Ref.~\cite{Brivio:2017bnu}), where it is shown in 
an example that when a generic (non-gauge) transformation is applied to a field, it is necessary to adjust the external  wave-function renormalisation factors in order to recover the original $S$-matrix from the transformed Lagrangian.

In other words, starting from an \ac{EFT} Lagrangian $\mathcal{L}_\text{EFT}$, in general  we cannot transform fields and pretend to use the transformed Lagrangian 
$\mathcal{L}'_\text{EFT}$ in a simple way as we would do with $\mathcal{L}_\text{EFT}$. If the fields are transformed with a non-gauge transformation, one would get different physical predictions from $\mathcal{L}_\text{EFT}$ and $\mathcal{L}'_\text{EFT}$.

In conclusion, we should ask ourselves if it is worth applying non-linear field 
transformations to transform away unpleasant operators from a basis: if this is done consistently, the result for a physical quantity will not change (but its computation will be cumbersome); if this is not done with care, the risk is to downgrade a rigorous theory to a phenomenological model, not suitable to inspect the nature of deviations.
\subsubsection*{Examples of phenomenological models}
To clarify the concept of phenomenological models, in the following we give some simple  examples. In the first, we explain how to build a phenomenological Lagrangian that provides, 
at \ac{LO}, rescaled \ac{SM} production and decay amplitudes as in the \kappaframework. 
The second example is provided by the Higgs Characterization framework as it was proposed 
in Ref.~\cite{Artoisenet:2013puc}.
\subparagraph*{\ac{SM} with rescaled Higgs couplings}
Let us consider the \kappaframework which (as explained in~\cref{sec:kappa}) was clearly proposed to address a type-(a) problem. A phenomenological Lagrangian for the framework can be obtained from the \ac{SM} Lagrangian (supplemented by effective, \ac{SM}-valued gluon-gluon-Higgs and photon-photon-Higgs couplings), with all the Higgs couplings modified by the introduction of multiplicative coefficients accordingly to~\cref{tab:kappaCouplings}. 

Indeed, using this Lagrangian to compute the \ac{LO} production 
cross sections and Higgs partial decay widths, it is possible to obtain the defining relations for the coupling scale factors given in Table 2 of 
Ref.~\cite{LHCHiggsCrossSectionWorkingGroup:2012nn}.
\begin{table}[h!]
\[
\begin{array}{cc}
\toprule
\text{\ac{SM} coupling} & \text{Rescaling factor} \\
\midrule
 W W \h & \kappa_W \\
 Z Z \h & \kappa_Z \\
 t t \h & \kappa_t \\
 b b \h & \kappa_b \\
 g g \h \,\text{(effective)} & \kappa_g \\
 A A \h \,\text{(effective)} & \kappa_\gamma \\
 \tau \tau \h & \kappa_\tau \\
\bottomrule
\end{array}
\]
\caption{Coupling factors in the \kappaframework, as proposed 
in Ref.~\cite{LHCHiggsCrossSectionWorkingGroup:2012nn} (see Table 2).}
\label{tab:kappaCouplings}
\end{table}
\subparagraph*{Higgs Characterization framework}
The Higgs characterization framework from Ref.\cite{Artoisenet:2013puc} can be thought of as the 
\ac{SM}, with the Higgs boson replaced by a resonance $X_J$ of mass $125\,\GeV$, where for the sake of generality the spin can take the values $J=0$, $1$ or $2$, supplemented by interaction terms coming from $\mathrm{dim} = 6$ operators. These terms come from hypothetical 
interactions of the Higgs-like resonance $X_J$ with a new heavy sector, which resides at 
the scale $\Lambda$. 

In the framework, all the $\mathrm{dim} = 6$ operators which do not 
include the scalar field $X_J$ are neglected. Moreover, all the interaction terms involving 
more than one $X_J$ field, or more than three fields are neglected. This leads to a very compact phenomenological Lagrangian, but one has to keep in mind that the predictions worked out within the framework are applicable consistently only when working at \ac{LO} with production or decay processes involving one and only one $X_J$ particle. 

The phenomenological Lagrangian for the case $J = 0$ consists, as given in the main reference, in
\begin{equation}
 \Lag_{\text{HC},0} = \Lag_{\SMm-\h} + \Lag_0 \spc
\end{equation}
where the fermionic interactions of the field $X_0$, related to the mass eigenstates of the $125\, \GeV$ resonance, are given by the Lagrangian
\begin{equation}
\label{eq:higgsCharFermion}
 \Lag_0^f = - \sum_{f = t,b,\tau} \bar \psi_f \left( c_\alpha \kappa_{\h ff} g_{\h ff} + 
i s_\alpha \kappa_{Aff} g_{Aff} \gamma_5 \right) \psi_f X_0 \spc
\end{equation}
and interaction terms with vector bosons are gathered in $\Lag_0^V$:
\begin{equation}
 \begin{split}
  \Lag_0^V =& \biggl\{
            c_\alpha \kappa_\SMm \biggl[ \frac{1}{2} g_{\h ZZ} Z_\mu Z^\mu + g_{\h WW} W_\mu^+ W^{- \mu} \biggr]
            \\ \quad &
            - \frac{1}{4} \biggl[ c_\alpha \kappa_{\h \gamma \gamma} g_{\h \gamma \gamma} A_{\mu \nu} A^{\mu \nu} + s_\alpha \kappa_{A \gamma \gamma} g_{A \gamma \gamma} A_{\mu \nu} \widetilde A^{\mu \nu} \biggr]
            \\ \quad &
            - \frac{1}{2} \biggl[ c_\alpha \kappa_{\h Z \gamma} g_{\h Z \gamma} Z_{\mu \nu} A^{\mu \nu} + s_\alpha \kappa_{A Z \gamma} g_{A Z \gamma} Z_{\mu \nu} \widetilde A^{\mu \nu} \biggr]
            \\ \quad &
            - \frac{1}{4} \biggl[ c_\alpha \kappa_{\h g g} g_{\h g g} G^a_{\mu \nu} G^{a,\mu \nu} + s_\alpha \kappa_{A g g} g_{A g g} G^a_{\mu \nu} \widetilde G^{a,\mu \nu} \biggr]
            \\ \quad &
            - \frac{1}{4} \frac{1}{\Lambda} \biggl[ c_\alpha \kappa_{\h Z Z} Z_{\mu \nu} Z^{\mu \nu} + s_\alpha \kappa_{A Z Z} Z_{\mu \nu} \widetilde Z^{\mu \nu} \biggr]
            \\ \quad &
            - \frac{1}{2} \frac{1}{\Lambda} \biggl[ c_\alpha \kappa_{\h W W} W^+_{\mu \nu} W^{-\mu \nu} + s_\alpha \kappa_{A W W} W^+_{\mu \nu} \widetilde W^{-\mu \nu} \biggr]
            \\ \quad &
            - \frac{1}{\Lambda} c_\alpha \biggl[ 
  \kappa_{\h \partial \gamma} Z_\nu \partial_\mu A^{\mu \nu} 
+ \kappa_{\h \partial Z} Z_\nu \partial_\mu Z^{\mu \nu} 
+ \bigl( \kappa_{\h \partial W} W_\nu^+ \partial_\mu W^{- \mu \nu} + \text{h.c.} \bigr) \biggr]
            \biggr\} X_0 \spp
 \end{split}
\end{equation}
Taking into account the fact that $X_0$ is the physical Higgs field, this phenomenological 
Lagrangian is very compact compared with a full \ac{EFT}. This is due to the selection of terms explained above. For instance, consider the fermion-Higgs operators of the Warsaw basis given in Table 2 of Ref.~\cite{Grzadkowski:2010es}: among all the fermion-Higgs interactions, only terms coming from the operators reported in the upper-right sector of the table (labelled as $\psi^2 \varphi^3$) survive the selection. The terms labelled as $\psi^2 X \varphi$ 
give four-point Higgs interactions and are neglected. In the operators $\psi^2 \varphi^2 D$, in order to have only three-point interactions involving one Higgs field, one doublet is replaced by the Higgs \ac{VEV} and the covariant derivative is replaced by the partial derivative, giving a vanishing contribution:
\begin{equation}
\begin{split}
\biggl( \varphi^\dagger i 
\overset{\leftrightarrow}{D}_\mu \varphi \biggr) \bigl( \bar \psi \gamma^\mu \psi \bigr)
&\longrightarrow
i \biggl( v 
\overset{\leftrightarrow}{D}_\mu X_0 + 
X_0 \overset{\leftrightarrow}{D}_\mu v \biggr) \bigl( \bar \psi \gamma^\mu \psi \bigr), \\
&\longrightarrow
i \biggl[ v \biggl(\partial_\mu X_0 \biggr) - 
v \biggl(X_0 \overset{\leftarrow}{\partial}_\mu \biggr) \biggr] 
\bigl( \bar \psi \gamma^\mu \psi \bigr) = 0 \spp
\end{split}
\end{equation}
In a similar way also the operator $Q_{\varphi u d}$ is killed by the selection and there are no
terms involving derivatives of $X_0$ in the Lagrangian $\Lag_0^f$ reported
in~\cref{eq:higgsCharFermion}.
\subsection{Tools}
The aim of the first part of this review was to present and discuss some of the 
techniques that can be used to observe and interpret possible deviations that might show up 
at the \ac{LHC} in the next years of data taking. 

Regarding the \acp{EFT} and the 
phenomenological models, we addressed how Lagrangians, that are suitable to understand 
possible \ac{NP} effects, can be built. Of course, as our practical way to observe 
elementary phenomena is through detector experiments, Lagrangians are not the full story. 

\begin{figure}
\begin{center}
\includegraphics[scale=0.5]{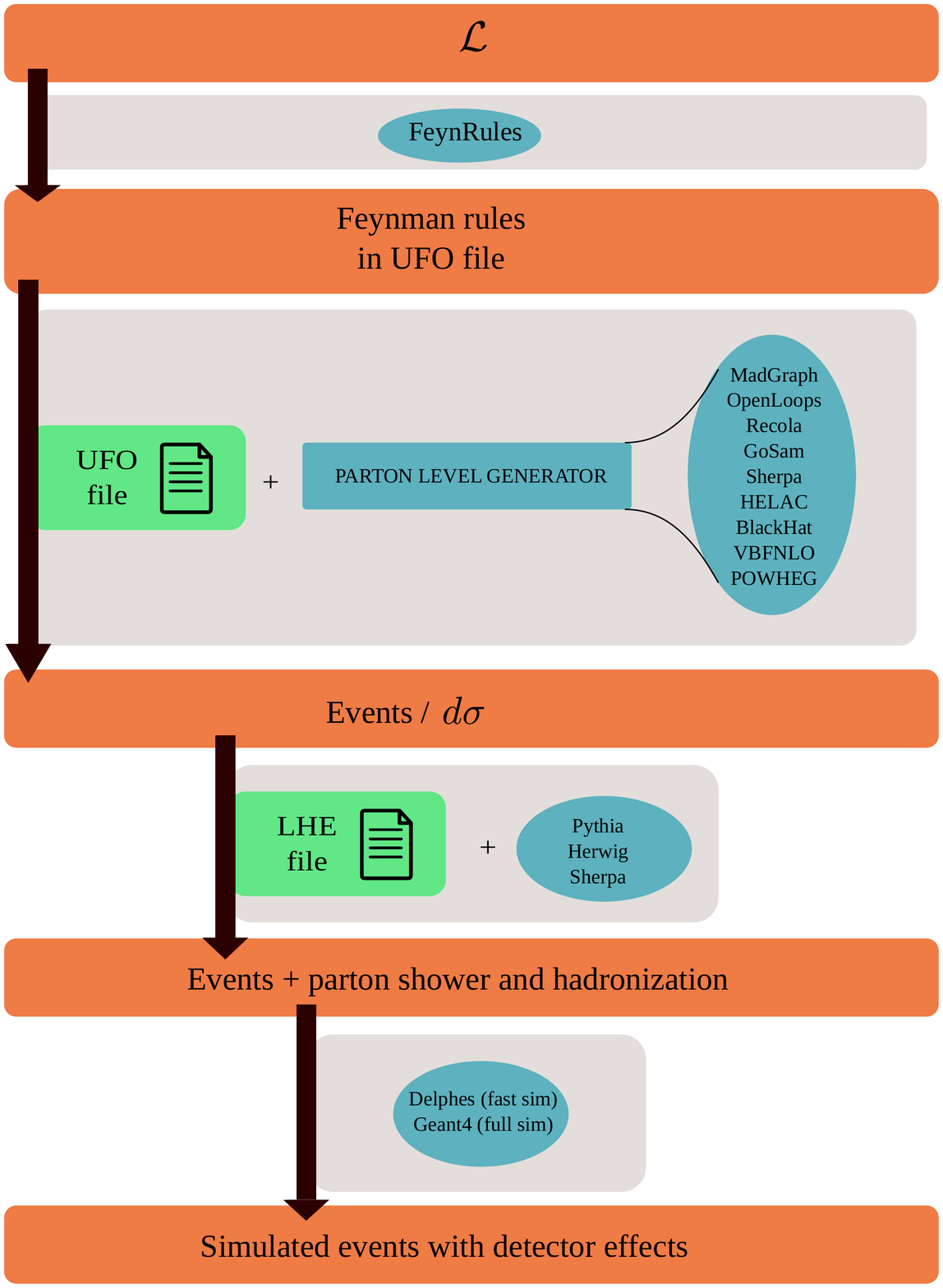}
\end{center}
\caption{Work flow for the simulation of events using \ac{MC} generators.}
\label{fig:simulationWorkFlow}
\end{figure}

To make a connection with the collected data it is necessary to have control of the hard scattering process, and also, we must be able to have a full simulation of the processes that we observe in the experiments. In~\cref{fig:simulationWorkFlow} we present schematically the steps that are necessary to have a full simulation, starting from the Lagrangian of the considered theory, and we report some of the available tools.
\begin{table}
\begin{center}
\scalebox{0.86}{
\begin{tabular}{lccll}
\toprule
Tool & LO & \ac{NLO} \ac{QCD} & Web page & References \\
\midrule
MadGraph	& $\checkmark$ & $\checkmark$ & \url{http://madgraph.physics.illinois.edu/} & \cite{Alwall:2014hca} \\
OpenLoops	&              & $\checkmark$ & \url{https://openloops.hepforge.org/} & \cite{Cascioli:2011va} \\
Recola		&              & $\checkmark$ & \url{https://recola.hepforge.org/} & \cite{Actis:2012qn,Actis:2016mpe,Denner:2014gla,Denner:2010tr,Denner:2005nn,Denner:2002ii} \\
GoSam		&              & $\checkmark$ & \url{https://gosam.hepforge.org/} & \cite{art:Cullen:2011ac,art:Cullen:2014yla} \\
Sherpa		& $\checkmark$ & & \url{https://sherpa.hepforge.org/trac/wiki} & \cite{Gleisberg:2003xi,Gleisberg:2008ta} \\
HELAC		& $\checkmark$ & $\checkmark$ & \url{http://helac-phegas.web.cern.ch/helac-phegas/} & \cite{Kanaki:2000ey,Papadopoulos:2000tt,Cafarella:2007pc} \\
\midrule
Tool & \multicolumn{2}{l}{Process} & Web page & References \\
\midrule
BlackHat	& \multicolumn{2}{l}{Multi-jet processes} & \url{https://blackhat.hepforge.org/} & \cite{art:Berger:2008sj} \\
VBFNLO		& \multicolumn{2}{l}{Vector boson fusion} & \url{https://www.itp.kit.edu/vbfnlo/wiki/doku.php} & \cite{Arnold:2008rz,Arnold:2011wj,Baglio:2014uba} \\
POWHEG		& \multicolumn{2}{l}{Various (see web page)} & \url{http://powhegbox.mib.infn.it/} & \cite{art:NasonPowheg2004,Frixione:2007vw,Alioli:2010xd} \\
\bottomrule
\end{tabular}
}
\end{center}
\caption{Web pages and references for some widely-used parton level generators available 
for automatic calculations.}
\label{tab:MCgeneratorList}
\end{table}

From the starting Lagrangian, the Mathematica package FeynRules, from Ref.~\cite{Alloul:2013bka} can be used to derive the Feynman rules for the theory. These can be exported in a \ac{UFO} file, as defined in Ref.~\cite{Degrande:2011ua} which, together with a parton level generator 
(see~\cref{tab:MCgeneratorList} for more details), can be used to generate events for the considered model.

The output of a parton level generator can be a differential distribution, but the generated events can also be stored in a \ac{LHE} file, a format that was introduced in Refs.~\cite{Alwall:2006yp,Butterworth:2010ym}, which in 
turn can be used to simulate showers and hadronisation of the final states using a 
dedicate software like Pythia, from Refs.~\cite{Sjostrand:2006za,Sjostrand:2014zea},  
Herwig, from Refs.\cite{Bahr:2008pv,Bahr:2008tf,Bahr:2008tx,
Bellm:2015jjp,Bellm:2017bvx,Gieseke:2011na,Arnold:2012fq,Bellm:2013hwb} or Sherpa, from Refs.\cite{Gleisberg:2003xi,Gleisberg:2008ta}). 

Further, detector effects can be simulated 
using Delphes, described in Ref.~\cite{deFavereau:2013fsa} or more commonly,
Geant4, from Refs.~\cite{Allison:2006ve,Agostinelli:2002hh,Allison:2016lfl}, which can be used respectively for fast (with simple smearing effects) or full (but time -consuming) simulations.
\subsection{Connection between $\kappa$s, POs and Wilson coefficients}\label{subsec:PO_k_Wilson}
To finalise, in this section the connection between $\kappa$s, \acp{PO} and Wilson coefficients is studied. An  example is given showing how Wilson coefficients collapse into a smaller number of more 
general objects which we call physical \acp{PO}, see Sect II.1.8 of Ref.~\cite{deFlorian:2016spz}. Then we study the strategy to simultaneously extract the values of such parameters from data, via a likelihood minimisation. Technical issues arise at this point, given the high dimensionality of the theory parameters space.
When going from \ac{SM} to \ac{SMEFT} the scattering amplitudes are modified as
\begin{equation}
\begin{split}
\mathcal{A}_{\rm SMEFT}^{\rm LO} = \sum_{i = 1,n} \mathcal{A}_{\rm SM}^{(i)} + ig_{_6}\kappa_c
\mathcal{A}_{\rm SMEFT}^{NLO} = 
\sum_{i = 1,n} \kappa_i \mathcal{A}_{\rm SM}^{(i)} + 
ig_{_6}\kappa_c + g_{_6} \sum_{i = 1,N} {\rm a}_i \mathcal{A}_{\rm n\!f}^{(i)} \spc
\end{split}
\end{equation}
with $g_{_6} = \left( \sqrt{2}G_F\Lambda^2 \right)^{-1}$.
The last term contains all the non-factorisable loop contributions. 
In there, we find the Wilson coefficients ${\rm a}_i$, and the factors $\kappa_i$ are linear 
combinations of the ${\rm a}_i$.

Now, to understand how to connect \acp{PO} with Wilson coefficients let us consider 
the process $h(P) \to \gamma_{\mu}(p_1)\gamma_{\nu}(p_2)$. The corresponding amplitude can be written as
\begin{equation}
\mathcal{A}_{\rm h\!A\!A}^{\mu \nu} = i \mathcal{I}_{\rm h\!A\!A}T^{\mu \nu} = 
-i \frac{2}{v_{F} M_{h}^{2}} \epsilon_{\gamma \gamma} T^{\mu \nu} \spc
\end{equation}
where $\epsilon_{\gamma \gamma}$ is a \ac{PO} which is based on quantities that can be 
extracted from Green's functions. This can be conveniently rewritten as
\begin{equation}
\mathcal{I}_{\rm h\!A\!A} = 
\frac{g_{R}^{3} s_{W}^{2}}{8 \pi^{2}} \sum_{I = W,t,b} 
\rho_{I}^{\rm h\!A\!A} \mathcal{I}_{\rm h\!A\!A;L\!O}^{I} + 
g_{F}g_{6} \frac{M^{2}_h}{M_W} {\rm a_{\rm \!A\!A}} + 
\frac{g^{3}_F g_{_6}}{\pi^{2}}\mathcal{I}_{\rm h\!A\!A}^{\rm n\!f} \spp
\end{equation}
Introducing $g_F^2 = 4 \sqrt{2} G_F M_W^2$ and $c_W = M_W / M_Z$ and deriving
\begin{equation}
\begin{split}
& \kappa_I^{\rm h\!A\!A}  =\frac{g_F^3 s^2_W}{8 \pi^2} \rho_I^{\rm h\!A\!A} \spc  \qquad \kappa_c^{\rm h\!A\!A} = g_F \frac{\mh^2}{M_W} {\rm a_{\rm A\!A}} \\
\end{split}
\end{equation}
and use \cref{rela}
where ${\rm a_{ij}}$ are Wilson coefficients. The $\rho$ factors are process dependent and are given by
\begin{equation}
\rho_{\rm I}^{\rm proc} = 1 + g_{_6} \Delta\rho_{\rm I}^{\rm proc} \spp
\end{equation}
In addition to this there are other non-factorizable contributions.
This means that the $\kappa$ factors can also be introduced at loop level, as combination 
of Wilson coefficients. But to do so it is needed to introduce process dependent and 
non-factorizable contributions, as described in Ref.~\cite{Ghezzi:2015vva}.
Therefore at LO we can see the relation between the intermediate \ac{PO} 
$\epsilon{\gamma \gamma}$ and a Wilson coefficient ${\rm a}_{AA}$:
\begin{equation}
\begin{split}
& \epsilon_{\gamma \gamma} = -\frac{1}{2} \frac{v_F}{m_h^2} \mathcal{I}_{{\rm h\!A\!A}}, \qquad \mathcal{I}_{{\rm h\!A\!A}} = 
\mathcal{I}_{{\rm h\!A\!A}}^{\rm SM} + g_Fg_{_6} \frac{m_h^2}{M_W} {\rm a}_{{\rm A\!A}} \spp \\
\end{split}
\end{equation}
Therefore with this formulation full consistency is achieved. 
We have some physical observable that is related to an \ac{EFT} and does not have the 
inconsistencies of the interim framework (as highlighted in~\cref{sec:kappa}).
Then the question to answer is whether or not it is possible to extract these parameters 
from data in a global and simultaneous fit. 

The main problem is the following: it is not clear if ATLAS/CMS can implement such a large number of independent parameters in the currently available fitting codes. They require speeds of recalculations with new input parameters of $<~1\,$ms on average on a single CPU. However, they could pre-compute and transfer the results of these calculations into a generalized $n\,$-dimensional polynomial which could 
then enter the experimental fits. 

Let us define for simplicity $P(X_1=\mu_1, X_2=\mu_2,\dots X_m=\mu_m)$ the probability of measuring $X_i$ with an $\mu_i$ expectation from the theoretical model (which can depend on some \acp{PO} or Wilson coefficients of a $\mathrm{dim} = 6$ EFT). The index $m$ runs over all the performed measurements, such as \ac{STXS}, as well as \acp{BR} and fiducial measurements.

For the sake of notation, let us call $X_i$ each measurement, $\sigma_i$ its error and $\rho_{ij}$ its correlation factor with another measurement $X_j$.
We also call $\mu_i$ the expected value of such a measurement. It depends on the Wilson coefficients of the underlying EFT or on \acp{PO}, namely  $\mu_i (k_1, \dots, k_p)$. Supposing each measurement follows a Gaussian distribution, which is justified by a generalized central limit theorem, the likelihood function becomes:
\begin{equation}\label{eq:prob_gaussian}
f_{X_1, \dots X_m}(\mu_1, \dots \mu_m)\propto \frac{1}{\Sigma} 
\text{exp}\left[ \left( X - \mu\right)^T\Sigma^{-1}
\left( X - \mu\right)\right], \qquad \text{with } \Sigma_{ij}=\sigma_i \sigma_j \rho_{ij} \spp
\end{equation}

All the errors and correlations between measurements are encoded in $\Sigma$.
Taking into account all \acp{STXS} bins from all stages and all branching fraction 
measurements, the order of $m$ is $\mathcal{O}(100)$.

The minimisation of such a function is challenging in many aspects. Firstly all the 
relations $\mu_i=\mu_i(k_1, \dots, k_p)$ are needed. These relations are most of the times not analytical and have to be computed through \ac{MC} simulations. The available tools which can do it were listed in~\cref{tab:MCgeneratorList}.

Various MC simulations need to be generated with different sets of input values 
for the parameters of the theoretical model implemented. From there, a continuous parametrisation of the observables of interest can be extrapolated. 
The finer the grid of MC simulation is, the more accurate is the extrapolation. If $p$ parameters are changed and each of them can take $n$ different values, the total  number of input simulations is $n^p$.

For example in $\mathrm{dim} = 6$ EFT, $p\approx59$, and in general $p \approx 50$. 
If every parameter takes $\sim 10$ values (\ac{SM} value, and some values around it), 
the number of simulations  grows up to $\approx10^{50}$ which is an unaffordable number.
 
Luckily, not all coefficients contribute to every observable, and, as explained in the example at the beginning of this section, often Wilson coefficients do combine into more general objects which are smaller in number. Typically less than $10$ coefficients enter in the calculation of a \ac{BR} or cross section.

On the other side, the number of values which parameters can take, can be minimised thanks to morphing techniques. Given an observable, polynomial function of parameters $\vec k$ (this is true if there 
is a polynomial dependence of the scattering amplitude on the theory parameters), 
such techniques allow to obtain the value of the observable in a new point of the 
parameters space, as a function of a previous computed values of the observable.
Labelling $\vec k_\text{i}$ the point of the parameters phase space where the observable 
has been calculated, it means:
\begin{equation}\label{eq:morphing}
O_{\text{output}}= 
O (\vec k_{\text{new}}) = \sum_{j=0}^{N_{\text{input}}} 
w\left(\vec k_{\text{new}}, \vec k_{\text{j}} \right) \cdot O (\vec k_{\text{j}}) \spc
\end{equation}
where $w\left(\vec k_{\text{new}}, \vec k_{\text{j}} \right)$ is a simple weight function, 
correlating the two parameters' phase space points $\vec k_{\text{new}}$ and $\vec k_{\text{j}}$.
This reduces the number of parameter variations to be simulated, making the number of MC samples to be produced to a reasonable number.

Once all the $\mu_i=\mu_i(k_1, \dots, k_p)$ relations have been obtained, the minimisation 
of the likelihood function has to be performed. 
Minimising a function requires multiple calls to it and if it is difficult to evaluate, the computing time to obtain the result can be large.
For a multivariate Gaussian distribution as the one introduced in~\cref{eq:prob_gaussian}, the order of the problem depends on the number of points in the space of parameters which have to be evaluated ($G\approx n^p$), as well as with the dimension of vector of measurements, $\text{dim}[\vec X] = \text{dim}[\vec \mu]= m$.
The order of the full calculation is found to be $G \cdot m!$ 
which in the case of $m=\mathcal(100)$,  makes it impossible to solve with the available tools, since $100!\approx 10^{150}$.

One possible solution to reduce the order of the problem could be to group observables which depend only on a subset of theory parameters, in order to factorise the problem into smaller dimension ones.
If $\vec X_a$ is a subset ($\text{dim}[\vec X_a] = m_a < \text{dim}[\vec X]$) of measurements for which only the subset $\vec k_a$ ($\text{dim}[\vec k_a]=p_a < \text{dim}[\vec k]$) of all parameters enter its theoretical calculation, the order of the problem would be reduced to:
\begin{equation}
G \cdot m!\approx n^p \cdot m! \ 
\longrightarrow \ \prod_a G_a \cdot m_a! 
\approx \prod_a n_a^{p_a} \cdot m_a! \spc \qquad \sum_ap_a=p \quad \spc 
\quad \sum_a m_a=m \spc
\end{equation}
by neglecting correlations between those observables.
In the case of \acp{STXS}, this would mean grouping all STXS \emph{bins} from different stages in terms of production modes.  Strong correlations anyway cannot be neglected, since they change not only the error on  the fitted parameters, but also their central value. 
Thus, if two bins from different production modes are found to be strongly correlated, factorisation is not possible.
To give a numerical example, imagine it is possible to split the whole set of measurements in $10$ subsets of dimension 10 each. 
To each of these subsets let us say that only $5$ parameters of the initial $50$ contribute. This translates into $m_a=10, p_a=5$ for each $a\in(1, 10)$. Let us also fix $n_a=10$. The order of the problem decreases from $10^{160}$ to $10^{13}$. 
The fit performed in this way is simultaneous for every subset of parameters, but not global. 

Another way to reduce the order of the likelihood minimisation is to extract one 
parameter at a time, while profiling all the others, and repeating the procedure for all of them. The idea of profiling has been introduced in~\cref{par:likelihood} and it is used in general in global fits up to now. This approximation makes the fit global, but not simultaneous, since every parameter is extracted separately.

In the case that no big deviation is found with respect to the \ac{SM}, an expansion of the likelihood function around 
$X_i=\mu_i^{\text{SM}}= \mu_i( k_1^{\text{SM}}, \dots, k_1^{\text{SM}})$ is also possible. In the case of no discovery, the condition
\begin{equation}\label{eq:likelihood_SMapprox}
X_i- \mu_i^{\text{SM}} < 5 \sigma_i  \qquad 
\end{equation}
holds for every $i$.

Summarising, the minimisation of the full multivariate likelihood in a global and simultaneous way is a difficult problem which can not be solved exactly with the current available tools. A number of approximations can be applied to the problem to make it easier and thus solvable. The validity of such approximations have to be checked carefully, since neglecting strong correlations can also change central values of the parameters being fitted.

\clearpage
\section{Summary}
After the discovery of the Higgs boson at the LHC we still have a conventional vision: some very different physics occurs at Planck scale, and the SM is just an effective field theory. What about the next SM? A new weakly coupled
renormalisable model?  A tower of EFTs? 

The SMEFT framework  is useful because one can set limits on the effective coefficients in a model-independent way. This is why SMEFT in the bottom-up approach is so useful:  we do not know what the tower of UV completions is (or if it exists at all) but we can formulate the SMEFT and perform calculations
with it without needing to know what happens at arbitrarily high scales.  

On the other hand interpreting such limits as bounds on UV models does require some assumption of the UV dynamics. Alternatively, is the SM close to a fundamental theory?

Understanding the Higgs boson properties is a pillar of the present paradigm. 
For a theory or hypothesis to count as scientific it ought to be falsifiable in principle. In that regard, the SM is a good falsifiable hypothesis. Nevertheless it is important to emphasise that the SM has withstood risky tests that it could have easily failed.

In this review we discussed the frameworks adopted during Run 1 at the LHC as well as some further improvements on the former. Understanding that the main accent should be put on observables (i.e. quantities related to
an $S\,$-matrix) is important, and mapping those observables to a Lagrangian is a truly subtle affair, that we must understand and that cannot really be demoted.

In our view the problem is not how to imagine wild scenarios, but
rather how to arrive at the correct scenario by making only small steps, without having to make unreasonable assumptions.

Finally, we want point out it is important to preserve the original data, not just our current interpretation of the results. The estimate of the missing higher order terms can change over time, modifying the lessons we have drawn from the data and projected into the parameters and implementation of the SMEFT. 

Furthermore, considering projections for the precision to be reached in LHC Run 2
analyses, LO results for interpretations of the data in the SMEFT are challenged by consistency concerns and are not sufficient if the cut-off scale is in the few TeV range. The assignment of a theoretical error for SMEFT analyses is always important.

\clearpage

\chapter{$\phistar$: A new variable for studying $\h \to \gamma\gamma$ decays}
\label{cha:phistar}
\section{Overview}
\label{sec:overview}

After the landmark discovery of a Standard Model-like Higgs boson~\cite{Aad:2012tfa,Chatrchyan:2012xdj} in 2012 the main thrust of the LHC physics program has shifted to measuring its precise properties.  The combination of experimental results from ATLAS and CMS using data taken in Run 1 at $\sqrt{s} = 7$ and 8~TeV~\cite{Khachatryan:2016vau} demonstrated that the properties are consistent with the predictions of the Standard Model (SM). At the same time, no convincing evidence for physics Beyond the Standard Model (BSM) was found. 

Since 2015, the LHC has been operating at the higher centre of mass energy of 13~TeV.  The cross sections for Higgs boson production are of course larger, but more importantly, the anticipated integrated luminosity for Run 2 is five times larger.  This increased data set will enable a more precise determination of the Higgs boson properties.
The current state of the art for predictions of production and decay of the Higgs bosons, and for the measurement of the Higgs boson properties is summarised in the Handbook of LHC Higgs cross sections: 4. Deciphering the nature of the Higgs sector
(YR4)~\cite{deFlorian:2016spz}.

\begin{figure}[htb]
  \centering{%
\includegraphics[width=0.2\textwidth]{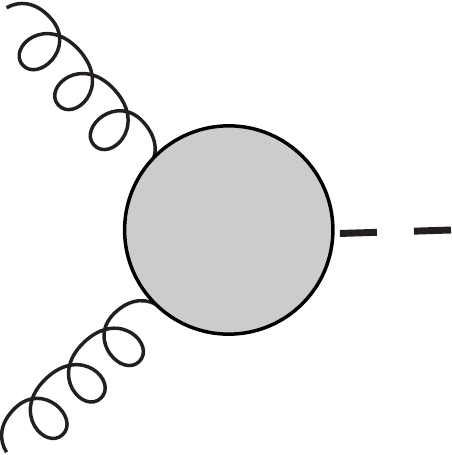}
  }
  \caption{Schematic diagram showing the production of a Higgs boson via gluon fusion.}
  \label{fig:blob}
\end{figure}

The main Higgs boson production mode takes place through gluon fusion, shown schematically in Fig.~\ref{fig:blob} where the shaded blob represents all possible interactions between gluons and the Higgs boson. Because of the very narrow width of the Higgs boson, the invariant mass of the gluon pair is concentrated around $\mh$.

\begin{figure}[t!]
  \centering{%
(a)~\includegraphics[width=0.2\textwidth]{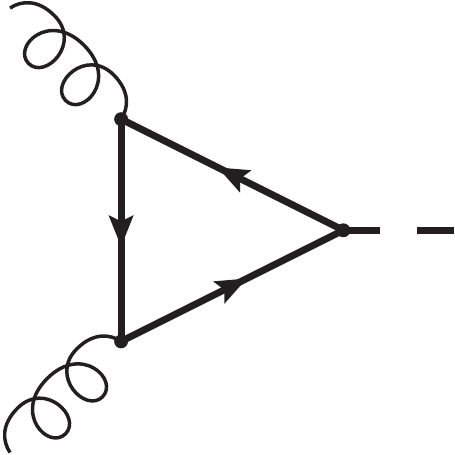}
\qquad 
(b)~
\includegraphics[width=0.2\textwidth]{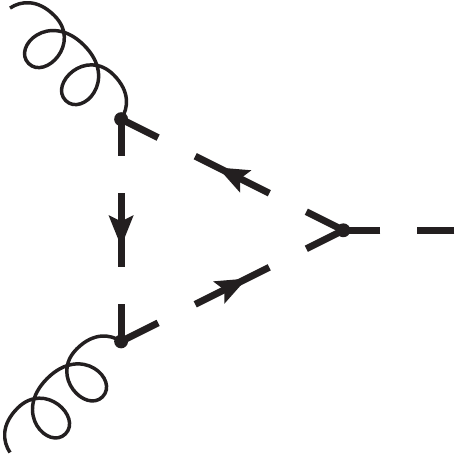}
\qquad 
(c)~\includegraphics[width=0.2\textwidth]{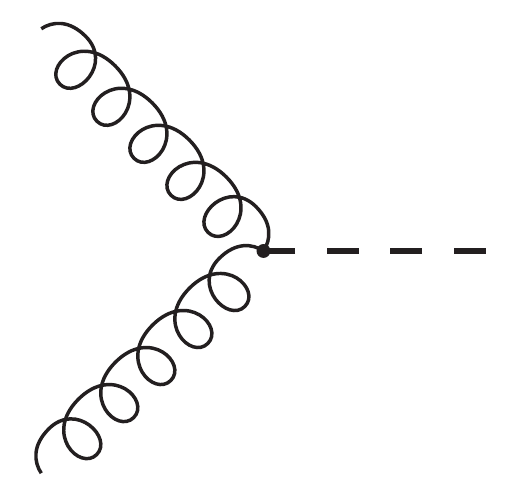}
  }
  \caption{Representative Feynman diagrams showing the production of a Higgs boson via gluon fusion for (a) fermion loops (b) scalar loops and (c) the effective interaction.}
  \label{fig:blobmediators}
\end{figure}

In the \ac{SM}, the Higgs boson couples directly to the quarks,
\begin{equation}
\label{eq:ffh}
{\cal L}_{\rm fermion} = -\sum_{q} \cq \frac{\mq}{v} \bar{q}q \h,
\end{equation}
and the interaction is simply mediated through the quark loop shown in Fig.~\ref{fig:blobmediators}(a) and $\cq = 1$.  The dominant contribution is from the top quark, but in certain kinematic regions the charm and bottom contributions are significant.   

On the other hand, new physical effects not described by the \ac{SM}, could also contribute to this process.  For example, a new heavy particle, $X$, of mass $m$, that couples to both gluons and the Higgs boson will change the rate according to the mass, spin, colour and couplings~\cite{Englert:2014uua}.  The rate for colour octet particles is eight times higher than for colour singlets.  As a concrete example, a massive colour triplet scalar particle, 
 $\tilde{q}$,
produces contributions as shown in Fig.~\ref{fig:blobmediators}(b).
   
In the limit where the particles circulating in the loop are heavy, $m \gg \mh$, the interaction is described by the effective interaction shown in Fig.~\ref{fig:blobmediators}(c),
\begin{equation}
\label{eq:eff}
{\cal L}_{\rm eff} = \cg \frac{\as}{12\pi}\frac{\h}{v} G^{a}_{\mu\nu} G^{a\,\mu\nu}.
\end{equation}
When the mass of the particle is produced by the Higgs mechanism, the heavy particle does not decouple in the $m \gg \mh$ limit and this leads to a contribution to $\cg$.

Because the kinematics of the $2\to 1$ process is so simple, different combinations of particles circulating in the loop simply affect the overall normalisation and measurements of the total cross section are not able to discriminate the mass scales of the various particles that mediate the $gg\h$ interaction.  This is a consequence of the Higgs low energy theorem that entangles short- and long-distance effects~\cite{Ellis:1975ap,Shifman:1979eb}. The normalisation is chosen such that (ignoring contributions from the light quarks and terms ${\cal O}(\mh^2/\mt^2$) ) the total inclusive rate simply depends on the squares of the sum of two terms proportional to $\ct$ and $\cg$ respectively~\cite{Grojean:2013nya},
\begin{equation}
\label{eq:sigmaincl}
\sigma_{\rm incl}(\ct,\cg) \sim \left(\ct + \cg\right)^2 \sigma^{\SM \, \LO (t)}_{\rm incl}.
\end{equation}
As a result, the measurement of the inclusive rate does not disentangle the effects of \ct and \cg.

In the \SM, $\cq=1$.  However, in the case of the top loop where $\mt > \mh$ it is common to make predictions for the gluon fusion cross section using the effective interaction of Eq.~\eqref{eq:eff}, i.e. setting $\cq = 0$ and $\cg = 1$. This is frequently called the Higgs Effective Field Theory (HEFT).
 
\begin{figure}[thb]
  \centering{%
\includegraphics[width=0.2\textwidth]{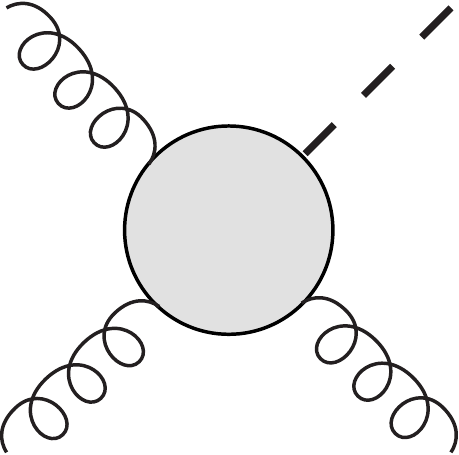}
  }
  \caption{Schematic diagram showing the production of a Higgs boson with transverse momentum via gluon fusion.}
  \label{fig:blobwithpt}
\end{figure}

Since \ct and \cg can be affected by new physics effects, it is clear that it is important to make measurements which allow an independent determination of $\cg$ and $\ct$. This requires a new energy scale much higher than the top-quark mass such that the low energy theorem no longer holds~\cite{Banfi:2013yoa} and the degeneracy of $\ct$ and $\cg$ breaks. This is done by recoiling the Higgs bosons against an additional jet as shown in Fig.~\ref{fig:blobwithpt}, and studying the process at high $\pth$~\cite{Harlander:2013oja,Azatov:2013xha,Grojean:2013nya,Dawson:2014ora,Langenegger:2015lra,Maltoni:2016yxb,Grazzini:2016paz,Deutschmann:2017qum}.
The QCD radiation acts as a probe to resolve the interaction, and at large $\pth$ there is sensitivity to particles with a mass $m ~\sim \pth$. 
As a consequence, Higgs boson production at large transverse momentum has been widely studied.

\begin{figure}[thb]
  \centering{%
(a)~
    \includegraphics[width=0.2\textwidth]{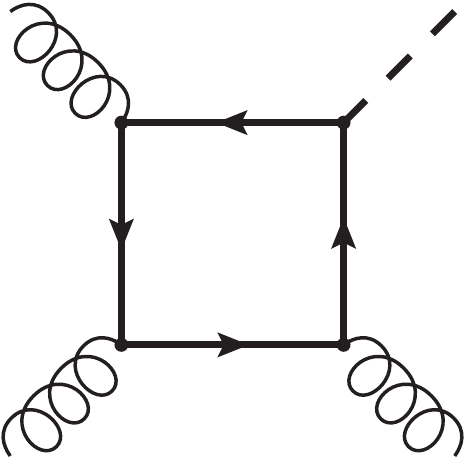}
\qquad
(b)~
    \includegraphics[width=0.2\textwidth]{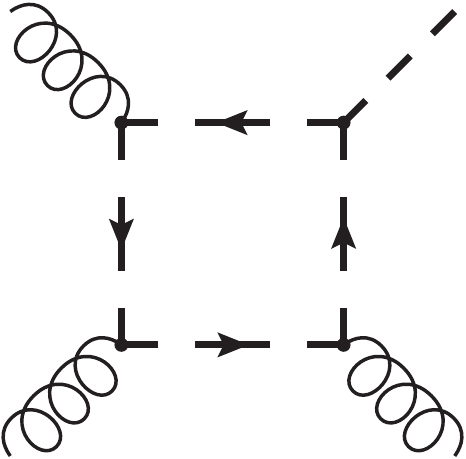}
\qquad
(c)~
  \includegraphics[width=0.2\textwidth]{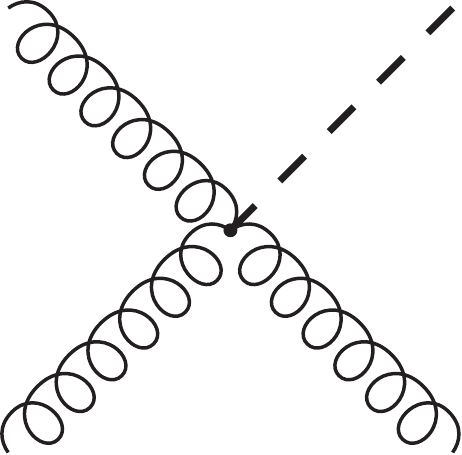}
  }
  \caption{Representative Feynman diagrams for (a) fermion loops (b) scalar loops and (c) the effective interaction for the production of a Higgs boson via gluon fusion at large transverse momentum.}
  \label{fig:blobwithptmediators}
\end{figure}

In the \SM, the Leading-Order (\LO) heavy quark contribution occurs at ${\cal O}(\as^3)$ and arises through loop diagrams~\cite{Ellis:1987xu,Baur:1989cm,Graudenz:1992pv,Harlander:2005rq} such as Fig.~\ref{fig:blobwithptmediators}(a). The ${\cal O}(\as^4)$ Next-to-Leading Order (\NLO) corrections
including finite top-mass effects involve complicated two-loop graphs~\cite{art:Bonciani:2016qxi} and are not currently known exactly, although efforts have been made to estimate their size~\cite{Harlander:2012hf,Neumann:2014nha,Neumann:2016dny}.   

The hadronic radiation probes the internal structure of the hard scattering and the effective interaction also receives radiative corrections as shown in Fig.~\ref{fig:blobwithptmediators}(c).
In this case, the \NLO corrections 
~\cite{Schmidt:1997wr,deFlorian:1999zd,Glosser:2002gm,Ravindran:2002dc} 
and ${\cal O}(\as^5)$ Next-to-Next-to-Leading Order (\NNLO) corrections~\cite{Boughezal:2013uia,Chen:2014gva,Boughezal:2015dra,Boughezal:2015aha,Caola:2015wna,Chen:2016zka}  
are known.   

Similarly, new heavy particles ($X$) will also receive different radiative corrections shown for example in Fig.~\ref{fig:blobwithptmediators}(b)
that depend sensitively on the mass of the particle and on the transverse momentum.  When the transverse momentum is large compared to $m$, the new physics degrees of freedom are resolved and can in principle be isolated from the \SM contributions.

Of course, the Higgs boson decays, and the transverse momentum can only be constructed from the decay products.   One of the cleanest decay channels is the rare $\h \to \gamma\gamma$ decay which has a small branching ratio of $\sim 2.3\cdot 10^{-3}$ so that current measurements are limited by statistics.   This will improve with time with the continued operation of the LHC.

The purpose of this report is to propose a new and complementary observable for studying Higgs boson production at large transverse momentum in the case where the Higgs boson decays to two photons.  

The $\phistar$ observable was originally proposed for the study of Drell-Yan production of lepton pairs~\cite{Banfi:2010cf}.  The $\ptz$ distribution is constructed from the transverse momentum of the leptons which relies on calorimetric measurements of the lepton energies.  
In contrast, the $\phistar$ observable relies on measurements of the directions in $\phi$ and $\eta$ of the two leptons rather than the energies.  
One can make use of both calorimeter and charged tracking information provided by the detector to determine the lepton directions. Ref.~\cite{Banfi:2010cf} demonstrated that the experimental resolution for $\phistar$ is much better than $\ptz$ particularly when $\ptz$ is small.  In the case at hand, we propose to extend the $\phistar$ variable to describe a two photon system, even though there is no tracking information about the photons. 

Our report is structured as follows.  In section~\ref{sec:observable}, we recall the definition and properties of the $\phistar$ variable itself. 
We show that although $\pth$ and $\phistar$ both vanish in the Born limit, they are never directly related.     
We see that the fermion loop and effective interactions behave differently for the $\phistar$ variable, and that this leads to clear differences between the leading order $\pth$ and $\phistar$ distributions.  In section~\ref{sec:photons} we make a first study of the experimental resolution for the $\phistar$ variable for photons. Even without the benefit of the charged track information, the experimental resolution for $\phistar$ is improved compared to the transverse momentum of the photon pair, particularly at low transverse momentum. The effect of the parton shower on the \pth and \phistar variables is discussed in section~\ref{sec:parton-shower}.

Once the viability of $\phistar$ as an experimental observable has been established we turn to more theoretical considerations in section~\ref{sec:theory}, where we discuss the general structure of the $gggH$ vertex and how to compute the loop diagrams that can contribute to this process. We show how such calculations are structured, using the one-loop \SM process as a pedagogic example. The notions of integral families, Integration-by-Parts identities, and master integrals are introduced and discussed, while the details of the calculation of the one-loop master integrals using differential equations are postponed to Appendix~\ref{sec:oneloopappendix}.  Details of the recently computed two-loop planar integrals~\cite{art:Bonciani:2016qxi} are discussed in Appendix~\ref{sec:twoloop} together with the more advanced techniques of integrand reduction, generalised polylogarithms, symbols and elliptic integrals.  
  
Armed with the one-loop amplitude in the \SM, section~\ref{sec:heavytoplimit} shows how to take the large mass limit  of the one-loop integrals.  This leads to a much simpler result that does not depend on \mt and we discuss for which values of $\pth$ this approximation breaks down.  The equivalent, much simpler, calculation in the Higgs Effective Theory (HEFT) is briefly discussed in section~\ref{sec:heavytopeffective}.

The uncertainties inherent in a perturbative calculation are discussed in section~{\ref{sec:theory-uncertainties}.  We discuss both the uncertainties that arise due to the truncation of the perturbative series, as well as the parametric uncertainties that arise from the choice of the values of the strong coupling, the heavy quark masses and the parton distribution functions (PDFs).
 
Higher order corrections are discussed in section~\ref{sec:higher-orders}.  
Due to the complexity of the two-loop master integrals (and especially the non-planar graphs), the full NLO corrections in the \SM are not yet available, and we only show results in the \HEFT.  Sections~\ref{sec:NLOinHEFT} and \ref{sec:NNLOinHEFT} make predictions for the \pth and \phistar distributions at \NLO and \NNLO respectively.  To simulate the anticipated effect of the heavy quark masses, we introduce two approximations - essentially retaining the exact \LO mass dependence in a multiplicative and additive way - and compare and contrast the \NLO and \NNLO predictions for the $\phistar$ and $\pth$ distributions. As is well known, at large \pth, the \HEFT breaks down leading to a large uncertainty in the \NLO and \NNLO predictions at large \pth.  However, the various approximations for the \phistar distribution are largely in agreement demonstrating that this is an observable that is much less sensitive to ultraviolet structure of the $ggg\h$ interaction and can be well described by the \HEFT.

The potential of the \phistar distribution as a probe of \BSM physics is discussed in section~\ref{sec:BSM}. In section~\ref{sec:BSMeffective} we study the effect of new physics effects on the \phistar distribution using a generic model in which a colour-triplet scalar particle is added to the \SM.  As expected from our analysis of the \SM predictions, the \pth distribution shows a clear effect, while the \phistar distribution does not.  As a more concrete example, we focus on the light stop scenario within the \MSSM proposed by the HXSWG~\cite{Carena:2013ytb} and study the feasibility of observing an effect in the \phistar distribution.   
Although the overall rate is affected by the presence of additional heavy fields, the shape of the the \phistar distribution is relatively insensitive to heavy particle thresholds.

Our findings are briefly summarised in section~\ref{sec:summary}.
 
\subsection{Default input parameters}
\label{sec:default}
Throughout our report, we fix the LHC centre-of-mass energy to be 13~TeV and, unless otherwise stated, use the following default input parameters: the Higgs boson mass, $\mh = 125$~GeV, the top quark mass, $\mt = 174$~GeV, the bottom quark mass, $\mb = 4.6$~GeV, the charm quark mass, $\mc = 1.42$~GeV.  We use the PDF4LHC15\_nnlo\_30 PDF set~\cite{Butterworth:2015oua} with $\as(\MZ) = 0.118$ and take the renormalisation and factorisation scales, $\mur$ and $\muf$ to be equal,
$$
\mu = \mur = \muf. 
$$
For the Born process, $pp \to H$, we set $\mu = \mh/2$ while for Higgs bosons produced with finite transverse momentum, we set $\mu = \MTh/2$ where $\MTh$ is the transverse mass of the Higgs boson,
\begin{equation}
\label{eq:transversemass}
\MTh = \sqrt{\mh^2 + \pthsq}.
\end{equation}
The Higgs boson decays into two photons with a branching ratio of 0.00235. The photons are required to have rapidity, $|\eta| < 2.5$, and be isolated, such that the sum of the hadronic energy within a cone of $R=0.4$ is less than 14 GeV. Jets are defined with the anti-$k_T$ algorithm with $R=0.4$.

\clearpage
\section{Observables for probing Higgs boson recoil}
\label{sec:observable}

In hadron collisions, the direction in which the two hadrons collide is extremely crowded and it is experimentally hard to distinguish what is a product of the collision itself and what is simply a remnant of the colliding hadrons. 

In order to isolate the products of the interaction, it is convenient to 
study the component of the momentum of the particle which is transverse to the beam axis, the transverse momentum,  
\begin{equation}
p_{T} = \sqrt{p_{x}^2 + p_{y}^2},
\end{equation}
as well as the momentum parallel to the beam direction which is characterised by the rapidity, 
\begin{equation}
\eta = \frac{1}{2}\log\left(\frac{E+p_z}{E-p_z}\right).
\end{equation}
For massless particles, $E = |p| = \sqrt{p_{x}^2 + p_{y}^2 +p_{z}^2} \equiv p_{T}\cosh \eta$, and the rapidity is also called the pseudorapidity.

In the {laboratory frame}, the four-momentum $p_i^\mu$ of a massless particle can be written in terms of the transverse momentum, $p_{Ti}$, the rapidity (or equivalently pseudorapidity), $\eta_i$ and the azimuthal angle $\phi_i$,
\begin{equation}
p_i^\mu = \left( p_{Ti}\cosh \eta_i , \vec{p}_{Ti}, p_{Ti}\sinh\eta_i \right),
\end{equation}
where the vector transverse momentum is,
\begin{equation}
\vec{p}_{Ti} = \left(p_{Ti}\cos \phi_i, p_{Ti} \sin \phi_i \right).
\end{equation}

\begin{figure}[b]
\centering
(a)
\raisebox{5mm}{\includegraphics[scale=0.4]{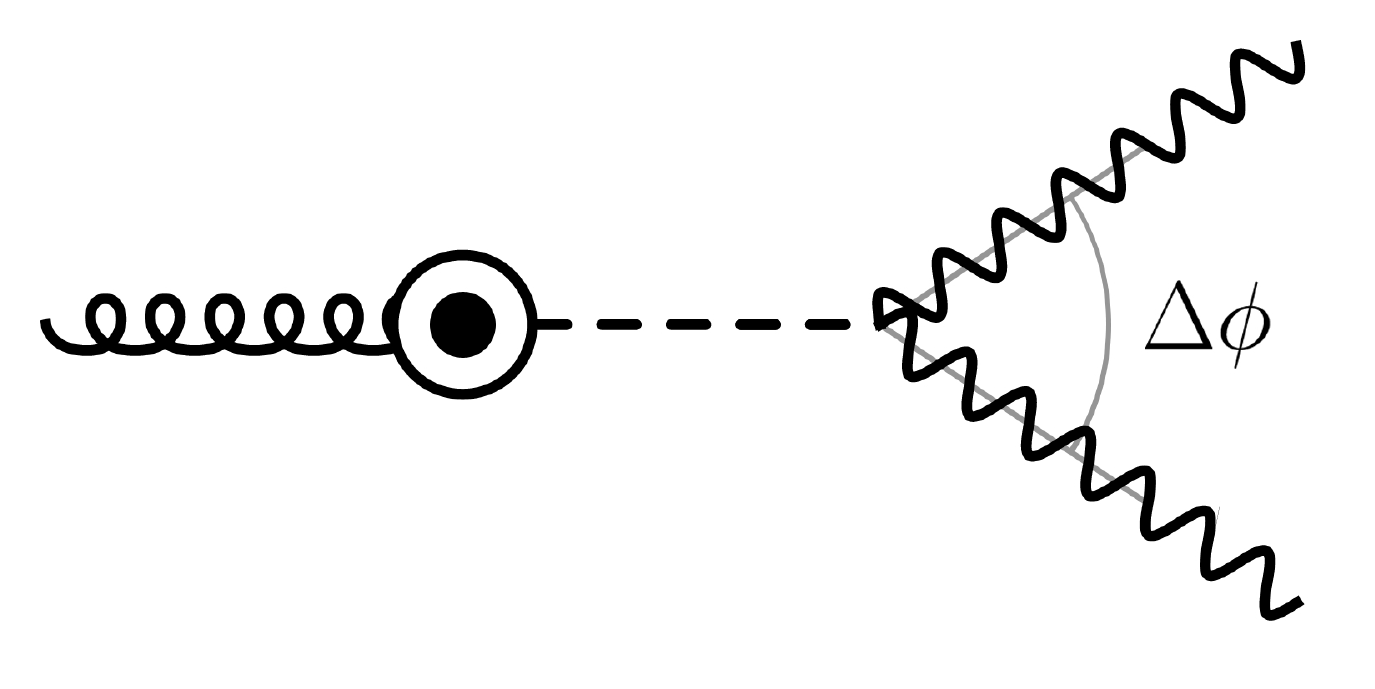}}
(b)
\raisebox{0mm}{\includegraphics[scale=0.4]{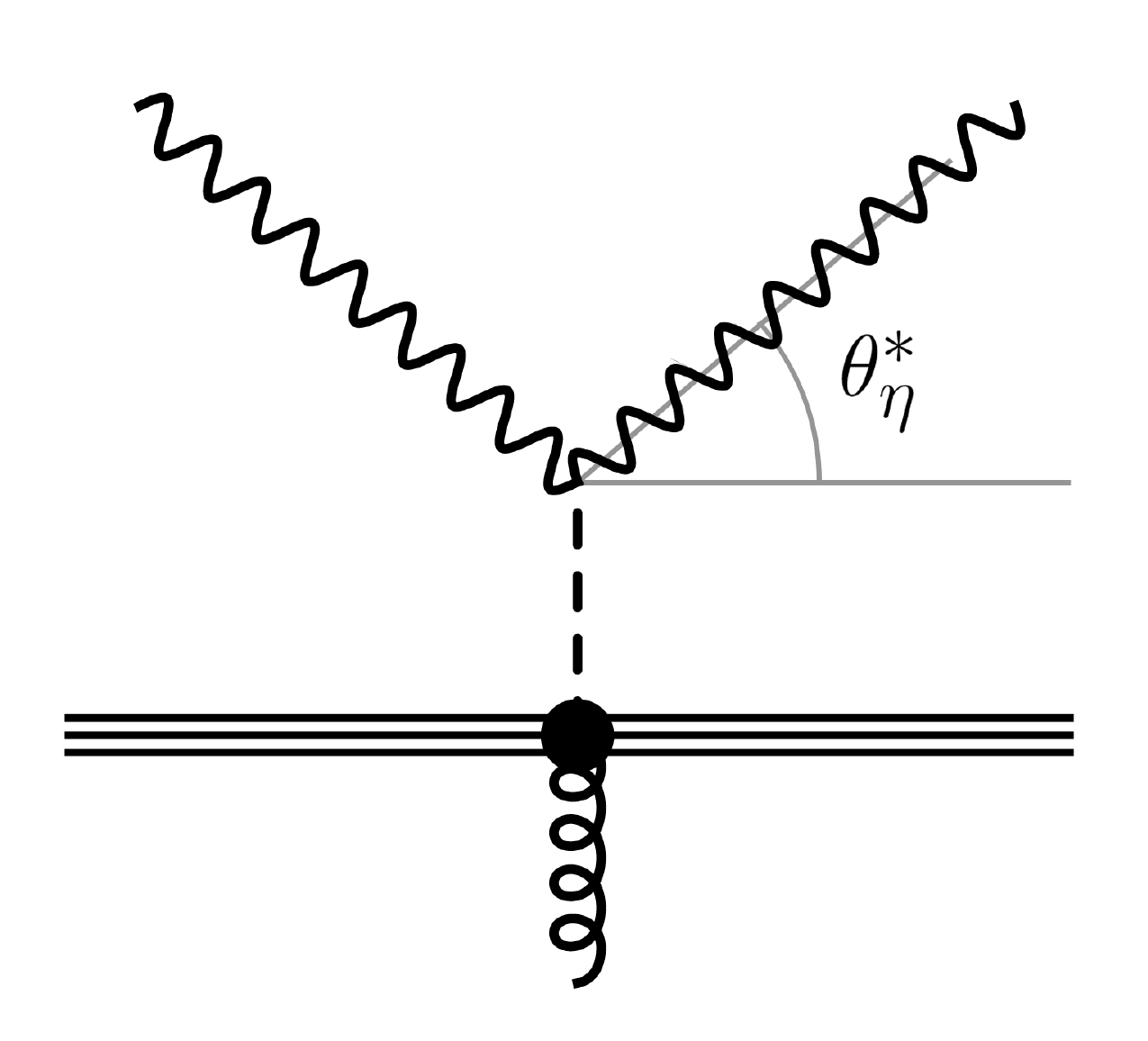}} 
\caption{Illustration of the two angles that appear in the definition of $\phistar$: (a) shows $\Delta \phi$ in the plane perpendicular to the beam axis. (b) shows $\thstar$ in the plane spanned by the beam axis and the Higgs boson, boosted along the beam axis so that the Higgs boson is purely transverse. }
\label{fig:phistarangles}
\end{figure}

A measurement of $p_T$ for a short-lived particle, is obtained by combining the $p_T$-measurements of its decay products. For small values of $p_T$, that procedure leads to a large experimental uncertainty on the reconstructed value. 
For a two particle system (such as Drell-Yan, or Higgs boson decay to photons), the transverse momentum of the pair is simply, 
\begin{equation}
\vec{p}_{T} = \vec{p}_{T1}+\vec{p}_{T2}.  
\end{equation}
As shown in Fig.~\ref{fig:phistarangles}(a), in the transverse plane, $\Delta \phi$ is the azimuthal angle between $p_{T1}$ and $p_{T2}$.   When there are only two particles in the final state, $\Delta\phi = \pi$, and $\vec{p}_{T}=\vec{0}$ while in the general case
$\Delta\phi < \pi$ and $\vec{p}_{T}\neq \vec{0}$.  The individual transverse momenta are related to the magnitude of the transverse momentum of the pair by,
\begin{eqnarray}
\label{eq:pt2def}
p_T^2 &=& p_{T1}^2 + p_{T2}^2 + 2 p_{T1} p_{T2} \cos \Delta\phi.
\end{eqnarray}
In the small $p_T$ region, $\cos\Delta\phi \sim -1$, and $p_{T1}  \sim p_{T2}$, so that there are large cancellations between the terms on the RHS of Eq.~\eqref{eq:pt2def}, which
ultimately leads to a poor experimental resolution.

\begin{figure}[t]
    \centering
(a)
    \begin{subfigure}[b]{0.45\textwidth}
        \includegraphics[width=\textwidth]{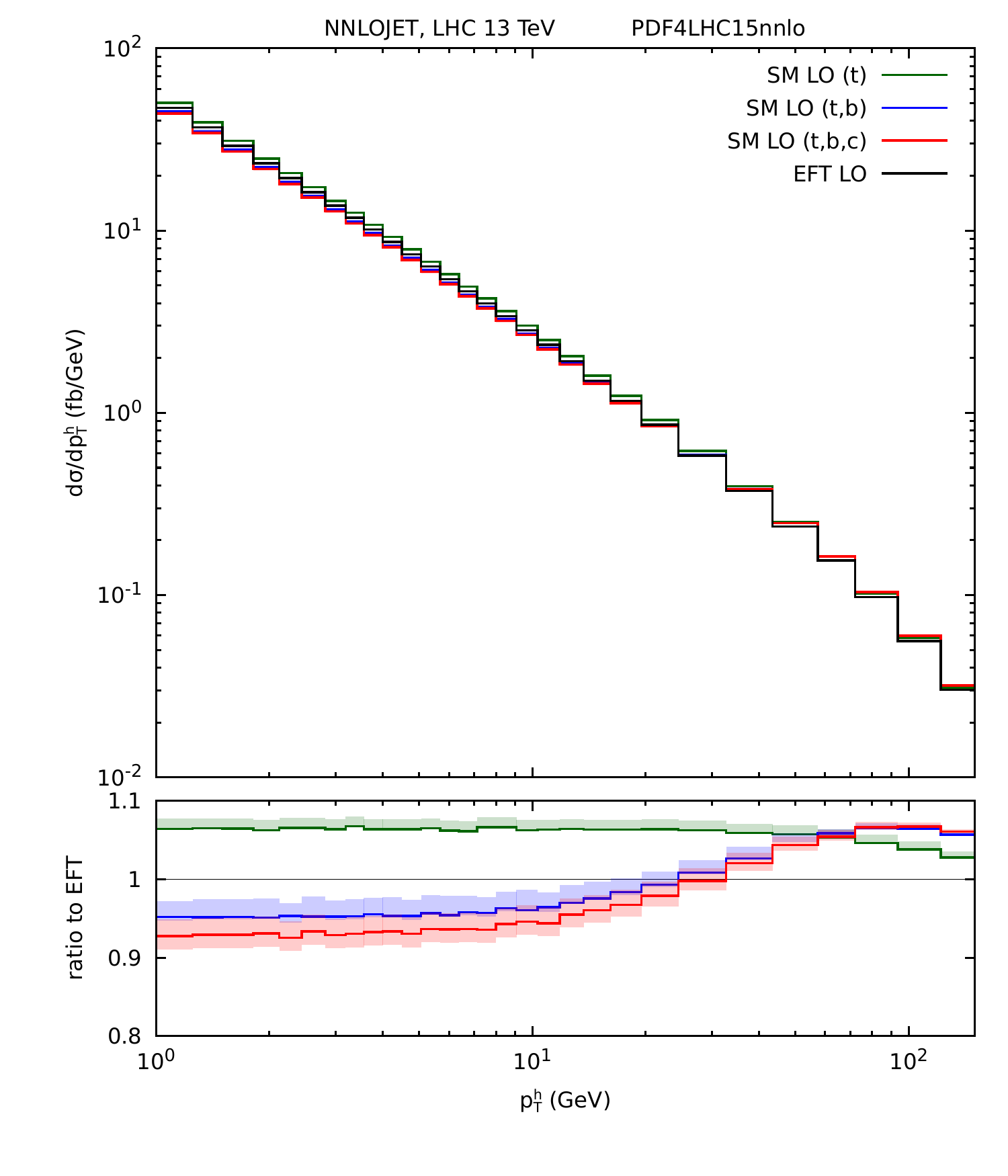}
    \end{subfigure}
(b)
    \begin{subfigure}[b]{0.45\textwidth}
        \includegraphics[width=\textwidth]{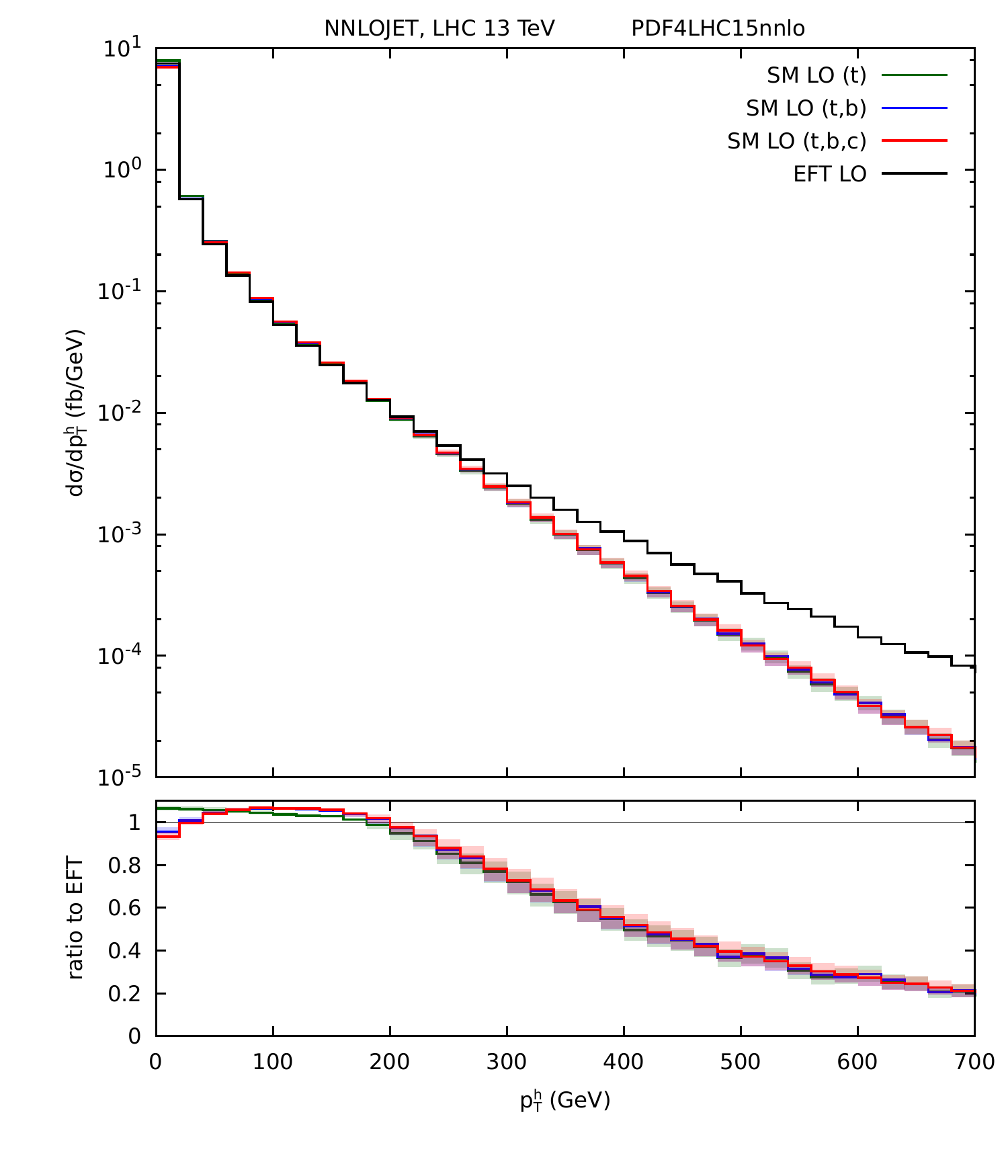}
    \end{subfigure}
    \caption{The Higgs boson transverse momentum distribution at $\sqrt{s}=13$~TeV at leading order for the \HEFT (black) and for the \SM (a) at low $\pth$ and (b) at high $\pth$.  The \SM predictions are shown for the top-quark alone (green), top and bottom quark loops (blue) and top, bottom and charm quark loops (red). The lower panels show the ratio normalised to the \HEFT result.}
    \label{fig:SMpth_13TeVcompare}
\end{figure}

As discussed in Sec.~\ref{sec:overview}, the main Higgs boson production mode is the gluon fusion process. Here we consider the rare but clean decay of the Higgs boson into two photons. The inclusive cross section for the two-photon decay of the Higgs boson produced via gluon fusion is dominated by low $\pth$ events and the inclusive cross sections for the \HEFT and \SM are similar. At leading order with the default parameters given in~\cref{sec:default} we find\footnote{We note that the higher order corrections are significant and have been computed through to next-to-next-to-next-to-leading order (N3LO)~\cite{Anastasiou:2015ema,Anastasiou:2016cez}},
\begin{align}
  \sigma_{\LO}^{\SM}(t,b,c)  & = 24  \pm 3  \text{ fb} \\
  \sigma_{\LO}^{\HEFT} &= 26  \pm 3 \text{ fb}.
\end{align}

Fig.~\ref{fig:SMpth_13TeVcompare} shows a comparison between the \HEFT and the \SM predictions for the $\pth$ distribution. 
As we can see Fig.~\ref{fig:SMpth_13TeVcompare}(a), for a Higgs boson with $\pth < \mt$ the \HEFT prediction almost perfectly reproduces the \SM result with a deviation of less than a 10\%. For $\pth \approx \mt$ the distributions start to differ as we begin to resolve the structure of the loop. At larger $\pth$, the two models lead to very different predictions demonstrating the potential of the $\pth$ distribution in resolving the details of the gluon-Higgs interaction.

\subsection{A new observable: \phistar}

To probe the low $p_T$ region, an alternative angular variable, $\phistar$, has been proposed\cite{Banfi:2010cf} which minimises the impact of these experimental uncertainties. This variable has been already studied for the case of $Z$ production decaying into two leptons \cite{Gehrmann-DeRidder:2016jns,Aad:2015auj}. It is our goal to extend these studies to Higgs boson production decaying into two photons.

The variable $\phistar$ is defined by:
\begin{equation}
    \phistar\equiv \tan\left(\frac{\phi_{\text{acop}}}{2}\right) \cdot \sin (\thstar) .
    \label{eq:phistardef}
\end{equation} 
In this definition, the acoplanarity angle is simply related to the angle between the particles in the transverse plane:
\begin{equation}
    \phi_{\text{acop}}\equiv \pi -\Delta\phi. 
    \label{eq:acop}
\end{equation}
As depicted in Fig.~\ref{fig:phistarangles}(b), $\thstar$ is the scattering angle of the two particles with respect to the proton beam in the reference frame in which the two particles are back to back in the $(r,\theta)$ plane. In this frame, the parent particle is purely transverse.
This frame is obtained by a longitudinal boost characterised by $\beta$ such that the pseudorapidities in the new frame, $\eta_i^\prime$ are related to $\eta_i$ by, 
\begin{equation}
\eta_i^\prime = \eta_i + \Delta \eta
\end{equation}
where
\begin{equation}
\Delta \eta = \frac{1}{2}\log\left(\frac{1-\beta}{1+\beta}\right) = -\left(\frac{\eta_1+\eta_2}{2}\right).
\end{equation}
The transverse momenta are unaffected by the boost so that in this frame,
\begin{eqnarray}
p_1^{\prime \mu} &=& \left( p_{T1}\cosh \left(\frac{\eta_1-\eta_2}{2}\right) , \vec{p}_{T1}, p_{T1}\sinh\left(\frac{\eta_1-\eta_2}{2}\right) \right),\\
p_2^{\prime \mu} &=& \left( p_{T2}\cosh \left(\frac{\eta_1-\eta_2}{2}\right) , \vec{p}_{Ti}, -p_{T2}\sinh\left(\frac{\eta_1-\eta_2}{2}\right) \right).
\end{eqnarray}
From here we immediately see that,
\begin{equation}
\sin(\thstar) = \frac{|p_T|}{E} = {\rm sech} \left(\frac{\eta_1-\eta_2}{2}\right),
\end{equation}
or equivalently,
\begin{equation}
    \cos(\thstar)= \frac{|p_z|}{E} = \tanh\left(\frac{\eta_1-\eta_2}{2}\right).
\end{equation}

The $\phistar$ variable measures the ``deviation from back-to-backness'' (acoplanarity) in the transverse plane and therefore 
vanishes at Born level where the azimuthal angle between the two leptons $\Delta\phi$ is exactly equal to $\pi$. 
Non-zero values of $\phistar$ are produced by the same mechanism that generates non-zero \pth, namely a recoil against hadronic emission from the partonic initial states.  As a consequence, the $\phistar$ distribution probes the same type of physics as the transverse momentum distribution. 
In the following we will investigate the relation between the two variables in further detail. 

\begin{figure}[t]
    \centering
(a)
    \begin{subfigure}[b]{0.45\textwidth}
        \includegraphics[width=\textwidth]{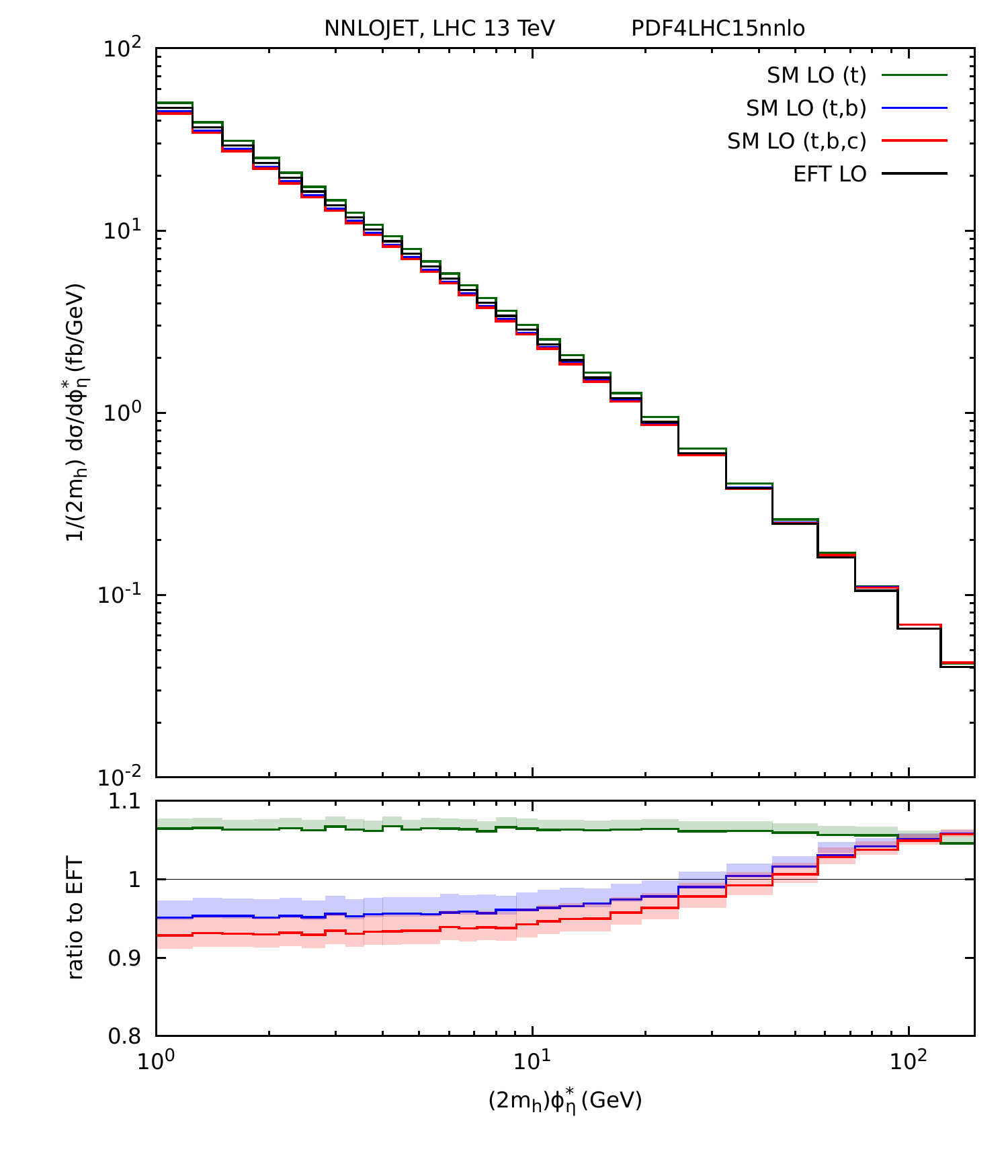}
    \end{subfigure}
(b)
    \begin{subfigure}[b]{0.45\textwidth}
        \includegraphics[width=\textwidth]{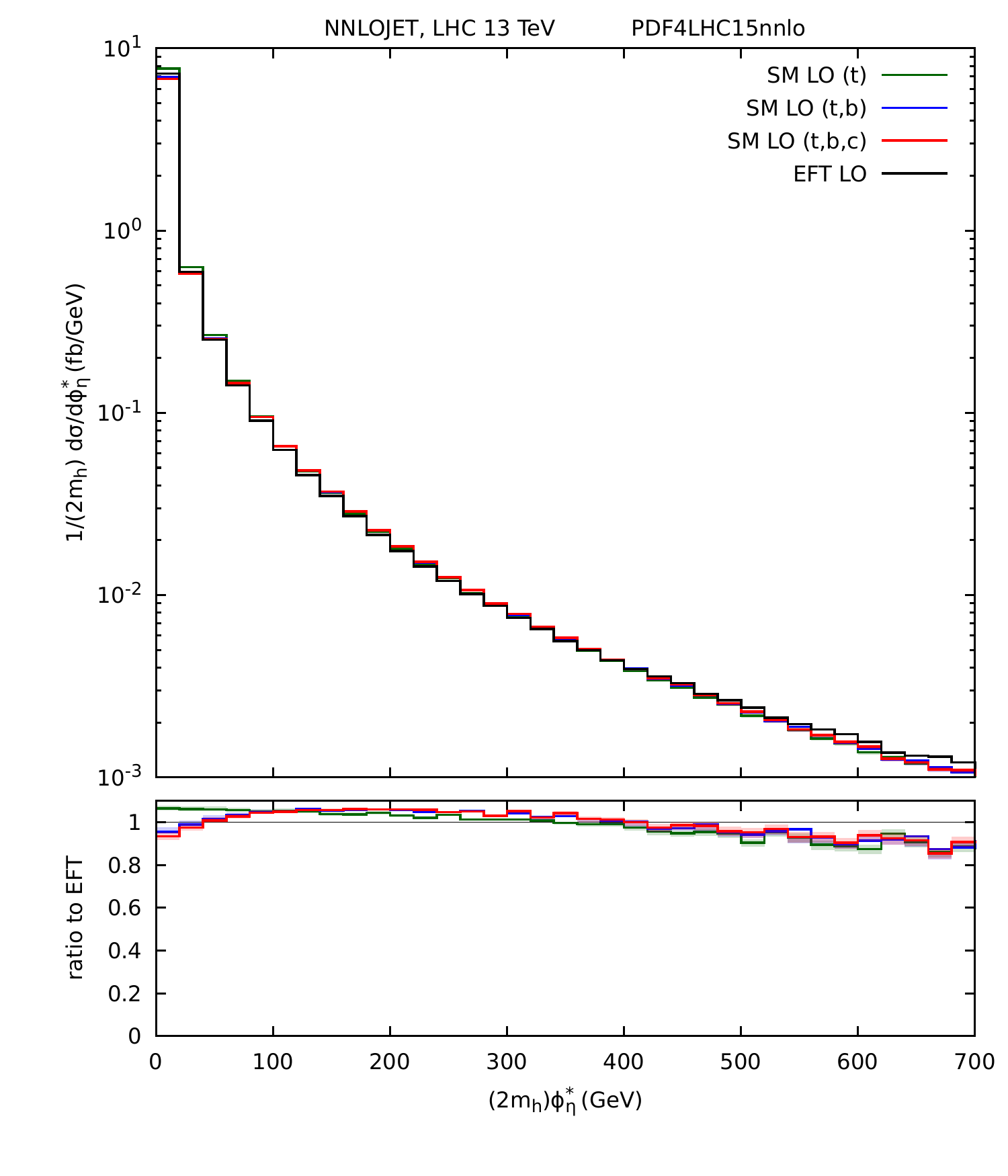}
    \end{subfigure}
    \caption{The Higgs boson $\phistar$ distribution at $\sqrt{s}=13$~TeV at leading order for the \HEFT (black) and for the \SM (a) at low $\phistar$ and (b) at high $\phistar$.  The \SM predictions are shown for the top-quark alone (green), top and bottom quark loops (blue) and top, bottom and charm quark loops (red). The lower panels show the ratio normalised to the \HEFT result.}
    \label{fig:SMphi_13TeVcompare}
\end{figure}

Fig.~\ref{fig:SMphi_13TeVcompare} compares the \HEFT and the \SM predictions for the $\phistar$ distribution for the Higgs boson decaying to two photons. As at low $\pth$, Fig.~\ref{fig:SMphi_13TeVcompare}(a), shows that for a Higgs boson with $2\mh\phistar < \mt$, the \HEFT prediction lies within 10\% of the \SM result. However, unlike the $\pth$ distribution, at higher values of $2\mh\phistar \approx \mt$ the agreement between the models persists.   On the one hand, this implies that the $\phistar$ distribution is less sensitive to the details of the gluon-Higgs interaction.  On the other hand, it means that the $\phistar$ distribution can be better predicted using the Higgs Effective Theory. 

\subsubsection{$\phistar$ in the low-\pth regime}

At very low \pth (or \phistar) the cross section is dominated by large Sudakov-type logarithms of the vanishing observable. As discussed earlier, \pth and \phistar vanish under the same circumstances. In the limit in which both variables vanish, we can find a relation between the two observables by looking at their dominant logarithmic behaviour~\cite{Bozzi:2005wk,Banfi:2009dy,Gehrmann-DeRidder:2016jns}:
\begin{eqnarray}
	  \frac{\pth}{d\sigma_{0}}\frac{d\sigma}{d\pth}  &=& - 8 C_{F} \frac{\alphaS}{\pi}\ln\left( \frac{\pth}{\mh}\right) + \mathcal{O}(\alphaS^{2}), \label{eq:logpth}
\\
	 \frac{\phistar}{d\sigma_{0}}\frac{d\sigma}{d\phistar}  &=& - 8 C_{F} \frac{\alphaS}{\pi}\ln\left(2\phistar\right) + \mathcal{O}(\alphaS^{2}), \label{eq:logphi} 
\end{eqnarray}
where $C_F = (N_c^2-1)/2/N_c$.
Although the two observables, \pth and \phistar, are not directly related, by equating the arguments of the logarithms in Eqs.~\eqref{eq:logpth} and \eqref{eq:logphi} we can derive an approximate relationship between the observables in the regions of the distributions that are dominated by Sudakov logarithms, 
\begin{equation}
  2 \mh \phistar \approx \pth.
  \label{eq:phistar_pt_relationship}
\end{equation}
We therefore systematically scale the $\phistar$ distributions by a factor of $2\mh$ throughout this report. 

\begin{figure}[t]
    \centering
(a)
    \begin{subfigure}[b]{0.45\textwidth}
        \includegraphics[width=\textwidth]{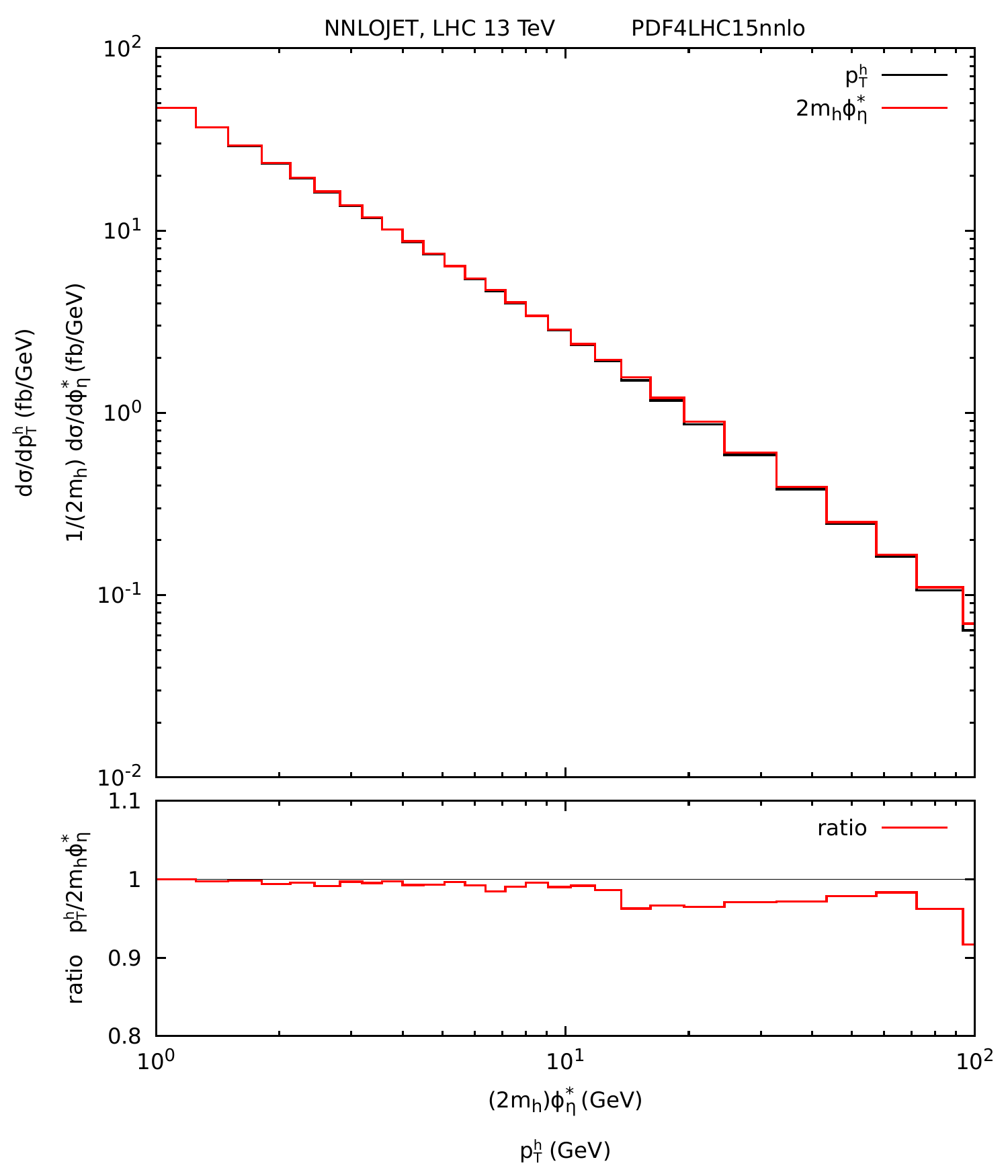}
    \end{subfigure}
(b)
    \begin{subfigure}[b]{0.45\textwidth}
        \includegraphics[width=\textwidth]{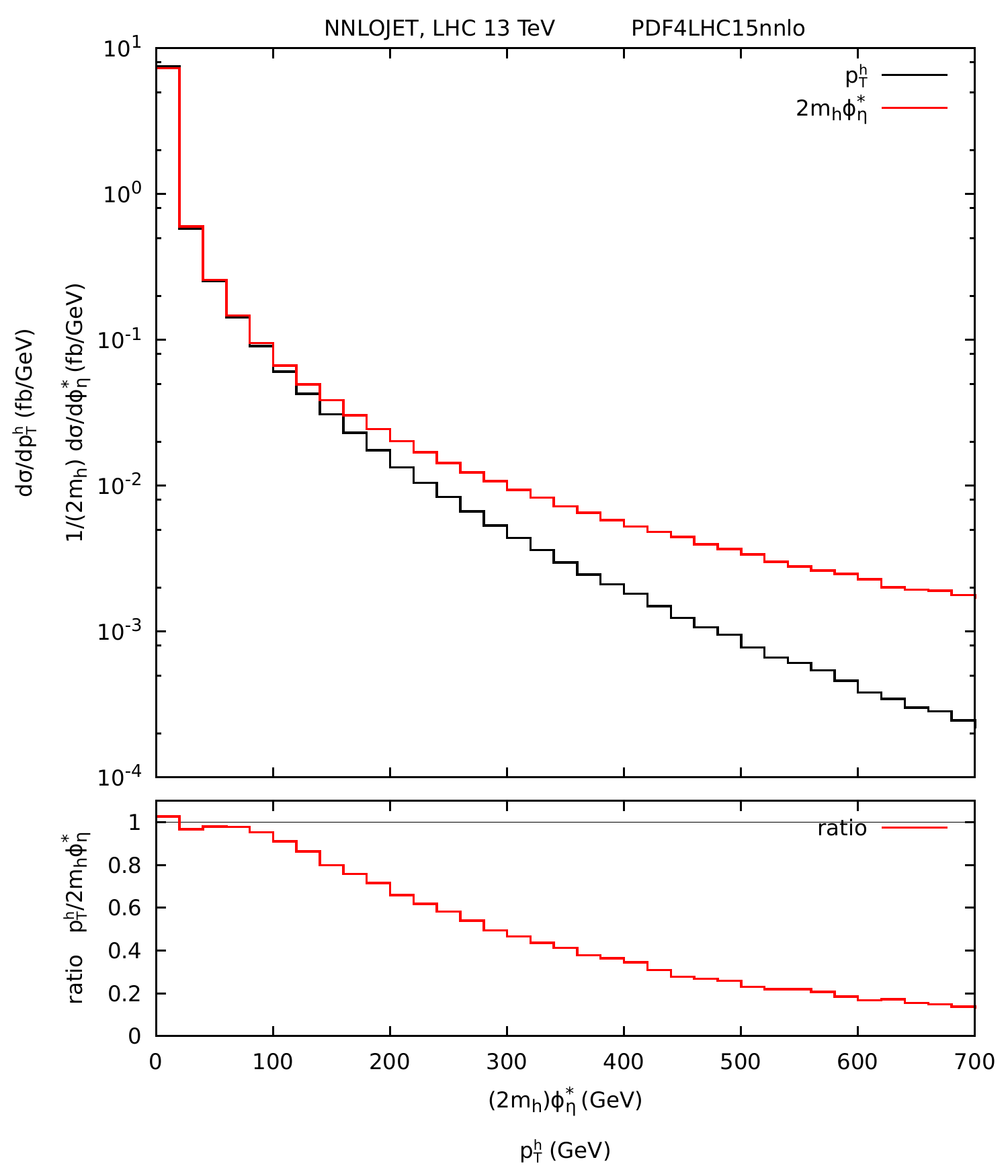}
    \end{subfigure}
    \caption{Comparison between the \LO \phistar (red) and \pth (black) distributions in the (a) low-\pth, low-\phistar and (b) high-\pth, high-\phistar regimes at 13 TeV.}
    \label{fig:phistar_pt_relationship}
\end{figure}

Fig.~\ref{fig:phistar_pt_relationship} shows the leading order \SM predictions for the \pth and \phistar distributions.  We see that at low-\pth, low-\phistar the two distributions are very similar.   However, at larger values of \pth and \phistar, the distributions are rather different.  This difference becomes apparent around $\pth \sim \mt$ and is because the \pth distribution is able to resolve the top quark loop.  In contrast, the \phistar distribution does not appear to resolve the top quark loop and from this point of view, behaves more like the inclusive cross section.

\clearpage
\section{Photons}
\label{sec:photons}
In this section we will study the experimental resolution of the $\phistar$ variable in two-photon events.  We will concentrate on the reconstruction of photons and electrons with the CMS detector. 
Like ATLAS, this enormous and incredibly complex detector is composed of different sub-detectors assembled in an onion-like configuration and designed to reconstruct all the visible particles produced during a proton-proton collision. Each sub-detector interacts with the particles produced by the collisions and is designed to obtain a particular piece of information about each particle. The specific combination of information collected from many sub-detectors allows the reconstruction of different particles types.  

\subsection{Photon reconstruction and identification at the LHC detectors}
Since we are concerned on the measurement of the Higgs boson when it decays into photons,  we will focus only on the inner CMS sub-systems: the tracker and the electromagnetic calorimeter (ECAL).  Within these sub-detectors, charged-particle trajectories are measured by the silicon pixel and strip tracker, with full azimuthal coverage within $|\eta| < 2.5$. A lead tungsten crystal ECAL and a scintillator hadron calorimeter (HCAL) surround the tracking volume and cover the region  $|\eta| < 3$. The ECAL cells have widths of 0.0174 in both $\eta$ and $\phi$. A detailed description of the CMS detector can be found in Ref.~\cite{Chatrchyan:2008aa} 

The photons and electrons are usually reconstructed in the same way at the LHC detectors, mainly by combining information from the electromagnetic calorimeter and the inner tracking detectors.  After being produced by a proton-proton collision, these particles travel through different sub-detectors before being stopped in the ECAL where they release all their energy as an electromagnetic shower. Photons and electrons are differentiated by the fact that electrons leave a charged track in the tracker that points towards the electromagnetic shower in the ECAL. 

Unlike electrons, whose momentum can be measured precisely by the tracker, the photon energy measurement depends entirely on the intrinsic detector resolution of the ECAL. A precise measurement of the transverse momentum of Higgs boson that decays into a pair of photons is then limited by the ECAL performance. 

As discussed in Sec.~\ref{sec:observable}, in order to probe the physics at low $\pth$ region ($\pth << \mh$), we propose the use of new observable, $\phi^*_{\eta}$, that could provide a precise measurement of the properties of $\h\to \gamma\gamma$ events by avoiding the ECAL intrinsic resolution and using the angular information instead.

\subsection{Event generation and detector simulation}
Monte-Carlo $H\to\gamma\gamma$ events are generated using the \texttt{Madgraph5\_aMC@NLO}~\cite{Alwall:2014hca} simulation framework with a centre-of-mass energy $\sqrt{s}=13~\mathrm{TeV}$ and a Higgs boson mass set to $\mh = 125~\mathrm{GeV}$. The events are generated at NLO accuracy including the matching and merging of parton multiplicities. The parton showering is taken care of by \texttt{Pythia8}~\cite{Sjostrand:2007gs} with the CUETP8M1~\cite{Skands:2014pea} tune. A fast detector simulation is performed on those events using \texttt{DELPHES 3}~\cite{Selvaggi:2014mya} with the CMS detector configuration. This framework supplies the relevant sub-sensors resolutions and reconstruction efficiencies of the CMS detector described earlier\footnote{Similarly a fast simulation with ATLAS can be performed.}.

In this study photon conversions into electron-positron pairs are neglected. The final photon and electron energy is obtained by applying the ECAL resolution function that depends on the intrinsic ECAL performances. True photons and electrons with no reconstructed track pointing to the ECAL are considered to be photons in the detector simulation \texttt{DELPHES 3}.

\subsection{Event selection }

We use the same fiducial volume as used for the differential cross section measurements by CMS~\cite{CMS-PAS-HIG-17-015}.  Events are accepted for further analysis if the diphoton invariant mass satisfies $|m_{\gamma\gamma} -125|<10~\rm GeV$ and the photons satisfy the following requirements:
\begin{itemize}
	\item The leading and sub-leading photons satisfy $p_{T}^{\gamma_1}/m_{\gamma\gamma} >1/3$ and $p_{T}^{\gamma_2}/m_{\gamma\gamma} >1/4$; 
	\item  Each photon of the candidate pair entering the analysis is required to have $|\eta| < 2.5$, excluding the region $1.4442 < |\eta| < 1.566$ that corresponds to the gap in the ECAL between the barrel and end-cap;
	\item Each photon must be isolated with an isolation value of $I_{\rm iso}< 0.1$ \footnote{The photon isolation is defined as the ratio between the transverse momentum of all the reconstructed objects within a radius of $\Delta R < 0.4$ and the photon candidate. This definition is slightly different from the usual definition used by the LHC experiments that use different types of isolation depending on the calorimeter and how the particles are reconstructed. More details can be found in Ref.~\cite{CMS:2009nxa}. }.
\end{itemize}

For each generated photon, the reconstructed photon candidate with the smallest $$\Delta R = \sqrt{(\eta_{rec} - \eta_{gen})^2 + (\phi_{rec} - \phi_{gen})^2}$$ is computed. When $\Delta R<0.2$ the generated photon is paired with a reconstructed photon.   

The distributions of  $\phi^*_{\eta}$ and $p_{T}^{\gamma\gamma}$ after the simulation of diphoton events are shown in Fig.~\ref{photons:phistar_pt_dist}. We see that for both variables, there is a very good correlation between {\tt truth} (the generated observable) and {\tt reco} (the reconstructed observable) apart from very small values of the observables.

The relationship between $\phistar$ and $p_{T}^{\gamma\gamma}$ for the diphoton pair in individual events is shown in Fig.~\ref{photons:phi_star_pt_relationship}. In order to make a fair comparison between these two observables we scale $\phistar$ by the factor of $2\times \mh$  as discussed in the previous section. 

\begin{figure}[!ht]
	\centering
(a)
	\includegraphics[width=0.45\textwidth]{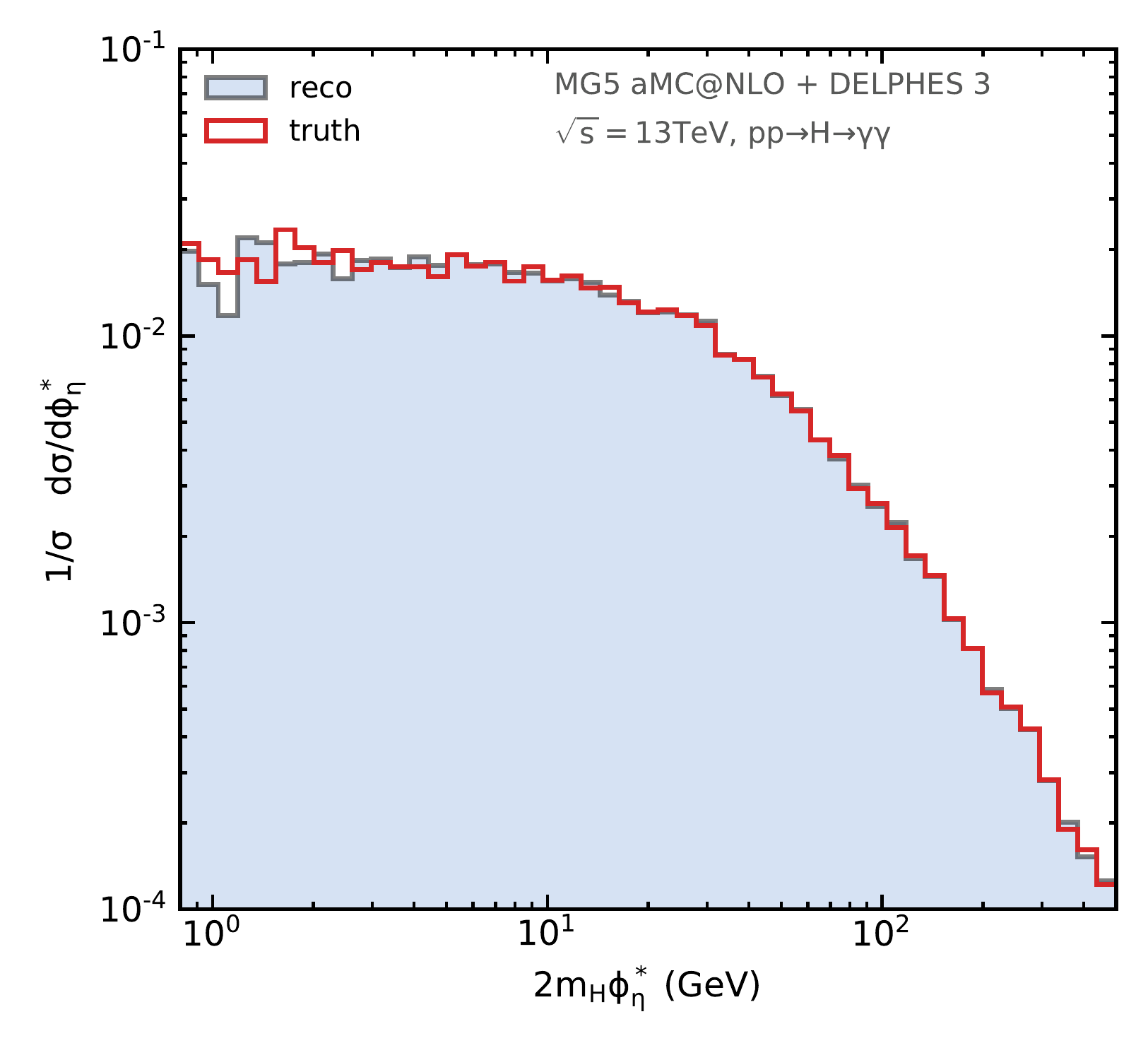}
(b)
	\includegraphics[width=0.45\textwidth]{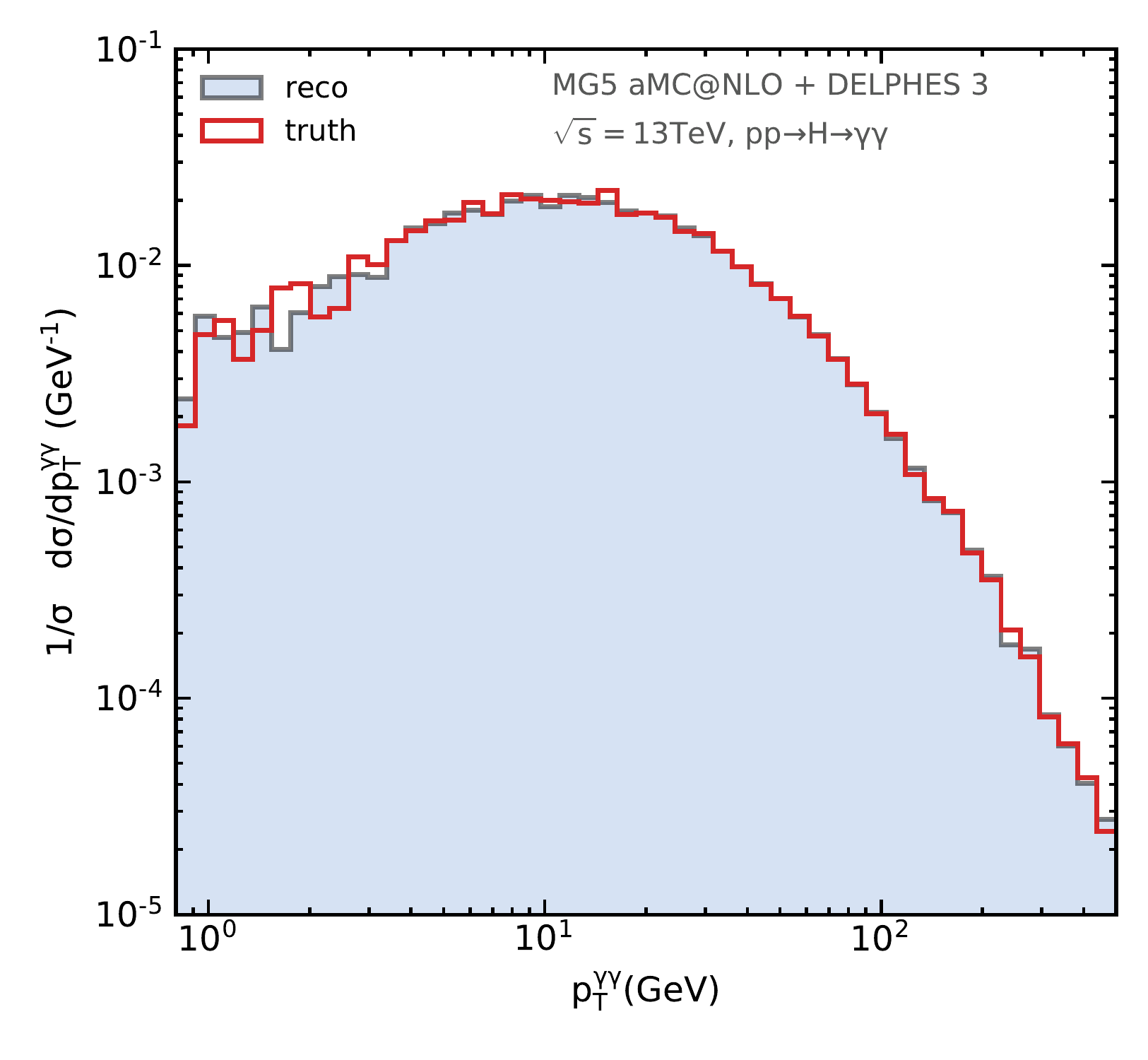}
	\caption{Distributions of (a) the transverse momentum $\h\to\gamma\gamma$  and (b) $\phistar$ at both generated ({\tt truth}) and reconstructed ({\tt reco}) levels. }
	\label{photons:phistar_pt_dist}
\end{figure}

\begin{figure}[!ht]
 	\centering
 	\includegraphics[height=0.45\textwidth]{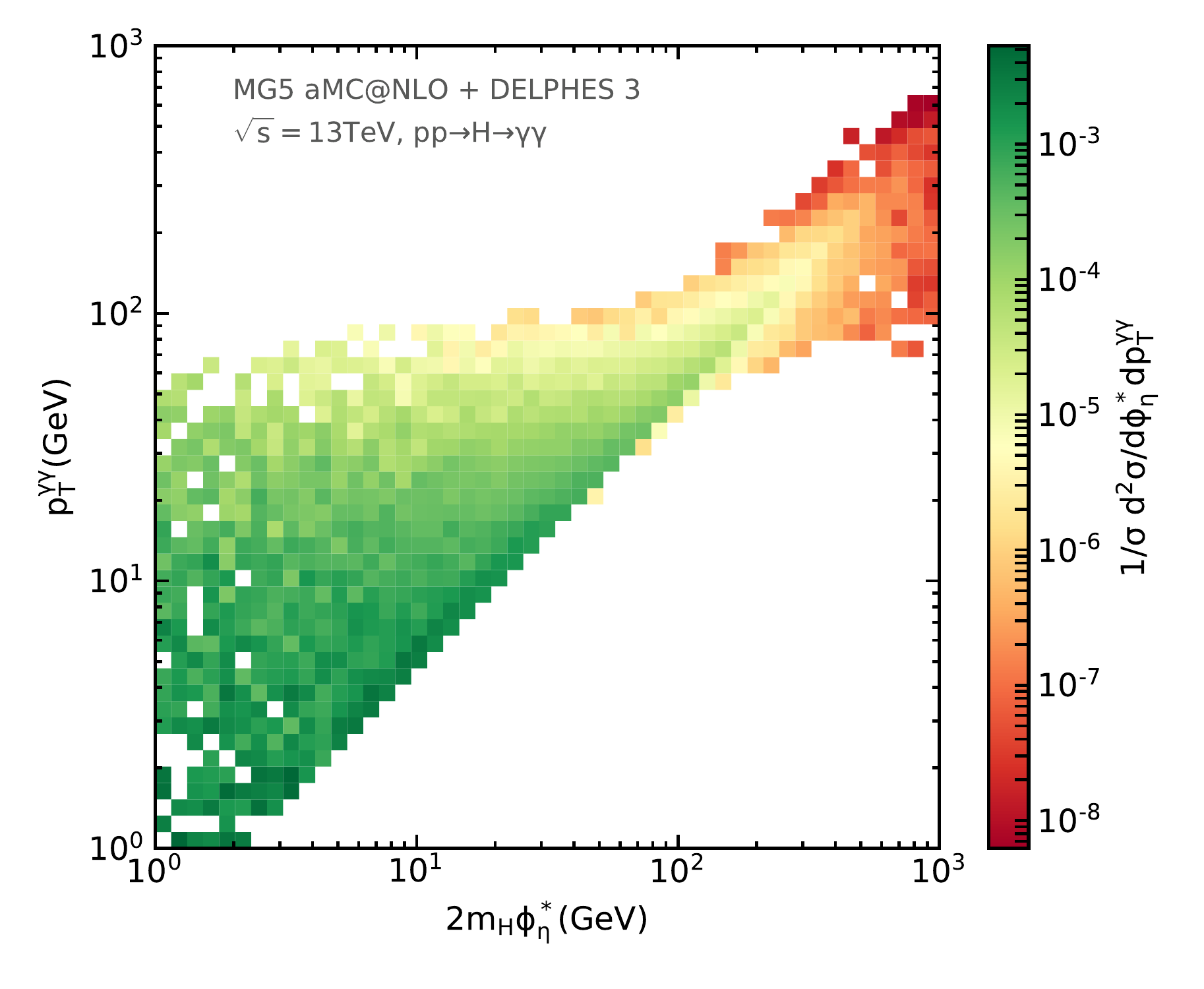}
 	\caption{Relationship between $p_{T}^{\gamma\gamma}$ and $\phistar$ for individual diphoton pairs. }
 	\label{photons:phi_star_pt_relationship}
\end{figure}

\subsection{Experimental resolution for diphoton scattering angle}

\begin{figure}[!ht]
 	\centering
(a)
 	\includegraphics[width=0.44\textwidth]{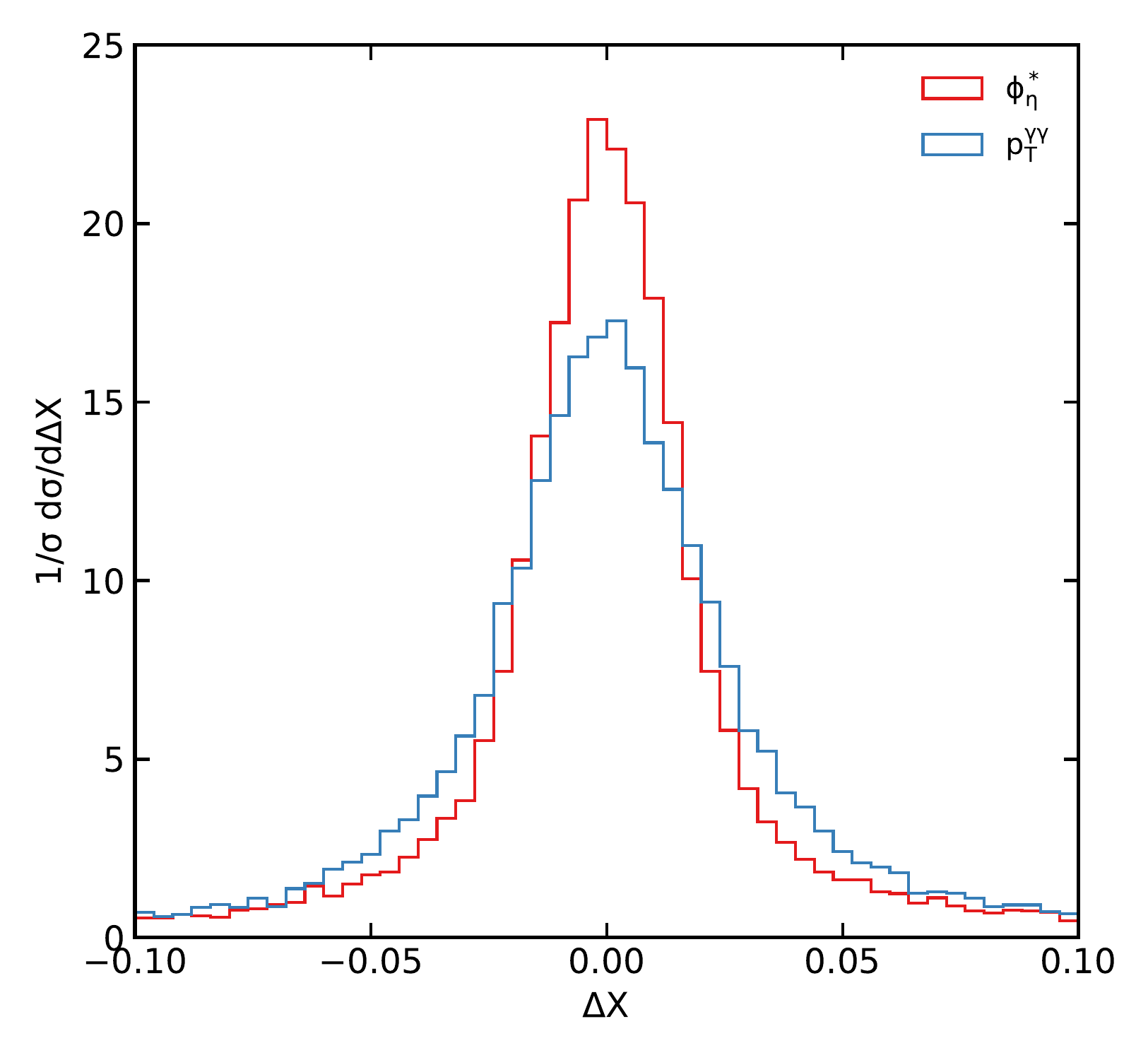}
(b)
	\includegraphics[width=0.46\textwidth]{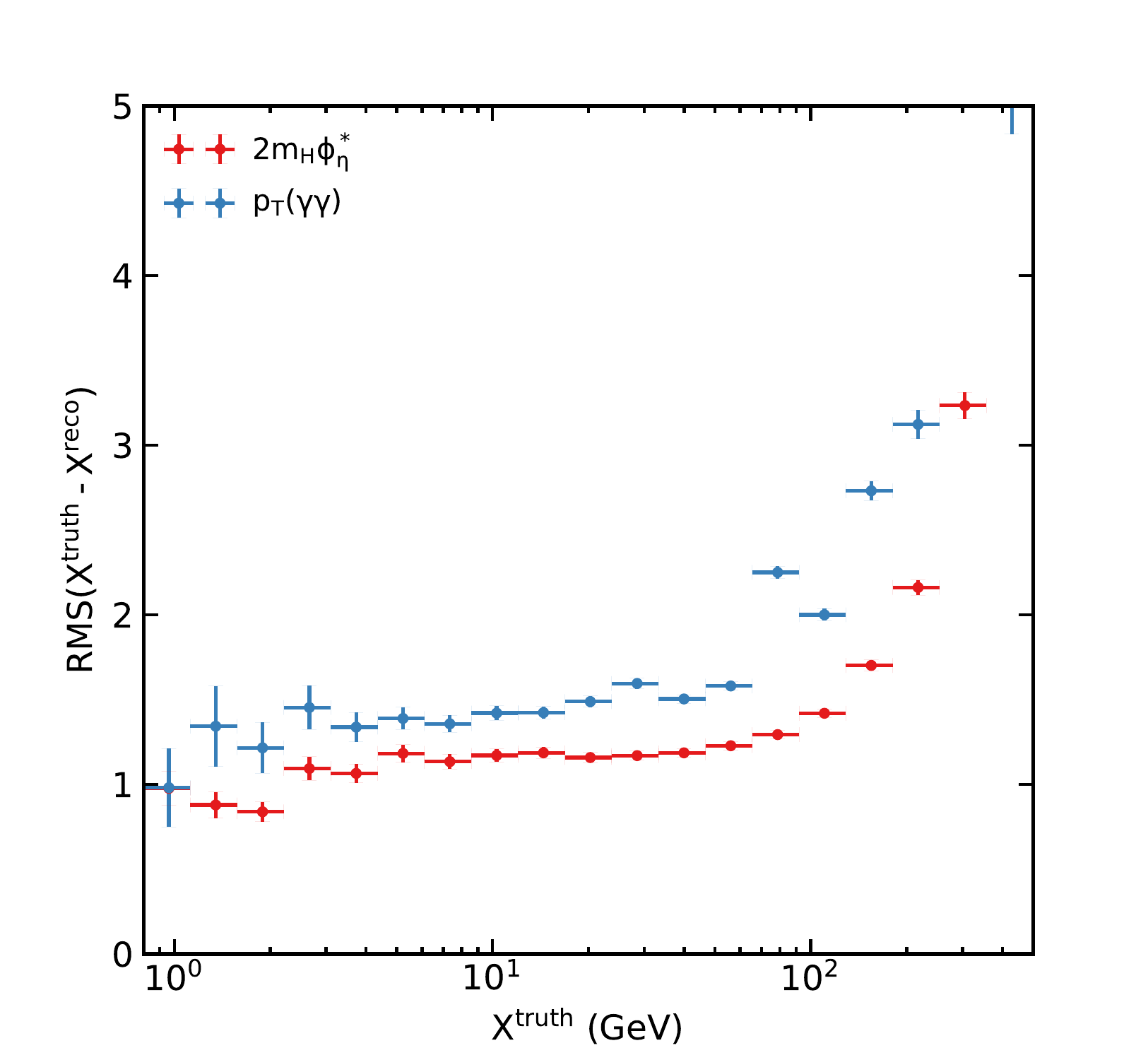}
 	\caption{(a) Relative resolution of $X = 2 \mh \phistar, ~p_{T}^{\gamma\gamma}$.
(b) Resolution of $X = 2 \mh \phistar, ~p_{T}^{\gamma\gamma}$. }
 	\label{photons:phi_star_resolution}
\end{figure}
Following the strategy as described in Ref.~\cite{Banfi:2010cf} in the context of the $Z$ transverse momentum measurement, we start by measuring the relative resolution of $\phistar$ and $p_{T}^{\gamma\gamma}$ in simulated diphoton events as shown in Figure~\ref{photons:phi_star_resolution}(a). Here we show the relative resolution for a given measured quantity $X$ ($X = 2 \mh \phistar, ~p_{T}^{\gamma\gamma}$) as a function of the relative difference $$\Delta X = (X^{\rm truth} - X^{\rm rec})/X^{\rm truth}.$$  As expected the $\phistar$ variable has a narrower relative resolution distribution than $p_{T}^{\gamma\gamma}$.  

The resolution is defined as the RMS value of $X^{\rm truth} - X^{\rm reco}$  
as a function of $X^{\rm truth}$ shown in Fig.~\ref{photons:phi_star_resolution}(b).  We conclude that the angular variable $2\mh \phistar$ has a better resolution in comparison with the simple calorimetric measurement of $p_{T}^{\gamma\gamma}$ in most of the range of $X^{truth}$. The improvement is roughly 30\% in the range $1~{\rm GeV} < X^{\rm truth} < 100~{\rm GeV}$.

Note that many systematic effects have been neglected in using the \texttt{DELPHES} simulation. One of the main sources of systematic uncertainties neglected is the determination of the hard scattering primary vertex in the presence of pile-up which could lead a mis-measurement of the scattering angles and thus the pseudo-rapidities of the two selected photons. The presence of pile-up also introduces additional systematic uncertainties in the isolation criteria that might further bias the identification of the two photons. 

\clearpage
\section{Parton Showers}
\label{sec:parton-shower}
Let us now investigate the effect of the  
parton shower (PS) when simulating the two observables discussed so far.
We concentrate on a very simple example, the production of a
Higgs boson through gluon fusion. To make the discussion even clearer,
we consider only the Higgs Effective Theory. Note that a complete
description of how a parton shower works is beyond the scope of this
report and we refer the reader to the useful references~\cite{Hoche:2014rga,Webber:1986mc,Schumann:2007mg,Catani:2001cc}.

The differential partonic cross section, ${\rm d} \hat{\sigma}$, for the Born (or leading order (LO))
hard scattering process, 
$gg \rightarrow \h$, can be written as
\begin{equation}
  {\rm d} \hat{\sigma}_{gg\rightarrow \h} = {\rm d} {\rm \Phi}_{2\rightarrow 1} \frac{|\overline{\mathcal{M}_{gg \to \h}}|^2}{F},
\end{equation}
where the flux factor $F = 4 p_1\cdot p_2 = 2 s_{12}$, 
$|\overline{\mathcal{M}_{gg \to \h}}|^2$ is the spin and colour summed and averaged matrix element and
${\rm \Phi}_{2\rightarrow 1}$ represents the one--particle phase space.
The explicit form for ${\rm d} {\rm \Phi}_{2\rightarrow 1}$, requires that
\begin{equation}
  {\rm d} \hat{\sigma}_{gg\rightarrow \h} \propto \delta^{(2)}(\vecpth),
\end{equation}
which follows from four--momentum conservation and implies that 
the value of $\pth$ is fixed to be identically zero.

\begin{figure}[thb]
  \centering{%
    \includegraphics[width=0.5\textwidth]{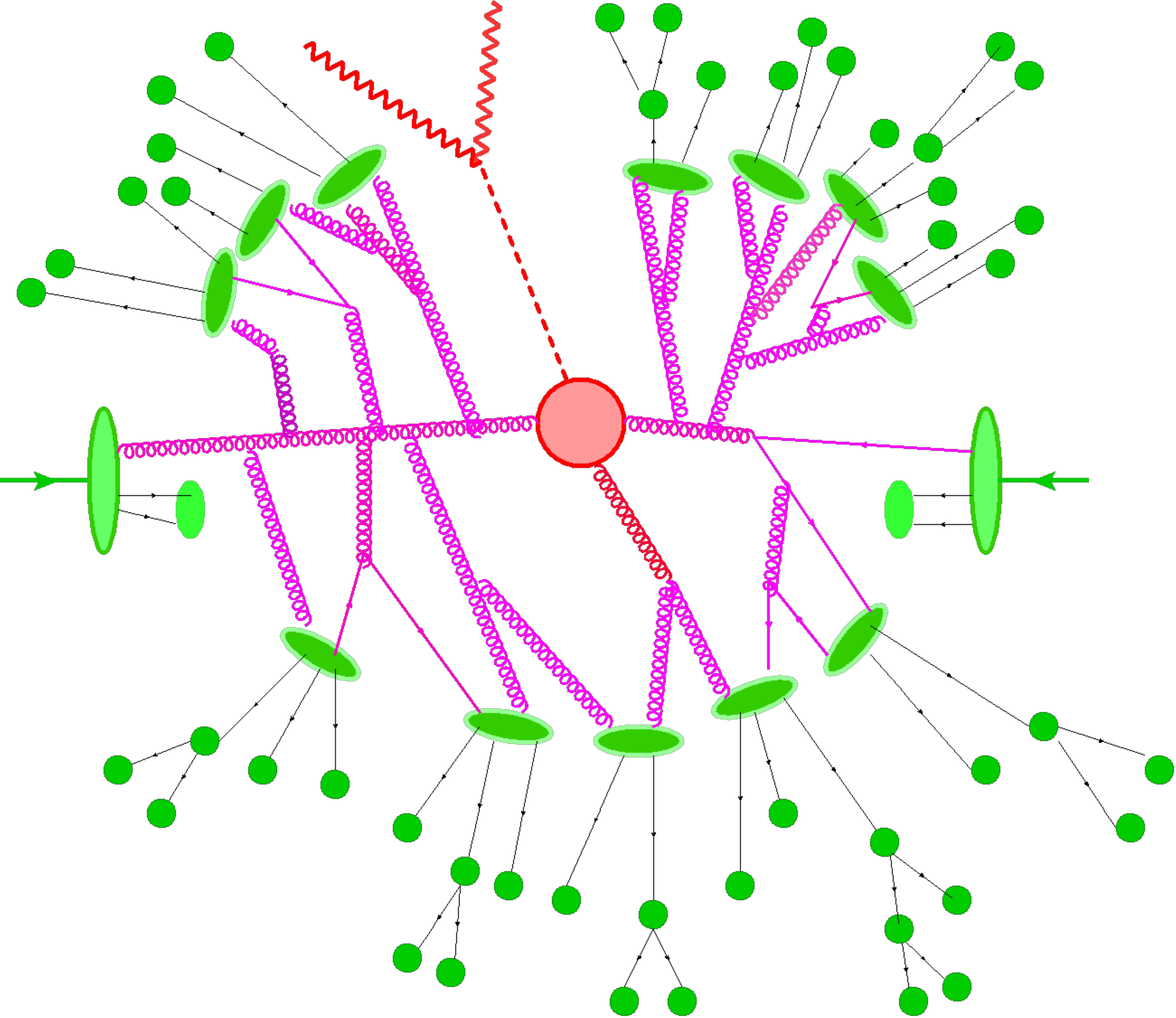}
  }
  \caption{Pictorial representation of how a large \pth Higgs boson event looks like from
    a Monte Carlo perspective. 
The hard interaction (big red blob) is followed by the decay of the Higgs boson to photons (red wavy lines)). Additional hard QCD radiation is produced through final state (red) and initial state (magenta) before the final-state partons hadronise (dark green blobs) and hadrons decay (light green blobs). 
  }
  \label{fig:event}
\end{figure}
In reality the two gluons
in the initial state are very energetic and will emit other particles,
gluons or quarks, which in turn will emit other particles and so on.
A schematic representation of an event in which a Higgs boson is produced at large transverse momentum 
is shown in Fig.~(\ref{fig:event}) showing emission from final state particles (red) and emission from the initial state (blue).
The emissions will most likely be soft or collinear
to the emitter, which allows us to make some approximations.
The parton shower simply does the job of simulating emission
after emission until the radiating particle runs out of energy,  which is the point when
partons hadronize (green lines and blobs in Fig.~\ref{fig:event}).

To see how this works consider the splitting
of a gluon into a pair of quarks where the splitting probability is
given by
\begin{equation}
  \mathcal{P}_{qq}(t,t^\prime)\,=\,\int_t^{t^\prime} \frac{{\rm d} t}{t}\biggl| \raisebox{-5.5mm}{ \includegraphics[scale=0.05]{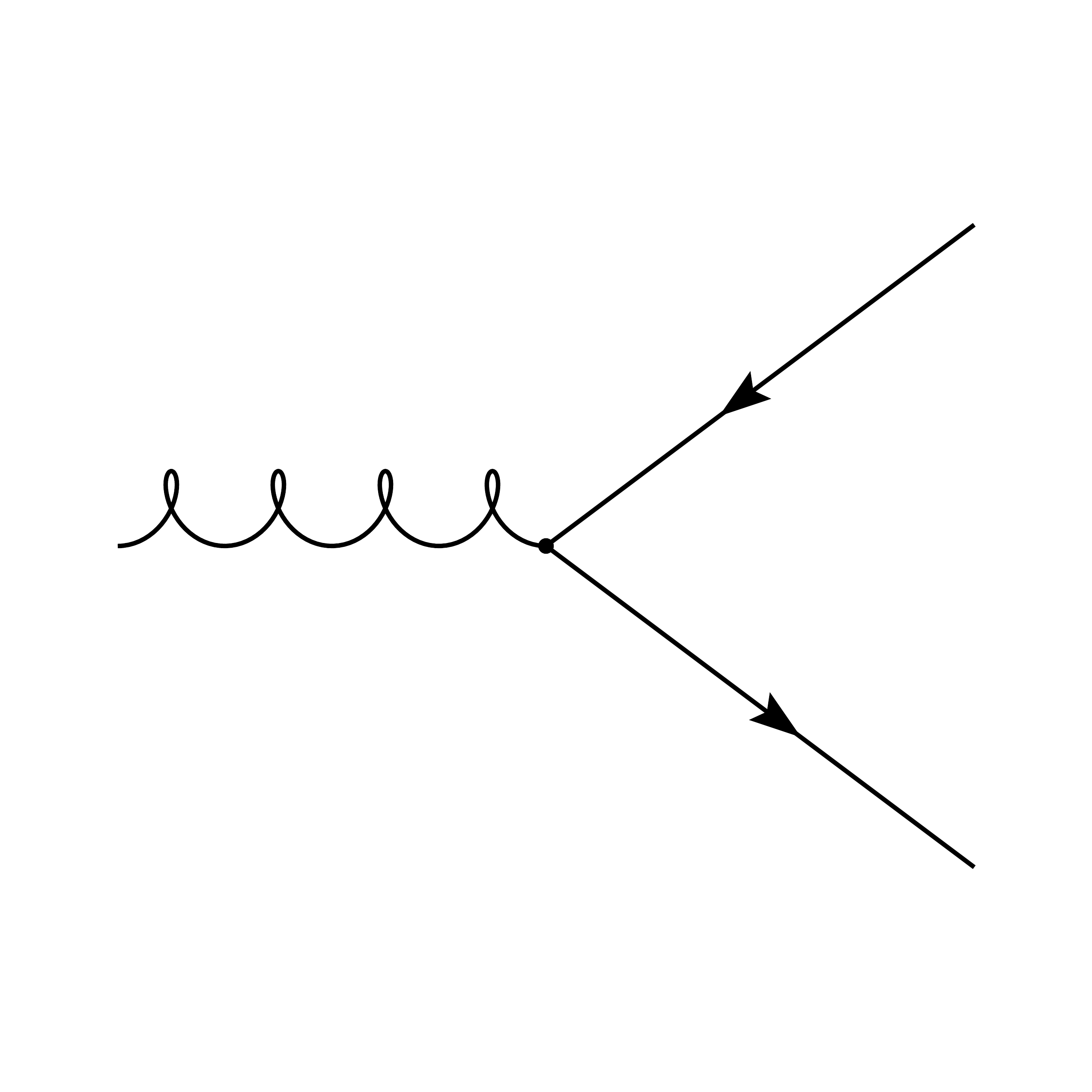}} \biggr|^2 \, \propto \, \frac{\alphas}{2\,\pi}  P_{qq}(z) \log\frac{t^\prime}{t}.
\end{equation}
Repeated splittings gives rise to a series that behaves like
 $$\frac{1}{m!}\left(\alpha_s \log\frac{t^\prime}{t}\right)^m, $$ which can symbolically be summed and exponentiated to give
\begin{equation}
  \Delta_{i}(t,t^\prime) = \exp{\left(-\sum_{j\in (q,g)}\mathcal{P}_{ij}(t,t^\prime)\right)}.
\end{equation}
$\Delta$ is called the Sudakov form factor, which represents
the unconditional branching probability for a parton not to undergo a branching
process between the two energy scales $t$ and $t^\prime$
\cite{Catani:1990rr,Ellis:1978sf,Amati:1978by}.
Note that the Sudakov form factor accurately represents the 
{\it leading logarithmic} (LL) behaviour of the splittings.
In practical implementations, the subleading logarithmic behaviour of parton showers
is not yet well understood, and varies from code to code. 
Keeping in mind that the outcome of the parton shower has leading order and leading
logarithmic accuracy, the main effect of showering a fixed order hard-process
is to add more particles to the final state, each carrying a transverse momentum
$\vec{k}_{iT}$ which is weighted according to $\Delta$.
The primary effect of adding more particles in the final state follows from
momentum conservation, and implies that
\begin{equation}
  \delta^{(2)}(\vecpth) \rightarrow \delta^{(2)}(\vecpth+\sum_{i=1}^N \vec{k}_{iT}),
\end{equation}
which shifts the peak of the distribution from zero to some non-zero value.
The energy that the shower can share among the additional final state
particles is limited by the partonic centre of mass energy
which in this case is of the order of $\mh$. Therefore, we should
expect that applying the parton shower to the LO $2 \rightarrow 1 $ process
will lead to a $\pth$ distribution that exhibits a broad peak that 
suddenly falls off for $\pth \approx \mh$.
\begin{figure}[tb]
  \centering{%
(a)    \includegraphics[width=0.45\textwidth]{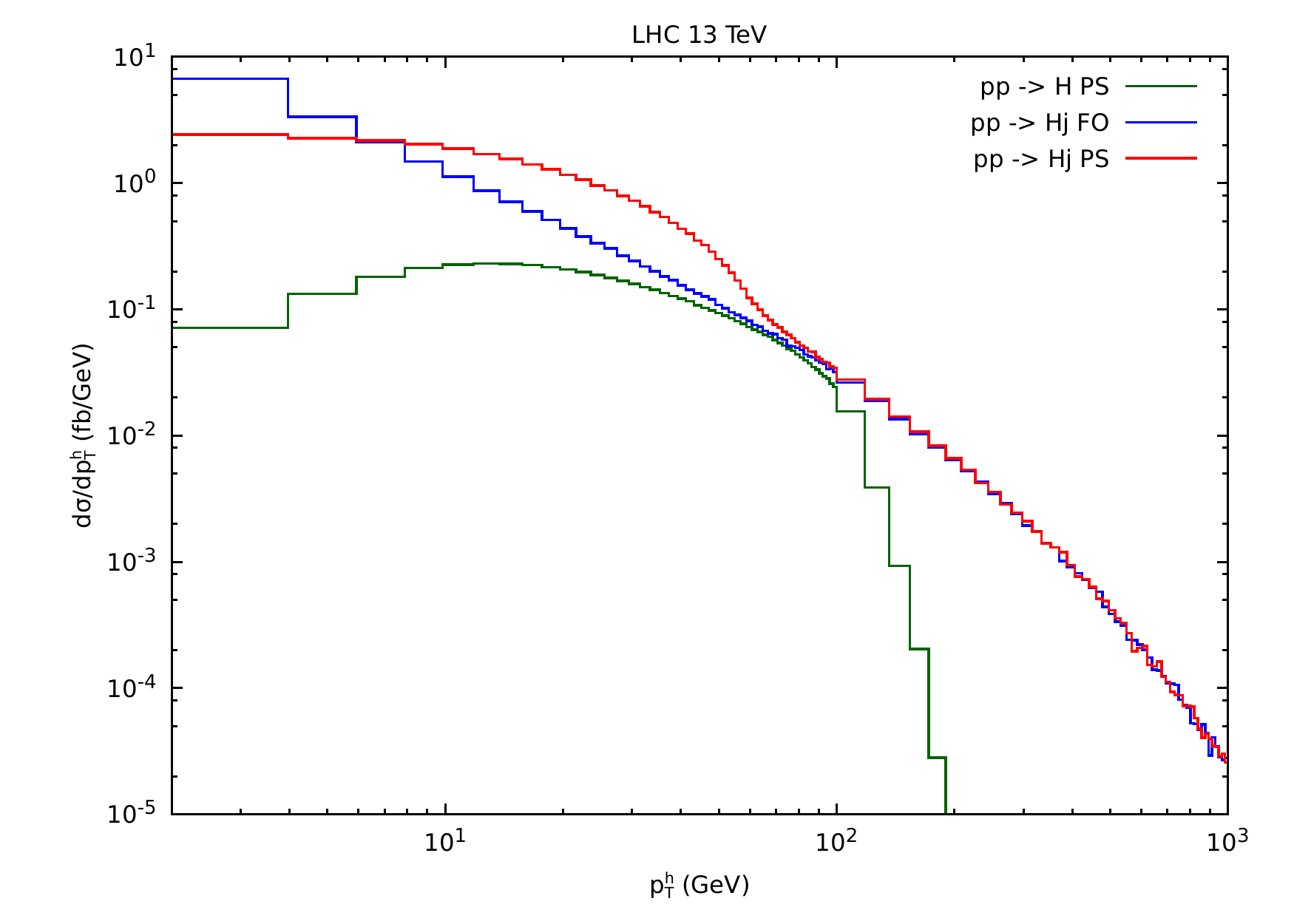}
(b)    \includegraphics[width=0.45\textwidth]{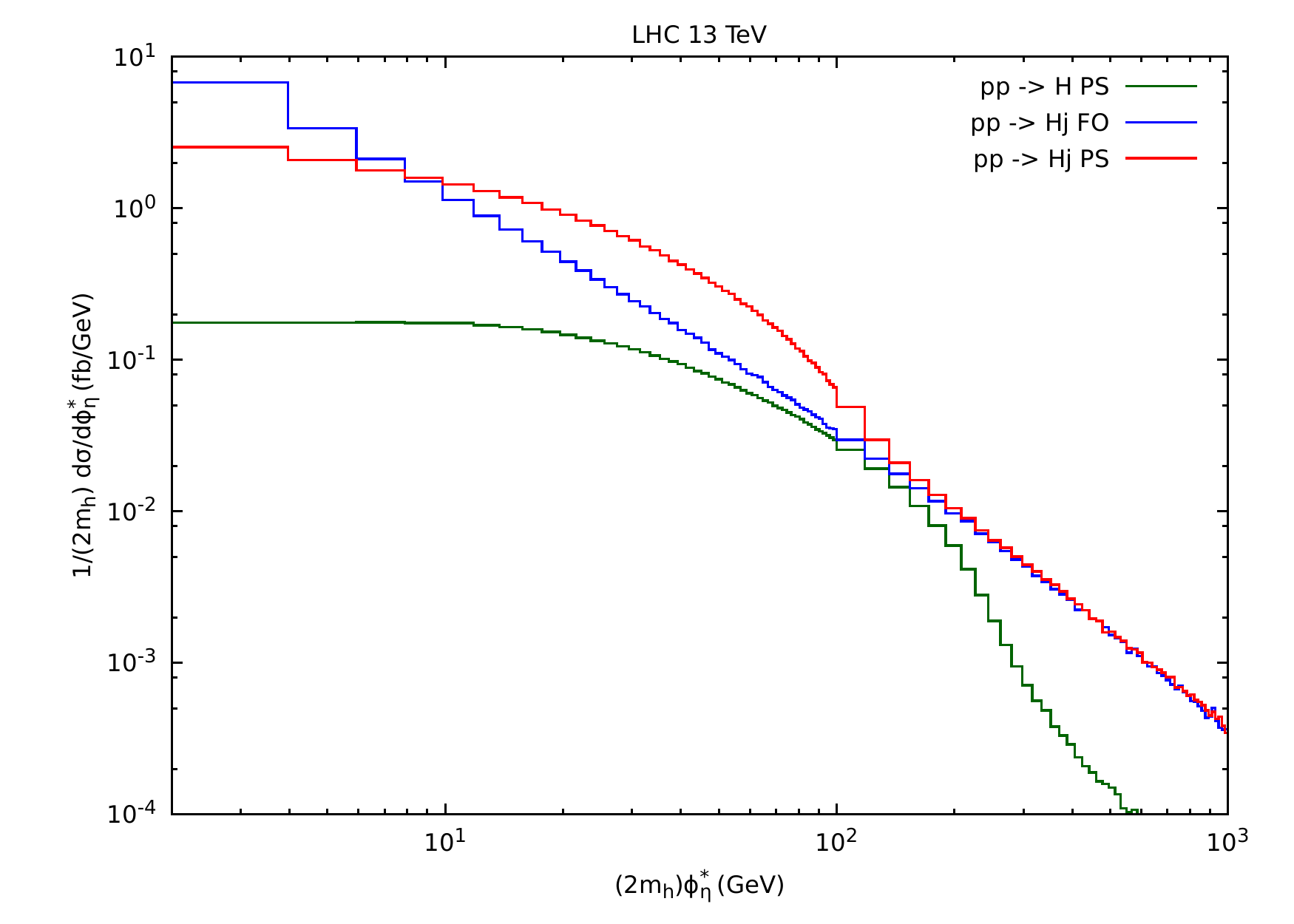}
  }
  \caption{The effect of the parton shower on (a) the $\pth$ and (b) $\phistar$ distributions. 
The curves show the effect of the parton shower on the $gg \to \h$ process, $pp \to \h$ PS (green), 
the fixed order $gg \to \h j$ process FO (blue) and the parton shower applied to the FO process, 
$pp \to \h j$ PS (red).}
  \label{fig:ps}
\end{figure}
This is indeed the behaviour that is shown in Fig.~(\ref{fig:ps})(a) by
the $pp\rightarrow \h$ PS prediction.

$\phistar$ is of course a different variable, but it shares the key feature that at the Born level, 
it is required to vanish.  The additional radiation in the parton shower produces a non-zero value for $\phistar$
and the $pp\rightarrow \h$ PS prediction shown Fig.~(\ref{fig:ps})(a) has the same gross features as the $\pth$ distribution, 
namely a broad shoulder accompanied by a tailing off at larger values of $\phistar$.

Although the shower generates many emissions, each emission is roughly a factor
of ten times less likely to happen than the previous one. 
The reason is that each emission comes at the cost
of an additional power of $\as$. Further, $\as$ is not
constant but a monotonically decreasing function, and for most
implementation of parton showers is evaluated at the $k_{T}$ of the
splitting. These two considerations together imply
that the first few emissions have a larger weight than the later emissions,
and splittings that exhibit a smaller $k_T$ are more likely than those
with a larger $k_T$. These two effects compete to determine the shape
of the distribution and the position of the peak.
Following these arguments, one could argue that a reasonable
approximation would simply be to go one order higher in $\as$
and include exactly the emission of one extra particle in the hard scattering itself.
In the case of $gg\rightarrow \h $, this process introduces a variety of new channels,
$gg \rightarrow \h g$, $q\bar q \rightarrow \h g$, $qg \rightarrow \h q$ that are discussed in detail in Sec.~\ref{sec:theory}.  
Here, we simply denote the sum of these
contributions by $gg\rightarrow \h j$ to reflect the fact that an additional parton $j$ 
is present in the final state. 

Let us first consider the case of $gg\rightarrow \h j$ at fixed order (FO),
{\it i.e.  with no parton shower}. As for the Born contribution,
we can write the differential cross section in terms of final state phase-space
and matrix elements,
\begin{equation}
  {\rm d} \hat{\sigma}_{gg\rightarrow \h j} = {\rm d} {\rm \Phi}_{2\rightarrow 2}\frac{ |\overline{\mathcal{M}_{gg \to \h j}}|^2}{F}\, \propto \frac{{\rm d}^2\vecpth}{\pth}
  \,\frac{{\rm d}^2\vec{k}_{jt}}{k_{jt}}\delta^{(2)}(\vecpth-\vec{k}_{jT}),
\end{equation}
where the $\delta$ function in this case simply forces the Higgs boson 
and the extra parton to be back-to-back and 
thus removes one of the two integration variables.
In contrast to the $gg\rightarrow \h$ case discussed earlier,   we find the
additional feature of the non trivial integration over $\pth$. Although
there is an extra particle recoiling against the Higgs boson, 
nothing prevents the extra parton to be very soft, or very collinear to the
beam axis. In both cases this gives rise to a logarithmic divergence.
What this means in practice is that events with $\pth\sim 0$~GeV will have a larger
weight ({\it i.e. a larger probability to be produced}).
This feature can clearly be seen as the $pp\rightarrow \h j$ FO prediction in Fig.~(\ref{fig:ps})(a).
In this case we indeed see a rise at low $\pth$ which diverges at $\pth=0$~GeV but is not included in the plot. We also see a rather hard distribution (compared to $pp\rightarrow \h$ PS) because the single hard emission is precisely modelled. One can see the same effects in the FO $\phistar$ distribution in  Fig.~(\ref{fig:ps})(b); a divergence at low $\phistar$ accompanied by a hard tail at high $\phistar$. 

Let us now investigate the effects of the parton shower on $gg\rightarrow \h j$.
As for the $2\rightarrow 1$ case, the main effect of the shower is to
shift and broaden the peak of the $\pth$ distribution. The main difference is that instead of having a sudden drop at $\pth \sim \mh$, the $\pth$ distribution is harder because more energy is available to the shower through the emission of an
extra particle in the hard process. We therefore
expect a transition between the low $\pth$ region where the shower dominates and distorts
the shape of the $pp\rightarrow \h j$ FO distribution and the high $\pth$ region where impact of the shower is negligible. This phenomenon is displayed as $pp\rightarrow \h j$ PS 
in Fig.~(\ref{fig:ps})(a). As one might expect, the same effects are evident in the $pp\rightarrow \h j$ PS  prediction for the $\phistar$ distribution in  Fig.~(\ref{fig:ps})(b); at low $\phistar$ the parton shower produces a smooth shoulder that merges onto the hard FO tail at high $\phistar$.



\clearpage
\FloatBarrier 
\section{Theoretical Predictions}
\label{sec:theory}

In this section we describe the calculation of the fixed order matrix elements relevant to the production of Higgs bosons with non-zero transverse momentum and non-zero $\phistar$.  We start by reviewing the general properties of the interaction shown in Fig.~\ref{fig:blobwithpt}.  In the Standard Model, this blob is mediated by heavy quark loops.  The one-loop matrix elements have been known for some time \cite{Ellis:1987xu,Baur:1989cm} but the two-loop matrix elements are not yet known exactly~\cite{art:Bonciani:2016qxi}.  We review the modern techniques that have been developed to tackle the complicated two-loop integrals that appear using the one-loop amplitude as an example.   
Finally, we show how the effective interaction can be recovered by taking the $\mt \to \infty$ limit.

\begin{figure}[hb]
  \centering{%
\includegraphics[width=0.2\textwidth]{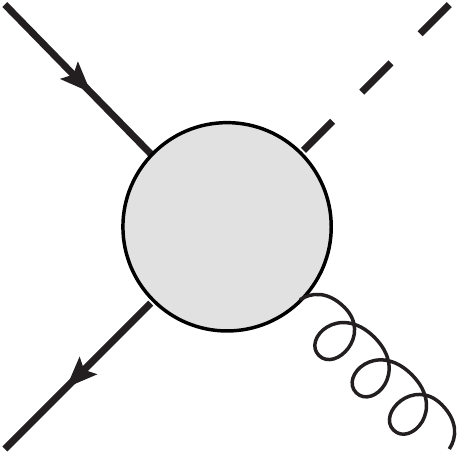}
  }
  \caption{Schematic diagram showing the production of a Higgs boson with transverse momentum from quark-antiquark annihilation.}
  \label{fig:blobwithptqq}
\end{figure}

Note that in addition to the gluonic process shown in Fig.~\ref{fig:blobwithpt}, there is also a process involving a quark-antiquark pair shown in Fig.~\ref{fig:blobwithptqq}
so that seven partonic channels contribute to the production of a Higgs boson plus a quark or gluon jet in proton-proton collisions:
\begin{align}
& qg \rightarrow q\h, \quad\quad\quad gq \rightarrow q\h, \quad\quad\quad \bar{q}g \rightarrow \bar{q}\h, \quad\quad\quad g\bar{q} \rightarrow \bar{q} \h, \nn \\
& q\bar{q}\rightarrow g\h, \quad\quad\quad \bar{q}q \rightarrow g\h, \quad\quad\quad gg \rightarrow g\h.
\label{eq:sevenchannels}
\end{align}

The calculational methodology is similar for all processes and in the following, we will mainly focus on the $gg \to g\h$ process.  

\subsection{General Properties}

\subsubsection{Tensorial Structure of the Amplitude}
\label{sec:tensor_dec}

The amplitude for the  $gg \rightarrow g\h$ process may be written in terms of the external polarisation vectors of the gluons and a third rank tensor $\M^{\mu \nu \tau}$ that describes the physics within the blob,
\begin{align}
\M_{gg \rightarrow g\h} = \M^{\mu \nu \tau} \, \varepsilon_{\mu}(p_1) \, \varepsilon_{\nu}(p_2) \, \varepsilon_{\tau}(p_3).
\label{eq:formfactor}
\end{align}
If one wishes to exploit the helicities of the external particles, then Eq.~\eqref{eq:formfactor} can be generalised,
\begin{align}
A_{gg \rightarrow gH}^{h_1h_2h_3} = \M^{\mu \nu \tau}_{h_1h_2h_3} \, \varepsilon^{h_1}_{\mu}(p_1) \, \varepsilon^{h_2}_{\nu}(p_2) \, \varepsilon^{h_3}_{\tau}(p_3),
\label{eq:formfactorhel}
\end{align}
where $h_i$ denotes the helicities of the gluons.
As we will see, all information required to evaluate  $\M^{\mu \nu \tau}_{h_1h_2h_3}$ for any combination of the helicities of the external gluons can be directly obtained from the general tensor  $\M^{\mu \nu \tau}$ 

The most general tensorial structure of $\M$ is formed by factors of the external momenta $p_i^{\mu}$,  the Lorentz invariant tensor $g^{\mu \nu}$ and the Lorentz invariant Levi-Civita tensor $\varepsilon^{\mu \nu \rho \sigma}$. The process $gg \rightarrow g\h$ has four external particles, but due to momentum conservation we may express the momentum of the Higgs boson, $p_4$, in terms of the momenta of the three gluons. We obtain the general expression for the tensor decomposed amplitude
\begin{align}
\M^{\mu \nu \tau} &= \sum_{ijk=1}^{3} F_{ijk} p_i^{\mu} p_j^{\nu} p_k^{\tau} + \sum_{i=1}^3 G_{i1} p_i^{\mu} g^{\nu \tau} + \sum_{i=1}^3 G_{i2} p_i^{\nu} g^{\mu \tau} + \sum_{i=1}^3 G_{i3} p_i^{\tau} g^{\mu \nu} + \sum_{i=1}^3 H_{i} p_i^{\sigma} \varepsilon^{\mu \nu \rho \sigma}
\label{eq:ffexpansion}
\end{align}
where $F_{ijk}$, $G_{ij}$ and $H_{i}$ are scalar functions called formfactors.
QCD is a parity invariant theory and therefore $H_i \equiv 0$. However, there may well be models in which this term is present. In this section, we are mainly concerned with the Standard Model and will neglect this parity violating term.
Eq.~\eqref{eq:ffexpansion} has $27$ $F_{ijk}$ terms and 9 $G_{ij}$ terms. It is important to note that while the values of the formfactors depend on the specifics of the physical theory, the number of loops etc. but the tensorial decomposition given by Eq.~\eqref{eq:ffexpansion} does not.

The terms in Eq.~\eqref{eq:ffexpansion} are not physically independent. This can be seen from the Ward identities imposed by the gauge invariance of the theory, which say that any amplitude that may be written as $\M^{\mu} \varepsilon_{\mu}(p)$ for a massless vector boson with momentum $p$, must satisfy $\M^{\mu} p_{\mu} = 0$. 

Imposing the Ward identity for each of the external gluons in turn and requiring that each of the (independent) second rank tensor coefficients vanish produces $30$ relations between the coefficients of Eq.~\eqref{eq:ffexpansion}, of which $22$ turn out to be independent. This leaves a gauge-invariant expression with $14$ independent terms. 

Of these $14$ terms $10$ turn out to be proportional to either $p_1^{\mu}$, $p_2^{\nu}$, or $p_3^{\tau}$. As polarization vectors have the property that $\varepsilon(p) \cdot p = 0$, these terms will yield no contribution to the amplitude of Eq.~\eqref{eq:formfactor}. Thus only four physical terms are left,
\begin{align}
\M^{\mu \nu \tau}_{\text{physical}} &= F_{212} T_{212}^{\mu \nu \tau} 
+ F_{332} T_{332}^{\mu \nu \tau}
+ F_{311} T_{311}^{\mu \nu \tau}
+ F_{312} T_{312}^{\mu \nu \tau},
\label{eq:mphysical}
\end{align}
where,
\begin{align} 
T_{212}^{\mu \nu \tau} =& ( s_{12} g^{\mu \nu} - 2 p_2^{\mu} p_1^{\nu} )( s_{23} p_1^{\tau} - s_{13} p_2^{\tau} ) / (2 s_{13}), \nn  \\
T_{332}^{\mu \nu \tau} =& ( s_{23} g^{\nu \tau} - 2 p_3^{\nu} p_2^{\tau} )( s_{13} p_2^{\mu} - s_{12} p_3^{\mu} ) / (2 s_{12}), \nn \\
T_{311}^{\mu \nu \tau} =& ( s_{13} g^{\tau \mu} - 2 p_1^{\tau} p_3^{\mu} ) ( s_{12} p_3^{\nu} - s_{23} p_1^{\nu} ) / (2 s_{23}), \label{eq:mphysicaltensors} \\
T_{312}^{\mu \nu \tau}  =& \Big( g^{\mu \nu} (s_{23} p_1^{\tau} - s_{13} p_2^{\tau}) + g^{\nu \tau} ( s_{23} p_2^{\mu} - s_{12} p_3^{\mu} ) + g^{\tau \mu} ( s_{12} p_3^{\nu} - s_{23} p_1^{\nu} ) \nn \\
& + 2 p_3^{\mu} p_1^{\nu} p_2^{\tau} - 2 p_2^{\mu} p_3^{\nu} p_1^{\tau} \Big) /2 \nn
\end{align}
with $s_{ij}=(p_i+p_j)^2$.

We notice a symmetry between the first three terms of Eq.~\eqref{eq:mphysical}. This is due to the cyclic permutation invariance of $\M$, e.g. $\M(p_1^{\mu},p_2^{\nu},p_3^{\rho}) = \M(p_2^{\nu},p_3^{\rho},p_1^{\mu})$, so we realize that only two of the terms of Eq.~\eqref{eq:mphysical} are truly independent. These can be chosen as e.g. $F_{311}$ and $F_{312}$, and are denoted the independent formfactors of the process.

The formfactors can be directly extracted from the tensor by introducing ``projectors'', 
$Q^{ijk}_{\mu \nu \tau}$~\cite{art:Gehrmann:2011aa} with the property that
\begin{align}
Q^{ijk}_{\mu \nu \tau} \, \M_{\text{physical}}^{\mu \nu \tau}  = F_{ijk}.
\label{eq:prodef}
\end{align}
It is natural to construct the $Q^{ijk}_{\mu \nu \tau}$ as linear combinations of the $T_{ijk}^{\mu \nu \tau}$ appearing in Eq.~\eqref{eq:mphysicaltensors} and to fix the coefficients such that Eq.~\eqref{eq:prodef} is satisfied. In general, these coefficients depend on the $s_{ij}$ and the dimensionality of space-time $d$. 
Explicit calculation yields, 
\begin{align}
Q^{\mu \nu \tau}_{212} &= \frac{1}{(d-3) s_{12}} \left( 
- \frac{d s_{13}}{s_{12}^2 s_{23}} T_{212}^{\mu \nu \tau}
+ \frac{d-4}{ s_{23}^2} T_{332}^{\mu \nu \tau}
+ \frac{d-4}{s_{12} s_{13}} T_{311} ^{\mu \nu \tau}
+ \frac{d-2}{s_{12} s_{23}} T_{312}^{\mu \nu \tau}
\right), \\
Q^{\mu \nu \tau}_{332} &= \frac{1}{(d-3) s_{23}} \left( 
   \frac{d-4}{s_{12} s_{23}} T_{212}^{\mu \nu \tau}
- \frac{d s_{12}}{ s_{13} s_{23}^2} T_{332}^{\mu \nu \tau}
+ \frac{d-4}{s_{13}^2} T_{311}^{\mu \nu \tau}
+ \frac{d-2}{s_{13} s_{23}} T_{312}^{\mu \nu \tau}
\right), \\
Q^{\mu \nu \tau}_{311} &= \frac{1}{(d-3) s_{13}} \left( 
   \frac{d-4}{s_{12}^2} T_{212}^{\mu \nu \tau}
+ \frac{d-4}{ s_{13} s_{23}} T_{332}^{\mu \nu \tau}
- \frac{d s_{23}}{s_{12} s_{13}^2} T_{311}^{\mu \nu \tau}
+ \frac{d-2}{s_{12} s_{13}} T_{312}^{\mu \nu \tau}
\right), \\
Q^{\mu \nu \tau}_{312} &= \frac{(d-2)}{(d-3) s_{12} s_{23} s_{13}} \left( 
   \frac{s_{13}}{s_{12}}  T_{212}^{\mu \nu \tau}
+ \frac{ s_{12}}{s_{23}}  T_{332}^{\mu \nu \tau}
+ \frac{ s_{23}}{s_{13}} T_{311}^{\mu \nu \tau}
+ \frac{d}{d-2}  T_{312}^{\mu \nu \tau}
\right),
\end{align}
where the symmetry between the first three projectors remains apparent.

For the quark channels, rather than three free Lorenz indices, the tensorial amplitude will have one Lorentz index along with a spinor and an anti spinor index. For that case a similar projector technique may be used, see e.g. Ref.~\cite{art:Gehrmann:2011aa}.

\subsubsection{From the Amplitude to the Cross-section}

The partonic cross-section is given in terms of the amplitude, $\M_{gg \rightarrow gH}$, by
\begin{align}
\hat{\sigma}_{gg \rightarrow g\h}(s_{12}) = \int \mathrm{d} \Phi_{2 \rightarrow 2} \frac{\overline{|\M_{gg \rightarrow gH}|}^2}{F}.
\end{align}
In four dimensions the 2-to-2 phase space for production of a Higgs boson plus jet is
\begin{equation}
\int \mathrm{d} \Phi_{2 \rightarrow 2} = \int_{-1}^{1} \mathrm{d} \cos \theta \int_{0}^{2 \pi} \mathrm{d} \phi \frac{1}{32 \pi^2} \left(1-\frac{\mh^2}{s} \right).
\end{equation}
The gluon channel contribution to the hadronic cross-section is given by
\begin{equation}
\sigma_{pp \rightarrow hj}(s) = \int_0^1 \mathrm{d} z_1 \int_0^1 \mathrm{d} z_2 \int_{s_{12}^{\mathrm{min}}}^s \mathrm{d} s_{12} \hat{\sigma}_{gg \rightarrow gh}(s_{12}) \delta(\hat{s}-z_1 z_2 s) f_g(z_1) f_g(z_2),
\end{equation}
where $f_g$ is the gluon PDF and $ s_{12}^{\mathrm{min}} = \mh^2$ is the minimum energy required to produce a Higgs boson on-shell.
Introducing the parton luminosity,
\begin{equation}
\frac{\mathrm{d} \mathcal{L}_{ij}}{\mathrm{d} \tau} = \int_\tau^1 \frac{\mathrm{d} x}{x} f_i(x) f_j \left( \frac{\tau}{x} \right) 
\end{equation}
we may write the hadronic cross-section as
\begin{equation}
\sigma_{pp \rightarrow hj}(s) = \int_{\tau_0}^1 \mathrm{d} \tau \frac{\mathrm{d} \mathcal{L}_{gg}}{\mathrm{d} \tau} \hat{\sigma}_{gg \rightarrow gh}(\tau s),
\end{equation}
where $\tau_0=\mh^2/s$.

In conventional dimensional regularisation (CDR) the spin and colour averaged absolute square of the amplitude is given by
\begin{equation}
\overline{|\M_{gg \rightarrow g\h}|}^2 = \frac{1}{(d-2)^2} \frac{1}{(N_c^2-1)^2} |\M_{gg \rightarrow g\h}|^2,
\end{equation}
where the we have divided by the number of polarisations, $(d-2)$, for each incoming gluons in order to average over their polarisations and by $N_c^2-1$ for each incoming gluon to average over their colours.
The absolute square of the matrix element is given in terms of the form factors by
\begin{align}
\label{eq:A2formfactors}
|\M_{gg \rightarrow g\h}|^2 &= 
\sum_{h_1,h_2,h_3}
(\varepsilon^{h_1}_\mu)^* \varepsilon^{h_1}_{\hat{\mu}} 
(\varepsilon^{h_2}_\nu)^* \varepsilon^{h_2}_{\hat{\nu}} 
(\varepsilon^{h_3}_\tau)^*\varepsilon^{h_3}_{\hat{\tau}} 
(\M^{\mu \nu \tau})^* \M^{\hat{\mu} \hat{\nu} \hat{\tau}} \\
&=  \frac{1}{4}
 \ \Big[\frac{(d-2) |F_{212}|^2 s_{12}^3 s_{23}}{s_{13}} + F_{212} F_{332}^* s_{12} s_{23}^2                 + F_{212} F_{311}^* s_{12}^2 s_{13}                 + (d-2) F_{212} F_{312}^* s_{12}^2 s_{23}  \nn \\
& + F_{332} F_{212}^* s_{12} s_{23}^2             + \frac{(d-2) |F_{332}|^2 s_{13} s_{23}^3}{s_{12}} +  F_{332} F_{311}^* s_{13}^2 s_{23}                + (d-2)  F_{332} F_{312}^* s_{13} s_{23}^2 \nn \\
& + F_{311} F_{212}^* s_{12}^2 s_{13}             + F_{311} F_{332}^* s_{13}^2 s_{23}                 + \frac{(d-2) |F_{311}|^2 s_{12} s_{13}^3}{s_{23}} + (d-2) F_{311} F_{312}^* s_{12} s_{13}^2 \label{eq:me2} \\
& +(d-2) F_{312} F_{212}^* s_{12}^2 s_{23}        + (d-2) F_{312} F_{332}^* s_{13} s_{23}^2            +(d-2) F_{312} F_{311}^*  s_{12} s_{13}^2          \nn \\ 
& + (3d-8) |F_{312}|^2 s_{12} s_{13} s_{23} \Big], \nn
\end{align}
where for the polarisation sums we have used the axial gauge expression
\begin{equation}
\sum_{pols} \left( \varepsilon^\mu_{i}(p_i) \right)^* \varepsilon^{\nu}_{i}(p_i) = 
- g^{\mu \nu} + \frac{p_i^\mu n_i^\nu + p_i^\nu n_i^\mu}{p_i \cdot n_i},
\end{equation}
with $n_i$ an arbitrary reference vector not collinear to $p_i$.  

Eq.~\eqref{eq:me2} looks rather cumbersome and a simpler expression for the absolute square of the matrix element can be written in terms of helicity amplitudes. We first fix the dimensionality of the space-time to four for all external particles, which corresponds to working in the 't Hooft-Veltman scheme (rather than CDR). In four dimensions we can then use the spinor helicity formalism to compute the helicity amplitudes, for a discussion of this topic see \cite{Xu:1986xb,Berends:1982ie,Berends:1983ez,Berends:1983ey,Dixon:1996wi}.

Each of the three gluons can have one of two helicities thus there are $2^3$ helicity amplitudes, though they are not all independent. 
Firstly, the helicity amplitudes are related by parity,
\begin{equation}
\M_{gg \rightarrow g\h}^{h_1 h_1 h_3} = - \M_{gg \rightarrow g\h}^{-h_1 -h_2 -h_3},
\end{equation}
this reduces the number of potentially independent helicity amplitudes to 4.
However, the remaining 4 are also related due to the permutation invariance of the amplitude
\begin{equation}
\M_{gg \rightarrow g\h}^{++-}(s_{12},s_{13},s_{23}) = \M_{gg \rightarrow g\h}^{+-+}(s_{13},s_{23},s_{12}) =  \M_{gg \rightarrow g\h}^{-++}(s_{23},s_{12},s_{13}).
\end{equation}
In terms of the form factors the two independent helicity amplitudes are given by
\begin{align}
\M_{gg \rightarrow g\h}^{+++} & = \frac{- s_{12} s_{23} s_{13} }{\sqrt{2} \langle 12 \rangle \langle 23 \rangle \langle 31 \rangle} \left( 
\frac{s_{12}}{2 s_{13}} F_{212} 
+ \frac{s_{23}}{2 s_{12}} F_{332}
+ \frac{s_{13}}{2 s_{23}} F_{311}
+ F_{312}
\right), \\
\M_{gg \rightarrow g\h}^{++-} & = \frac{F_{212} s_{12} \langle 2 3 \rangle [21]^2}{2 \sqrt{2} \langle 12 \rangle [31]}.
\end{align}
We may write now the absolute square of the matrix element as a sum of absolute squares of the 8 helicity amplitudes
\begin{equation}
|\M_{gg \rightarrow g\h}|^2 = \sum_{h_1, h_2, h_3 = \pm} |\M_{gg \rightarrow g\h}^{h_1 h_2 h_3}|^2.
\label{eq:A2helamps}
\end{equation}
When using this expression care should be taken to average over the correct number of transverse polarisations of the gluon, now 2 rather than $d-2$.

Unlike the formfactors themselves, Eqs.~\eqref{eq:A2formfactors} and \eqref{eq:A2helamps} are quite general and do not depend on the details of the interaction.

\subsection{The Standard Model at one-loop\label{sec:oneloop}}
\label{sec:SMoneloop}

\begin{figure}
\begin{center}
    \begin{subfigure}[b]{0.30\textwidth}
        \includegraphics[width=\textwidth]{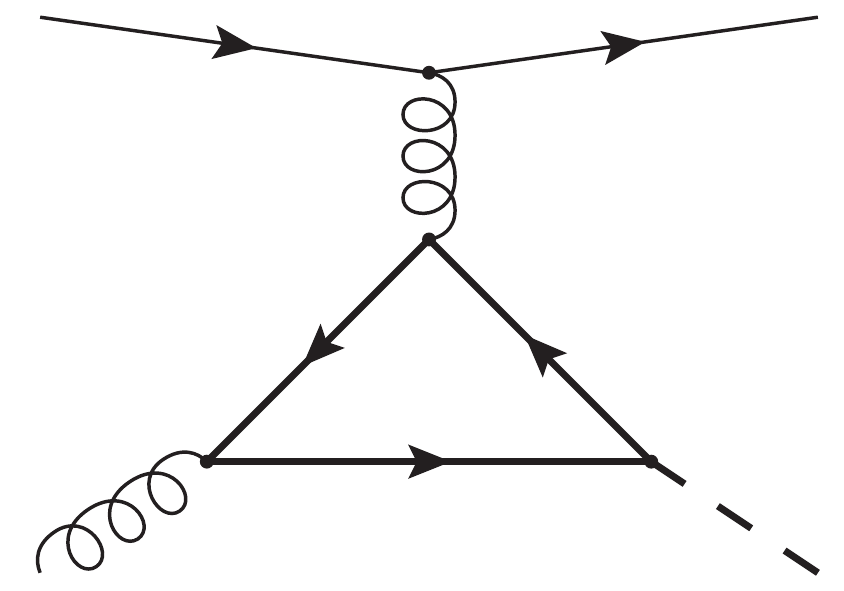}
    \end{subfigure}
    \quad
        \begin{subfigure}[b]{0.30\textwidth}
        \includegraphics[width=\textwidth]{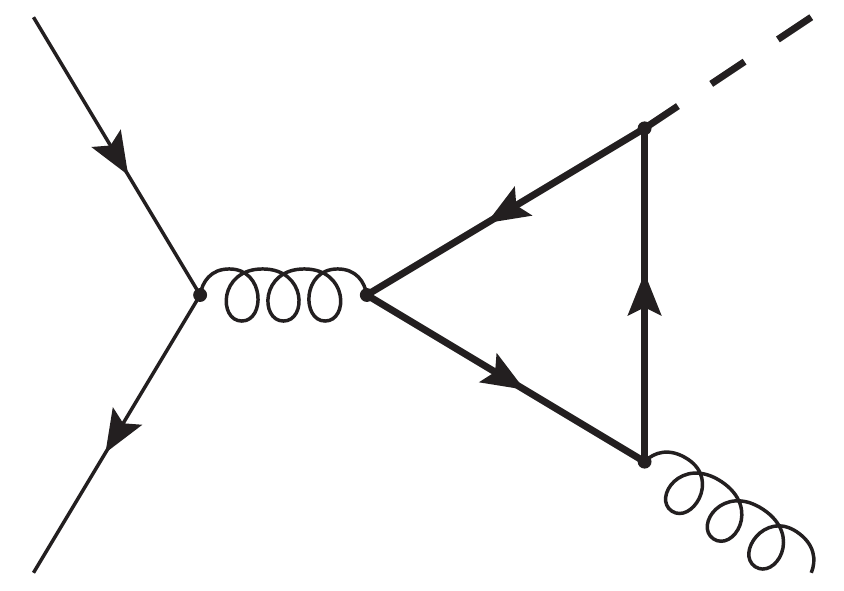}
    \end{subfigure}
    \quad
     \begin{subfigure}[b]{0.30\textwidth}
        \includegraphics[width=\textwidth]{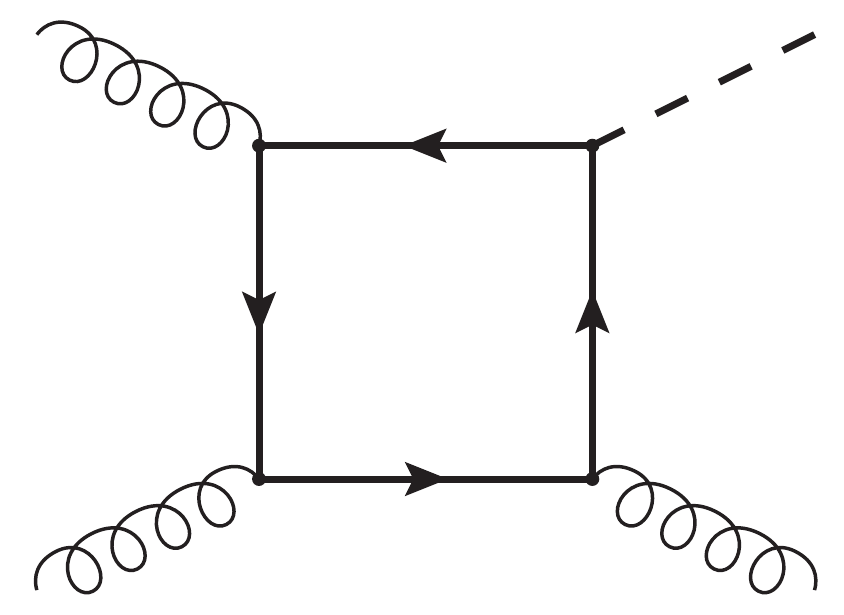}
    \end{subfigure}
    \caption{Example Feynman diagrams for the channels $qg\rightarrow q\h$, $q\bar{q}\rightarrow g\h$ and $gg\rightarrow g\h$ at leading order in the Standard Model.}
    \label{fig:allchannels1loopsm}
\end{center}
\end{figure}
The \LO\ contribution to Higgs boson plus jet production contains Feynman diagrams with one loop. Examples of the Feynman diagrams in the $qg\rightarrow q\h$, $q\bar{q}\rightarrow g\h$ and $gg\rightarrow g\h$ channels are shown in Fig.~\ref{fig:allchannels1loopsm}.
The channels involving a quark in the initial or final state are all related by crossing symmetries and so it is sufficient to calculate the matrix element for any one of these channels and obtain the others by crossing. The result for all partonic channels was first calculated in 1988~\cite{Ellis:1987xu} and later analysed in detail for Higgs boson plus jet production, including expansions in the limit of large top-quark mass~\cite{Baur:1989cm}.

\begin{figure}
    \centering
(a)
    \begin{subfigure}[b]{0.45\textwidth}
        \includegraphics[width=\textwidth]{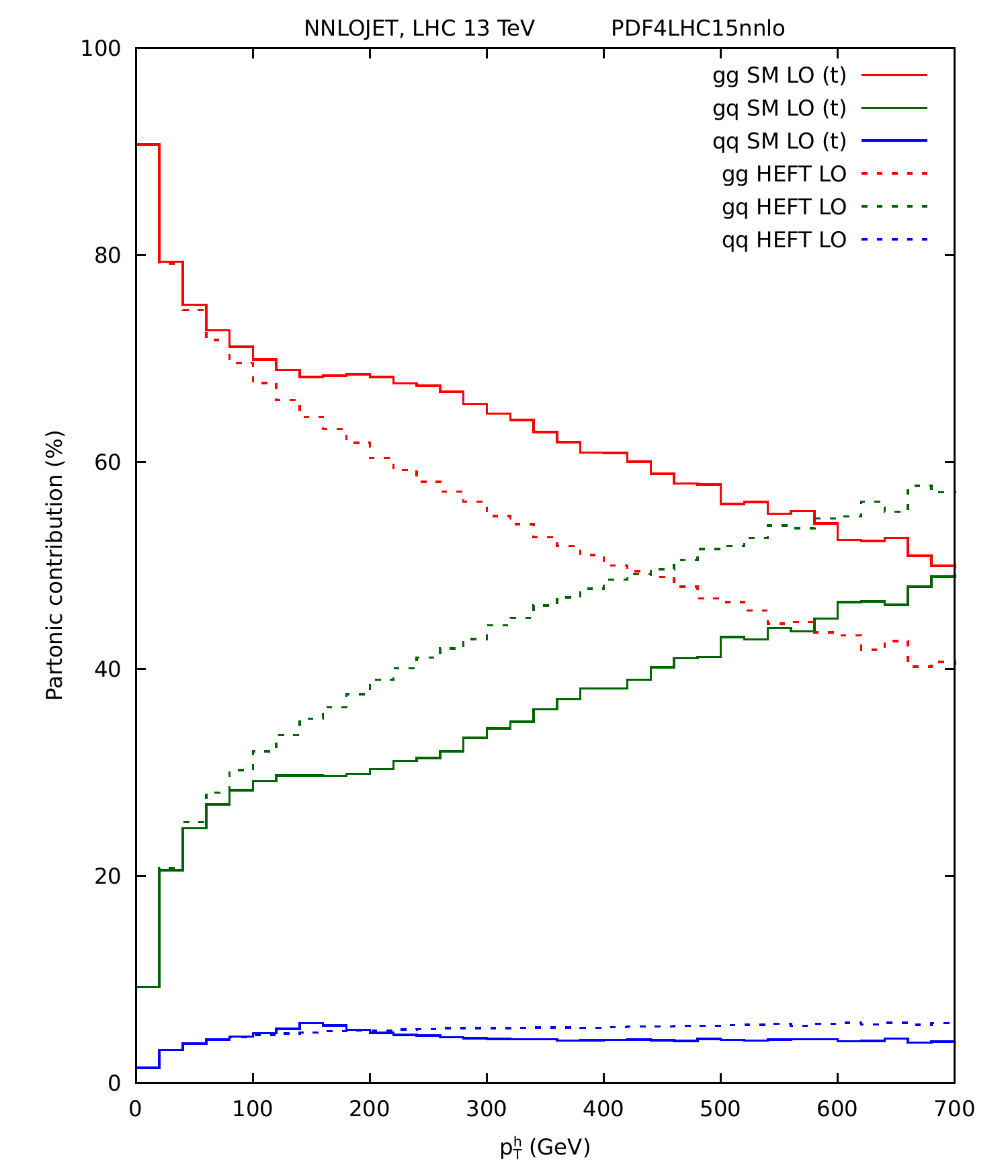}
    \end{subfigure}
(b)
    \begin{subfigure}[b]{0.45\textwidth}
        \includegraphics[width=\textwidth]{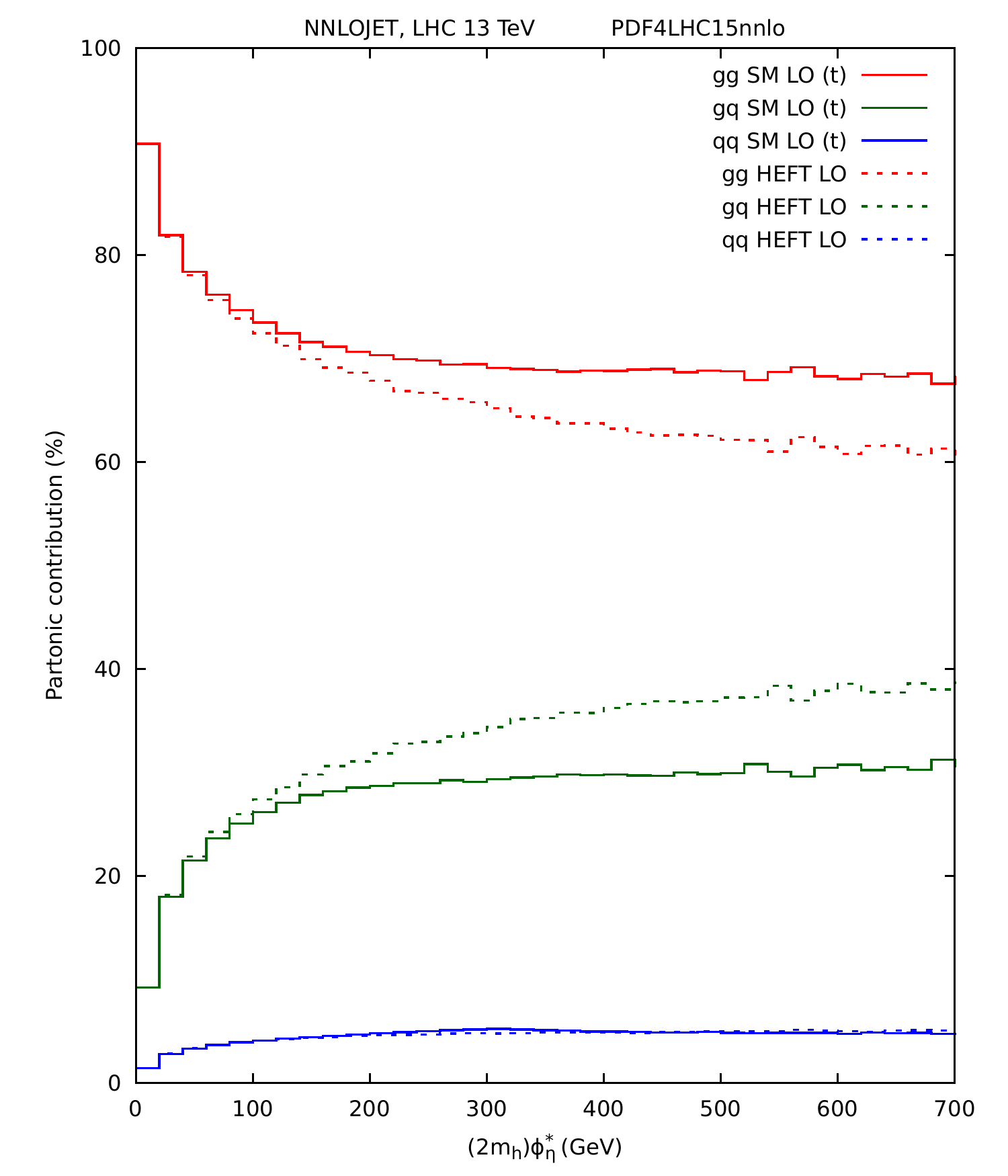}
    \end{subfigure}
    \caption{The percentage contributions from the gluon-gluon (red), gluon-quark (green) and quark-quark (blue) channels to the  (a) \pth and (b) \phistar distributions at $\sqrt{s}=13$~TeV in the Standard Model with a top quark loop.  The breakdown using the Higgs Effective Theory is shown dashed.}
    \label{fig:channels}
\end{figure}

Fig~\ref{fig:channels} shows the percentage contribution of the $gg$, $qg \equiv qg+\bar{q}g+gq+g\bar{q}$ and $qq \equiv q\bar{q} + \bar{q}q + qq + \bar{q}\bar{q}$ partonic channels for the \pth and \phistar distributions. We see that for the LHC the gluon-gluon channel is dominant due to the high gluon luminosity, but that the quark-gluon channel is also important. We clearly see the effect of the $gg \to t\bar{t}$ threshold that enhances the $q\bar q$ channel around $\pth \sim \mt$. We note that dominance of the gluon-gluon channel persists to higher values of 2\mh\phistar.

We will recompute this channel using techniques that are also applicable to multi-loop calculations. For this channel, there are 12 diagrams contributing at leading order: the two diagrams shown in Fig.~\ref{fig:gggh1loopsm} plus the 3! permutations of the gluons.
\begin{figure}
\begin{center}
    \begin{subfigure}[b]{0.30\textwidth}
        \includegraphics[width=\textwidth]{phistar/figures/hj1loop/gggh/diag1}
    \end{subfigure}
    \qquad
    \qquad
    \begin{subfigure}[b]{0.30\textwidth}
        \includegraphics[width=\textwidth]{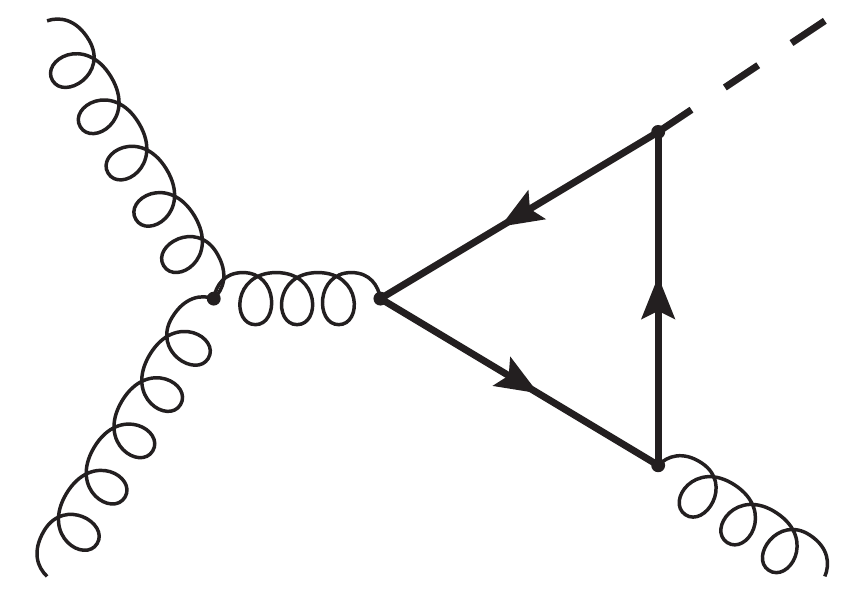}
    \end{subfigure}
    \caption{Feynman diagrams (2 out of 12) for the process $gg\rightarrow gH$ at leading order in the Standard Model.}
    \label{fig:gggh1loopsm}
\end{center}
\end{figure}

In general we can write an $L$-loop amplitude as
\begin{align}
A_{L-\text{loop}} = \sum_{i \in \text{diagrams}}  \int \prod_{l=1}^L  \frac{\id^d k_l}{i \pi^{d/2}} \frac{N_i(\{k\})}{\prod_{j \in i} D_j(\{k\})}
\label{eq:Lloopdiagrams}
\end{align}
Here $D_j$ denote the denominators stemming from the propagators of the Feynman diagrams, i.e. terms of the form $(k+p)^2-m^2$, while $N_i$ contains the rest of the terms in the Feynman diagram, i.e all gamma matrices, colour factors, polarization vectors, spinors, and all other factors provided by the Feynman rules. Most importantly, $N_i$ is a scalar.  This means that after performing the traces, all of the loop momenta appear either contracted with one of the physical momenta, $k_i\cdot p_j$, or one of the external polarisations, $k_i\cdot \varepsilon_j$, or with another loop momentum $k_i\cdot k_j$.  Here we will adopt the strategy of projectors described in Sec.~\ref{sec:tensor_dec} so that the polarisation vectors are removed and the loop momentum only appears in terms like $k_i\cdot p_j$ and $k_i\cdot k_j$.

There are several public codes available for generating expressions for Eq.~\eqref{eq:Lloopdiagrams}, see for example, Refs.~\cite{art:Nogueira:1991ex, art:Hahn:1998yk, art:Hahn:2000kx, art:Shtabovenko:2016sxi}. These programs typically generate algebraic expressions for the Feynman diagrams (according to some model) without performing any algebra. This means that the numerators of the diagrams in general will contain long strings and traces of (Dirac) gamma matrices, from external fermions and fermions in loops respectively, along with products of the colour factors $T^a_{ij}$ and $f_{abc}$ from vertices containing coloured particles. Several algebraic manipulation programs enabling the reduction of such factors exist, see for example, Refs.~\cite{art:Hahn:1998yk, art:Vermaseren:2000nd}. 

\subsubsection{Integral Families}
\label{sec:integralfamilies}

Finding a systematic and homogeneous way of expressing Feynman integrals, is crucial in order to identify relations between them. For a process with $P$ external particles in (integer) $D$ dimensions, only $E = \min(P-1, D)$ of these external momenta will be independent. $P$ is reduced by one due to momentum conservation relating the momenta, while in a $D$-dimensional space, any vector can be expressed as a linear combination of $D$ basis vectors. For an $P$-point process with $L$ loops, it is not hard to realize that the number of inherently different scalar products of the type $k_i\cdot p_j$ or $k_i\cdot k_j$ is given by
\begin{align}
N &= L (L+1)/2 + LE
\end{align}
where the first term comes from contracting the loop momenta with themselves, and the second from contraction the loop momenta with the external ones.



For $H+j$ production $E=3$ so that at one-loop $N=4$. This is exactly the number of propagators in the box-diagram (seen as $g_9$ on Fig.~\ref{fig:precanonical}) and reflects the fact that all propagators of integrals contributing to the one-loop process can be written as a subset of those of the box. We choose the propagators of the box diagram to be
\begin{align}
D_1 &= k_1^2 - \mt^2, \nn \\
D_2 &= (k_1-p_1)^2 - \mt^2, \nn \\
D_3 &= (k_1-p_1-p_2)^2 - \mt^2, \nn \\
D_4 &= (k_1-p_1-p_2-p_3)^2 - \mt^2,
\label{eq:boxpropagators}
\end{align}
and we introduce the {\it integral family}
\begin{equation}
I_{\alpha_1,\alpha_2, \alpha_3 , \alpha_4} = \int \frac{\mathrm{d}^d k_1}{i \pi^{d/2}} \frac{1}{D_1^{\alpha_1} D_2^{\alpha_2} D_3^{\alpha_3} D_4^{\alpha_4}}.
\end{equation}
Here the indices $\alpha_1, \ldots, \alpha_4 \in \mathbb{Z}$ encode the power of the propagators appearing in a particular integral. As we will see, there are nine master integrals, which can be obtained by setting one or more of the $\alpha_i$ equal to zero.

In this case, the four independent scalar products are related to the propagator factors through,
\begin{align}
k_1\cdot k_1 &= D_1+\mt^2,\nn \\
2 k_1\cdot p_1 &= D_1-D_2+p_1^2,\nn \\
2 k_1\cdot p_2 &= D_2-D_3+(p_1+p_2)^2-p_1^2,\nn \\
2 k_1\cdot p_3 &= D_3-D_4+(p_1+p_2+p_3)^2-(p_1+p_2)^2.  
\end{align}
The general strategy is to rewrite scalar products involving loop momentum appearing in the numerator as propagator factors.  Sometimes these propagator factors will cancel against one of the propagators of the Feynman diagram, reducing the corresponding $\alpha_i$ by one. Any propagator factors remaining in the numerator simply have a negative value for the corresponding $\alpha_i$.

To express all integrals appearing in the problem we will also need ``crossed'' integral families, these are related to the above integral family by permuting the external momenta. 

For the two-loop process $N=9$, but as we will see later, the most complicated Feynman integrals have at most seven propagators, so only seven of the nine scalar products can be expressed in terms of the propagators actually present in the problem. The remaining two are denoted irreducible scalar products (or ISPs) and they are inherently present in the problem. To identify and apply such relations systematically, goes under the name of integrand reductions, and is described in further detail in Appendix~\ref{sec:IntegrandReduction}.

\subsubsection{Integration-By-Parts identities}
\label{sec:IBPs}

One way of obtaining relations between Feynman integrals within a particular integral family is through the application of Integration-By-Parts (IBP) relations. The general form of the IBP relation states that
\begin{align}
\int \! \frac{\id^d k}{i \pi^{d/2}} \, \frac{\partial}{\partial k^{\mu}} v^{\mu} F(k) &= 0,
\label{eq:IBP}
\end{align}
where $v$ can be any Lorentz vector and where $F(k)$ contains the Feynman integrand along with the remaining integrations for a multi-loop integral~\cite{Grozin:2011mt}.

Eq.~\eqref{eq:IBP} works by letting the differential operator work on $F(k)$, where it will yield a sum of different terms, and thus expose a relationship between the corresponding Feynman integrals. We  illustrate this with the example of the one-loop massless ($p_1^2=p_2^2=0$) triangle. We define
\begin{align}
J_{a_1,a_2,a_3} &= \int \! \frac{\id^d k}{i \pi^{d/2}} \, \frac{1}{((k-p_1)^2)^{a_1} (k^2)^{a_2} ((k+p_2)^2)^{a_3}},
\end{align}
where the normal triangle integral of course corresponds to $J_{1,1,1}$. Applying Eq.~\eqref{eq:IBP} to $J_{1,1,1}$ with $v = k$ (and using the fact that $\partial k^{\mu}/\partial k^{\mu}  = d$), immediately yields the relation
\begin{align}
(d-4) J_{1,1,1} - J_{2,0,1} - J_{1,0,2} &= 0,
\end{align}
showing that the triangle integral $J_{1,1,1}$ may be re-expressed solely in terms bubble-type integrals (i.e. integrals with only two propagators), and the use of a second IBP will show that the triangle $J_{1,1,1}$ may be reduced solely to the plain bubble integral $J_{1,0,1}$, the relation being
\begin{align}
J_{1,1,1} &= \frac{-2 (d-3)}{s_{12} (d-4)} J_{1,0,1}.
\end{align}

Choosing different values for $v$, built from either internal or external momentum, and using different starting (or seed) integrals produces different IBP relations. The complete set of linearly independent IBP relations is finite, but nevertheless, it is conjectured that these IBP relations are sufficient to deduce all relations between Feynman integrals~\cite{art:Lee:2008tj}\footnote{Except relations imposed by symmetries of the diagrams, and relations imposed by the finite dimensionality of the space spanned by the external momenta.}.

The integrals that cannot be eliminated by the IBP reduction are known as {\it master integrals}. While the number of master integrals is fixed by the full set of IBP relations, there is a lot of freedom in choosing which set of integrals to use. In general it is convenient to pick a set as {\it simple} as possible, which may be defined firstly as integrals with a minimal set of propagators (so choosing the bubble over the triangle), and then as integrals with as few propagators with powers higher than one, or alternatively with the smallest number of numerator factors.

An algorithm which systematizes the IBP relations according to such a criterion of simplicity, exists and is known as the Laporta algorithm~\cite{art:Laporta:2001dd}, and several implementations of that algorithm are publicly available~\cite{art:Anastasiou:2004vj, art:Smirnov:2008iw, art:vonManteuffel:2012np, art:Lee:2012cn, art:Smirnov:2014hma,Maierhoefer:2017hyi}.

Even with the implementations mentioned above, the integral reductions are a bottleneck in current multi-loop Feynman integral calculations. Therefore several attempts have been made at finding alternative methods to achieve the reduction to master integrals. Examples of this include attempts at simplifying the IBPs in order to avoid integrals with squared propagators in the intermediate results~\cite{art:Gluza:2010ws, art:Ita:2015tya, art:Larsen:2015ped}, or at reducing directly to master integrals by identifying characteristic ``master contours'' in complex phase space~\cite{art:CaronHuot:2012ab, art:Sogaard:2013fpa}. None of these developments will be described further here.

The number of master integrals for a given multi-loop process is  highly dependent on the number of loops and mass-scales and can range from a few to a number in the hundreds. For the two-loop planar $gg \rightarrow g\h$ there are 125 master integrals - too many for them to be listed here, see Ref.~\cite{art:Bonciani:2016qxi} for the full list and an accompanying figure. 

Beyond one-loop, more than one integral family is needed when it is not possible to write all the integrals as having propagators which are a subset of the propagators of one diagram. For the planar contribution to $gg \rightarrow g\h$ two families are needed (one corresponding to diagrams where the massive quark goes all the way around the diagram, and one where it stays on one side), but for the practical calculation four different families were used~\cite{art:Bonciani:2016qxi}.

\subsubsection{One-Loop Master Integrals}
\label{sec:oneloopmasters}

Use of the IBP identities shows that there are 9 master integrals (other integrals related to these by crossing symmetries also appear in the amplitude) defined by 
\begin{align}
\label{eq:precanonical}
    &g_1(\mt^2) = I_{2,0,0,0}, & & \nn \\
    &g_2(s_{12},\mt^2) = I_{2,0,1,0}, &g_3(s_{23},\mt^2) =I_{0,2,0,1}, & &g_4(\mh^2,\mt^2) = I_{2,0,0,1}, & \nn \\
    &g_5(s_{12},\mt^2) = I_{1,1,1,0},  &g_6(s_{23},\mt^2) =I_{0,1,1,1},  & \nn \\
    &g_7(s_{12},\mh^2,\mt^2) = I_{1,0,1,1}, & \quad g_8(s_{23},\mh^2,\mt^2) = I_{1,1,0,1}, \nn \\
    &g_9(s_{12},s_{23},\mh^2,\mt^2) = I_{1,1,1,1}, &   
\end{align}
and 
shown in Fig.~\ref{fig:precanonical}. For practical calculations, these integrals need to be evaluated, either analytically or numerically (see for example Ref.~\cite{Borowka:2015mxa} and references theirin).  The most common method for analytic evaluation of integrals is by identifying the differential equations (in terms of the external parameters, $s_{ij}$) and solving them~\cite{art:KOTIKOV1991158, art:Remiddi:1997ny,art:Gehrmann:1999as, art:Gehrmann:2000xj}.  In Appendix~\ref{sec:oneloopappendix} we show how this can be done.

\begin{figure}[htb]
  \centering
  \begin{subfigure}[b]{0.25\textwidth}
    \centering
    \includegraphics[valign=m,width=0.5\textwidth]{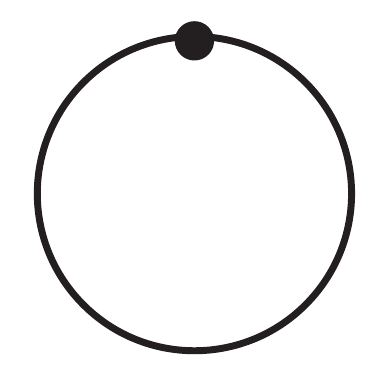}
    \caption{$g_1$}
  \end{subfigure}
  \begin{subfigure}[b]{0.25\textwidth}
    \includegraphics[valign=m,width=\textwidth]{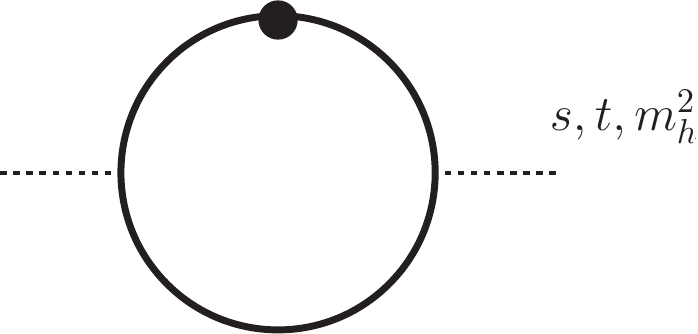}
    \caption{$g_2, g_3, g_4$}
  \end{subfigure}
  \begin{subfigure}[b]{0.25\textwidth}
    \includegraphics[valign=m,width=\textwidth]{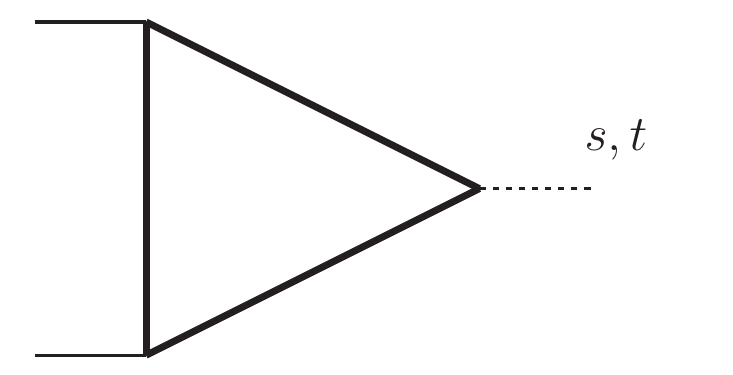}
    \caption{$g_5, g_6$}
  \end{subfigure}
  \begin{subfigure}[b]{0.25\textwidth}
    \includegraphics[valign=m,width=\textwidth]{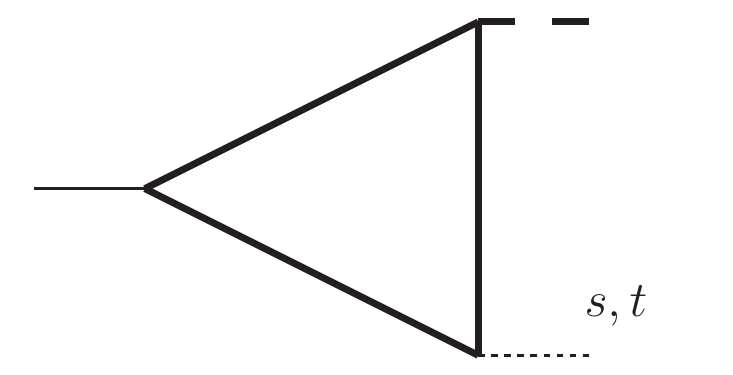}
    \caption{$g_7, g_8$}
  \end{subfigure}
  \begin{subfigure}[b]{0.25\textwidth}
    \includegraphics[valign=m,width=\textwidth]{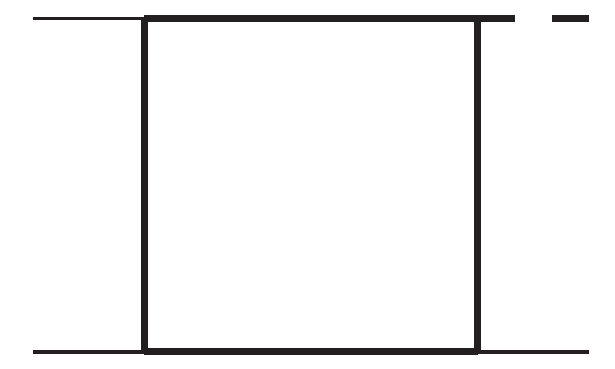}
    \caption{$g_9$}
  \end{subfigure}
  \caption{The nine one-loop master integrals for the $gg \to g\h$ process.}
  \label{fig:precanonical}
\end{figure}

\subsubsection{The One-Loop Formfactors}
\label{sec:oneloopformfactors}

By applying the projectors, as described in Sec.~\ref{sec:tensor_dec}, each diagram's contribution to each form factor can be computed. For example, the contribution of the triangle diagram shown in \ref{fig:gggh1loopsm} to $F_{312}$ is:
\begin{align*}
F_{312} \supset f^{abc} C_\epsilon \frac{e g_{ht} g_s^3}{(d-3)} \left[ (d-4) \frac{2 \mt}{s_{12}^2 s_{23}} I^{(132)}_{1,-1,0,1} + d \frac{4 \mt}{s_{12}^2 s_{23}} I^{(132)}_{1,1,-1,0} + (d-4)\frac{-4 \mt}{s_{12}^2 s_{23}} I^{(132)}_{1,0,-1,1}  + \ldots \right],
\end{align*}
where $C_\epsilon = \pi^{d/2}/(2\pi)^d \Gamma(1-\epsilon)^2 \Gamma(1+\epsilon)/\Gamma(1-2\epsilon) = 1/(16\pi^2) + \mathcal{O}(\epsilon)$ and the ellipsis denotes a further 50 terms.

Using the IBP identities we can relate each integral appearing in the amplitude to our selected master integrals. Inserting these relations into the expression for our amplitude we obtain:
\begin{align*}
F_{212}(s_{12},s_{13},s_{23}) & = f^{abc} C_\epsilon \frac{2 \mt e g_{ht} g_s^3}{(d-3) (d-2)} \left[
\frac{8 (d-4) \left(s_{12}^2-s_{13} s_{23}\right)}{s_{12} s_{23} (s_{12}+s_{13}) (s_{12}+s_{23})} A(\mh^2,\mt^2) \right. \\
&-\frac{4 (d-4)}{s_{12} s_{23}} B(s_{12},\mh^2,\mt^2) \\
& -\frac{4 s_{13} \left((d-4) s_{12}^2+2 d s_{12} s_{23}+d s_{23}^2\right)}{s_{12}^2 s_{23} (s_{12}+s_{23})^2} B(s_{13},\mh^2,\mt^2) \\
& -\frac{4 \left((d-4) s_{12}^2+2 d s_{12} s_{13}+d s_{13}^2\right)}{s_{12}^2 (s_{12}+s_{13})^2} B(s_{23},\mh^2,\mt^2) \\
& -\frac{(d-4) \left(s_{13}^2+s_{23}^2\right)}{s_{13} s_{23}^2} C(s_{12},s_{13},s_{23},\mt^2) \\
& -\frac{s_{13}^2 \left((d-4) s_{12}^2+d s_{23}^2\right)}{s_{12}^3 s_{23}^2} C(s_{13},s_{23},s_{12},\mt^2) \\
& -\frac{s_{23} \left((d-4) s_{12}^2+d s_{13}^2\right)}{s_{12}^3 s_{13}} C(s_{23},s_{12},s_{13},\mt^2) \\
& +\frac{2 (d-3) (s_{13}+s_{23}) \left((d-2) s_{12}-8 \mt^2\right)}{s_{12}^2 s_{23}} g_7(s_{12},\mh^2,\mt^2) \\
& +\frac{2 (d-3) \left((d-2) s_{12} \left(s_{12}^2-s_{23}^2\right)-8 \mt^2 \left(s_{12}^2-2 s_{12} s_{23}-s_{23}^2\right)\right)}{s_{12}^2 s_{23} (s_{12}+s_{23})} g_7(s_{13},\mh^2,\mt^2) \\
& +\frac{2 (d-3) \left((d-2) s_{12} \left(s_{12}^2-s_{13}^2\right)-8 \mt^2 \left(s_{12}^2-2 s_{12} s_{13}-s_{13}^2\right)\right)}{s_{12}^2 s_{23} (s_{12}+s_{13})} g_8(s_{23},\mh^2,\mt^2) \\
& +\frac{(d-2) \left((d-4) s_{12} s_{23} + (d-3) s_{12} s_{13} -4 \mt^2 s_{13}\right)}{s_{12} s_{13}} g_9(s_{12},s_{23},\mh^2,\mt^2)  \\
& +\frac{(d-2) s_{13} \left((d-4) s_{12} s_{13}+(d-3) s_{12} s_{23}-4 \mt^2 s_{23}\right)}{s_{12} s_{23}^2} g_9(s_{12},s_{13},\mh^2,\mt^2) \\
& \left. +\frac{(d-2) s_{13} \left(-(d-3) s_{12}^2+d s_{13} s_{23}+12 \mt^2 s_{12}\right)}{s_{12}^3} g_9(s_{23},s_{13},\mh^2,\mt^2)  \right], \\
F_{311}(s_{23},s_{12},s_{13}) &= F_{332}(s_{13},s_{23},s_{12}) = F_{212}(s_{12},s_{13},s_{23}), \\
F_{312}(s_{12},s_{13},s_{23}) & = f^{abc} C_\epsilon \frac{2 \mt e g_{ht} g_s^3}{(d-3) (d-2)} \left[
 \left(\frac{8 (d-4)}{s_{12} s_{13}+s_{13} s_{23}}\right) A(\mh^2) \right.\\
& +\frac{4 \left((d-2) s_{13}^2+2 d s_{13} s_{23}+(d-2) s_{23}^2\right)}{s_{13} s_{23} (s_{13}+s_{23})^2} B(s_{12},\mh^2,\mt^2)  \\
& + \frac{(d-2) s_{12} \left(s_{13}^2+s_{23}^2\right)}{s_{13}^2 s_{23}^2} C(s_{12},s_{13},s_{23},\mt^2) \\
& + \frac{4 (d-3) \left((d-2) (s_{13}+s_{23})-8 \mt^2\right)}{s_{12} (s_{13}+s_{23})} g_7(s_{12},\mh^2,\mt^2) \\
& \left. + \frac{(d-2) \left(-(d-2) s_{12} s_{23}+(d-3) s_{13}^2-4 \mt^2 s_{13}\right)}{s_{13}^2} g_9(s_{12},s_{23},\mh^2,\mt^2) \right] \\
& + (s_{12}\rightarrow s_{13}, s_{13} \rightarrow s_{23}, s_{23}\rightarrow s_{12}) + (s_{12}\rightarrow s_{23}, s_{13} \rightarrow s_{12}, s_{23}\rightarrow s_{13}). \label{eq:1loopresult}
\end{align*}
The abbreviations $A, B, C$ are
\begin{align*}
A(\mh^2,\mt^2) &= g_1(\mt^2) + (4 \mt^2 - \mh^2) g_4(\mh^2,\mt^2), \\
B(x,\mh^2,\mt^2) &= (4 \mt^2 - x) g_{2,3}(x,\mt^2) - (4 \mt^2 -\mh^2) g_4(\mh^2,\mt^2), \\
C(x,y,z,\mt^2) &= (d-2) g_{5,6}(x,\mt^2) - (d-2) \frac{y+z}{x} g_{7,8}(x,\mh^2,\mt^2),
\end{align*}
here $g_{i,j}$ means either $g_i$ or $g_j$ depending on the argument $x$. Note that some integrals which are related to the master integrals by crossing, e.g., $g_7(s_{13},\mh^2,\mt^2)$ also appear in the formfactors.

\subsubsection{Heavy Top-Quark Limit \label{sec:heavytoplimit}}

One interesting limit is obtained by considering that the top-quark mass is very much larger than the Higgs boson mass and the other scales present in the integral ($s_{12},s_{13},s_{23}$).

Starting from the form factors presented in Sec.~\ref{sec:oneloopformfactors} we can straightforwardly expand the coefficient of each master integral in inverse powers of $\mt^2$. The expansion of the master integrals is straightforward given the analytic results (see Appendix~\ref{sec:oneloopappendix}),
\begin{align}
g_1(\mt^2) &= (-1)^2 \Gamma(2-d/2) \left( \frac{1}\mtsq \right)^{2-d/2}, \\
g_2(s_{12},\mt^2) &= (-1)^3 \Gamma(3-d/2) \left( \frac{1}\mtsq \right)^{3-d/2} \left( \frac{1}{2} + \frac{s_{12}}\mtsq \frac{(3-d/2)}{12}   + \mathcal{O}\left(\frac{1}{\mt^4} \right)\right),\\
g_5(s_{12},\mt^2) &= (-1)^3 \Gamma(3-d/2) \left( \frac{1}\mtsq \right)^{3-d/2} \left( \frac{1}{2} + \frac{s_{12}}\mtsq \frac{(3-d/2)}{24}   + \mathcal{O}\left(\frac{1}{\mt^4} \right)\right),\\
g_7(s_{12},\mh^2,\mt^2) &= (-1)^3 \Gamma(3-d/2) \left( \frac{1}\mtsq \right)^{3-d/2} \left( \frac{1}{2} + \frac{(s_{12}+\mh^2)}\mtsq \frac{(3-d/2)}{24}   + \mathcal{O}\left(\frac{1}{\mt^4} \right)\right), \label{eq:g7expansion}\\
g_9(s_{12},s_{23},\mh^2,\mt^2) &= (-1)^4 \Gamma(4-d/2) \left( \frac{1}\mtsq \right)^{4-d/2} \left( \frac{1}{6} + \mathcal{O}\left(\frac{1}\mtsq \right)\right).
\end{align}

The series expansion in some kinematic invariant or mass of the master integrals can also be obtained without knowing the result of the integral using the technique of {\it expansion by regions} \cite{Beneke:1997zp,Smirnov:1998vk,Smirnov:1999bza}.  Let us demonstrate the procedure by applying it to one of the multi-scale triangle integrals $g_7(s,\mh^2,\mt^2)$. In the case at hand we wish to expand in inverse powers of the top-quark mass. All propagators are top-quark propagators and therefore there is only one {\it region}, which is equivalent to the one-loop integral itself. We begin by Feynman parametrising the integral,
\begin{align*}
g_7(s_{12},\mh^2,\mt^2) &= (-1)^3 \Gamma(3-d/2) \int_0^\infty  \left(\prod_{i=1}^3 \mathrm{d} x_i \right) \delta(1- \sum_{i=1}^3 x_i) \frac{\mathcal{U}^{3-d}}{\mathcal{F}^{3-d/2}}, \\
\mathcal{U} &= x_1+x_2+x_3, \\
\mathcal{F} &= -s_{12} x_2 x_3 - \mh^2 x_1 x_3 + \mt^2 (x_1+x_2+x_3)^2.
\end{align*}
Introducing a scaling parameter $\rho$ via the substitution $\mt^2 = \bar{m}_t^2/\rho, s_{12}=\bar{s}_{12}, \mh^2 = \bar{m}_{\h}^2$ and factoring out the overall dependence on $\rho$ we obtain
\begin{align*}
g_7(s_{12},\mh^2,\mt^2)  &= (-1)^3 \Gamma(3-d/2) \int_0^\infty  \left(\prod_{i=1}^3 \mathrm{d} x_i \right) \delta(1- \sum_{i=1}^3 x_i) \rho^{3-d/2} \frac{\mathcal{U}^{3-d}}{\mathcal{F}^{3-d/2}}, \\
\mathcal{U} &= x_1+x_2+x_3, \\
\mathcal{F} &= \rho(-\bar{s}_{12} x_2 x_3 - \bar{m}_{\h}^2 x_1 x_3) + \bar{m}_t^2 (x_1+x_2+x_3)^2.
\end{align*}
Series expanding in $\rho$ gives
\begin{align*}
g_7(s_{12},\mh^2,\mt^2)  &= (-1)^3 \Gamma(3-d/2) \int_0^\infty \left(\prod_{i=1}^3 \mathrm{d} x_i \right) \delta(1- \sum_{i=1}^3 x_i) \rho^{3-d/2}  \\
& \left[ \frac{(x_1+x_2+x_3)^{3-d}}{(\bar{m}_t^2(x_1+x_2+x_3)^2)^{3-d/2}} \right. \\
&\left.  - \rho (3-d/2) (-\bar{s}_{12}x_2 x_3 - \bar{m}_{\h}^2 x_1 x_3) \frac{(x_1+x_2+x_3)^{3-d}}{(\bar{m}_t^2(x_1+x_2+x_3)^2)^{4-d/2}} + \mathcal{O}(\rho^2)  \right].
\end{align*}
All scales can be factored out of each integral and this greatly simplifies their evaluation,
\begin{align*}
\int_0^\infty  \left(\prod_{i=1}^3 \mathrm{d} x_i \right) \delta(1- \sum_{i=1}^3 x_i) \frac{1}{(x_1+x_2+x_3)^{3}} & = \frac{1}{2}, \\
\int_0^\infty  \left(\prod_{i=1}^3 \mathrm{d} x_i \right) \delta(1- \sum_{i=1}^3 x_i) \frac{x_2 x_3} {(x_1+x_2+x_3)^{4}} & = \frac{1}{24}, \\
\int_0^\infty  \left(\prod_{i=1}^3 \mathrm{d} x_i \right)  \delta(1- \sum_{i=1}^3 x_i) \frac{x_1 x_3}{(x_1+x_2+x_3)^{4}} & = \frac{1}{24}. 
\end{align*}
Inserting the results for these integrals into the expansion of $g_7(s_{12},\mh^2,\mt^2)$ and rewriting $\bar{m}_t^2, \bar{s}_{12}, \bar{m}_{\h}^2$ in terms of the original variables, the dependence on $\rho$ cancels and we obtain the result of Eq.~\eqref{eq:g7expansion}.

Expanding each of the form factors in the $\mt \rightarrow \infty$ limit we obtain:
\begin{align}
&F_{212} = N \frac{1}{s_{23}},&
&F_{332} = N \frac{1}{s_{13}},&
&F_{311} = N \frac{1}{s_{12}},&
&F_{312} = N \left( \frac{1}{s_{23}} + \frac{1}{s_{13}} + \frac{1}{s_{12}} \right),&
\end{align}
where
\begin{equation}
N = f^{abc} C_\epsilon  \frac{8}{3} \frac{e g_{ht} g_s^3}{\mt} = f^{abc} C_\epsilon  \frac{8}{3} \frac{g_s^3}{v}.
\end{equation}
Inserting the result for each of the form factors into Eq.~\eqref{eq:me2} we obtain the result for the squared amplitude in the limit $d\rightarrow 4-2\epsilon$,
\begin{equation*}
|\M_{gg \rightarrow g\h}|^2 = 4  N_c \left(N_c^2-1\right) \frac{\as^3}{\pi v^2} \frac{ (\mh^8 + s_{12}^4 + s_{13}^4 + s_{23}^4) (1-2\epsilon) + \epsilon/2 (\mh^4 +s_{12}^2 +s_{13}^2 +s_{23}^2)^2 }{9 s_{12} s_{13} s_{23}}.
\end{equation*}

\begin{figure}[htpb]
  \centering
  \includegraphics[width=0.7\textwidth]{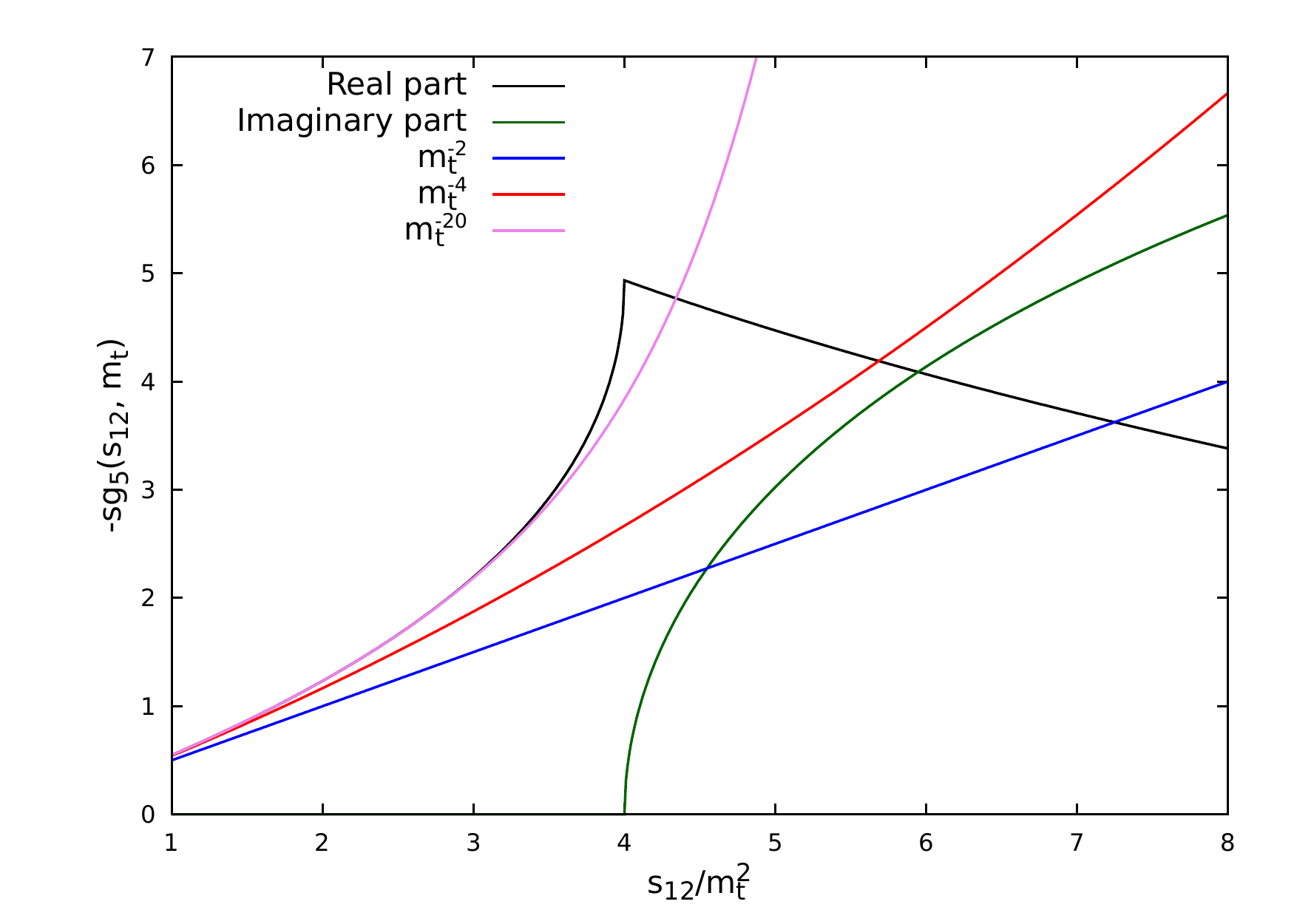}
  \caption{The real and imaginary part of the triangle integral $g_5(s_{12},\mt^2)$ together with the result obtained in the large top-mass expansion keeping up to the 1st, 2nd and 10th terms in the series.}
  \label{fig:triangleexpand}
\end{figure}

We must now assess in what range of the kinematic invariants $s_{12}$ and $s_{13}$ this expansion is valid. To gain some intuition, let us first consider a single integral $g_5(s_{12},\mt^2)$ appearing in the amplitude. The integral has two massless on-shell legs and one leg off-shell with mass $s_{12}$, the internal propagators are all massive with mass $\mt$. In a compact notation the result for this integral is
\begin{align*}
&-s_{12}\ g_5(s_{12},\mt^2) = -\frac{1}{2} \ln^2 \left( -\frac{1-\beta}{1+\beta} \right),& &\beta = \sqrt{ 1- \frac{4\mt^2}{s_{12}} }.&
\end{align*}
The integral has a branch point at $s_{12} = 4 \mt^2$ which corresponds to the top-quark propagators going on-shell. We can see in Fig.~\ref{fig:triangleexpand} that for $s_{12} \ge 4 \mt^2$ the integral develops an imaginary part. Since the integral depends only on $s_{12}$ and $\mt$ the heavy top-quark mass expansion for this integral corresponds to an expansion in $s_{12}/4\mt^2$ centred at 0 with a radius of convergence equal to 1. The first few terms are given by
\begin{equation}
-s_{12}\ g_5(s_{12},\mt^2)  = \frac{s_{12}}{2\mt^2} + \frac{s_{12}^2}{24 \mt^4} + \mathcal{O}\left( \frac{s_{12}^3}{\mt^6} \right).
\end{equation}
Fig.~\ref{fig:triangleexpand} also shows the large mass expansion result keeping the $n$ terms of the series in $(s_{12}/\mt^2)$ with $n=1,2,10$.  We can see that above the top-quark threshold the series expansion does not converge to the true result. 

\begin{figure}
    \centering
    \begin{subfigure}[b]{0.49\textwidth}
        \includegraphics[width=\textwidth]{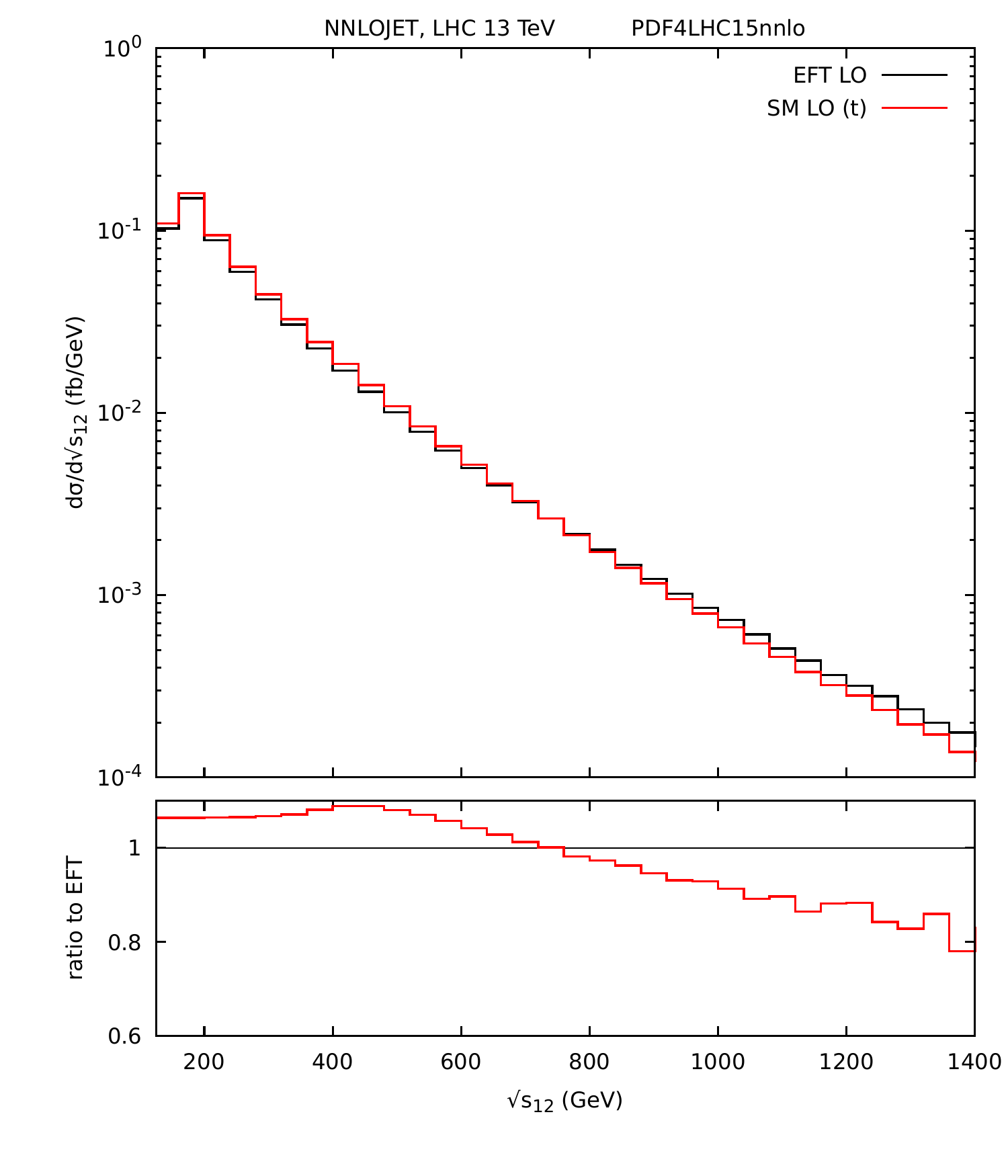}
    \end{subfigure}
    \caption{The partonic invariant mass distribution for the Standard Model with a single top quark compared with the Higgs Effective Theory for the gluon induced partonic channel.  The lower panel shows the ratio.}
    \label{fig:invmasspt}
\end{figure}

Given the break down of the heavy-top quark expansion of the loop integrals for $s_{12} \ge 4 \mt^2$, upon which the amplitude and cross-section depend, we expect that the discrepancy between the heavy top-quark limit and the full calculation grows above this threshold.  Fig.~\ref{fig:invmasspt} shows that for $\sqrt{s_{12}} <  2\mt$ the ratio of the \SM prediction retaining the full top mass dependence to the \HEFT prediction is roughly constant. 
Since $s_{12}$ is greater than $\mh^2$ this implies that at the lowest values of $s_{12}$ the expansion parameter is,
$$
\frac{s_{12}}{4 \mt^2} \sim \frac{\mh^2}{4\mt^2} \sim 0.13.
$$
Relative to the leading term, we have neglected terms of this order, and indeed, we see that the \SM and \HEFT predictions differ by ${\cal O}(10\%)$.
Somewhat above $\sqrt{s_{12}} \sim  2\mt$ the \SM prediction is further enhanced relative to the \HEFT prediction as the top-quark can become on shell. 
When the partonic invariant mass $\sqrt{s_{12}}$ is much greater than $2\mt$ the \SM begins to fall relative to the \HEFT prediction leading to a larger discrepancy far above the threshold.

In the $gg \to g\h$ process, the energy $\Eh$ of the Higgs boson in the partonic centre of mass frame is given by,
\begin{equation}
\Ehsq =  \frac{(s_{12}-\mh^2)^2}{4s_{12}}.
\label{eq:higgsenergy}
\end{equation}
The Higgs boson transverse momentum is 
\begin{equation}
\pth = \Eh \sin\theta,
\end{equation} 
where $\theta$ is the angle between the Higgs boson and the beam axis in this frame. It is further related to the Mandelstam variables by,
\begin{equation}
\pthsq = \frac{s_{13} s_{23}}{s_{12}}. 
\label{eq:higgsptmandelstam}
\end{equation}
Since $\Eh \geq \pth$, solving Eq.~\eqref{eq:higgsenergy} for $s_{12}$ yields a lower bound in terms of \pth,
\begin{equation}
s_{12} = \left(\Eh + \sqrt{\Ehsq+\mh^2}\right)^2 \geq \left(\pth+ \sqrt{\pthsq+\mh^2}\right)^2.
\label{eq:s12bound}
\end{equation}

For a fixed $s_{12}$, the largest $\pth$ is obtained when the bound in Eq.~\eqref{eq:s12bound} is saturated.  Setting $s_{12} = 4 \mt^2$ therefore tells us for which values of $\pth$ the threshold starts to open,
\begin{equation}
\pth > \frac{4 \mt^2-\mh^2}{4  \mt}. 
\end{equation}
For example, with $\mh = 125\ \GeV $, $\mt = 174\ \GeV $ we obtain $\pth > 152\ \GeV$.
Above this value of $\pth$ 
all contributions to the differential cross-section come from above the top-quark threshold
where we can not reliably trust the heavy top-quark expansion. The $\pth$ distribution for the heavy-top quark limit and for the Standard Model is shown in Fig.~\ref{fig:SMpth_13TeVcompare} and, as expected, at large $\pth$ we can see a very large deviation between the exact and approximate results. 

At lower values of $\pth$ there may be a contribution to the differential cross-section from above the top-quark threshold (since $\Eh \geq \pth$). However, the parton luminosity strongly prefers smaller invariant masses so that the majority of the cross section comes from within the radius of convergence of the heavy top-quark expansion and the approximate results are reliable. This can be also seen in Fig.~\ref{fig:SMpth_13TeVcompare} where the exact and approximate results are in good agreement at low to medium $\pth$.

The discrepancy is less marked for the \phistar distribution shown in Fig.~\ref{fig:SMphi_13TeVcompare}
since a particular value of \phistar gets contributions from a range of \pth.  Again, since the \pth distribution is peaked at lower \pth, this means that most of the \phistar distribution is dominated by the low \pth region. 
 
Another approach to studying the high $\pth$ behaviour of Higgs boson plus (multi) jet production is the use of leading-log high energy resummation techniques. It has been shown shown that for large $\pth$ the differential partonic cross section scales as $1/\pthsq$ in the SM, whilst in the large top-quark mass limit it scales only as $1/\pth$~\cite{Bagnaschi:2015qta,Caola:2016upw}.

\subsection{Heavy Top-Quark Effective Theory \label{sec:heavytopeffective}}

The one-loop amplitude in the limit of large top-quark mass is extremely simple in comparison to the result retaining the full top mass dependence. From its simplicity, one might guess that there is a much simpler way of obtaining the amplitude directly in the large quark mass limit. Indeed, this turns out to be the case. Rather than computing the full amplitude and then taking the limit $\mt \rightarrow \infty$, it is much simpler to first take the limit at the level of the Lagrangian. This can be done by integrating out the top-quark field.  The result is that the Lagrangian now contains an effective coupling between the Higgs and two or more gluons:
\begin{equation*}
\mathcal{L}_{eff} \supset  \cg  \frac{\as}{12\pi} \frac{\h}{v} G^a_{\mu \nu} G^{a \  \mu \nu}  
\end{equation*}
where the effective coupling $\cg$ is computed by matching the effective theory to the full theory. This involves computing some quantity that depends on the effective coupling in the effective theory and in the full theory and then demanding that the two quantities are equivalent. Thus, we could in principle compute the matching using our full result for Higgs plus jet or by computing a simpler quantity, for example the top-quark contribution to the gluon self energy~\cite{Chetyrkin:2005ia}. 
The effective coupling $\cg$ is known to four-loops~\cite{Chetyrkin:2005ia,Chetyrkin:2016uhw} and, to ${\cal O}(\as^2)$ is given by the perturbative expansion,
\begin{equation*}
c_g =  1 + \frac{11}{4} \frac{\as}{\pi}  + {\cal O}(\as^2).
\end{equation*}

\begin{figure}
\begin{center}
    \begin{subfigure}[b]{0.25\textwidth}
        \includegraphics[valign=m,width=\textwidth]{phistar/figures/hj1loop/gggh/diag1}
    \end{subfigure}
{\Large $\rightarrow$}
        \begin{subfigure}[b]{0.20\textwidth}
        \includegraphics[valign=m,width=\textwidth]{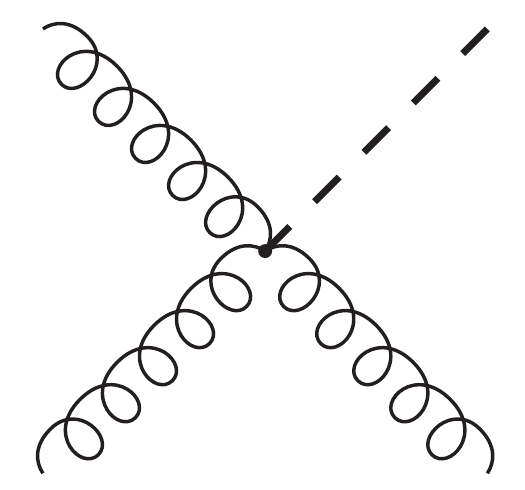}
    \end{subfigure}
    \\
    \begin{subfigure}[b]{0.25\textwidth}
        \includegraphics[valign=m,width=\textwidth]{phistar/figures/hj1loop/gggh/diag2}
    \end{subfigure}
{\Large $\rightarrow$}
        \begin{subfigure}[b]{0.20\textwidth}
        \includegraphics[valign=m,width=\textwidth]{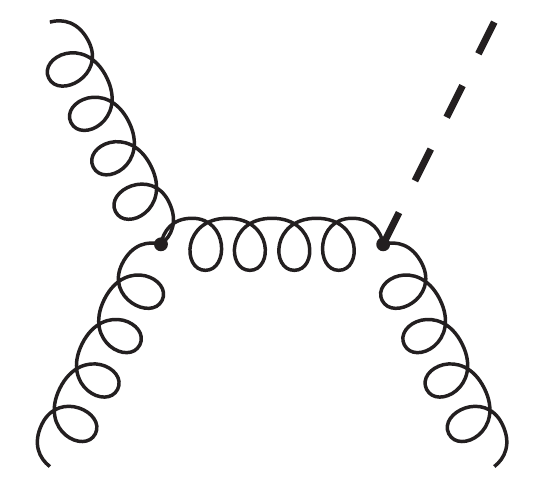}
    \end{subfigure}
    \caption{Schematic picture of the relation between the \SM Feynman Diagrams and those of the \HEFT, where the Higgs boson has an effective coupling to gluons.}
    \label{fig:smtoheft}
\end{center}
\end{figure}
The interactions in the effective theory are schematically related to those in the full theory, as illustrated in 
Fig.~\ref{fig:smtoheft}.

\begin{figure}
\begin{center}
    \begin{subfigure}[b]{0.20\textwidth}
        \includegraphics[valign=m,width=\textwidth]{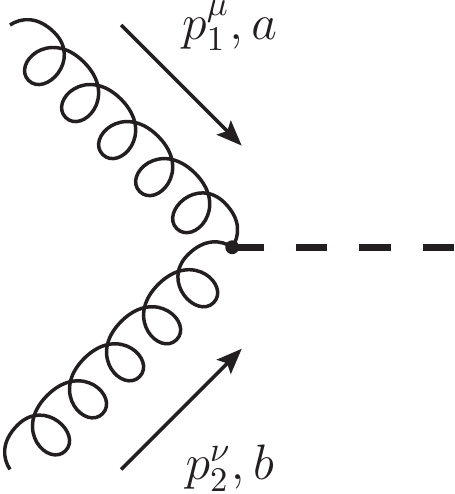}
    \end{subfigure}
$=-ic_g \delta^{ab} H^{\mu \nu}(p_1,p_2)$,\\
\vspace{0.15cm}
        \begin{subfigure}[b]{0.20\textwidth}
        \includegraphics[valign=m,width=\textwidth]{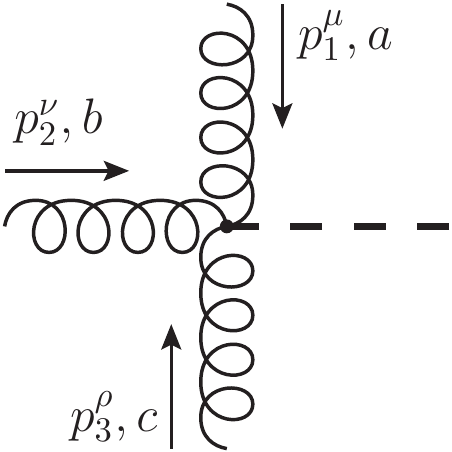}
    \end{subfigure}
    $=-c_g  V^{\mu \nu \rho}_{abc}(p_1,p_2,p_3)$,\\
    \begin{subfigure}[b]{0.20\textwidth}
        \includegraphics[valign=m,width=\textwidth]{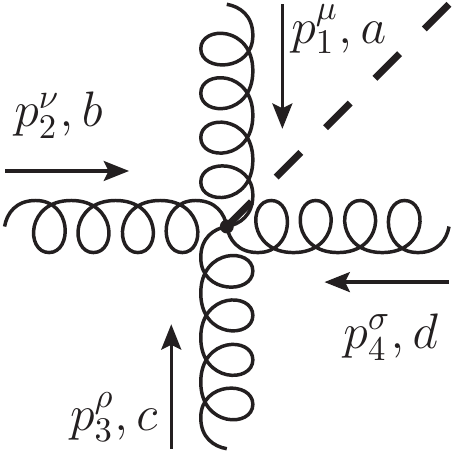}
    \end{subfigure}
  $=- c_g X^{\mu \nu \rho \sigma}_{abcd}$,

    \caption{The \HEFT Feynman rules involving the effective coupling of the Higgs boson to gluons.}
    \label{fig:heftfeynmanrules}
\end{center}
\end{figure}

The effective Lagrangian generates vertices involving the Higgs boson and two or more gluons. From the Lagrangian we can derive the Feynman rules shown in Fig.~\ref{fig:heftfeynmanrules}, where
\begin{align*}
  H^{\mu \nu}(p_1,p_2) = & g^{\mu \nu} p_1 \cdot p_2 - p_1^\nu p_2^\mu , \\
  V^{\mu \nu \rho}_{abc}(p_1,p_2,p_3)  = & g_s f^{abc}
  \left(  g^{\mu \nu}(p_1^{\rho} - p_2^{\rho}) \right.
  + g^{\nu \rho}(p_2^{\mu} - p_3^{\mu})
  \left. + g^{\rho \mu}(p_3^{\nu} - p_1^{\nu}) \right), \\
  X^{\mu \nu \rho \sigma}_{abcd} = & -ig_s^2
  \left(  f^{abe}f^{cde}   ( g^{\mu \rho}g^{\nu \sigma} - g^{\mu \sigma}g^{\nu \rho}) \right.\\
  & + f^{ace}f^{bde}  ( g^{\mu \nu}g^{\rho \sigma} - g^{\mu \sigma}g^{\nu \rho}) \\
  & \left. + f^{ade}f^{bce} ( g^{\mu \nu}g^{\sigma \rho} - g^{\mu \rho}g^{\nu \sigma}) \right) .
\end{align*}

\begin{figure}
\begin{center}
    \begin{subfigure}[b]{0.20\textwidth}
        \includegraphics[valign=m,width=\textwidth]{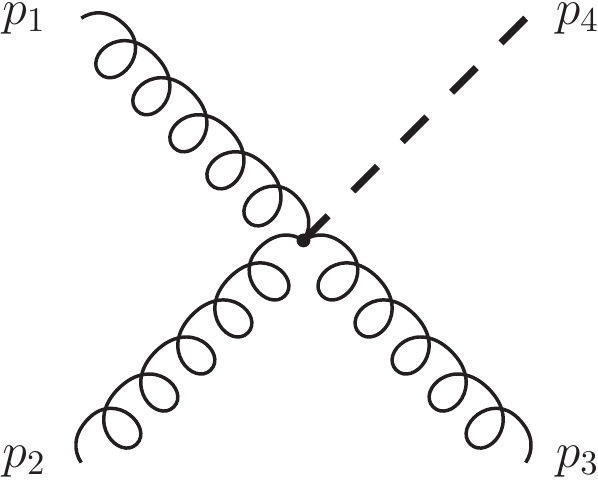}
    \end{subfigure}
        \begin{subfigure}[b]{0.20\textwidth}
        \includegraphics[valign=m,width=\textwidth]{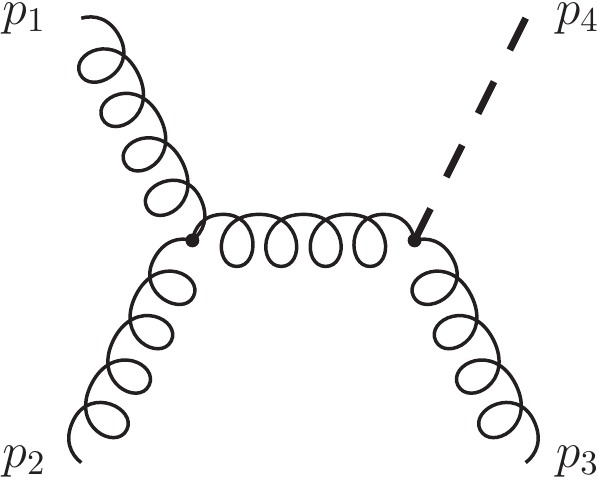}
    \end{subfigure}
    \begin{subfigure}[b]{0.20\textwidth}
        \includegraphics[valign=m,width=\textwidth]{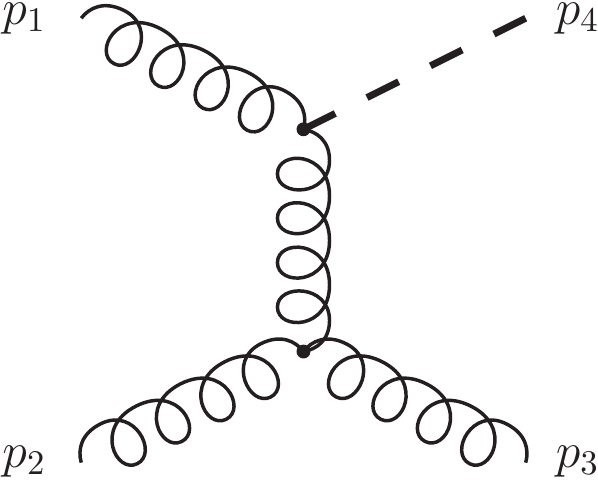}
    \end{subfigure}
    \begin{subfigure}[b]{0.20\textwidth}
        \includegraphics[valign=m,width=\textwidth]{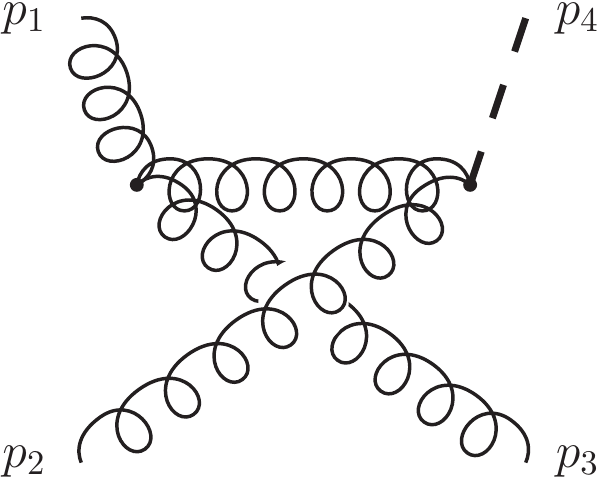}
    \end{subfigure}
    \caption{The Feynman diagrams for the process $gg \rightarrow gH$ at leading order in the HEFT.}
    \label{fig:feynmanEFTtree}
\end{center}
\end{figure}

In the effective theory, four diagrams contribute to the $gg \rightarrow g\h$ partonic channel at leading order, shown in Fig.~\ref{fig:feynmanEFTtree}. We can compute these diagrams by decomposing the amplitude as described in \ref{sec:tensor_dec} and using projectors to extract the formfactors, or (since they are tree level diagrams) simply by squaring them. The result for the squared amplitude in the limit $d\rightarrow 4-2\epsilon$ is
\begin{equation*}
|\M_{gg \rightarrow g\h}|^2 = 4  N_c \left(N_c^2-1\right) \frac{\as^3}{\pi v^2} \frac{ (\mh^8 + s_{12}^4 + s_{13}^4 + s_{23}^4) (1-2\epsilon) + \epsilon/2 (\mh^4 +s_{12}^2 +s_{13}^2 +s_{23}^2)^2 }{9 s_{12} s_{13} s_{23}}.
\end{equation*}
in agreement with the first term of the expansion of the full theory result in the limit of large top mass.

Note that in contrast to the full theory, the Feynman rules of the effective theory allow for Higgs plus jet production at tree-level. Concretely, the ${\cal O}(\as^3)$ calculation was one-loop in the full theory but tree-level in the effective theory, the ${\cal O}(\as^4)$ calculation is two-loop in the full theory but only one-loop in the effective theory and so on. This fact makes computing a given perturbative order vastly simpler in the effective theory than in the full theory. However, the effective theory is yet simpler still, since the top-quark has been integrated out of this theory the integrals in the effective theory contain no top-quark propagators, this means that there are no integrals with massive internal lines at any order in QCD in the effective theory.

\subsection{Theory Uncertainties}
\label{sec:theory-uncertainties}

In order to fully assess the potential of $\phistar$ as an observable, 
we must determine whether it leads to smaller theoretical
uncertainties than $\pth$ and find out whether the overall conclusions
regarding its usefulness may be affected by arbitrary choices in the
theory settings.

The theoretical uncertainties can be loosely classified as:
\begin{itemize}
	\item Missing higher order uncertainties (MHOU): Observables
		in high energy quantum field theory can only be calculated as
		a perturbative expansion in the coupling constants, of which
		only the first few terms are known exactly. Usually the terms 
		which have not been computed are
		estimated though \emph{scale variations}.

	\item Parametric uncertainties: These arise from the fact that some
		theoretical parameters
		parameters cannot be calculated from the theory. These include, for example, 
		the PDF uncertainties, the value of the strong coupling
		constant and the treatment of the heavy quark masses.
\end{itemize}

A particular difficulty with MHOUs is that there currently isn't
a well established procedure to interpret them in a statistically
consistent way, because nothing is known a priori about the
distribution of the higher order coefficients. Some refinements can be
made if we make some assumptions about the coefficients of the
perturbative expansion. For example, that they are bounded from above
and distributed uniformly in a given range~\cite{Cacciari:2011ze}, or
that they exhibit a similar behaviour to that of known perturbative
series~\cite{David:2013gaa}.

On the other hand the parametric uncertainties are obtained by
procedures that involve propagating statistical uncertainties from the
experimental data and  simulations, where the statistical distribution
of the parameters are under control. However, the parameters
themselves are typically affected by MHOUs, and therefore the
distinction above is only indicative.

We now discuss the various sources of theory uncertainty in
the Leading Order computation described in Section~\ref{sec:theory} and
the approach we have taken to estimate their impact, while making
connections to the results in Sections~\ref{sec:parton-shower} and
\ref{sec:higher-orders} where appropriate.

\subsubsection{Scale uncertainties \label{sec:theory-uncertainties-scale-variation}}

MHOUs are typically estimated by varying the renormalisation and
factorization scales that enter the calculation. This procedure
(commonly called {\it scale variations}) can capture some
of the effect of the missing higher orders by exploiting the fact that
the dependency on the scales is formally a higher order effect that cancels at
all orders (see e.g. Chapter 12 of Ref.~\cite{peskin1995introduction}).

Conventionally, the following procedure is followed:
\begin{itemize}
\item Select a {\it central scale.} Typically one chooses a scale
	that minimizes the scale dependence (see e.g.~\cite{Buehler2013}),
	since this choice is likely to also minimize the logarithmic
	contributions in the higher order terms.
\item A {\it scale uncertainty} is assigned by varying the
	factorization and renormalisation scales following some
	prescription and repeating the calculation. A common choice is to
	use the same value for $\mur$ and $\muf$,
	$\mu=\mur=\muf$, and call the ``scale uncertainty'' the range
	$\frac{1}{2}(\sigma_{2\mu}-\sigma_{\frac{1}{2}\mu})$, where
	$\sigma_{2\mu}$ and $\sigma_{\frac{1}{2}\mu}$ are the values of
	the observable calculate setting both the renormalisation and
	factorization scale to 2 and $\frac{1}{2}$ times the central scale
	respectively.  The Higgs Cross Section Working group recommends
	a more involved 7 point variation procedure~\cite{deFlorian:2016spz}.
\end{itemize}
We note that while finding a sensible value for the central scale is
simple for single scale problems, it is a known fact that for more
complicated process selecting an {\it optimal} scale is non-trivial 
and can lead to important phenomenological differences, bigger
than the scale variations discussed above. For example in, Ref.~\cite{Currie:2017ctp}
various choice of scales for the jet cross sections
are discussed and it is shown that the scale choice can have an effect
on the description of the data that is more significant than the
error band estimated by scale variations. Consequently, while
currently there doesn't exist an agreed upon procedure among the
community, it is recommended that several scale choices are
considered, possibly also assigning an uncertainty associated to the
choice of central scale.
%

To estimate the scale uncertainties, we have obtained the $\pth$ and
$\phistar$ distributions for our central scale choice,
$$
\mur=\muf = \frac{1}{2} \MTh
$$
and for both scales multiplied simultaneously by either
$\frac{1}{2}$ or 2. Additionally, we have considered a fixed scale
$\mur=\muf=\mh$. While this is not a good choice at large $\pth$~\cite{Greiner:2016lsq}, 
it allows us to set limits on the effect of a particularly unfortunate choice, 
and furthermore it is interesting
to observe the different behaviour of \phistar and \pth. The results
are presented in Fig.~\ref{fig:scale_variations}. By fluctuating about the
central scale by a factor two we obtain a scale
uncertainty of approximately  $\pm 30\%$, which contains the NLO prediction~\cite{Greiner:2016lsq}.
Because high values of \phistar are affected by the contributions coming from
low \pth, they change less when the scale is set to $\mh$.

\begin{figure}
    \centering
(a)    \begin{subfigure}[b]{0.45\textwidth}
        \includegraphics[width=\textwidth]{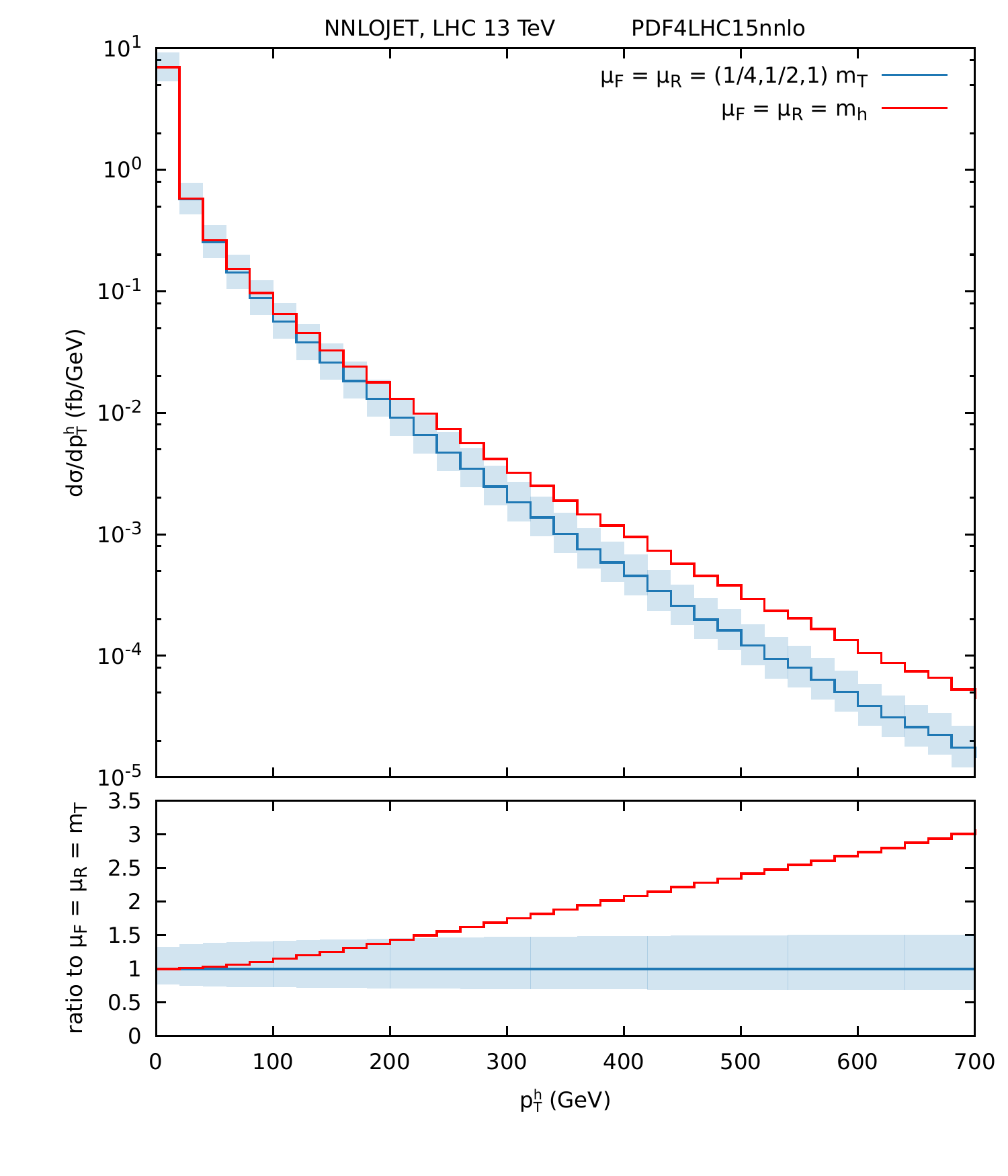}
    \end{subfigure}
(b)    \begin{subfigure}[b]{0.45\textwidth}
        \includegraphics[width=\textwidth]{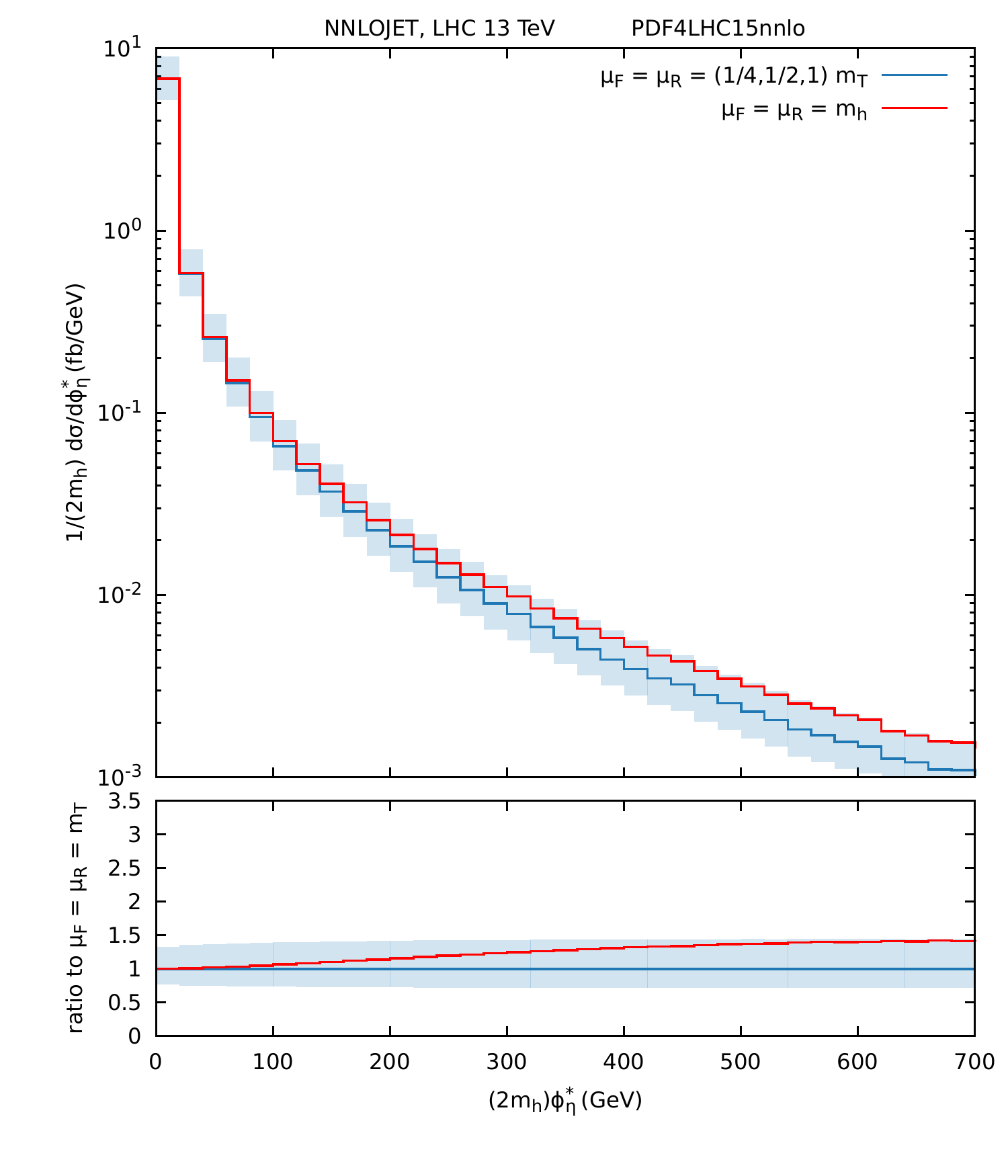}
    \end{subfigure}
	\caption{(a) $\pth$ and (b) $\phistar$ distributions with
    scale variations. The band shows the scale uncertainty estimation
	obtained by multiplying the central choice of scales
	($\mur=\muf=m_T/2$) by
	0.5 and 2. The red line shows the effect of fixing both scales to
	  $\mh$.
	}	
\label{fig:scale_variations}
\end{figure}

\subsubsection{$\alphaS$ uncertainties}

The value of the strong coupling constant, $\alphaS$, is not
a fundamental input of the theory and needs to be estimated from data.
Several methods exist for determining $\alphaS$ at various scales
(including Lattice calculations, determinations from parton densities
and structure functions, global electroweak fits, and $\tau$ decays~\cite{Olive:2016xmw}).
These are related though the running of the coupling constant to yield
a value of $\alphaS(\mz)$ in the $\MSbar$ scheme. Note that it is only
possible to relate the value of the coupling constant obtained at some
arbitrary scale to $\alphaS(\mz)$ though QCD running up to some
finite perturbative order. Similarly, the dependence on $\alphaS$ in the high
energy processes that are used for the coupling constant
determinations is only known in perturbation theory.  Consequently it
is affected by MHOUs, which need to be estimated for each measurement.

One would typically repeat the calculation with a matching PDF set for
the two values corresponding to the upper and lower fluctuations
around the global average.  Currently, the recommended values are~\cite{deFlorian:2016spz}.
\begin{equation}
	\alphaS(\mz)=0.118\pm 0.001
\end{equation}
with the prescription of using $\alphaS(\mz)=0.118$ as the central
value and
$\alphaS(\mz)=0.117$ and $\alphaS(\mz)=0.119$ as upper and
lower variations respectively.

The PDF4LHC PDF set we are using comes with the slightly more conservative choice of
\begin{equation}
	\alphaS(\mz)=0.118\pm 0.0015
\end{equation}
and we have computed the $\pth$ and $\phistar$ distributions using the  
PDF4LHC15\_nnlo\_asvar PDF sets~\cite{Butterworth:2015oua}.
The results are presented in
Fig.~\ref{fig:as_variations}. For both distributions, we 
obtain uncertainties of less than $\pm 5$\% across the whole range of the observable.    

\begin{figure}
    \centering
(a)    \begin{subfigure}[b]{0.45\textwidth}
        \includegraphics[width=\textwidth]{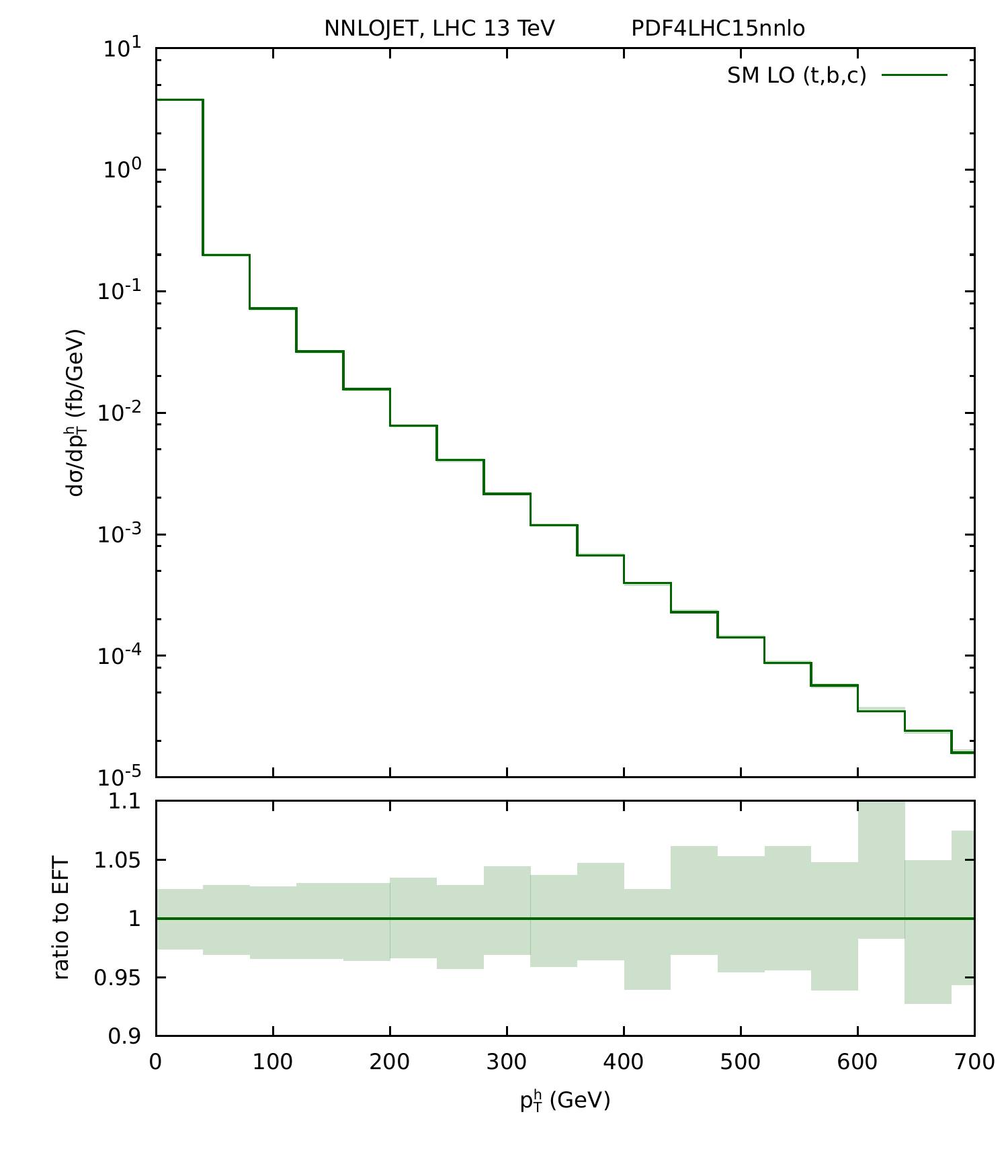}
    \end{subfigure}
(b)    \begin{subfigure}[b]{0.45\textwidth}
        \includegraphics[width=\textwidth]{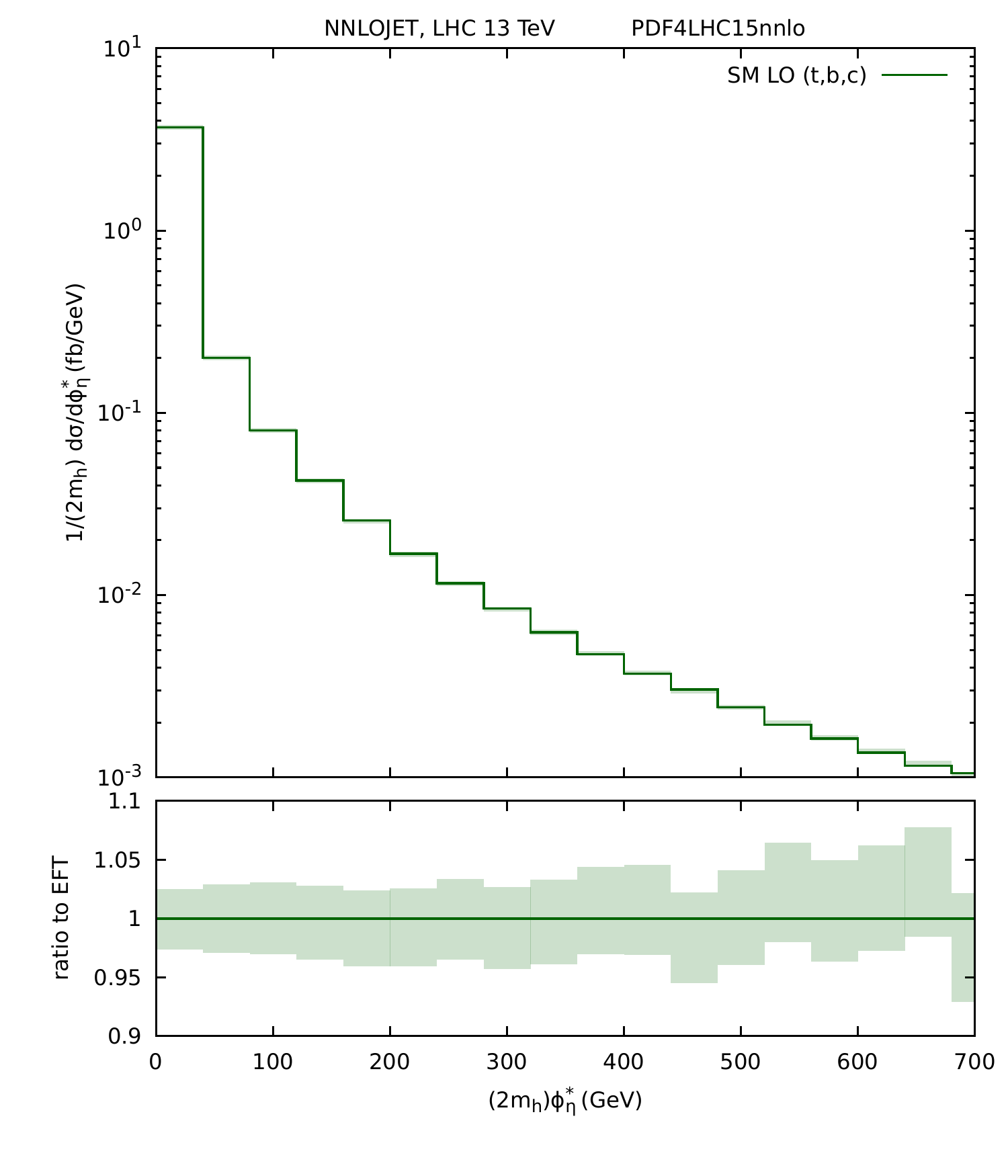}
    \end{subfigure}
	\caption{(a) $\pth$ and (b) $\phistar$ distributions with
    $\alphaS$ variations. As described in the main text, we have used
    the $\as(\mz)= 0.118$ as central reference (solid line) and the
    predictions with $\as(\mz)= 0.1165$ and $\as(\mz)= 0.1195$ 
    to compute the uncertainty band.}
	\label{fig:as_variations}
\end{figure}

\subsubsection{Heavy quark mass uncertainties}

The precise formulation of the renormalisation scheme used to define
the masses of the heavy quarks can have an important impact on the
numerical results. At leading order, the different schemes simply reduce to
different numerical values of the quark masses. Consequently, we
estimate the associated uncertainty as coming from the difference between the
results for these different schemes.  These differences (of $\sim
15\%$) are considerably larger than the uncertainties associated with
the determination of the masses within each of the schemes (of $\sim 3\%$), 
which we therefore neglect.

At higher orders, this uncertainty is significantly reduced notably due to the
appearance of the renormalisation
scheme dependent counterterms which compensate the mass differences (see
Ref.~\cite{Schmidt:2015cea} and references therein).

In order to assess the uncertainty related to the definition of the
heavy quark masses, we have computed the $\pth$ and $\phistar$ distributions
with three variations of the top and bottom masses, but otherwise using the default
setup described in Sec.~\ref{sec:default}. We have selected ranges
consistent with the differences arising from renormalisation schemes
quoted in the PDG 2016 report~\cite{Olive:2016xmw}: For the top quark
we have used $160$, $174$ and $188$~\GeV, and for the bottom quark we
have used $4.2$, $4.6$ and $5$~\GeV. For the top mass, this range
contains the pole and $\MSbar$ masses as well as the results from
the direct measurement of the $t\bar{t}$ cross section. For the bottom
mass, the range contains the results from the renormalon subtracted
and $\MSbar$ determinations. 
We display the results in Fig.~\ref{fig:hq_mass_variations}. 
In the low $\pth$ and $\phistar$ regions, the impact of varying the quark
masses is negligible compared to impact of the parton shower (see
Sec.~\ref{sec:parton-shower}).  

In fact, the effects of changes in the
bottom mass are at the percent level over the whole range of \pth and \phistar, 
and are therefore negligible at the level of precision of this calculation.
The value of the top mass becomes a relevant sources of uncertainty at
large $\pth$ as it leads to a $\sim 10\%$ increased uncertainty. On
the other hand, the predictions for large values of $\phistar$ are
affected much less by the changes in $\mt$, with changes below $3\%$.

\begin{figure}
    \centering
(a)    \begin{subfigure}[b]{0.45\textwidth}
        \includegraphics[width=\textwidth]{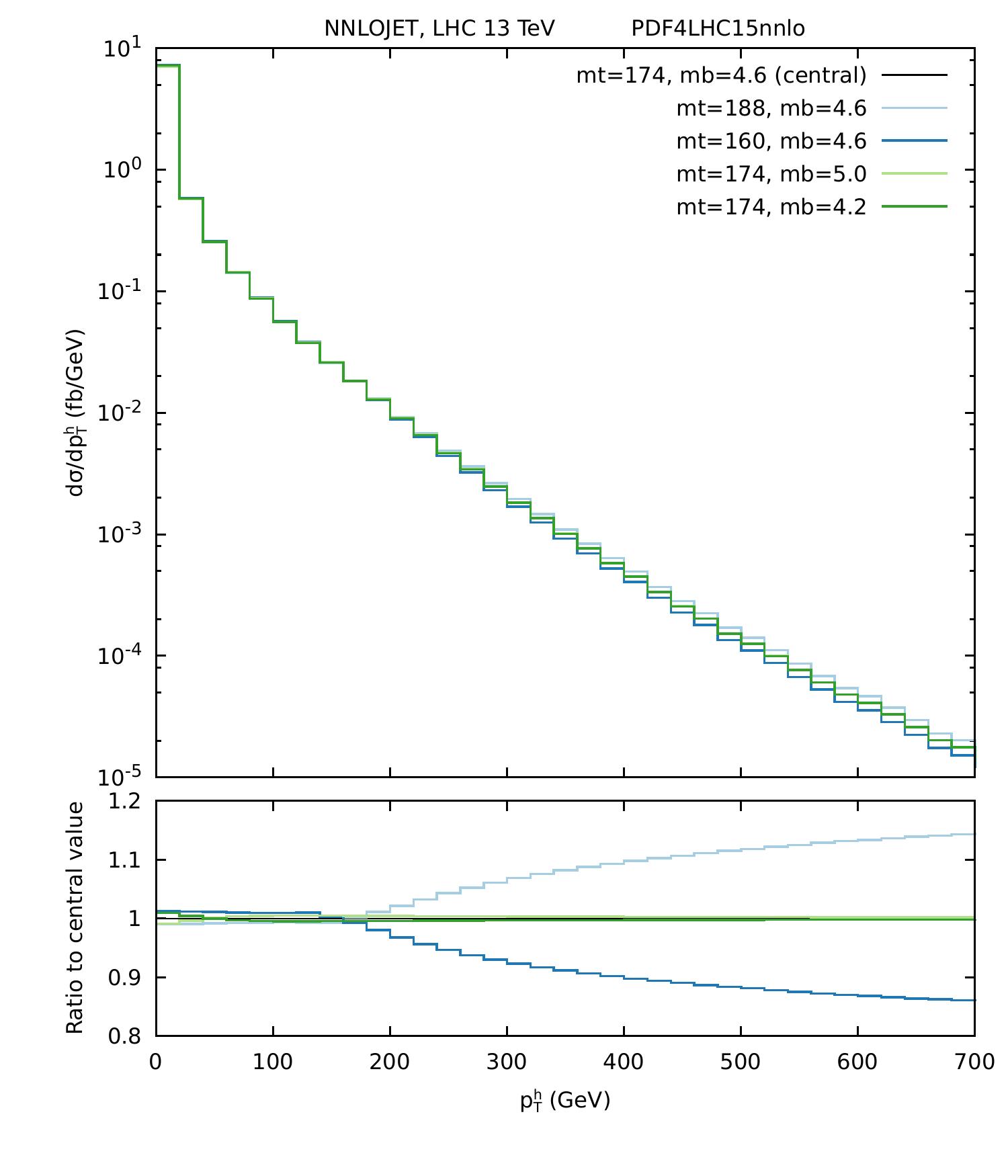}
    \end{subfigure}
(b)    \begin{subfigure}[b]{0.45\textwidth}
        \includegraphics[width=\textwidth]{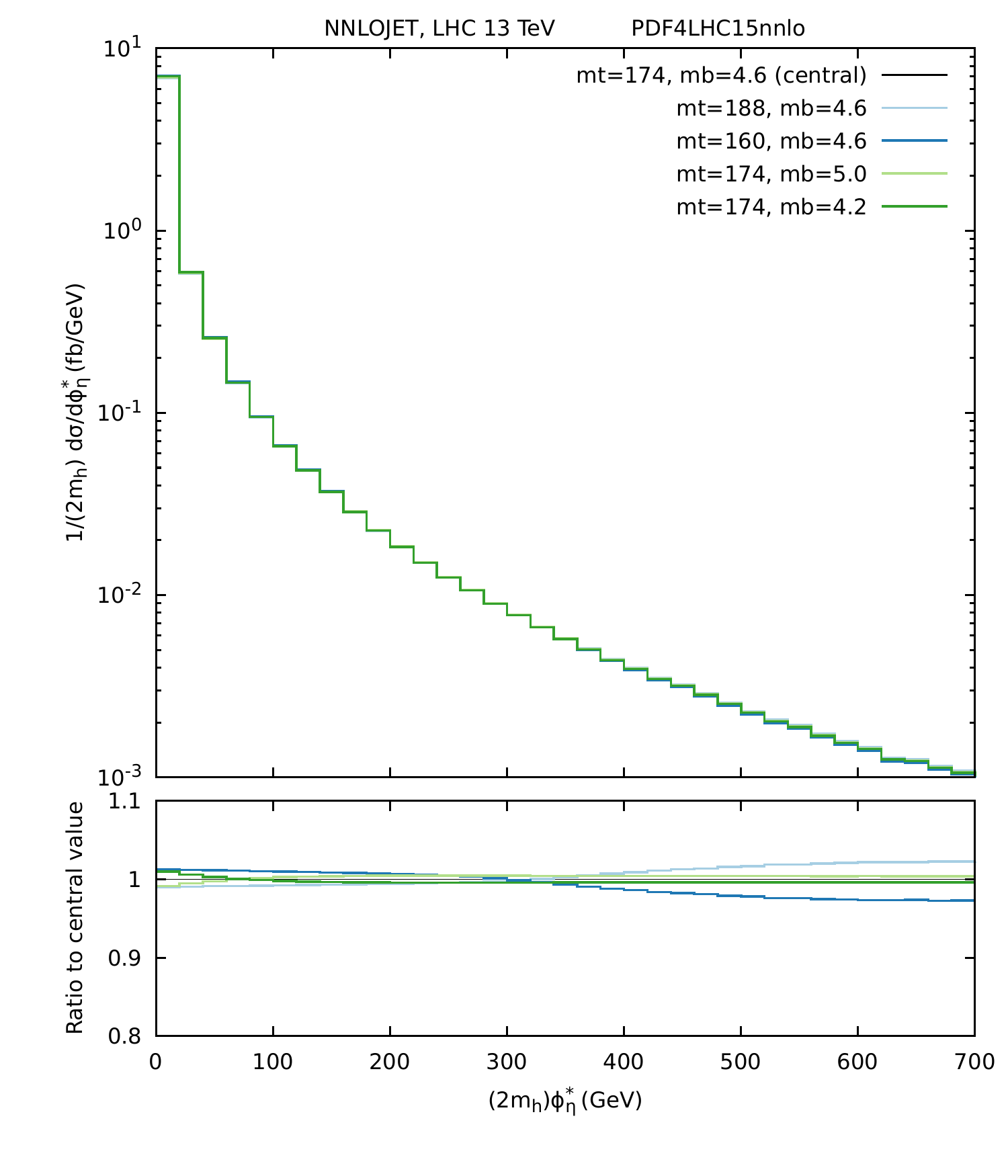}
    \end{subfigure}
	\caption{(a) $\pth$ and (b) $\phistar$ distributions with
    variations in the heavy quark masses. The lines show upwards
	(dark) and downwards (light) variations
    of \mt at fixed $\mb$ (blue) and variations of \mb at fixed \mt (green).}
	\label{fig:hq_mass_variations}
\end{figure}

\subsubsection{PDF uncertainties}

The structure of the proton cannot be calculated in perturbation theory
and must be estimated by matching the theoretical knowledge of the
short range interaction partonic interaction to the experimental results.
The proton structure is characterized by the so called {\it Parton
Distribution Functions} (PDFs). See Ref.~\cite{Forte:2010dt} and references
therein for a more detailed discussion.

Several groups produce global PDF fits, which differ in the way
the PDFs are parametrized, the fitting algorithm, the dataset that is
used to produce the fit, theoretical choices such as the treatment of
the perturbative evolution of the heavy quarks.

{\it PDF uncertainties} are computed by first convolving the partonic
matrix element with variations of the central PDF determination, and
then following a specific prescription defined for each PDF set to
obtain the final value of the uncertainty. 
Ref.~\cite{Butterworth:2015oua} contains a recent review of the
recommended settings and usages.

Note that PDFs themselves depend heavily on the strong coupling
constant. Normally PDF fits are provided for different values
$\alphaS$, which must be used consistently with the $\alphaS$
variations discussed above. The usual prescription is to compute only
the central PDF value for the two variations of the strong coupling,
without also estimating the PDF errors.

Additionally the PDFs are affected by MHOU and parametric
uncertainties such as the value of the heavy quark masses (although
the dependence is moderate  and in the case of the charm mass
can be reduced by explicitly fitting the charm PDF
\cite{Ball:2016neh}.  These are currently not
included in the definition of ``PDF uncertainty''. 


We have used the PDF4LHC15\_nnlo\_30 set to estimate the PDF uncertainty for 
the \pth and \phistar distributions shown in Fig.~\ref{fig:pdf_variations}. 
Because these observables probe slightly 
different parton-parton luminosities, see Fig.~\ref{fig:channels}, the detailed 
behaviour is different for the two observables.
For the \pth distribution, we obtain uncertainties that are typically $\pm (2-3)\%$ 
but growing at larger \pth. The \phistar distribution is dominated by the regions where
the PDF's are better known with a correspondingly smaller uncertainty of $\pm(1-2)\%$.

\begin{figure}
    \centering
(a)    \begin{subfigure}[b]{0.45\textwidth}
        \includegraphics[width=\textwidth]{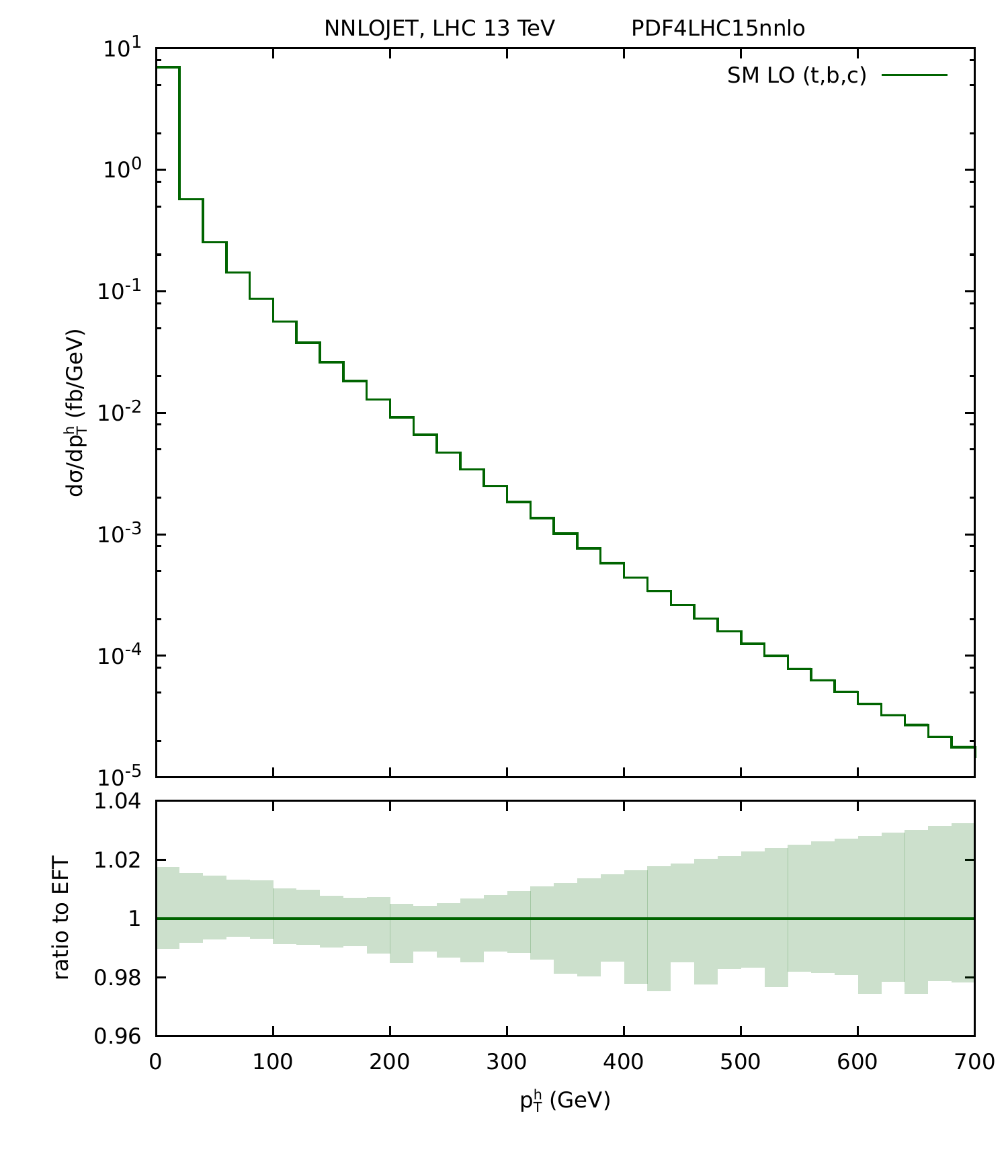}
    \end{subfigure}
(b)    \begin{subfigure}[b]{0.45\textwidth}
        \includegraphics[width=\textwidth]{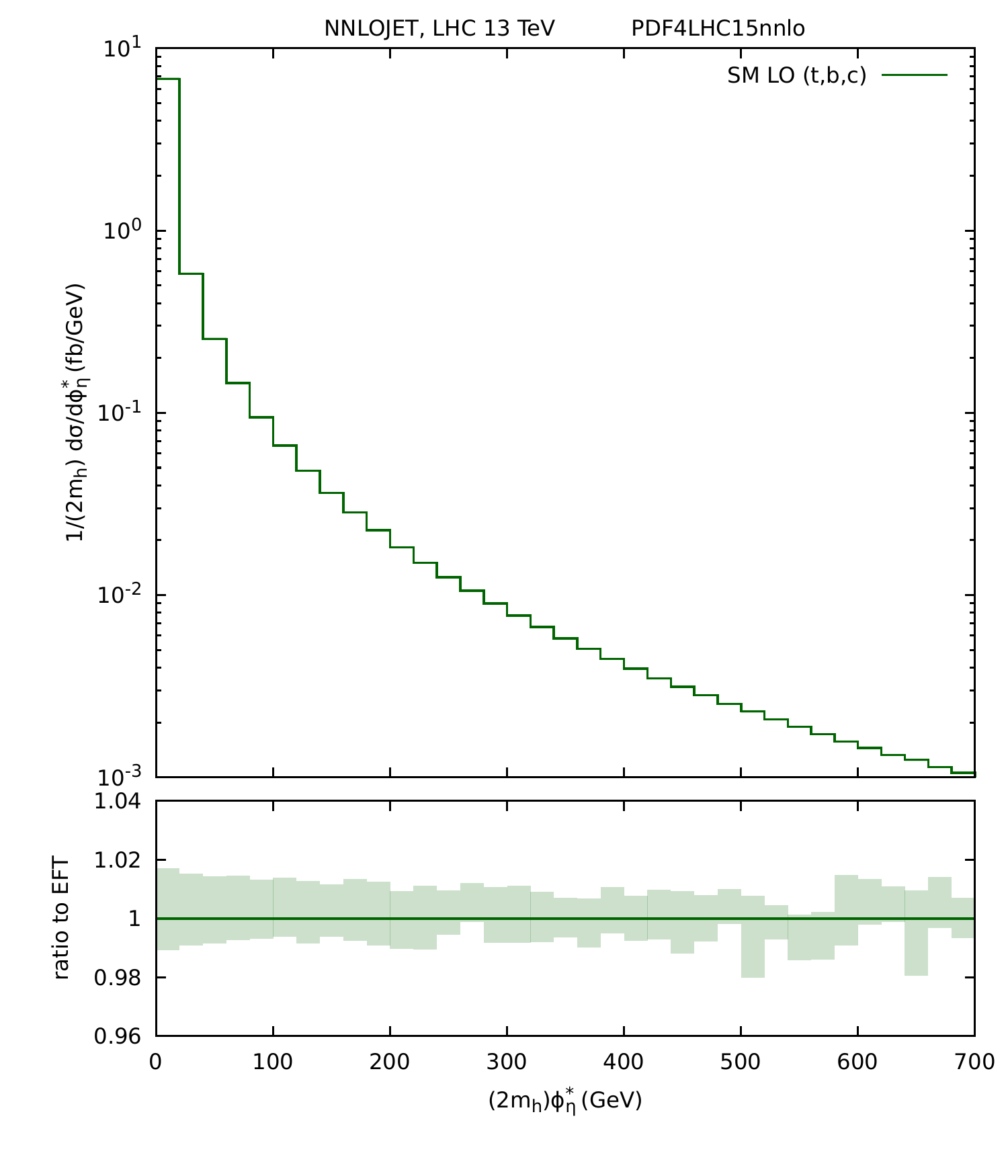}
    \end{subfigure}
	\caption{(a) $\pth$ and (b) $\phistar$ distributions with
    the associated PDF uncertainty.}
	\label{fig:pdf_variations}
\end{figure}

We conclude that the main sources of theoretical uncertainties in this
process are those due to missing higher orders, which motivates the
effort of improving the accuracy of the calculation as we discuss next.

\subsection{Higher Orders}
\label{sec:higher-orders}

In order to make precise predictions in perturbative quantum field theories, one has 
to go beyond the leading order. In QCD, the perturbation is done around small values 
of of the strong coupling constant, or rather small values of $\as$ which is 
proportional to the square thereof. At the energy scale of the LHC, the value of 
$\as$ is slightly smaller than $0.1$ which means that a first estimate of the 
uncertainty introduced by leaving out the next-to leading order (NLO) contribution 
is approximately $10\%$, that of leaving out the NNLO will be $1\%$ and so on.

Another related source of uncertainty is that fixed-order calculations introduce a 
``characteristic energy scale'' $\mu$ in the problem, and the predicted cross 
section depends on the value assigned to that parameter. Introducing higher order 
terms makes the dependence of the result on that parameter smaller, and thus 
decreases the uncertainty from that source as described in 
Sec.~\ref{sec:theory-uncertainties-scale-variation}.

\subsubsection{NLO corrections in the Standard Model}
At NLO there are two conceptually different contributions, denoted virtual 
corrections and real corrections respectively. The former comes from diagrams 
with one loop more than the leading order diagrams, and the latter from diagrams 
with one additional radiated parton integrated over the enlarged phase-space:
\begin{align}
\sigma_{gg \rightarrow gH}^{\text{NLO}} &= \raisebox{-5mm}{\includegraphics[scale=0.25]{phistar/figures/hj1loop/forNigel/gggH_topbox}} \! \times \raisebox{-5mm}{\includegraphics[scale=0.25]{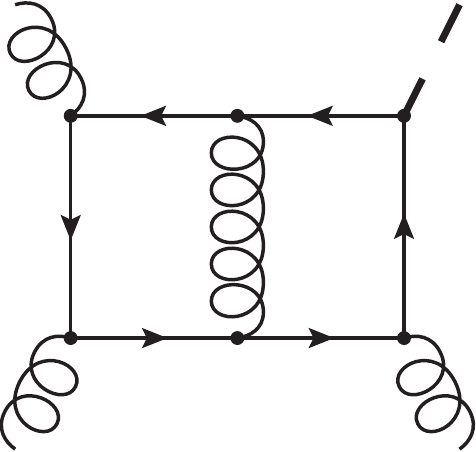}}\raisebox{3.5mm}{\Large *} \! + \; \text{cc.} \; + \, \int \! \id \Phi \, \Bigg| \raisebox{-5mm}{\includegraphics[scale=0.25]{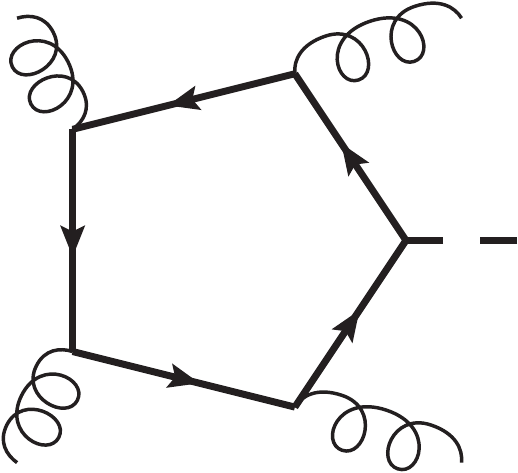}} \Bigg|^2
\label{eq:realvirtual}
\end{align}

The terms combine in such a way that divergences from the additional particle 
in the real radiation becoming ``soft'' (e.g. having small values of the momentum), 
cancel with some of the divergences (the so-called IR divergences) coming from 
the loop in the virtual correction.

At the next order in the perturbative expansion (NNLO) there will be three kinds 
of terms, one with two extra loops (the virtual-virtual corrections), one with 
an extra loop and an extra particle (the real-virtual corrections), and one 
with two extra radiated particles (the real-real corrections). The cancellations 
between the associated divergences take place in a way that is conceptually 
similar to the NLO case but in practice much more complicated both in terms 
of calculation the loop corrections, performing the phase-space integrals, 
and deriving the cancellation of the divergences.

This growth in the number of terms, continues at each order in the perturbative 
expansion. In the following we will focus on the virtual corrections, but have 
in mind that all the terms have to be included in order to obtain a finite contribution.

\begin{figure}[t!]
\centering
\includegraphics[width=0.20\textwidth]{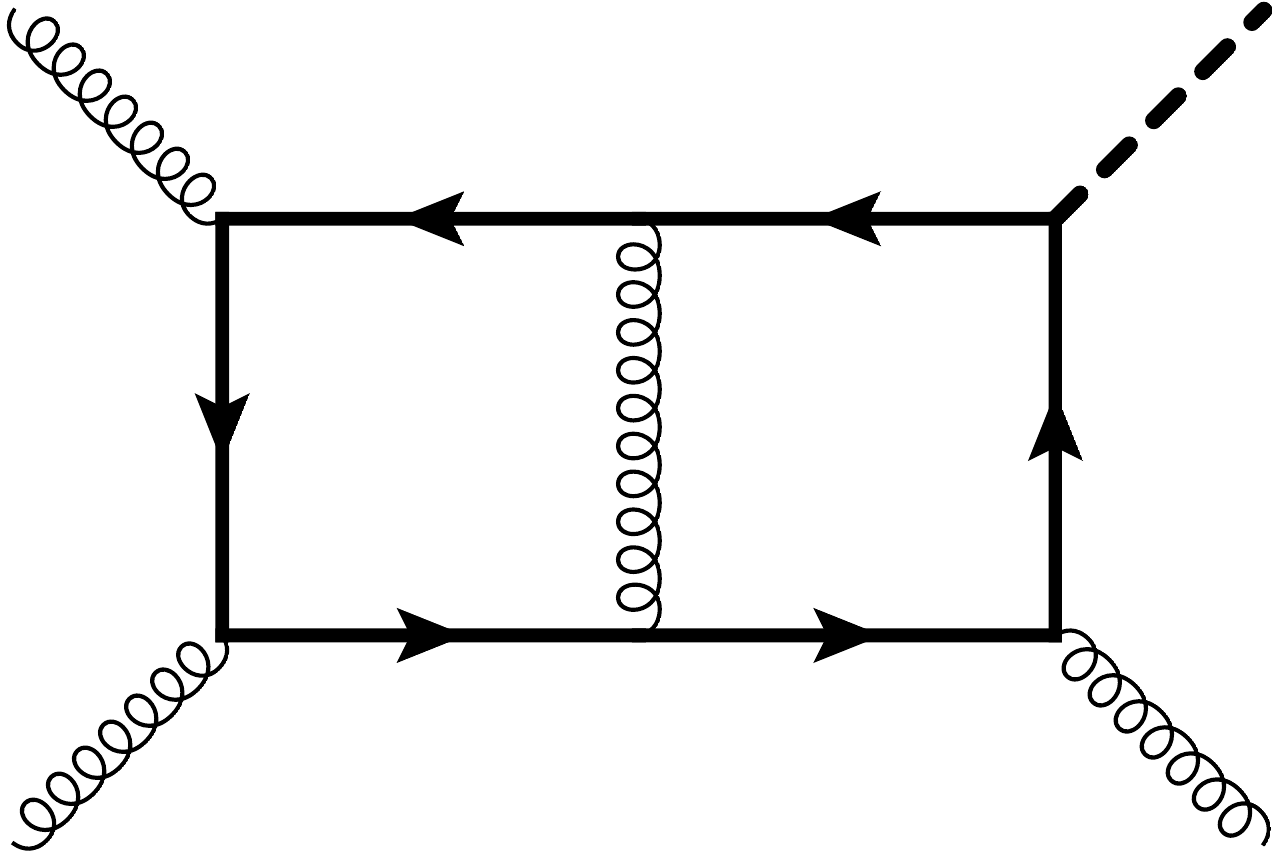} \hspace{0.03\textwidth}
\includegraphics[width=0.20\textwidth]{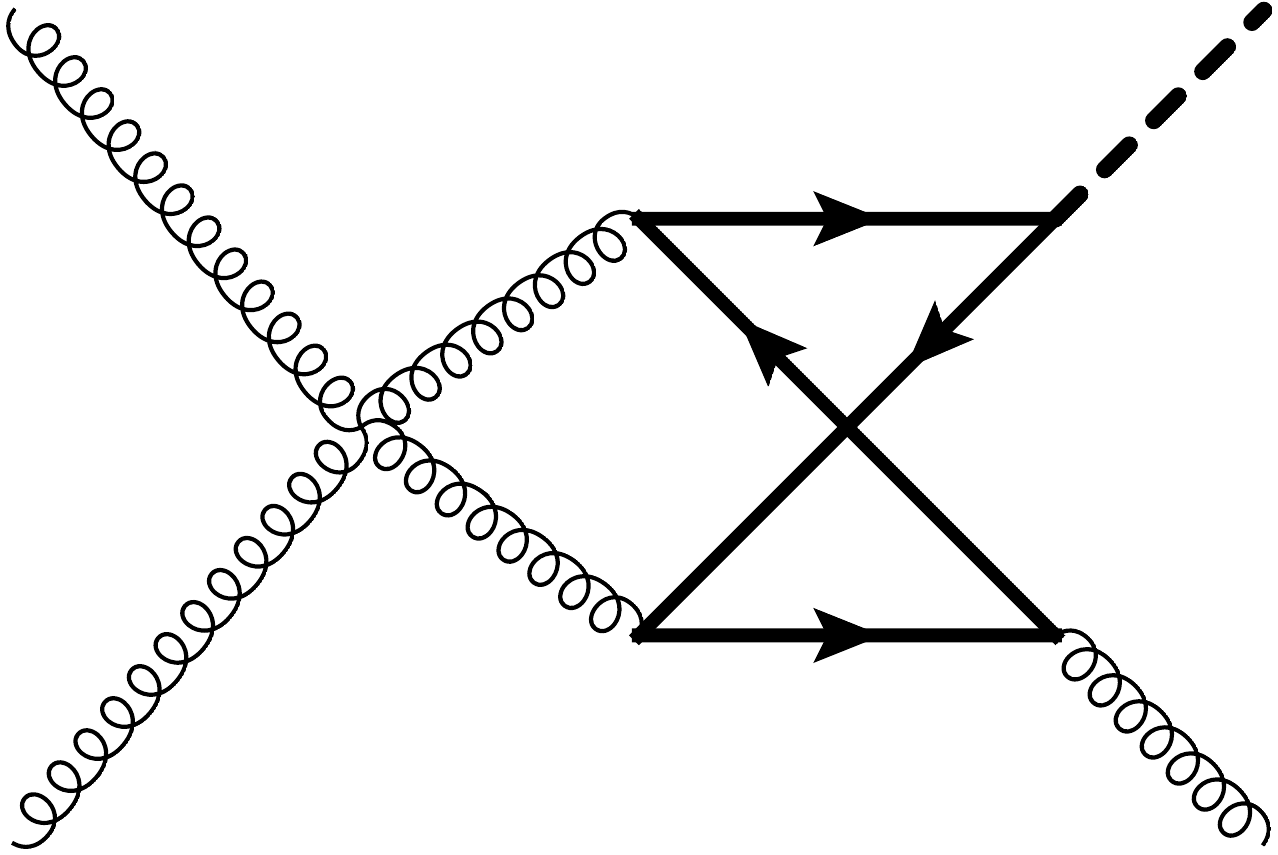} \hspace{0.03\textwidth}
\includegraphics[width=0.20\textwidth]{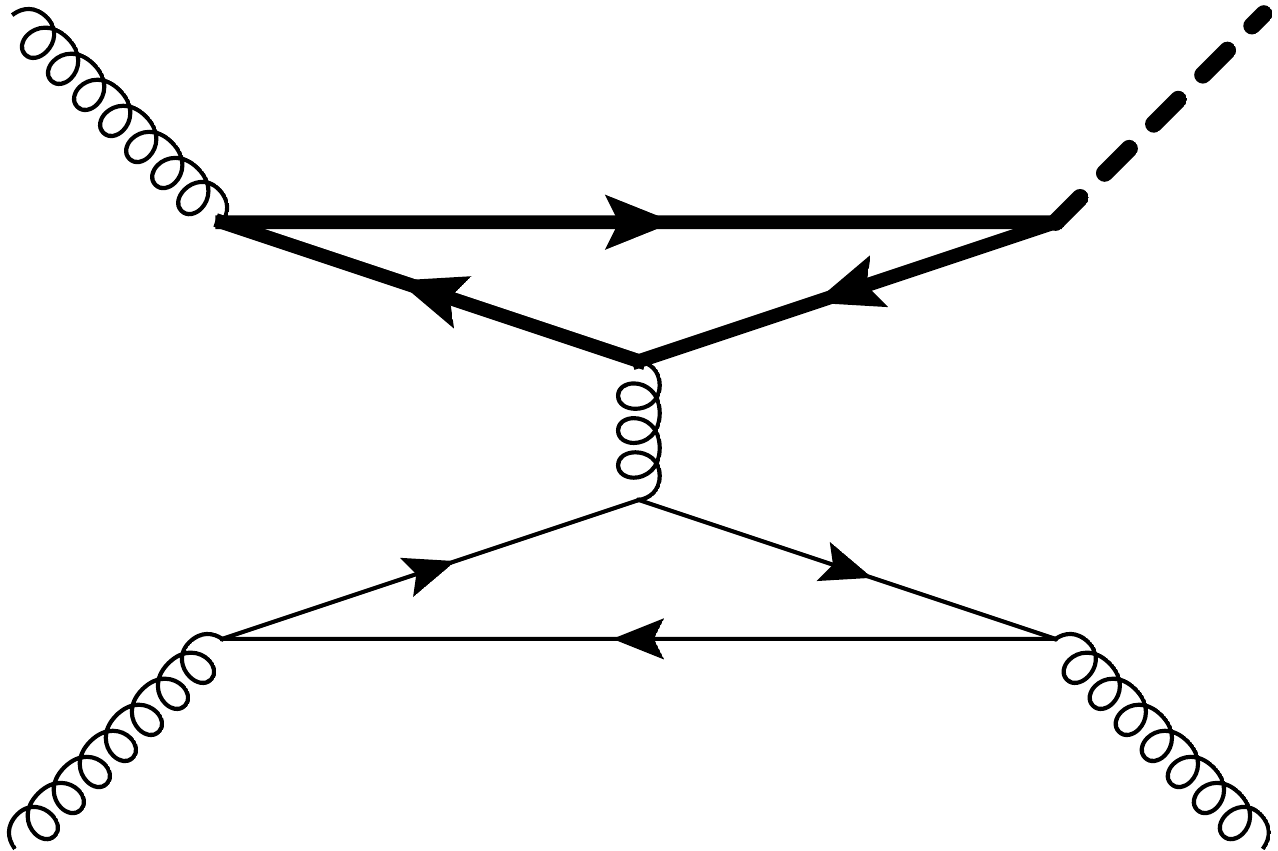} \hspace{0.03\textwidth}
\includegraphics[width=0.20\textwidth]{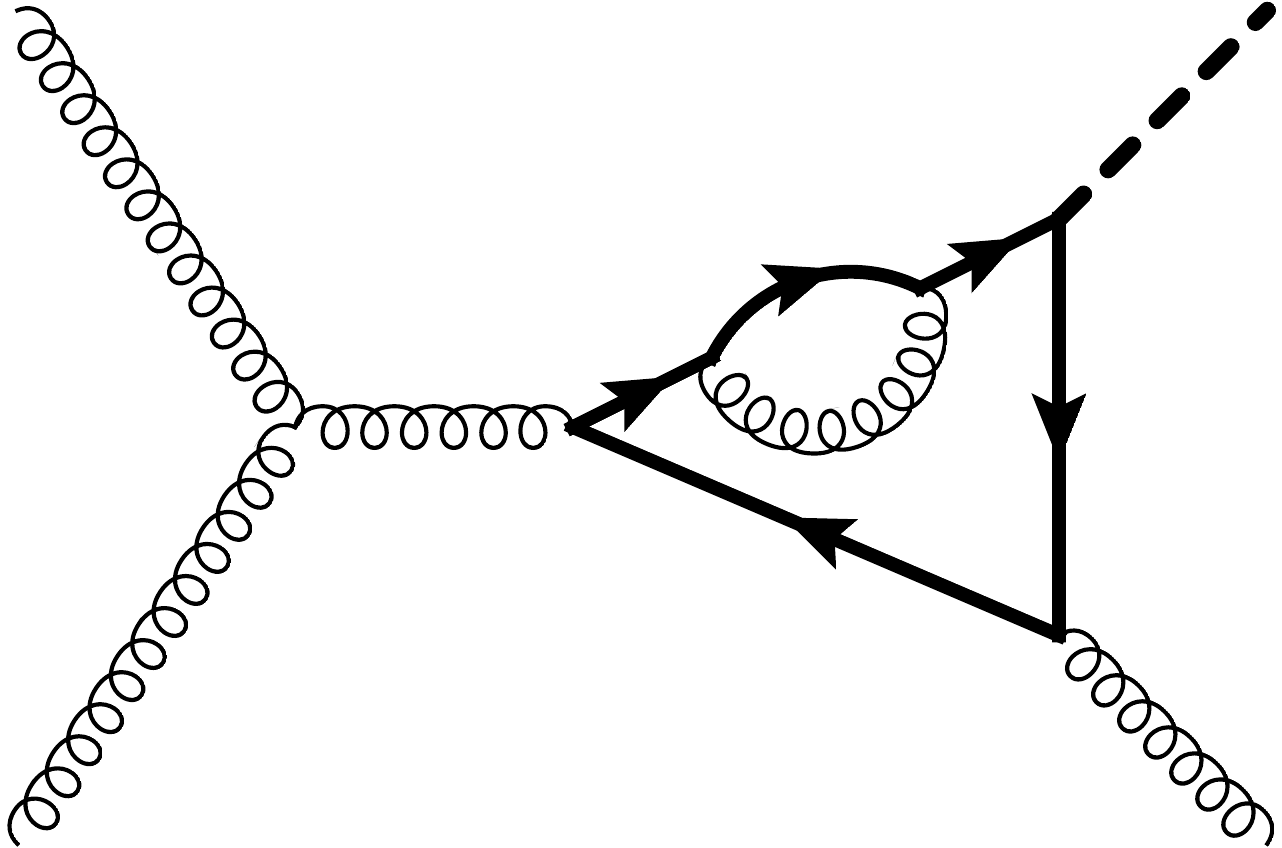}
\caption{Four of the 286 Feynman diagrams contributing to the two-loop QCD correction to $gg \rightarrow gH$.}
\label{fig:feynmantwoloop}
\end{figure}

For Higgs boson $+$ jet production where the leading order term was given by Feynman 
diagrams with one loop, the next-to leading order virtual correction is given 
by two-loop diagrams, and that computation is at the very edge of what is 
possible with the current computational technology~\cite{art:Bonciani:2016qxi, 
art:Neumann:2016dny, art:Melnikov:2016qoc}.

$gg \rightarrow g\h$ contains 286 Feynman diagrams, and $q \bar{q} \rightarrow g\h$ 
and $qg \rightarrow q\h$ each an additional 61. The numerator of each diagram will 
initially contain between 100 and 1000 terms, but this number will grow immensely 
in the intermediate steps of the gamma matrix algebra and the colour factor algebra. 
But while these numbers are large, this part of the calculation is no challenge 
for a computer. The real challenge is in computing the two-loop Feynman integrals. 

(Semi)analytical expressions for the planar subset of the integrals, were 
recently computed in Ref.~\cite{art:Bonciani:2016qxi}. A more thorough discussion 
of the techniques used for that computation will be described in 
Appendix~\ref{sec:twoloop}. The final result of that computation takes up 
about 500 MB, and this fact alone should motivate the search for simpler methods 
to approximate the NLO and the higher contributions.

\subsubsection{NLO corrections in the Higgs Effective Theory}
\label{sec:NLOinHEFT}

For a fully consistent description of the mass effects at high transverse momentum, 
one would like to have the NLO (and ultimately also NNLO) predictions with exact 
mass dependence. However, as discussed in the previous section, owing to the 
complexity of the two-loop virtual amplitudes, these are not available at present. 

\begin{figure}[htb]
  \centering{%
\includegraphics[width=0.2\textwidth]{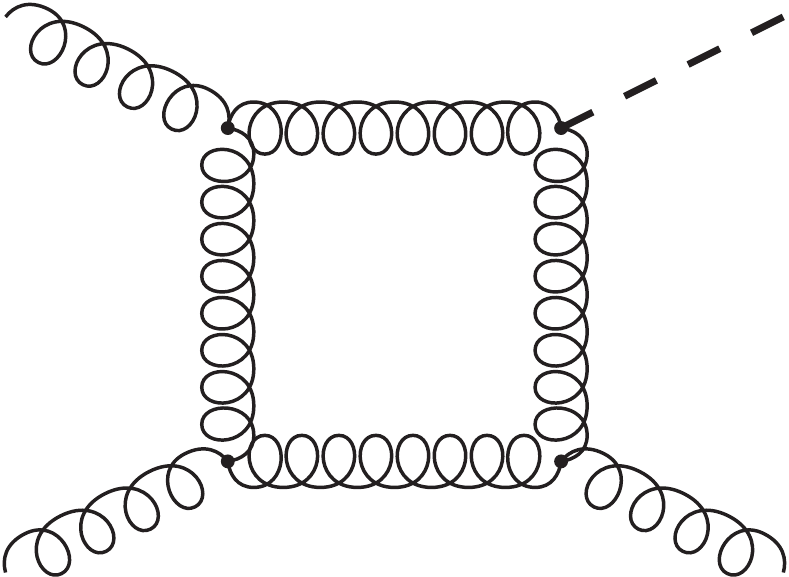}
  }
  \caption{One of the Feynman diagrams contributing to $gg \to g\h$ at one-loop in the HEFT.}
  \label{fig:heftoneloop}
\end{figure}
The NLO corrections have been computed in the \HEFT~\cite{Schmidt:1997wr,
deFlorian:1999zd,Ravindran:2002dc}.  There are 60 one-loop diagrams, of which 
the most complicated is the box integral shown in Fig.~\ref{fig:heftoneloop}. 
As noted earlier, these one-loop integrals are simpler than those in the \SM at 
\LO - the external particles are the same, but the particles circulating in the 
loop are massless gluons. 

As also discussed earlier, at large values of \pth, the \HEFT approach is 
guaranteed to fail.  
We therefore introduce two approximate approaches to estimating the effect of 
a finite top quark mass.
For the inclusive Higgs boson production cross section, it has been observed that the top 
quark mass corrections at NLO~\cite{Spira:1995rr,Harlander:2005rq,Aglietti:2006tp,
Anastasiou:2009kn} can be 
well-approximated  by re-weighting the NLO \HEFT cross section by the ratio of \SM 
and \HEFT predictions at leading order:
\begin{equation}
\label{eq:Rdef}
R = \sigma_{{\rm LO}}^{\SM} / \sigma_{{\rm LO}}^{{\HEFT}}
\end{equation}
where $\sigma_{{\rm LO}}^{\SM}$ includes the exact mass dependence of top quark 
loops. 
The numerically smaller contributions from charm and bottom quarks to 
the inclusive Higgs boson cross section can be accounted for in the same form by 
including the relevant quark loops in the numerator.
In the following, the re-weighting factor $R$ will always include charm, bottom 
and top loops in $\sigma_{{\rm LO}}^{\rm SM}$, while normalising to 
$\sigma_{{\rm LO}}^{{\HEFT}}$ with infinite top quark mass. In the \HEFT, all 
other quarks are treated massless, and their Yukawa couplings are set to zero.

This inclusive re-weighting factor can be generalised to the transverse momentum 
distribution (which is also inclusive in all hadronic radiation) as
\begin{equation}
R(\pth) = \left(\frac{\dsigma_{{\rm LO}}^{\SM}}{\dpth}\right) \Bigg/ 
\left(\frac{\dsigma_{{\rm LO}}^{{\HEFT}}}{\dpth}\right).
\label{eq:rmass}
\end{equation}
Multiplying the higher order \HEFT predictions bin-by-bin with this factor, 
yields the \EFTtimes\ approximation
\begin{equation}
\label{eq:MtimesNLO}
\frac{\dsigma_{{\rm NLO}}^{\EFTtimes}}{\dpth} \equiv R(\pth) 
\left(\frac{\dsigma_{{\rm NLO}}^{{\HEFT}}}{\dpth}\right)  
\end{equation}
which correctly captures the leading logarithms in the quark mass 
corrections~\cite{Bagnaschi:2015qta,Caola:2016upw} at all orders, while failing 
in general to describe subleading 
logarithms and non-logarithmic terms. The computation of subleading mass 
corrections at NLO~\cite{Harlander:2012hf,Dawson:2014ora,Dawson:2015gka} 
also suggests the applicability of 
the \EFTtimes\ procedure. 
To quantify the uncertainty associated with  this re-weighting procedure, 
we consider also
the additive \EFTplus\ prediction obtained by substituting only the LO \HEFT 
contribution by the full LO mass-dependence,
\begin{equation}
\label{eq:MplusNLO}
\frac{\dsigma_{\NLO}^{\EFTplus}}{\dpth} \equiv  \left(\frac{\dsigma_{\NLO}^{{\HEFT}}}{\dpth}\right) + \left (R(\pth) -1\right) \left(\frac{\dsigma_{\LO}^{{\HEFT}}}{\dpth}\right).
\end{equation}

The inclusion of quark mass effects at \LO damps the \pth spectrum. 
Consequently, in the \EFTplus prediction at large \pth, the
harder higher order \HEFT corrections dominate over the softer LO contribution 
with exact mass dependence. Even if the yet unknown \NLO corrections to the 
exact mass dependence turn out to be numerically large, there is no reason for 
them to increase substantially with transverse momentum. The \EFTplus approximation 
is therefore an overestimate of the hardness of the mass-corrected transverse 
momentum spectrum, and can be considered to be an upper bound on the actual exact mass dependence.

The \EFTtimes prediction reweights the full spectrum with the softness of the 
LO mass dependence of the (\h + 1)-parton process. A recent study~\cite{Greiner:2016awe} 
demonstrated that the mass-dependent suppression (with respect to the 
\HEFT prediction) of large transverse momentum configurations is less strong 
for the (\h + 2)-parton and (\h + 3)-parton processes than it is for the 
(\h + 1)-parton process. Consequently, \EFTtimes could be considered 
as a lower bound on the exact mass dependence.

\begin{figure}
    \centering
(a)
    \begin{subfigure}[b]{0.45\textwidth}
        \includegraphics[width=\textwidth]{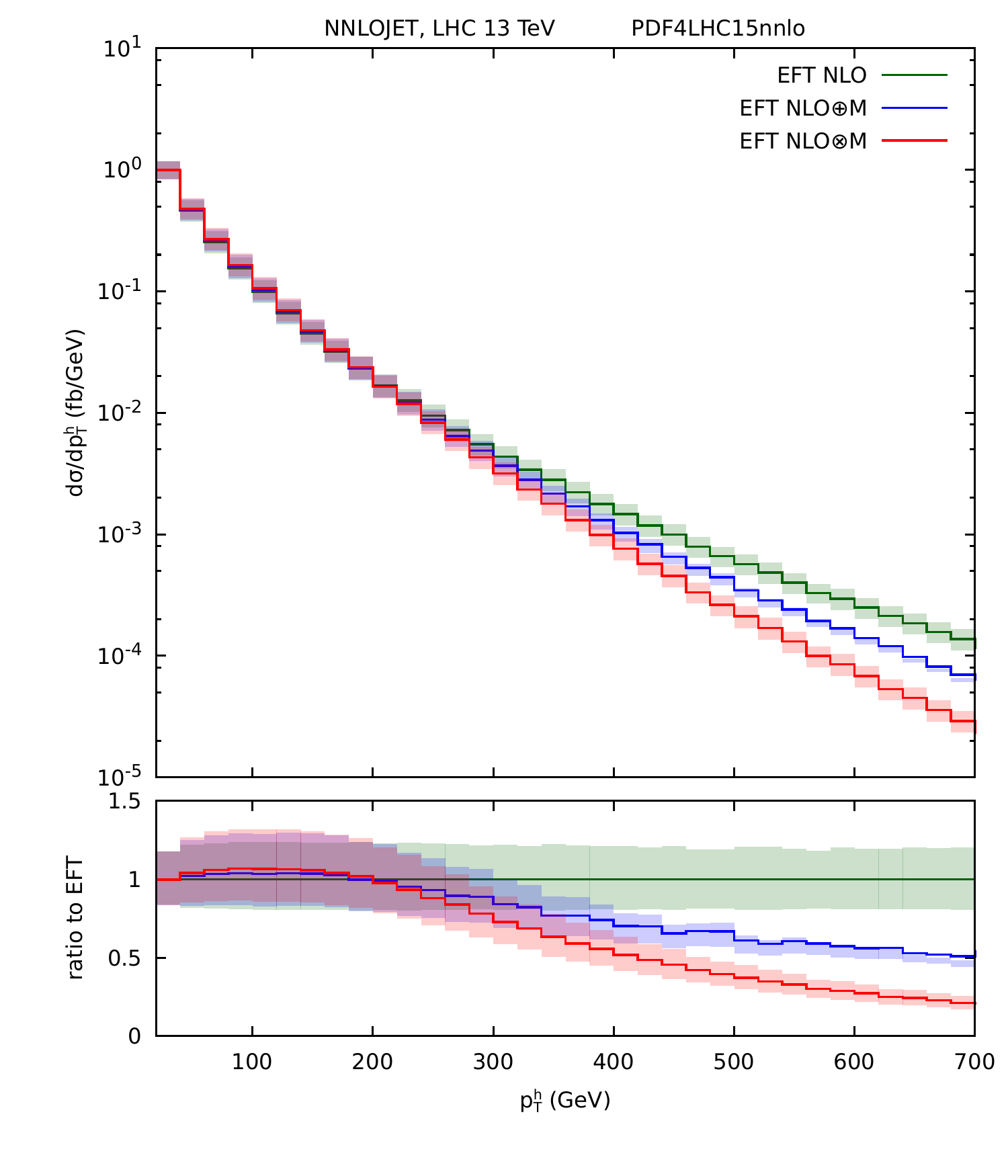}
    \end{subfigure}
(b)
    \begin{subfigure}[b]{0.45\textwidth}
        \includegraphics[width=\textwidth]{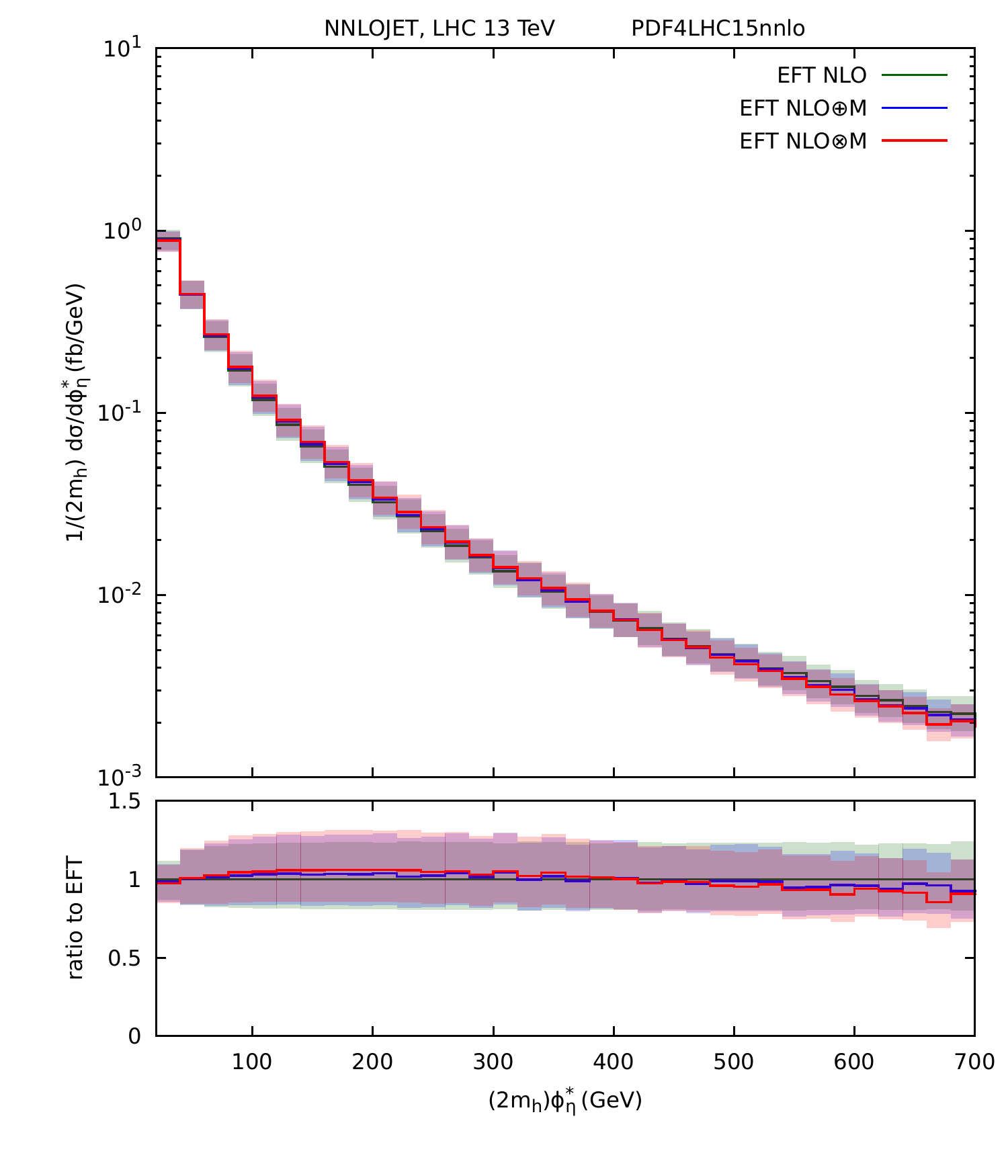}
    \end{subfigure}
    \caption{The Higgs boson (a) \pth and (b) \phistar distributions at $\sqrt{s}=13$~TeV for the NLO \HEFT (green), NLO $\EFTplus$ (blue) and NLO $\EFTtimes$ (red) approximations discussed in the text. The lower panels show the ratio normalised to the NLO Higgs Effective Theory result.}
    \label{fig:phistar_observable_pt_13TeVcompareNLO}
\end{figure}

Fig.~\ref{fig:phistar_observable_pt_13TeVcompareNLO} shows the NLO predictions for the \pth and \phistar distributions in the three approximations, \HEFT, \EFTplus and \EFTtimes.  We see that below $\pth \sim 200$~\GeV, all three approximations lead to a similar prediction for the \pth distribution.  Above $\pth \sim 200$~\GeV, the three are clearly different with the \EFTtimes prediction being softer than the \EFTplus distribution.  As discussed above, the \SM\ \NLO result is likely to lie between the \EFTplus and \EFTtimes distributions and the uncertainty on the \pth distribution is thus much larger than the scale uncertainty for any individual prediction. 

On the other hand, each bin in the \phistar distribution samples a wide range of $\pth$ 
leading to rather similar predictions for all three approximations with at most a $\pm$5\% difference, while the scale uncertainty of {\cal O}($\pm$ 20\%) is much larger.  If the  \SM\ \NLO result does actually lie between the \EFTplus and \EFTtimes distributions, then this implies that the
NLO \HEFT is good estimator of the \phistar distribution. 
   
\subsubsection{NNLO corrections in the Higgs Effective Theory}
\label{sec:NNLOinHEFT}

The NNLO ${\cal O}(\as^5)$ corrections have also been computed in the \HEFT~\cite{Boughezal:2013uia,Chen:2014gva, Boughezal:2015dra,Boughezal:2015aha,Caola:2015wna,Chen:2016zka}. 
There are 1305 two-loop $gg \to g\h$ and 328 two-loop $q\bar{q} \to g\h$ diagrams, and a representative diagram is shown in Fig.~\ref{fig:hefttwoloop}. 
\begin{figure}[tbh]
  \centering{%
\includegraphics[width=0.2\textwidth]{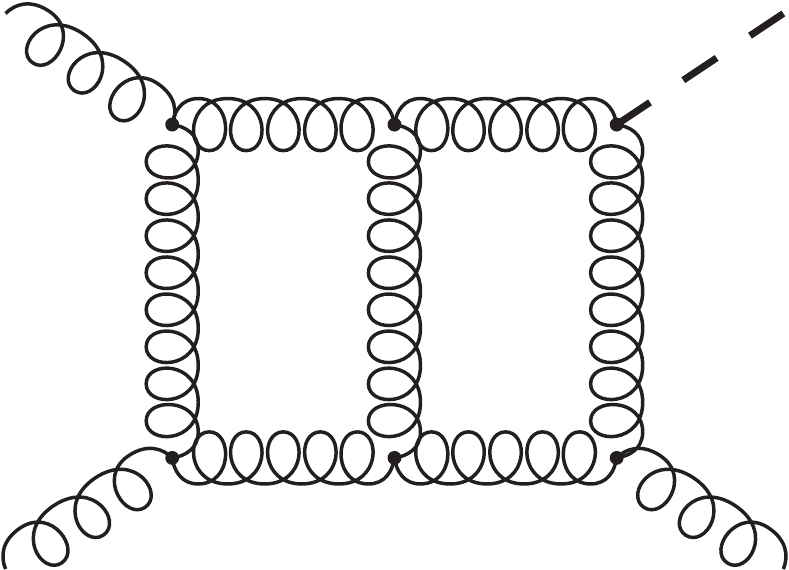}
  }
  \caption{One of the Feynman diagrams contributing to $gg \to g\h$ at two-loops in the HEFT.}
  \label{fig:hefttwoloop}
\end{figure}

The two-loop matrix elements were computed in Ref.~\cite{art:Gehrmann:2011aa} using the 
projector method discussed in Sec.~\ref{sec:theory}.  The two-loop master integrals are 
the same as those for $\gamma^* \to q\bar qg$~\cite{Gehrmann:2000zt,Gehrmann:2001ck} 
continued into the relevant phase space region~\cite{Gehrmann:2002zr} and were computed 
using the differential equation method~\cite{art:Gehrmann:1999as}.  These expressions are 
in terms of generalised harmonic polylogarithms which can be evaluated numerically~\cite{Gehrmann:2001pz,Gehrmann:2001jv}. 
The one-loop $gg \to gg\h$, $q\bar q \to gg\h$ and $q\bar q \to Q\bar{Q}\h$ helicity 
amplitudes have been computed analytically using unitarity methods~\cite{Glover:2008ffa,Badger:2009hw,Badger:2009vh}.
The implementation of the NNLO corrections within NNLOJET employs the antenna subtraction 
method to isolate the infrared singularities~\cite{GehrmannDeRidder:2005cm,GehrmannDeRidder:2004tv,GehrmannDeRidder:2005aw,
GehrmannDeRidder:2005hi,GehrmannDeRidder:2007jk,Daleo:2006xa,Daleo:2009yj,Gehrmann:2011wi,
Boughezal:2010mc,GehrmannDeRidder:2012ja,Currie:2013vh}.  The \h+jet and \pth distributions 
have been computed within the NNLOJET framework~\cite{Chen:2014gva,Chen:2016zka}.

The \NNLO predictions in the Higgs Effective Theory are shown in 
Fig.~\ref{fig:phistar_observable_pt_13TeVall}.   
We see that for both \pth and \phistar spectra, the \NLO effects are sizeable compared to 
\LO\ (typically 80-100\%), but that for $\pth, 2\mh\phistar > 100$~\GeV, the \NNLO effects 
are small compared to \NLO.  For both distributions, the scale uncertainty is reduced from {\cal O}(30\%) at \LO to {\cal O}(20\%) at \NLO, and reduced further to {\cal O}(10\%) at \NNLO.

\begin{figure}[tbh]
    \centering
(a)
    \begin{subfigure}[b]{0.45\textwidth}
        \includegraphics[width=\textwidth]{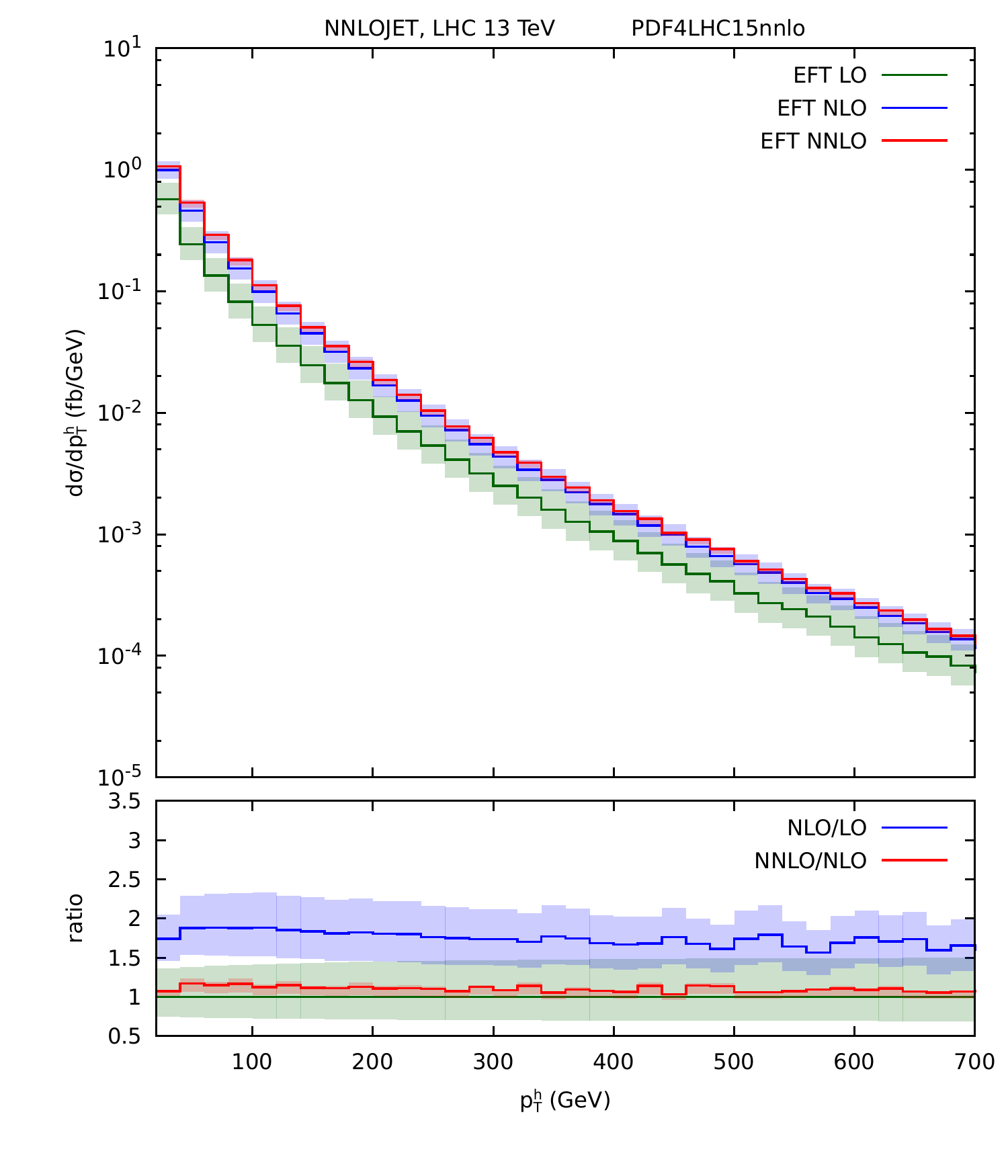}
    \end{subfigure}
(b)
    \begin{subfigure}[b]{0.45\textwidth}
        \includegraphics[width=\textwidth]{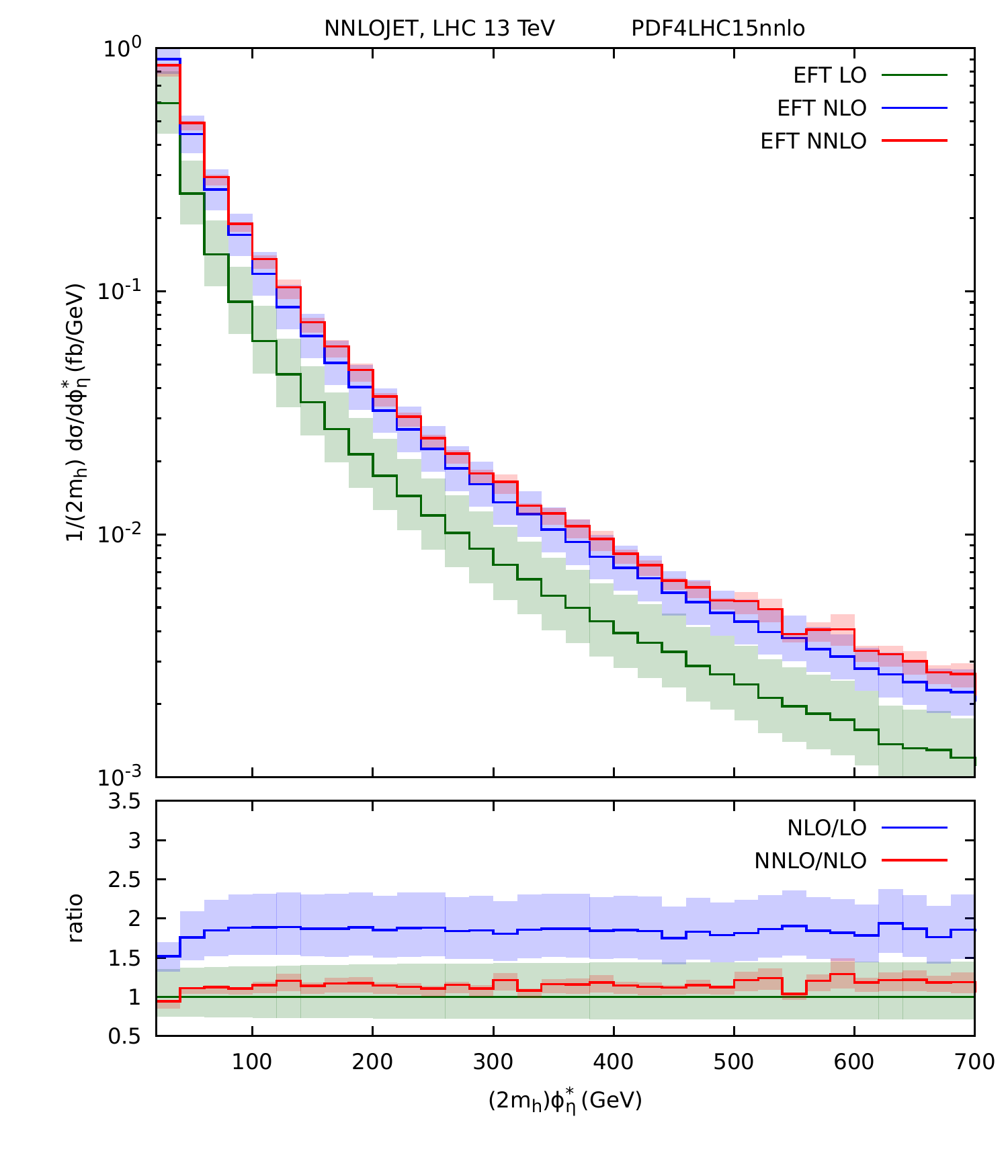}
    \end{subfigure}
    \caption{The Higgs boson (a) \pth and (b) \phistar distributions at $\sqrt{s}=13$~TeV 
for the \HEFT at \LO (green), \NLO (blue) and \NNLO (red).  The lower panels show the 
ratio of \NLO/\LO (blue) and \NNLO/\NLO (red).}
    \label{fig:phistar_observable_pt_13TeVall}
\end{figure}

Following the discussion at NLO, we modify the \HEFT prediction either by scaling the 
differential cross section by the \SM LO result,
\begin{equation}
\label{eq:MtimesNNLO}
\frac{\dsigma_{\NNLO}^{\EFTtimes}}{\dpth} \equiv R(\pth) 
\left(\frac{\dsigma_{\NNLO}^{{\HEFT}}}{\dpth}\right),  
\end{equation}
or by simply replacing the LO \HEFT contribution by the LO result with the exact mass dependence,
\begin{equation}
\label{eq:MplusNNLO}
\frac{\dsigma_{\NNLO}^{\EFTplus}}{\dpth} \equiv  \left(\frac{\dsigma_{\NNLO}^{{\HEFT}}}{\dpth}\right) + \left (R(\pth) -1\right) \left(\frac{\dsigma_{\LO}^{{\HEFT}}}{\dpth}\right).
\end{equation}

The NNLO predictions for the \pth and \phistar distrbutions are shown in 
Fig.~\ref{fig:phistar_observable_pt_13TeVcompareNLO}. At small $\pth$, the \EFTtimes 
and \EFTplus approximations are very similar and lead to a small enhancement compared to the pure \HEFT.  
However, as at LO and NLO, with increasing $\pth$, the non-pointlike nature of the heavy quark 
loop becomes resolved leading to a softening of the \pth spectrum as compared to the \HEFT prediction.  
This suppression is more severe for the \EFTtimes\ approximation than for the \EFTplus\ 
approximation.
We observe that at large \pth the scale uncertainty for the \EFTplus\ approximation is much larger than that for the \EFTtimes\ prediction~\cite{Chen:2016zka}. 
Even so, the difference between the two predictions is larger than the scale uncertainty of either
and could reasonably be considered as an estimate of the current theoretical 
uncertainty at large \pth. 

As at \NLO, the \phistar distribution exhibits a rather different behaviour and the
three approximations again lie within a few percent of each other.
Employing the same logic as for the \pth distribution, i.e., that the \EFTplus\ and \EFTtimes\ distributions span the likely range of the prediction including top mass effects, 
we deduce that the \phistar distribution 
can be reliably estimated by the \HEFT prediction with a residual scale uncertainty 
of about 10\%.

\begin{figure}
    \centering
(a)
    \begin{subfigure}[b]{0.45\textwidth}
        \includegraphics[width=\textwidth]{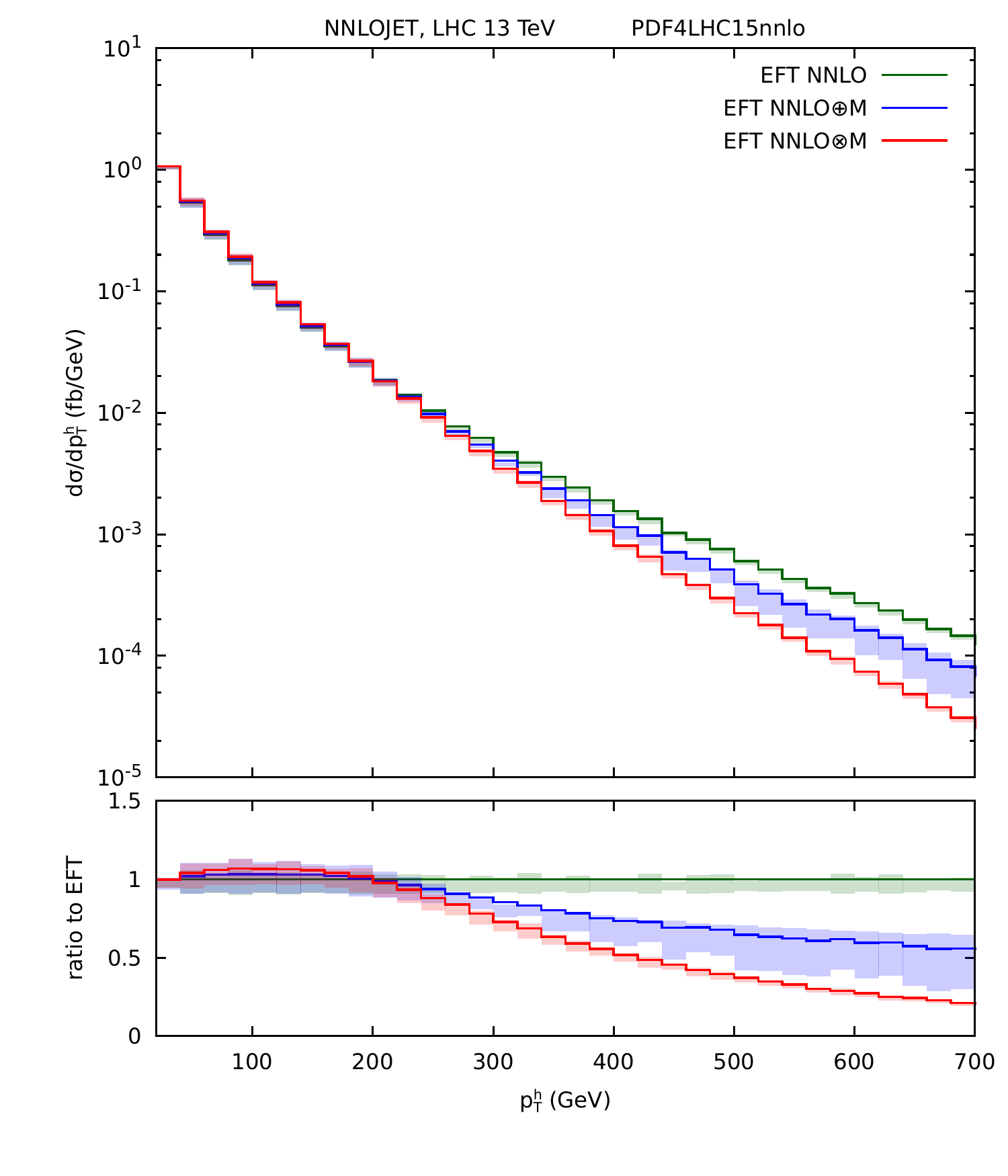}
    \end{subfigure}
(b)
    \begin{subfigure}[b]{0.45\textwidth}
        \includegraphics[width=\textwidth]{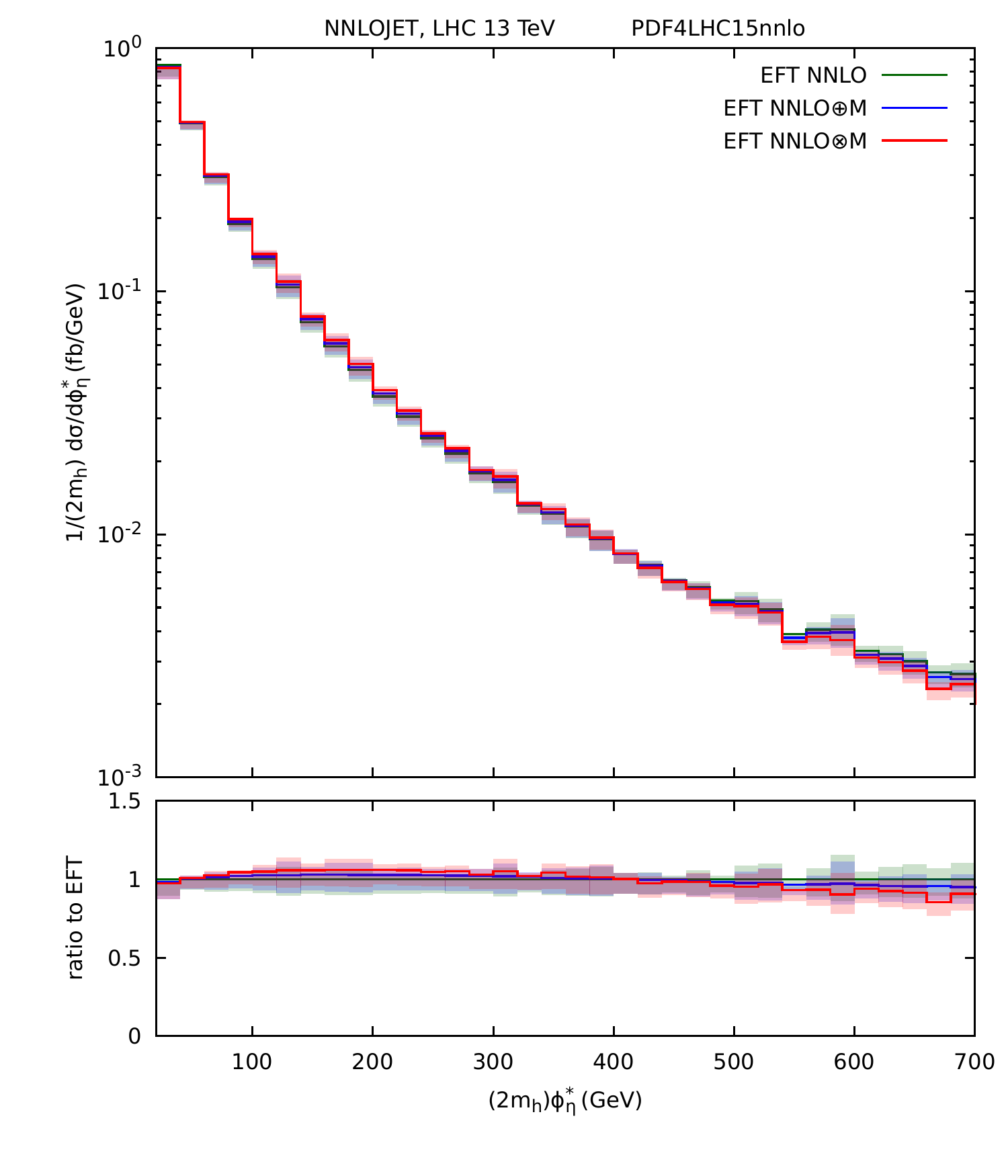}
    \end{subfigure}
    \caption{The Higgs boson (a) \pth and (b) \phistar distributions at $\sqrt{s}=13$~TeV for 
the NNLO HEFT (green), NNLO $\EFTplus$ (blue) and NNLO $\EFTtimes$ (red) approximations 
discussed in the text. The lower panels show the ratio normalised to the NNLO Higgs Effective Theory result.}
    \label{fig:phistar_observable_pt_13TeVcompareNNLO}
\end{figure}

\clearpage
\section{Beyond the Standard Model}
\label{sec:BSM}

The Standard Model has been extremely compatible with almost all the measurements made in particle physics experiments over the last few decades with no convincing hints of additional physics. However, despite its success so far, the Standard Model suffers a number of theoretical shortcomings and also falls short of explaining several experimental observations. This suggests that it cannot be the complete theory of nature, but rather serves as an effective theory of the energy regime we are currently able to probe. One of the theoretical failings of the SM concerns the stablity of the Higgs mass, which requires large cancellations between its bare mass and the quantum corrections from the particles that couple to it. New Physics models that address this issue predict new particles whose contributions to the loops can soften the unnatural fine tuning. These new particles are typically heavy resonances that couple strongly to the Higgs and can either be scalars, as in the case of supersymmetric models, or fermionic top parters, predicted by many composite Higgs models.

\subsection{Searches for BSM Physics} 
\label{sec:BSMsearches}

The new particle spectrum can affect the production and decay rates of SM particles, in particular the Higgs boson. New massive coloured particles can participate in the loop induced gluon fusion process and alter the Higgs boson production cross section. Heavy top partners may also mix with the top quark and induce a modified Yukawa coupling, which would also affect the gluon fusion cross section\footnote{This would also change the experimentally difficult high-multiplicity $t\bar{t}h$ final state.}.  It is therefore important to resolve the details of the gluon-Higgs interaction. Precise measurements of this vertex can reveal the dynamics of the particles in the loop, and quantify possible deviations from the SM predictions.

We can parametrise the missing information about the exact nature of the gluon-Higgs interaction in terms of effective operators. If the new particles in the loop are much heavier than the Higgs, they will modify the coefficient $\cg$ of the effective interaction of Eq.~\eqref{eq:eff}. 
If there are no new particles participating in the loop as in the Standard Model then $\cg = 0$. Similarly, modifications to the Yukawa coupling 
arising from, for instance, heavy top partners that mix with the top quark}
can adjust the top Yukawa interaction \eqref{eq:ffh} by a factor $\ct$ (compared to the \SM value of $\ct = 1$). 

As discussed earlier, to a good approximation, the gluon fusion cross section is dominated by contributions from top-quark loops and from the effective interaction and is proportional to 
$\left(\ct + \cg\right)^2$ (see Eq.~\eqref{eq:sigmaincl}).
As a result, the measurement of the inclusive rate does not disentangle the effects of \ct and \cg. In a large class of BSM models such as Composite Higgs or Little Higgs models, the change in the top-quark Yukawa coupling  may also compensate for the effects of the new particles such that the inclusive cross section is SM-like and there are no traces of the new particle spectrum observed at the LHC. Cancellations of this nature may also happen in SUSY models such as the MSSM, where the deviations of the inclusive cross section from the observed SM-like values vanish for certain values of the trilinear coupling of the Higgs boson and a pair of scalar top quarks (stops), $A_t$~\cite{Haber:1984zu}.

\begin{figure}
  \begin{center}
      \includegraphics[width=0.20\textwidth]{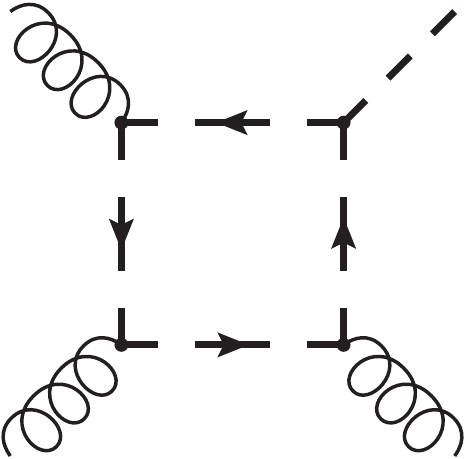}                 \hspace{0.03\textwidth}
      \includegraphics[width=0.20\textwidth]{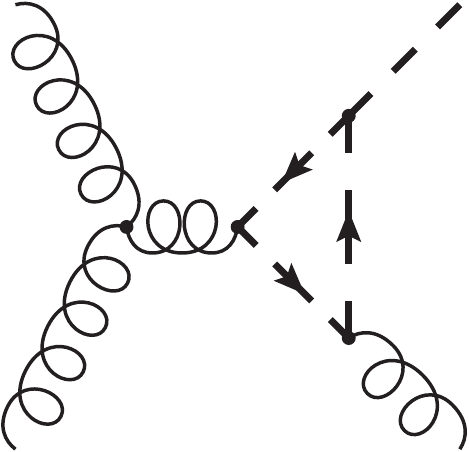}                \hspace{0.03\textwidth}
      \includegraphics[width=0.20\textwidth]{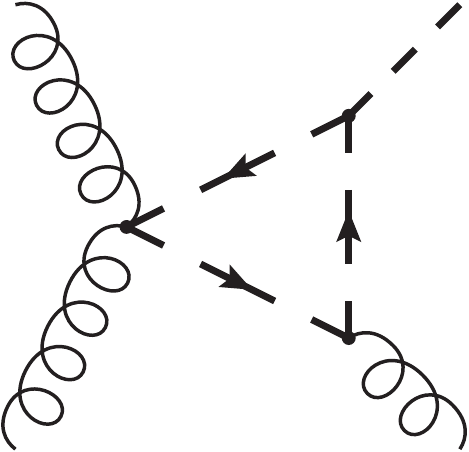}       \hspace{0.03\textwidth}
      \includegraphics[width=0.20\textwidth]{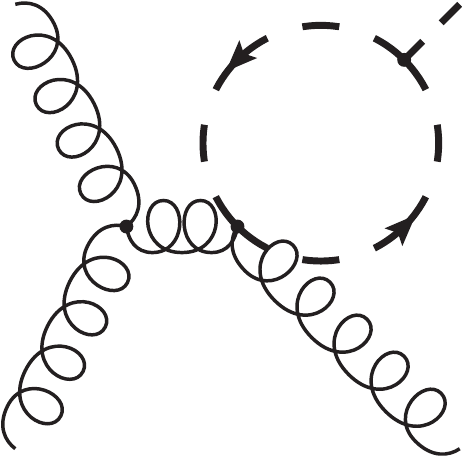}
    \caption{Feynman diagrams for the process $gg\rightarrow gh$ mediated by scalar particles at leading order.}
    \label{fig:BSMloop}
  \end{center}
\end{figure}

In order to independently determine $\cg$ and $\ct$ 
a new energy scale much higher than the top-quark mass is necessary such that 
the low energy theorem no longer holds~\cite{Banfi:2013yoa}. In this case,  
the degeneracy of $\ct$ and $\cg$ can be lifted. This is achieved by recoiling the Higgs bosons against an additional jet, and studying the process at high $\pth$~\cite{Harlander:2013oja,Azatov:2013xha,Grojean:2013nya,Dawson:2014ora,Langenegger:2015lra,Maltoni:2016yxb,Grazzini:2016paz,Deutschmann:2017qum}. 

As a concrete example, in BSM models such as the MSSM, heavy squark loops will also contribute to the production of boosted Higgs bosons as shown in Fig.~\ref{fig:BSMloop}.
Lets assume that the mass of the squark $m >> \mh$.  In this case, one could use the effective interaction to compute the contribution to the inclusive Higgs boson cross section.  At low \pth, this would also be a good description.  However, because the effective interaction is a dimension-5 operator, the amplitude grows with energy, $E/m$, and the cross section will grow like $E^2/m^2 \sim \pth^2/m^2$. Eventually this growth will violate unitarity.  Long before that happens, the approximation $m \rightarrow \infty$ loses validity, and the differential cross section becomes sensitive to the masses in the loop, thereby revealing possible new physics which was otherwise invisible in both the inclusive cross section and in the low $\pth$ region.   
One can now understand why it is important to study the $\pth$ spectrum of the Higgs to find deviations from SM predictions.


\subsection{A Simple Model}
\label{sec:BSMeffective}

To see more explicitly how 
BSM effects can be disentangled
we construct a simple but generic BSM model that might contain either heavy fermions or heavy scalars or both. We consider a model where boosted Higgs boson production cross section receives contributions from a complex colour triplet scalar loop with mass $m_s$ as well as contributions involving the top-quark and some heavy unknown particles whose interaction with the Higgs boson is described by the effective vertex of Eq.~\eqref{eq:eff}.
The scalar-scalar-higgs interaction is given by,
\begin{equation}
{\cal L}_{\rm scalar} = \cqt \frac{2m_s^2}{v} S^\dagger S\h,
\end{equation}
while the interactions of the scalar with the gluons are determined by gauge invariance and the covariant derivative.
We can write the full matrix element for the Higgs boson coupling to gluons as
\begin{equation}\label{eq:BBSM_matrix_element}
\mathcal{M}({\ct}, \cqt, \cg)  = \ct \MF +  \cqt \MS + \cg \MUV,
\end{equation}
where $\MF$ is the one-loop amplitude taking into account the full top-quark mass dependence,  
$\MS$ is the amplitude with the full scalar mass dependence and $ \MUV$ is the amplitude arising from the effective interaction.
The coefficients $\ct$ and $\cqt$ scale any effects that the new heavy particles may have on the couplings of the top-quark and the scalar to the Higgs boson.
This simple model can be straightforwardly generalised to contain contributions from the charm and bottom quarks and multiple scalar particles (as in SUSY models), however for simplicity we will only consider the top-quark and a single generic scalar colour triplet. 

The amplitude for the gluon fusion production of a Higgs boson is,
\begin{equation}
\label{eq:Mggmodel}
\mathcal{M}^{gg} = \ct \AF + \cqt \AS + \cg \AUV,
\end{equation}
where the individual contributions can be summarised as,
\begin{eqnarray}
\AF & = & \frac{1}{v} A_{1/2}(\tau_t),\\
\AS & = & \frac{1}{v} A_{0}(\tau_s),\\
\AUV & = & \frac{1}{v} A_{\infty},
\end{eqnarray}
where $\tau_{i} = 4m_i^2/\mh^2$.
The well known loop functions are (see for example Ref.~\cite{Englert:2014uua}),
\begin{eqnarray}
A_0 (\tau) &=& \ - \tau [1-\tau f(\tau)],\\
A_{1/2} (\tau) &=& \ 2\tau[1+(1-\tau)f(\tau)],
\end{eqnarray}
with 
\begin{eqnarray}
f(\tau) =& \ \begin{cases}
{\arcsin}^2 \frac{1}{\sqrt{\tau}} , &  \tau \geq 1 \\
-\frac{1}{4} \left( \log \frac{1+ \sqrt{1-\tau}}{1-\sqrt{1-\tau}} - i\pi \right)^2, & \tau < 1
\end{cases},
\end{eqnarray}
and
\begin{eqnarray}
A_{\infty} &=& \frac{4}{3}.
\end{eqnarray}

If the mass of $m_{X_i}$ is large enough, we can simplify the loop functions by taking $\tau \to \infty$ such that 
\begin{eqnarray}
A_0(\tau) &\to & \frac{1}{4} A_{\infty}\\
A_{1/2}(\tau) &\to & A_{\infty}.
\end{eqnarray}
This UV behaviour means that we can simplify Eq.~\eqref{eq:Mggmodel} when the scalar mass is much bigger than \mh by the replacements,
\begin{equation}
\cqt \to 0, \qquad \cg \to \cg + \frac{\cqt}{4}.
\end{equation}
Similarly, the \HEFT is obtained by taking the limit where the top-quark mass is much larger than \mh, and amounts to the replacements,
\begin{equation}
\ct \to 0, \qquad \cg \to \cg + \ct.
\end{equation}
This model therefore includes both the top-quark contribution to the \SM ($\ct=1$, $\cqt=0$ and $\cg=0$) and the \HEFT ($\ct=0$, $\cqt=0$ and $\cg=1$). 
If both top-quark and scalar masses are large compared to the Higgs boson mass, then we can set,
\begin{equation}
\ct \to 0, \qquad \cqt \to 0, \qquad \cg \to \cg + \ct + \frac{\cqt}{4}.
\end{equation}

In this model, because $\mt > \mh$ and we assume that $m_s > \mt$, the lowest order gluon fusion cross section is sensitive to the combination of couplings,
\begin{equation}
\sigma_{\LO} \sim \left({\ct} + \frac{1}{4}\cqt + \cg\right)^2 \sigma_{\LO}^\SM,
\end{equation}
and measurements of the total cross section cannot discriminate between models that have different values of \ct, \cqt and \cg .

As discussed earlier, QCD radiation probes the internal structure of the effective interaction and the 
effects of various new physics contributions can be disentangled by studying boosted Higgs bosons. 
The amplitudes due to fermion loops and/or the effective interaction have been discussed earlier.  
The scalar contribution was computed in Refs.~\cite{Graudenz:1992pv,Spira:1995rr} and the matrix elements implemented in the numerical code
{\tt SusHi}~\cite{Harlander:2012pb}.  We have ported the routines directly from {\tt SusHi} to NNLOJET. 

Following Ref.~\cite{Schlaffer:2014osa,Harlander:2016hcx}, we write the contribution to the cross section of the simple model above \pthcut and normalised to the top-quark contribution as
\begin{align}\label{eq:BBSM_norm}
\frac{\sigma(\pthcut)}{\sigma^{SM}(\pthcut)} =& \frac{\int_{\pthcut}^{\infty} {\rm d}  \pth \dOmega \ |\ct \MF +  \cqt \MS + \cg \MUV|^2}{\int_{\pthcut}^{\infty} \dpth \dOmega \ |\MF|^2} \nonumber \\
=& \left({\ct} + \frac{1}{4}\cqt + \cg\right)^2 \nonumber \\
&+ \delta_{ts}(\pthcut) \ \ct \cqt \ +  \delta_{tg}(\pthcut) \ \ct \cg \ + \delta_{sg}(\pthcut) \ \cqt \cg  \nonumber \\
&+ \epsilon_{s} (\pthcut) \ \cqt^2 \ + \epsilon_{g} (\pthcut) \ \cg^2
\end{align}
where 
\begin{align}\label{eq:delta_eps}
\delta_{tg}(\pthcut) =& \ \frac{2 \ \int_{\pthcut}^{\infty} \dpth \dOmega \ \text{Re}( \MF\MUV^{*}) }{\int_{\pthcut}^{\infty} {\rm d}  \pth \dOmega \ |\MF|^2 } - 2, \\
\delta_{ts}(\pthcut) =& \ \frac{2 \ \int_{\pthcut}^{\infty} \dpth \dOmega \ \text{Re}( \MF\MS^{*}) }{\int_{\pthcut}^{\infty} {\rm d}  \pth \dOmega \ |\MF|^2 } - \frac{1}{2},\\
\delta_{sg}(\pthcut) =& \ \frac{2 \ \int_{\pthcut}^{\infty} \dpth \dOmega \ \text{Re}( \MS\MUV^{*}) }{\int_{\pthcut}^{\infty} {\rm d}  \pth \dOmega \ |\MF|^2 } - \frac{1}{2},\\
\epsilon_{g}(\pthcut) =& \  \frac{\int_{\pthcut}^{\infty} \dpth \dOmega \ |\MUV|^2}{\int_{\pthcut}^{\infty} {\rm d}  \pth \dOmega \ |\MF|^2} - 1,\\
\epsilon_{s}(\pthcut) =& \  \frac{\int_{\pthcut}^{\infty} \dpth \dOmega \ |\MS|^2}{\int_{\pthcut}^{\infty} {\rm d}  \pth \dOmega \ |\MF|^2} - \frac{1}{16}.
\end{align}
Analogous expressions can be derived for the normalised cross sections above $\phistarcut$. 

\begin{figure}[t]
\centering
(a)
    \begin{subfigure}[b]{0.45\textwidth}
\includegraphics[width=\textwidth]{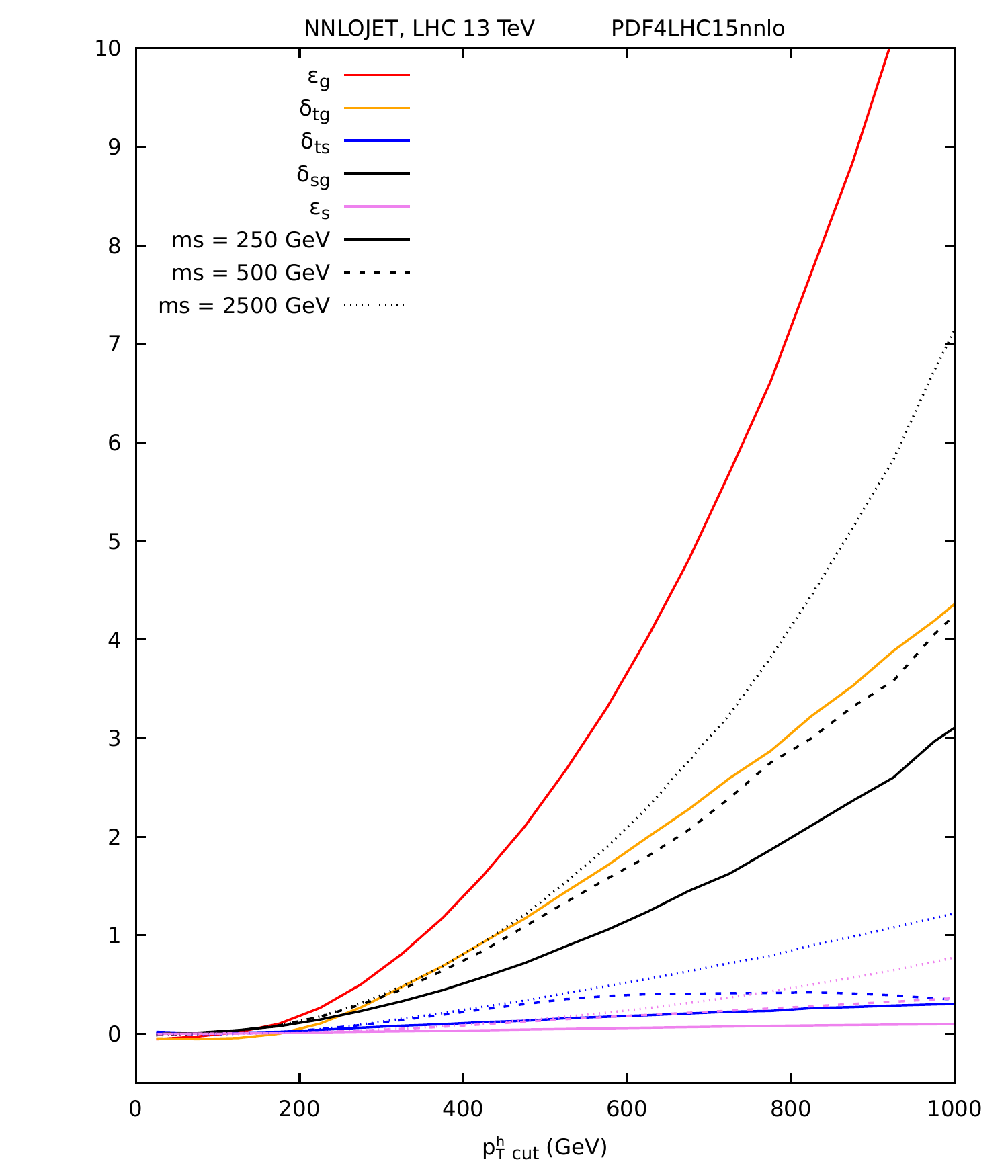}
    \end{subfigure}
(b)
    \begin{subfigure}[b]{0.45\textwidth}
\includegraphics[width=\textwidth]{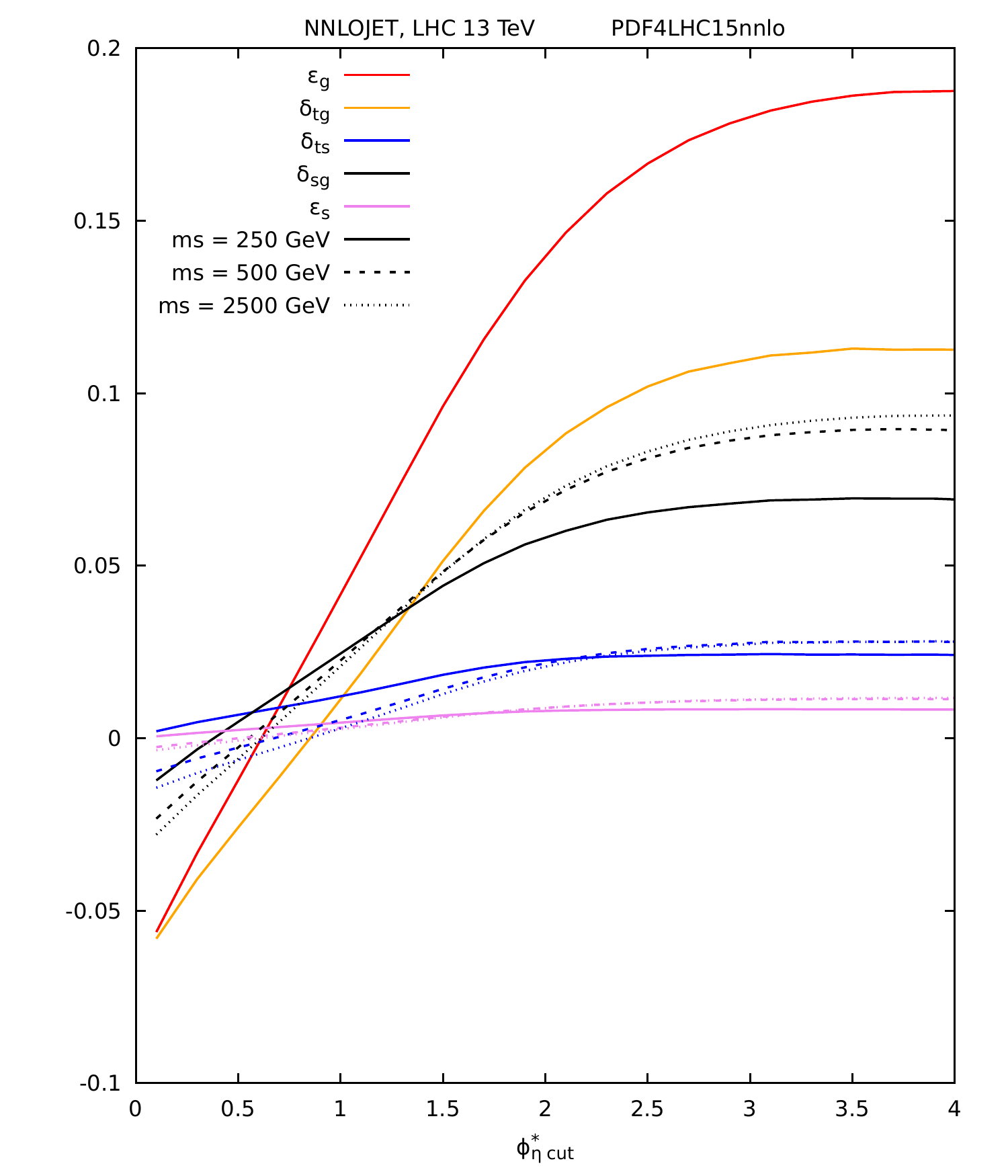}
    \end{subfigure}
\caption{Coefficients $\epsilon_i$ and $\delta_{ij}$ as a function of (a) $\pthcut$ and (b) $\phistarcut$ for $m_s$ = 250 GeV (solid), $m_s$ = 500 GeV (dashed) and $m_s$ = 2500 GeV (dotted).}
\label{fig:EpsilonDelta}
\end{figure}
For small values of \pthcut or \phistarcut, the coefficients $\epsilon_i$, $\delta_{ij}$ are all small. In this limit, we are simply probing the gross effect of the gg\h coupling and we recover the total cross section which is proportional to $\left({\ct} + \frac{1}{4}\cqt + \cg\right)^2$.
However, at larger values of \pthcut, the top-quark and scalar loops become resolved and the $\epsilon_i$ and $\delta_{ij}$ grow.  This is demonstrated in Fig.~\ref{fig:EpsilonDelta}(a) where the various coefficients are shown as a function of \pthcut for $m_s=250$ GeV, $m_s=500$ GeV and $m_s=2500$ GeV. In all cases, the value of the coefficient increases with $\pthcut$. 
As a consistency check, we observe that for a very large scalar quark mass, $m_s=2500$~GeV, the hierarchy predicted by absorbing the effect of the scalar into the effective interaction,
\begin{equation}
\delta_{sg} \sim \frac{1}{2} \epsilon_{g}, \qquad
\delta_{ts} \sim \frac{1}{4} \delta_{tg}, \qquad
\epsilon_{s} \sim \frac{1}{16} \epsilon_{g},
\end{equation}
is satisfied.
We also see that for smaller values of $m_s$, and particularly when $\pthcut$ is large enough to resolve the mass of the particle circulating the loop, the effects are smaller.

In contrast, as shown in Fig.~\ref{fig:EpsilonDelta}(b), the corresponding coefficients as a function of $\phistarcut$  are much smaller and are less sensitive to the masses of the new scalars or the contribution of the effective vertices.
In particular, increasing the value of $\phistarcut$ does not lead to larger values of these parameters.

This comparison demonstrates that the variable $\phistar$, which we have shown to have various theoretical and experimental merits over the more traditional variable $\pth$, fails to be a better variable for discriminating between new physics models. 

\subsection{MSSM: light stop scenario}
\label{sec:lightstops}
In supersymmetric models,  stops contribute to the gluon fusion process through loop diagrams and their effect depends mainly on their masses and mixing~\cite{Dawson:1996xz,Dermisek:2007fi}. It has been shown that even within the current limits on squark masses, the total cross section could be suppressed by a factor $\sim 15 \%$ (citation here) in the context of the Minimal Supersymmetric Standard Model (\MSSM). 
In this section we therefore carry out a numerical study of the impact of the new states predicted in the \MSSM on the Higgs boson $\pth$ and $\phistar$ distributions.

To illustrate the size of the possible effects, we consider the {\it light stop} \MSSM benchmark point proposed by the HXSWG~\cite{Carena:2013ytb}. 
This benchmark point features relatively light stops. However, the couplings of the squarks to the lightest higgs boson have opposite signs so that if both squarks have the same mass, they destructively interfere and give a null effect. Therefore, the stop masses $m_{\tilde{t}_1}$ and $m_{\tilde{t}_2}$ should be somewhat different to give rise to a sizeable effect. The parameters for the HXSWG light stop scenario are:
\begin{eqnarray}
& \mu=350 \text{ GeV}, \quad \tan\beta = 20, \quad M_3=1500  \text{ GeV},
\\
& M_A=600  \text{ GeV}, M_2= 500  \text{ GeV}, \quad M_{\tilde{l}_3}=1000  \text{ GeV} \\
&  A_t= X_t + \mu/\tan\beta; \quad X_t=2 \MSUSY,
\end{eqnarray}
%
where $\MSUSY$ is the SUSY scale, given by the mass terms of the third generation squarks;
\begin{eqnarray}
M_{\tilde{t}_{L}}=M_{\tilde{t}_{R}}=M_{\tilde{b}_{L}}=M_{\tilde{b}_{R}}= \MSUSY.
\end{eqnarray}
The stop masses are obtained by diagonalising the stop mixing matrix,
\begin{eqnarray}
\label{eq:stopmasses}
M_{\tilde{t}}^2 &=& \left ( 
\begin{array}{cc}
M_{L}^2 & \mt X_t \\
\mt X_t & M_{R}^2
\end{array}
\right)
\end{eqnarray}
where,
\begin{eqnarray}
\label{eq:stopL}
M_L^2 &=& M_{\tilde{t}_{L}}^2 + \mt^2 +\left(\frac{1}{2}-\frac{2}{3} \sin^2\theta_W \right) \cos(2\beta)  \mz^2\\
\label{eq:stopR}
M_R^2 &=& M_{\tilde{t}_{R}}^2 + \mt^2 +\frac{2}{3} \sin^2\theta_W \cos(2\beta) \mz^2
\end{eqnarray}
and $\sin^2\theta_W = 1 -\mw^2/\mz^2$.
The physical stop masses are given by,
\begin{eqnarray}
\left ( 
\begin{array}{cc}
m_{\tilde{t}_1}^2 & 0 \\
0 & m_{\tilde{t}_2}^2 \nn \\
\end{array}
\right)
&=& \left ( 
\begin{array}{cc}
\cos\theta_t & \sin\theta_t \\
-\sin\theta_t & \cos\theta_t \nn \\
\end{array}
\right)
M_{\tilde{t}}^2
\left ( 
\begin{array}{cc}
\cos\theta_t & -\sin\theta_t \\
\sin\theta_t & \cos\theta_t \nn \\
\end{array}
\right)
\end{eqnarray}
where
\begin{eqnarray}
\label{eq:tantheta}
\tan 2\theta_t = \frac{2m_t X_t}{M_L^2-M_R^2},
\end{eqnarray}
so that,
\begin{eqnarray}
\label{eq:masst1}
m_{\tilde{t}_1}^2 &=& \frac{M_L^2+M_R^2}{2} -
\frac{1}{2} \left( (M_L^2-M_R^2)^2 +4\mt^2 X_t^2 \right)^{1/2}, \\
\label{eq:masst2}
m_{\tilde{t}_2}^2 &=& \frac{M_L^2+M_R^2}{2} +
\frac{1}{2} \left( (M_L^2-M_R^2)^2 +4\mt^2 X_t^2 \right)^{1/2}.  
\end{eqnarray}

The couplings of the lightest Higgs boson to stops are given by,
\begin{eqnarray}
g_{h\tilde{t}_1\tilde{t}_1} &=&
\frac{\cos(\alpha)}{\sin(\beta)}
- \sin(\alpha+\beta) \left[\frac{1}{2}\cos(\theta_t)^2 - \frac{2}{3}\sin^2\theta_W \cos(2\theta_t)\right] \frac{\mz^2}{\mt^2}\nonumber \\
&&
-\frac{1}{2\mt}\sin(2\theta_t) \left[\frac{\cos(\alpha)}{\sin(\beta)} A_t +\frac{\sin(\alpha)}{\sin(\beta)}\mu\right],\\
g_{h\tilde{t}_2\tilde{t}_2} &=&
\frac{\cos(\alpha)}{\sin(\beta)}
- \frac{1}{2}\sin(\alpha+\beta) \left[\frac{1}{2}\sin(\theta_t)^2 +  \frac{2}{3}\sin^2\theta_W \cos(2\theta_t)\right] \frac{\mz^2}{\mt^2}\nonumber \\
&&
+\frac{1}{2\mt}\sin(2\theta_t) \left[\frac{\cos(\alpha)}{\sin(\beta)} A_t +\frac{\sin(\alpha)}{\sin(\beta)}\mu\right].
\end{eqnarray}
In the specific MSSM benchmark points considered here, we are in the \emph{decoupling limit} where the lightest CP-even state $h$ has SM couplings to the SM particles~\cite{Haber:1995be,Djouadi:1996pb,Djouadi:2005gj} and the mixing angle $\alpha$ is related to $\beta$ by,
\begin{equation}
\sin(\beta-\alpha) = 1.
\end{equation}
This produces a much simpler expression for the couplings, 
\begin{eqnarray}
\label{eq:kappat1}
g_{h\tilde{t}_1\tilde{t}_1} &=& 1 +\cos(2\beta) \left(\frac{1}{2}\cos(\theta_t)^2-\frac{2}{3} \sin^2\theta_W \cos(2\theta_t)\right)  \frac{\mz^2}{\mt^2} 
-\frac{1}{2}\sin(2\theta_t)\, \frac{X_t}{m_t}, \\
\label{eq:kappat2}
g_{h\tilde{t}_2\tilde{t}_2} &=& 1 +\cos(2\beta) \left(\frac{1}{2}\sin(\theta_t)^2+\frac{2}{3} \sin^2\theta_W \cos(2\theta_t)\right)  \frac{\mz^2}{\mt^2}
 +\frac{1}{2}\sin(2\theta_t)\, \frac{X_t}{m_t}.
\end{eqnarray}
These couplings are normalised to the Lagrangian interaction,
$$
{\cal L}_{\rm int} = i\frac{2 m_t^2}{v} g_{h\tilde{t}_i\tilde{t}_i}
$$
so that the relationship between the $g_{h\tilde{t}_i\tilde{t}_i}$ and $\kappa_{\tilde{t}_i}$ for $i=1,2$ is  
\begin{equation}
\kappa_{\tilde{t}_i} = \frac{\mt^2}{m_{\tilde{t}_i}^2} g_{h\tilde{t}_i\tilde{t}_i}.
\end{equation}

For the benchmark points, inserting the values for $\MSUSY$ and the default parameters of Section~\ref{sec:default} into Eqs.~\eqref{eq:masst1}--\eqref{eq:kappat2}, we find the input parameters relevant for Higgs production shown in Table~\ref{tab:SUSYparameters}.
\begin{table}[h]
\begin{center}
\begin{tabular}{c|c|c|c|c|c}
BP & \MSUSY ~(\GeV) & $m_{\tilde{t}_1}$ (\GeV) & $m_{\tilde{t}_2}$ (\GeV) & $g_{h\tilde{t}_1\tilde{t}_1}$ & $g_{h\tilde{t}_2\tilde{t}_2}$\\ \hline 
1 & 500 & 323 & 672 & -1.94 & 3.80 \\
2 & 600 & 423 & 773 & -2.52 & 4.38 \\
3 & 800 & 624 & 972 & -3.67 & 5.53\\ \hline
\end{tabular}
\caption{\label{tab:SUSYparameters} Input parameters for the three light stop MSSM benchmark points, BP1, BP2 and BP3.}
\end{center}
\end{table}

For these SUSY parameters, Eqs.~\eqref{eq:stopL}--\eqref{eq:kappat2} simplify considerably, and lead to;
\begin{eqnarray}
\theta_t &\sim &\frac{\pi}{4},\\
m_{\tilde{t}_1} &\sim& \MSUSY -\mt,\\
m_{\tilde{t}_2} &\sim& \MSUSY +\mt,\\
g_{h\tilde{t}_1\tilde{t}_1} &\sim& 1 - \frac{\MSUSY}{\mt},\\
g_{h\tilde{t}_2\tilde{t}_2} &\sim& 1 + \frac{\MSUSY}{\mt}.
\end{eqnarray}
Essentially this amounts to ignoring terms proportional to $\mz^2/\mt/\MSUSY$.  

Using the results of the previous section, we can also estimate the size of the stop contribution to the gluon fusion cross section.  Both stops make a contribution to the gluon fusion amplitude, $\delta M^{gg}$, where
\begin{eqnarray}
\delta M^{gg} / M^{gg}_{\SM} &=& \frac{\kappa_{\tilde{t}_1}}{4} + \frac{\kappa_{\tilde{t}_2}}{4} \nonumber \\
\label{eq:exactMSSMgg}&=& 
 \frac{\mt^2}{4m_{\tilde{t}_1}^2}g_{h\tilde{t}_1\tilde{t}_1} 
+\frac{\mt^2}{4m_{\tilde{t}_2}^2}g_{h\tilde{t}_2\tilde{t}_2}\\
\label{eq:approxMSSMgg}&\sim &
-\frac{\mt^2}{2 (\MSUSY^2-\mt^2)}.
\end{eqnarray}
For the benchmark points BP1, BP2 and BP3, the gluon fusion cross section is changed by 
-14.9\% (-13.3\%), -9.9\% (-9.0\%) and -5.3\% (-4.9\%) respectively.  The numbers in brackets are those obtained through use of the approximate formula~\eqref{eq:approxMSSMgg}. 
 
The sbottom masses are given by a similar equation to \eqref{eq:stopmasses}, but because $\mb << \MSUSY$, the two sbottom masses are both close to $\MSUSY$ and their contributions largely cancel.

\begin{figure}[ht]
\centering
(a)
    \begin{subfigure}[b]{0.45\textwidth}
\includegraphics[width=\textwidth]{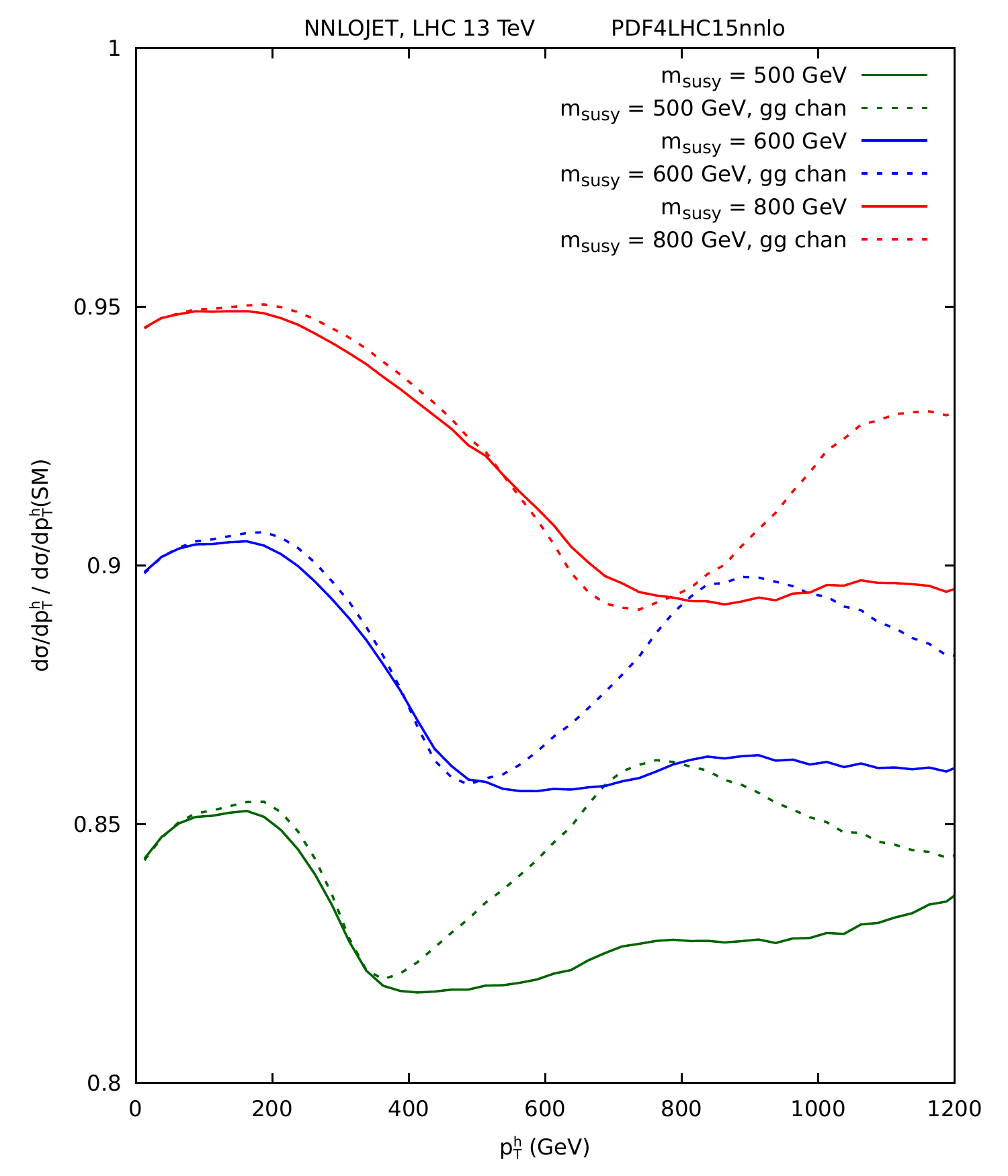}
    \end{subfigure}
(b)
    \begin{subfigure}[b]{0.45\textwidth}
\includegraphics[width=\textwidth]{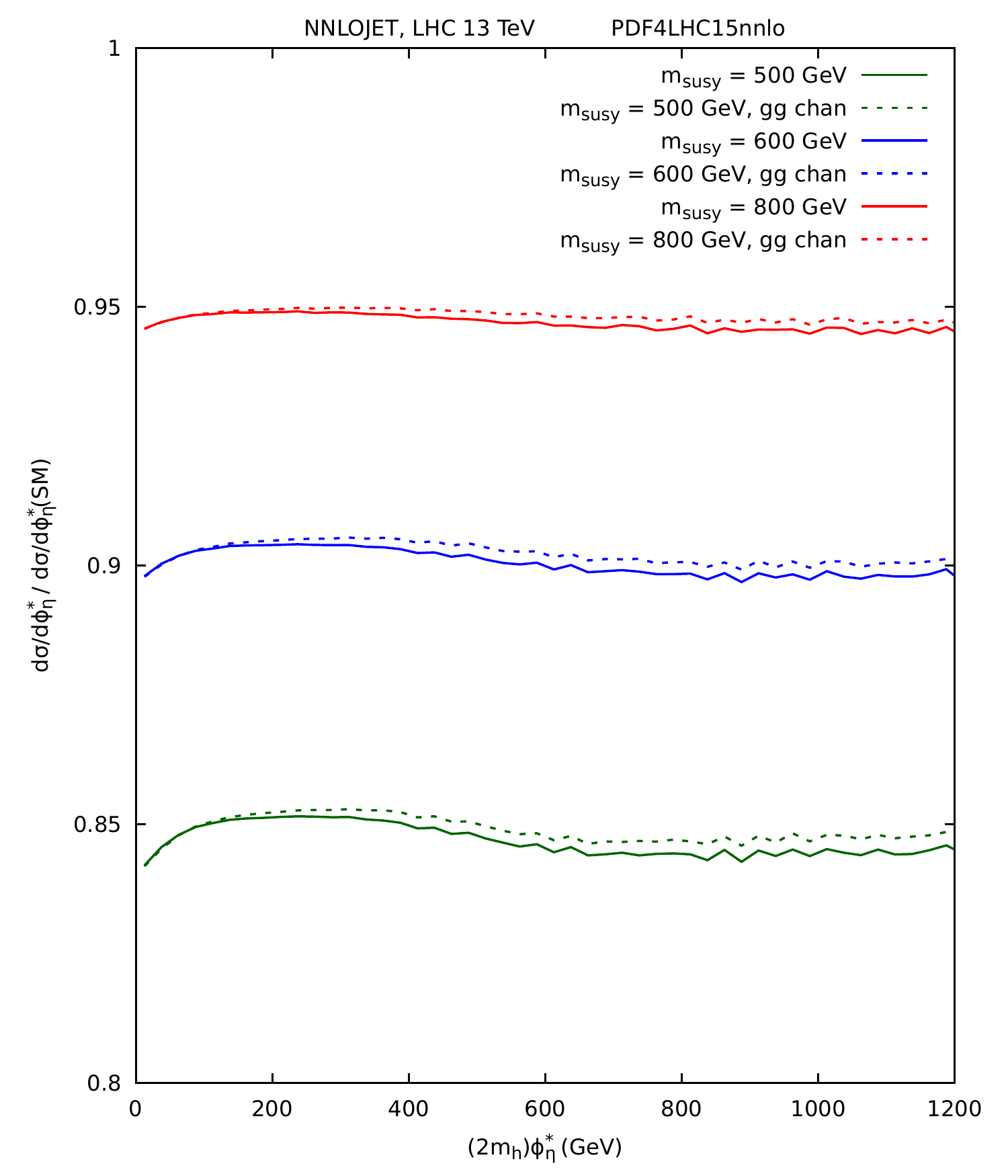}
    \end{subfigure}
\caption{\label{fig:MSSM13TeVdistributions} The ratio of the Higgs boson (a) \pth and (b) \phistar distributions at $\sqrt{s}$ = 13 TeV for the three benchmark points, $\MSUSY=500$~GeV (green), $\MSUSY=600$~GeV (blue) and $\MSUSY=800$~GeV (red) relative to the LO \SM prediction computed with top, bottom and charm quark loops. The dotted lines show the contributions coming from the gluon-gluon initial state.}

\end{figure}

The \pth and \phistar distributions for the three benchmark points characterised by different values of \MSUSY are shown in Fig.~\ref{fig:MSSM13TeVdistributions}.  These distributions are obtained with the NNLOJET code and the default input parameters as specified in Sec.~\ref{sec:default} and are normalised to the \LO\ \SM distribution with the exact mass dependence of the top-, botton- and charm-quark loops as shown in Figs.~\ref{fig:SMpth_13TeVcompare} and \ref{fig:SMphi_13TeVcompare}. The dominant gluon fusion contribution is shown as dashed lines.  
As expected from the earlier discussion, the stop quarks suppress the total cross section, by around 15\%, 10\%, and 5\% in  BP1, BP2 and BP3 respectively.  This is evident in the low \pth and low \phistar where most of the cross section is concentrated.\footnote{The small additional suppression is due to the charm and bottom quark loops.}

Figure~\ref{fig:MSSM13TeVdistributions}(a) shows that the \pth distribution is sensitive to the masses of the stop quarks.  As discussed in Section~\ref{sec:heavytoplimit}, for a heavy particle of mass $M$, the thresholds start to open when 
$$
\pth \sim M \left( 1-\frac{\mh^2}{4M^2}\right).
$$
The first peak for each of these plots occurs when $\pth$ is close to the top mass. Even though the cross section is normalised to the SM value, the effects of the top quark threshold is not completely washed away because of the interference between the top and squark loops. 
There is dip at around $m_{{\tilde t}_1} \sim \MSUSY-m_t$ where the first stop quark threshold starts to open, and a smaller peak at around $m_{{\tilde t}_2} \sim \MSUSY + m_t$ where the second stop quark threshold opens. These effects are much more evident in the gluon-gluon initiated channel than in the quark-gluon contribution. 

In complete contrast, and as expected from the discussion in the section~\ref{sec:BSMeffective}, the \phistar distribution shown in Figure~\ref{fig:MSSM13TeVdistributions}(b) is essentially insensitive to the heavy particle thresholds.

Even though Figure~\ref{fig:MSSM13TeVdistributions} is a naive leading order estimate which ignores higher order contribution that could significantly affect the cross sections, we expect that the sensitivity of the \pth distribution, and the corresponding insensitivity of the \phistar distribution will be preserved.   Therefore our recommendation is that:
\begin{itemize}
\item[a)] the \pth distribution is exploited for new physics searches
\item[b)] the \phistar distribution is used for precise measurements of the Higgs boson parameters.
\end{itemize}

\clearpage
\section{Summary}
\label{sec:summary}

In this part of the report, we proposed a new observable for the study of Higgs bosons that decay into 
two photons, \phistar.  This variable is complementary to the transverse momentum of 
the Higgs, since it relies on the measurement of the directions in $\phi$ and $\eta$ of 
the two photons rather than the energies.

When the Higgs boson is produced at rest in the transverse plane, i.e. $\pth = 0$, 
then \phistar also vanishes.  However, for finite values of $\pth$ and \phistar, the two variables are unrelated.

We made a rudimentary study of the experimental resolution for photon pairs coming from 
Higgs boson decays using a fast detector simulation of the CMS detector.   While a number 
of systematic effects, notably pileup, were ignored, our study showed that $2\mh \phistar$ 
has a better resolution compared to the simple calorimetric measurement 
of $p_{T}^{\gamma\gamma}$. This means that although there is no tracking information about the photons, \phistar is neverthless a viable observable and has the potential to give 
additional information on the production of boosted Higgs bosons.  

The \phistar distribution behaves very differently to the (well-studied) \pth distribution. 
In particular, as \phistar grows, it does not appear to resolve the short distance properties 
of the interaction producing the Higgs.  This is in complete contrast to the \pth distribution 
in which finite mass effects start to become important (typically for $\pth > \mt$ in the \SM, 
and for $\pth > M$ for new particles of mass $M$). 

The normalisation of the \phistar distribution is sensitive to the combination of particles that are involved in the $gg \to \h$ interaction (just as for the inclusive Higgs boson cross section), 
but the shape of the \phistar distribution is less sensitive to new physics effects.

On the other hand, this has the positive feature that the \phistar distribution can be more reliably calculated, 
even in the Higgs Effective Theory.  
We showed explicitly that different ways of estimating the top-quark mass dependence of the spectrum beyond 
Leading-Order  (the  \EFTplus and \EFTtimes approximations) produced essentially identical 
results to the \HEFT at \NLO and at \NNLO.  We also found that relative to \NLO, the \NNLO corrections are small,  and that the scale uncertainty is significantly reduced.  
We take this as a strong indication that the \NNLO\ \HEFT prediction for
the \phistar distribution is close to what might be found if the full top mass dependence would be available.  The \phistar distribution is therefore an ideal observable for extracting detailed information about the Higgs boson.

Armed with a new and precisely measurable observable, we anticipate that measurements of boosted Higgs bosons in the two photon decay channel could provide useful additional information about the interactions of the Higgs boson.

\clearpage

\appendix
\section{Calculation of the One-Loop Master Integrals}
\label{sec:oneloopappendix}

In this Appendix, we will demonstrate how to compute the one-loop master integrals that appeared in Sec.~\ref{sec:oneloopmasters}.  

\subsection{The canonical basis}
\label{sec:canonical}

As discussed in Sec.~\ref{sec:oneloopmasters}, the integral family defined by the four propagators of Eq.~\eqref{eq:boxpropagators},
\begin{align}
D_1 &= k_1^2 - \mt^2, \nn \\
D_2 &= (k_1-p_1)^2 - \mt^2, \nn \\
D_3 &= (k_1-p_1-p_2)^2 - \mt^2, \nn \\
D_4 &= (k_1-p_1-p_2-p_3)^2 - \mt^2,\nn
\end{align}
can be written in terms of the 9 master integrals $g_1 \ldots g_9$ described in Eq.~\eqref{eq:precanonical}.  
However, we are free to choose a convenient basis of master integrals, and the best choice depends on the method to be used to solve the integrals. 
Here we will use the method of differential equations since that is a powerful method for computing 
more complicated two-loop integrals. 
This method works by identifying the differential equations that each integral satisfies~\cite{art:KOTIKOV1991158, art:Remiddi:1997ny, art:Gehrmann:1999as, art:Gehrmann:2000xj} and solving the equations iteratively.

In this case, a good choice of basis for this method is the 
so-called {\it canonical} basis where the differential equations for each integral is in a very particular form discussed below. The question then is how to identify the canonical basis.  
The set of integrals given in Eq.~\eqref{eq:precanonical} is {\it not} the canonical basis.  
However, it is a suitable starting point for finding the canonical basis and is known as a {\it pre-canonical} basis.

Several methods are known for finding the canonical basis, see for example \cite{art:Gehrmann:2014bfa,art:Henn:2013pwa,art:Argeri:2014qva,art:Lee:2014ioa,art:Meyer:2016slj}. Here we will use a method inspired by Magnus and Dyson series as described in~\cite{art:Argeri:2014qva}. The 9 pre-canonical integrals obey a set of 4 partial differential equations in $ s \in \{s_{12}, s_{23}, p_4^2, \mt^2\}$,
\begin{equation}
\partial_s \ \vec{g} = B_s \ \vec{g}, \label{eq:differentialequation}
\end{equation}
where each of the $B_s$ are $9 \times 9$ matrices.
Additionally, the integrals obey a so-called {\it scaling} equation, 
\begin{equation}
\partial_{\varphi} \vec{g}  = B_{\varphi} \ \vec{g},
\end{equation}
where
\begin{equation}
\partial_{\varphi} = \left[ - \frac{\alpha}{2} + s_{12} \partial_{s_{12}} + s_{23} \partial_{s_{23}} + p_4^2 \partial_{p_4^2} \right].
\end{equation}
Here, $B_{\varphi}$ is another $9 \times 9$ matrix and $\alpha= md + 2s - 2r$ is the mass dimension of an $m$-loop integral with $r$ propagators in the denominators and $s$ propagators in the numerator. The ``scaling'' equation can be used as a consistency check of the partial differential equations in the other variables.

A differential equation is said to be in canonical form if it can be written as
\begin{equation}
\mathrm{d} \vec{f} = \epsilon (\mathrm{d} \tilde{A}) \vec{f}. \label{eq:canonicalform}
\end{equation}
The matrix $(\mathrm{d} \tilde{A})$ should be in so-called ``$\mathrm{d} \log$ form'', this means that it should be possible to write each of its elements as a sum of derivatives of logarithms with rational number coefficients. 

To obtain the canonical form for our differential equation, our goal will be to find a matrix $T$ with entries consisting of algebraic functions of the invariants, masses and $d$ such that $\vec{f} = T \vec{g}$ obeys \eqref{eq:canonicalform}.
In order to achieve this it is useful to understand how the differential equation transforms under a basis change.
Inserting $\vec{f} = T \vec{g}$ into~\eqref{eq:differentialequation} and using the product rule, we obtain the basis change formula,
\begin{equation}
\partial \vec{f} = T^{-1} \left( BT - \partial(T) \right) \vec{f}. \label{eq:basischange}
\end{equation}

\subsection{Solving the differential equations}
\label{sec:solveDEs}

It is often useful to write the differential equations in terms of dimensionless variables. To do this we can factor a dimensionful parameter, which we choose as $\mt^2$, from each integral in order to make it dimensionless.  
The integrals are then functions only of three dimensionless invariants, which we choose as
\begin{align}
&x_1 = \frac{s_{12}}{\mt^2}, &
&x_2 = \frac{p_4^2}{\mt^2}, &
&x_3 = \frac{s_{23}}{\mt^2}. &
\end{align}
The 3 partial derivatives with respect to the dimensionful variables $s_{12}, s_{23}, p_4^2$ are related to 3 partial derivatives with respect to the dimensionless invariants $x_1,x_2,x_3$ via a Jacobian matrix, 
\begin{equation}
\begin{pmatrix} \partial_{x_1}  \\ \partial_{x_2} \\ \partial_{x_3}  \end{pmatrix}
=
\begin{pmatrix} 
\frac{\partial x_1}{\partial s_{12}}  & \frac{\partial x_2}{\partial s_{12}} & \frac{\partial x_3}{\partial s_{12}} \\
\frac{\partial x_1}{\partial p_4^2}  & \frac{\partial x_2}{\partial p_4^2} & \frac{\partial x_3}{\partial p_4^2} \\
\frac{\partial x_1}{\partial s_{23}}  & \frac{\partial x_2}{\partial s_{23}} & \frac{\partial x_3}{\partial s_{23}} \\
\end{pmatrix}^{-1}
\begin{pmatrix} \partial_{s_{12}}  \\ \partial_{p_4^2} \\ \partial_{s_{23}}   \end{pmatrix}.
\end{equation}
Using this we first exchange our partial differential equations with respect to dimensionful variables for those in terms of dimensionless variables.
The scaling equations provide only information on the mass dimension of the integrals and can be used to determine the power of $\mt^2$ to be factored from each integral. The 9 pre-canonical integrals now obey 3 partial differential equations in $x \in \{ x_1, x_2, x_3\}$,
\begin{equation}
\partial_x \vec{g} = B_x \vec{g},
\end{equation}
with $B_x$ a $9 \times 9$ matrix for each variable, their entries consist of rational functions of the dimensionless invariants, the mass $\mt^2$ and $d$. 

We first try to rescale each of our integrals by $\epsilon^{a_i}$ with $a_i \in \mathbb{Z}$ such that the off-diagonal terms of our matrix have leading term $\epsilon$. The differential equation in terms of the new integrals can be obtained by using the basis change formula~\eqref{eq:basischange} with
\begin{equation}
T= \mathrm{diag}(\epsilon^{a_1},\ldots,\epsilon^{a_9}). \label{eq:epsilonrescale}
\end{equation}
In the case at hand we find that choosing $a_1,a_2,a_3,a_4 = 0$ and $a_5, a_6, a_7, a_8, a_9 = 1$ allows us to obtain a differential equation in the desired form.

Next, we note that each of our $B_x$ matrices depend linearly on $\epsilon$, 
\begin{equation}
B_x(\vec{x},\epsilon) = B_x^{(0)}(\vec{x}) + \epsilon B_x^{(1)}(\vec{x}).
\end{equation}
This simplifies finding the canonical basis. Cases without this property are also discussed in~\cite{art:Argeri:2014qva}. 
We change the basis of integrals via Magnus series $\Omega[B_x^{(0)}]$ obtained using $B_x^{(0)}$ as kernel,
\begin{equation}
\vec{f}(\vec{x},\epsilon) = e^{\Omega[B_x^{(0)}]} \vec{g}(\vec{x},\epsilon).
\end{equation}
Due to rescaling the integrals by $\epsilon$ in~\eqref{eq:epsilonrescale} the matrices $B_x^{(0)}$ are diagonal and the Magnus series is truncated at first order,
\begin{equation}
\Omega_1[B_x^{(0)}] = \int_{x_0}^{x} \mathrm{d} \tau_1 \ B_x^{(0)}(\tau_1).
\end{equation}
Repeating this basis change for each variable we finally obtain an $\epsilon$ factorised form for each of the partial derivative matrices $B_x$. The canonical form of the differential equation can be constructed from these transformed matrices.

The relation between the canonical master integrals, $\vec{f}$, and the pre-canonical master integrals, $\vec{g}$, is given by taking the product of all basis change matrices, the result is
\begin{align}
f_1 &= g_1, \\
f_2 &= \mt^2 \sqrt{4-x_1} \sqrt{x_1} \  g_2, \\
f_3 &= \mt^2 \sqrt{4-x_3} \sqrt{x_3} \  g_3, \\
f_4 &= \mt^2 \sqrt{4-x_2} \sqrt{x_2} \ g_4, \\
f_5 &= \epsilon \mt^2 x_1 \ g_5, \\
f_6 &= \epsilon \mt^2 x_3 \ g_6, \\
f_7 &= \epsilon \mt^2 (x_2-x_1) \ g_7, \\
f_8 &= \epsilon \mt^2 (x_2-x_3) \ g_8 \\
f_9 &= \epsilon \mt^4 \sqrt{4 x_2-4 x_1- 4 x_3 + x_1 x_3} \sqrt{x_1} \sqrt{x_3} \ g_9.
\end{align}


In the canonical form, the differential equation system may be written as
\begin{align}
d \bar{f} &= (d \tilde{A}) \bar{f},
\end{align}
where the coupling matrix $\tilde{A}$ is
\begin{align}
\tilde{A} &= \epsilon \left[ \begin{tabular}{ccccccccc}
0 & 0 & 0 & 0 & 0 & 0 & 0 & 0 & 0 \\
$E_2(x_1)$ & $E_1(x_1)$ & 0 & 0 & 0 & 0 & 0 & 0 & 0 \\
$E_2(x_3)$ & 0 & $E_1(x_3)$ & 0 & 0 & 0 & 0 & 0 & 0 \\
$E_2(x_2)$ & 0 & 0 & $E_1(x_2)$ & 0 & 0 & 0 & 0 & 0 \\
0 & $E_2(x_1)$ & 0 & 0 & 0 & 0 & 0 & 0 & 0 \\
0 & 0 & $E_2(x_3)$ & 0 & 0 & 0 & 0 & 0 & 0 \\
0 & $E_2(x_1)$ & 0 & $-E_2(x_2)$ & 0 & 0 & 0 & 0 & 0 \\
0 & 0 & $E_2(x_3)$ & $-E_2(x_2)$ & 0 & 0 & 0 & 0 & 0 \\
0 & $E_3(x_1,x_3)$ & $E_3(x_3,x_1)$ & $E_4$ & $E_5$ & $E_5$ & $E_5$ & $E_5$ & $E_6$
\end{tabular} \right]
\end{align}

and where the entries of the matrix are given as

\begin{align}
\label{eq:atildeol}
E_1(x_i) &= - \log(4 - x_i) \nn \\
E_2(x_i) &= - 2 \log \Big( \sqrt{4 - x_i} + \sqrt{-x_i} \Big) \nn \\
E_3(x_i,x_j) &= 2 \log( x_2 - x_1) + 2 \log( x_2 - x_3) \nn \\
& \;\; - 4 \log \Big( \sqrt{4 - x_i} \sqrt{-x_j} + \sqrt{4 (x_2 - x_1 - x_3) + x_1 x_3} \Big) \nn \\
E_4 &= - 2 \log(x_2-x_1) - 2 \log(x_2-x_3) \nn \\
& \;\; + 4 \log \Big( \sqrt{4 - x_2} \sqrt{-x_1} \sqrt{-x_3} + \sqrt{-x_2} \sqrt{4 (x_2 - x_1 - x_3) + x_1 x_3} \Big) \nn \\
E_5 &= - \log(x_2-x_1-x_3) + 2 \log \Big( \sqrt{-x_1} \sqrt{-x_3} + \sqrt{4 (x_2 - x_1 - x_3) + x_1 x_3} \Big) \nn \\
E_6 &= \log(x_2-x_1-x_3) - \log \big(4 (x_2-x_1-x_3) + x_1 x_3 \big)
\end{align}

In order to integrate the system of differential equations, we have to decide on a boundary condition. In the phase-space point $x_1=x_2=x_3=0$ (corresponding to $s_{12}=s_{23}=\ppf=0, \mtsq =1$), the Feynman integrals themselves are finite\footnote{this can be seen e.g. from the Schwinger parametrization.} and as all the members of the canonical basis except $f_1$ have prefactors that are zero in that point, they all vanish with the exception of $f_1$ - the tadpole integral - which is given as
\begin{align}
f_1|_{x_1=x_2=x_3=0} = 1 + \frac{\pi^2}{12} \eps^2 + \kur{O}(\eps^3)
\end{align}

As all the members of the canonical basis have uniform weight\footnote{Uniform weight means that the $n$th term in the $\eps$ expansion consists solely of functions with weight $n$, as explained in appendix \ref{sec:GPLs}, and additionally that the prefactors of these functions have to given solely in terms of rational numbers, so without any kinematical dependence.} we know that the $\eps^0$ term is purely numerical, and thus we deduce from the boundary point that all the members of the canonical basis are $0$ at the $\eps^0$ order, with the exception of $f_1$. To find the expressions at the next order in $\eps$ we will use the differential equation. In order to simplify the solution procedure we will make a change of variables from $x_i$ to the new dimensionless variables $z_i$, known as Landau variables, defined as
\begin{align}
x_i = - \frac{(z_i-1)^2}{z_i}.
\end{align}
These variables have the property of making some of the square-roots present in Eq.~\eqref{eq:atildeol} factorize, as\footnote{One might worry about whether the right hand side of Eq.~\eqref{eq:squarerootsplit} contains a factor of $i \pi$ from the branch cut of the logarithm. But any such constants can be discarded as the matrix $\tilde{A}$ only contributes through its derivative.}
\begin{align}
- 2 \log \Big( \sqrt{4 - x_i} + \sqrt{-x_i} \Big) = - 2 \log(2) - \log(z_i).
\label{eq:squarerootsplit}
\end{align}
Inserting these variables and doing the derivative with respect to $z_1$, we get the differential equation
\begin{align}
\frac{d f_2^{(1)}}{d z_1} = -\frac{1}{z_1},
\end{align}
while the derivative of the first-order terms of the remaining canonical master integrals with respect to $z_1$ is zero.
Integrating the system up, gives $f_2^{(1)} = -\log(z_1) + c(z_2,z_3)$, and as the derivative with respect to the remaining variables vanishes along with the value at the boundary point, we conclude that $c(z_2,z_3)=0$, so after substituting back, we get the result
\begin{align}
f_2^{(1)} &= - 2 \log \Big( \sqrt{4 - x_1} + \sqrt{-x_1} \Big) + 2 \log(2).
\end{align}

Following the same procedure with $z_2$ and $z_3$, gives the expressions
\begin{align}
f_3^{(1)} &= - 2 \log \Big( \sqrt{4 - x_3} + \sqrt{-x_3} \Big) + 2 \log(2), \nn \\
f_4^{(1)} &= - 2 \log \Big( \sqrt{4 - x_2} + \sqrt{-x_2} \Big) + 2 \log(2),
\end{align}
while the remaining canonical masters give zero, also at the first order in $\eps$.

Continuing to the second order in $\eps$, the derivatives with respect to $z_1$ become
\begin{align}
\frac{d f_2^{(2)}}{d z_1} &= - \Big( \frac{1}{z_1} - \frac{2}{z_1+1} \Big) \log(z_1), \nn \\
\frac{d f_5^{(2)}}{d z_1} &= \frac{1}{z_1} \log(z_1) ,\nn \\
\frac{d f_7^{(2)}}{d z_1} &= \frac{1}{z_1} \log(z_1),
\end{align}
with all the remaining $z_1$ derivatives except the last, that of the box, being zero.

Integrating up gives
\begin{align}
f_2^{(2)} &= \Li_2(-z_1) - \tfrac{1}{2} \log^2(z_1) + \log(z_1) \log(1+z_1) + c_2(z_2,z_3), \nn \\
f_5^{(2)} &= \tfrac{1}{2} \log^2(z_1) + c_5(z_2,z_3), \nn \\
f_7^{(2)} &= \tfrac{1}{2} \log^2(z_1) + c_7(z_2,z_3),
\end{align}
in terms of $z_1$ and the remaining integration constant.

It may at this point be worth mentioning that while in this specific case the integrals could be performed rather easily, in general that is not the case. The function class known as generalized polylogarithms (with is discussed in further detail in Appendix~\ref{sec:GPLs}) can be of great help with this, as the way it is definited identifies it as a natural way of expressing such integrals. It is however not needed for this case, and will not be mentioned any further.

As the derivatives of $f_2^{(2)}$ and $f_5^{(2)}$ with respect to the remaining variables are zero, we deduce that $c_2$ and $c_5$ are genuine constants, and then the boundary point allow us identify $c_2 = -\Li_2(-1) = \pi^2/12$ and $c_5=0$. For $c_7$ on the other hand, we get that
\begin{align}
\frac{d c_7}{d z_2} & = \frac{d f_7^{(2)}}{d z_2} = -\frac{1}{z_2} \log(z_2) \Leftrightarrow \nn \\
c_7 &= - \tfrac{1}{2} \log^2( z_2 ) + b(z_3)
\end{align}
and following the same procedure once again, we conclude that $b(z_3)=0$.

Going through the same procedure with the remaining canonical master integrals in the remaining variables, and substituting the original dimensionless variables, the $x_i$, back, we get in total
\begin{align}
f_1 &= 1 + \tfrac{\pi^2}{12} \eps^2 + \kur{O}(\eps^3) ,  \\
f_2 &= H_1(x_1) \eps + H_2(x_1) \eps^2 + \kur{O}(\eps^3) ,  \\
f_3 &= H_1(x_3) \eps + H_2(x_3) \eps^2 + \kur{O}(\eps^3) ,  \\
f_4 &= H_1(x_2) \eps + H_2(x_2) \eps^2 + \kur{O}(\eps^3) ,  \\
f_5 &= H_3(x_1) \eps^2 + \kur{O}(\eps^3),  \\
f_6 &= H_3(x_3) \eps^2 + \kur{O}(\eps^3) ,  \\
f_7 &= H_4(x_1,x_2) \eps^2 + \kur{O}(\eps^3) ,\\
f_8 &= H_4(x_3,x_2) \eps^2 + \kur{O}(\eps^3),  
\end{align}
with
\begin{align}
H_1(x_i) &= 2 \Big( \log(2) - \log \! \big( \sqrt{4 - x_i} + \sqrt{-x_i} \big) \Big) ,\nn \\
H_2(x_i) &= 2 \Big( \log ( 4 - x_i ) + \log \! \big( \sqrt{4 - x_i} - \sqrt{-x_i} \big) - \log(2) \Big) ,\nn \\
& \;\; \times \Big( \log \! \big( \sqrt{4 - x_i} - \sqrt{-x_i} \big) - \log(2) \Big) + 2 \, \Li_2 \! \bigg( \frac{\big( \sqrt{4 - x_i} - \sqrt{-x_i} \big)^2}{-4} \bigg) + \frac{\pi^2}{6} ,\nn \\
H_3(x_i) &= \tfrac{1}{2} H_1(x_i)^2 ,\nn \\
H_4(x_i,x_2) &= H_3(x_i) - H_3(x_2).
\end{align}

Only the box integral $f_9$ remains to be calculated. The transformation to Landau variables does not factorize the remaining square root, the $\sqrt{4 (x_2 - x_1 - x_3) + x_1 x_3}$. This makes this case a good illustration of an alternative method of solving a canonical system of differential equations - that of ``symbols''. For introductions to {\it symbols}, see Refs.~\cite{art:Goncharov:2010jf, art:Duhr:2012po} or Appendix~\ref{sec:symbols} where the subject is introduced more thoroughly than it will be here. The main idea is to associate a certain algebraic object - the symbol - to the various functions (logarithms and polylogarithms) that may appear in the result. Additionally such a {\it symbol} may be associated with the differential equation, and the procedure then consists of making a linear fit of the set of posible functions, to the differential equation. The result of such a fit, will be very close to the desired solution, and the difference can, in this case, be found solely from the differential equation along with a numerical fits.

Following this method, the result for the box $f_9$ may be expressed as
\begin{align}
\frac{f_9}{\eps^2} &= -4 \log(\alpha_{1,1})^2+4 \log(\alpha_{1,2})^2+\log(-\alpha_{u})^2-4 \log(\alpha_{2}) \log(x_2-x_1) \nn \\
& -4 \log(\alpha_{2}) \log(x_2-x_3)-2 \log(x_1) \log(x_2-x_3)+\log(x_2-x_3)^2 \nn \\
& +4 \log(\alpha_{1,2}) \big( -2 \log(\alpha_{4})+\log(x_2-x_1)+\log(x_2-x_3) \big) \nn \\
& + \log(\alpha_{1,1}) \big( 8 \log(\alpha_{3,1})-4 \log(-x_3) \big) + 4 \log(\alpha_{2}) \log(-x_3) + \log(-x_3)^2 \nn \\
& + 2 \log(-\alpha_{u}) \big( -\log(x_1)+\log(x_2-x_1)+\log(x_2-x_3)-2 \log(-x_3) \big) \nn \\
& + 2 \bigg[ \pi^2+ 4 i \pi \log(2)+2 \log(2)^2-8 \log(2) \log(\alpha_{1,3})+4 \log(\alpha_{1,3})^2 \nn \\
& + 4 \log(2) \log(\alpha_{3,3})-2 \log(\alpha_{3,3})^2+4 \log(2) \log(\alpha_{4})-4 \log(2) \log(x_1) \nn \\
& + 2 \log(\alpha_{2}) \log(x_1)+2 \log(\alpha_{3,3}) \log(x_1) -4 \log(2) \log(\alpha_{3,1})\nn \\
& + i \pi \Big(-2 \log(\alpha_{2})-2 \log(\alpha_{3,3})+\log(x_1)-\log(x_2-x_1)+2 \log(-x_3) \Big) \\
& - 2 \Li_2 \bigg( \frac{-4}{\alpha_{1,1}^2} \bigg) + 2 \Li_2 \bigg( \frac{-4}{\alpha_{1,3}^2} \bigg) - 2 \Li_2 \bigg( \frac{4 \alpha_{u}}{\alpha_{2}^2} \bigg) + \Li_2 \bigg( \frac{16 \alpha_{u}}{\alpha_{1,2}^2 \alpha_{2}^2} \bigg) + \Li_2 \bigg( \frac{\alpha_{1,2}^2 \alpha_{u}}{\alpha_{2}^2} \bigg) \nn \\
& - 4 \Li_2 \bigg( \frac{\alpha_{1,3} \sqrt{-x_1}}{\alpha_{3,3}} \bigg) - 2 \Li_2 \bigg( \frac{\alpha_{1,1}^2 (x_1-x_2)}{\alpha_{3,1}^2} \bigg) + \Li_2 \bigg( \frac{x_2-x_1}{\alpha_{u}} \bigg) + 2 \Li_2 \bigg( \frac{4 (x_2-x_1)}{\alpha_{3,1}^2} \bigg) \nn \\
& + \Li_2 \bigg( \frac{x_1}{x_2-x_3} \bigg) - 2 \Li_2 \bigg( \frac{4 (x_2-x_3)}{\alpha_{3,3}^2} \bigg) + 4 \Li_2 \bigg( \frac{\alpha_{1,1} (x_1-x_2)}{\alpha_{3,1} \sqrt{-x_3}} \bigg) + 2 \Li_2 \bigg( \frac{\alpha_{1,3}^2 (x_3-x_2)}{\alpha_{3,3}^2} \bigg) \bigg] , \nn
\end{align}
where we have used the abbreviations
\begin{align}
\alpha_u &\equiv x_2 - x_1 - x_3 ,\nn \\
\alpha_{1,i} &\equiv \sqrt{4-x_i} + \sqrt{-x_i} ,\nn \\
\alpha_2 &\equiv \sqrt{-x_1} \sqrt{-x_3} + \sqrt{4(x_2-x_1-x_3) + x_1 x_3} ,\nn \\
\alpha_{3,1} &\equiv \sqrt{4-x_1} \sqrt{-x_3} + \sqrt{4(x_2-x_1-x_3) + x_1 x_3} ,\nn \\
\alpha_{3,3} &\equiv \sqrt{-x_1} \sqrt{4-x_3} + \sqrt{4(x_2-x_1-x_3) + x_1 x_3} ,\nn \\
\alpha_{4} &\equiv \sqrt{-x_1} \sqrt{-x_3} \sqrt{4-x_2} + \sqrt{-x_2} \sqrt{4(x_2-x_1-x_3) + x_1 x_3}.\nn \\
\end{align}

\clearpage
\section{Two-loop Feynman Integrals \label{sec:twoloop}}

In this appendix we will lay out the steps needed to compute the two-loop contributions to Higgs boson $+$ jet production.  We will follow closely the approach used in Ref.~\cite{art:Bonciani:2016qxi}.


We begin with a description of how to get expressions for the Feynman diagrams or the numerators thereof. Then follows a discussion of the reduction of the Feynman integrals to a minimal set - the master integrals. Next we discuss how to evaluate those integrals using the differential equation method enhanced by the use of canonical bases and symbols, in order to obtain an expression in terms of generalized polylogarithms, which at the two-loop level are reducible to $\Li_n$ and $\Litt$. We conclude with a discussion of cases where these approaches are insufficient and bigger classes of functions, such as elliptic integrals, are needed.

\subsection{Extracting the numerator}

In general we can write any loop-level amplitude as
\begin{align}
A_{L-\text{loop}} = \sum_{i \in \text{diagrams}}  \int \frac{\prod_l^L \id^d k_l}{(i \pi^{d/2})^{L}} \frac{N_i(\{k\})}{\prod_{j \in i} D_j(\{k\})}
\label{eq:lloopdiagrams}
\end{align}
where as usual $D_j$ denotes the denominators stemming from the propagators of the Feynman diagrams, i.e. terms of the form $(k+p)^2-m^2$, while $N_i$ contains the rest of the terms in the Feynman diagram, i.e all gamma matrices, colour factors, polarization vectors, spinors, and all other factors provided by the Feynman rules.

As discused earlier, the Feynman rules from which to generate Eq.~\eqref{eq:lloopdiagrams} have been implemented in several codes~\cite{art:Shtabovenko:2016sxi, art:Hahn:1998yk, art:Hahn:2000kx, art:Nogueira:1991ex}. In general, the numerators of the diagrams will contain long strings and traces of (Dirac) gamma matrices, from external fermions and fermions in loops respectively, along with products of the colour factors $T^a_{ij}$ and $f_{abc}$ from vertices containing coloured particles.

Several computer implementations of the algebra enabling the reduction of such factors exist\cite{art:Hahn:1998yk, art:Vermaseren:2000nd}, the authors of ref. \cite{art:Bonciani:2016qxi} used a private implementation in FORM. The bottle-neck for such reductions are the gamma Matrix identities, the trace of a product of $2n$ gamma matrices has in general $(2n-1)!!$ terms, so finding ways to minimize that number can save significant computational time.

The are other methods which may be used to minimize the work which has to be done in order to extract the numerator of Eq.~\eqref{eq:lloopdiagrams}, though such methods were not found necessary for the calculations in ref. \cite{art:Bonciani:2016qxi}. Without going into detail, the numerators of diagrams consisting purely of three-point vertices will be identical to products of three-point tree-level amplitudes, except that the propagating particles are off shell. So, by mathematically putting these particles on shell with a procedure known as {\em generalized unitarity cuts}~\cite{art:Cutkosky:1960spc, art:Britto:2004nc}, the coefficients of those diagrams can be extracted solely from tree-level diagrams without the need for performing any kind of gamma algebra. The coefficients of the remaining diagrams may be extracted by similar methods using the knowledge of the three-point vertex diagrams as {\em subtraction terms}. This method, known as the {\em OPP method}~\cite{art:Ossola:2006us}\cite{art:Forde:2007mi, art:Badger:2008cm, art:Giele:2008ve}, have been used for the complete automation of one-loop-calculations~\cite{art:Ossola:2007ax, art:Berger:2008sj, art:Ellis:2008qc, art:Mastrolia:2010nb, art:Cullen:2011ac, art:Bevilacqua:2011xh, art:Badger:2012pg, art:Cullen:2014yla}, and similar method may be extended to higher loops\cite{art:Mastrolia:2011pr, art:Kosower:2011ty, art:Badger:2012dp, art:Badger:2012dv, art:Zhang:2012ce, art:Kleiss:2012yv, art:CaronHuot:2012ab, art:Badger:2013gxa, art:Sogaard:2013fpa, art:Badger:2015lda, art:Dunbar:2016aux, art:Badger:2016ozq}.

\subsection{Integrand reduction \label{sec:IntegrandReduction}}

After applying projectors (as described in section~\ref{sec:tensor_dec}), we can express the matrix element as $M \define M^{\mu \nu \tau} \kur{P}_{\mu \nu \tau}$, where
\begin{align}
M &= \sum_{i \in \text{diagrams}}  \int \frac{\prod_l^L \id^d k_l}{(i \pi^{d/2})^{L}} \frac{N_i(\{k\})}{\prod_{j \in i} D_j(\{k\})},
\label{eq:projectedintegral}
\end{align}
as in Eq.~\eqref{eq:lloopdiagrams}. After performing all gamma- and colour algebra, the $k$ dependence of the numerators $N_i$ will be solely true scalar products of the form $k_i \cdot k_j$, $k_i \cdot p_j$, or $k \cdot \omega$ where the vector $\omega$ is defined to be perpendicular to all the physical momenta $\omega^{\mu} \propto \varepsilon^{\nu_1 \nu_2 \nu_3 \mu} {p_1}_{\nu_1} {p_2}_{\nu_2} {p_3}_{\nu_3}$.

The renormalizability of the theory imposes that maximally four powers of each loop-momentum is allowed to appear in the numerators. That constraint limits the number of different Feynman integrals which have to be performed to a large but finite number, but it is still clearly desirable to find a way to minimize that number.

One way of doing so is called {\em Integrand Reduction}. This first consist of re-expressing the amplitude as a sum over {\em topologies} defined as distinct sets of propagators
\begin{align}
M &= \sum_{i \in \text{topologies}}  \int \frac{\prod_l^L \id^d k_l}{(i \pi^{d/2})^{L}} \frac{\Delta_i(\{k\})}{\prod_{j \in i} D_j(\{k\})},
\end{align}
where $\Delta$ denotes the irreducible numerator, which means that all terms in $N$ which are proportional to some of the propagators $D$ have been ``cancelled out'' leaving only the irreducible part $\Delta$. A simple example would be a term in the denominator of Eq.~\eqref{eq:projectedintegral} which contains the factor $k_i \cdot p_j$ while the denominator contains e.g. the propagators $k_i^2$ and $(k_i-p_j)^2 - m^2$. We might then perform a form of partial fractioning
\begin{align}
C \frac{k \cdot p}{(k^2) \, ((k-p)^2 - m^2)} &= \frac{C}{2} \left( \frac{p^2-m^2}{(k^2) \, ((k-p)^2 - m^2)} - \frac{1}{k^2} + \frac{1}{(k-p)^2 - m^2} \right)
\end{align}
showing how that term is reducible to an integral with a constant numerator plus terms in lower topologies. For multi-loop integrals performing the integrand reduction solely using ``partial fractioning'' is not always practical, one will miss relations unless one is very careful. There are ways to systematize this reduction which ensure that all relations are captured. These methods utilize insights from the mathematical field of algebraic geometry, and this is not the place to describe these developments. See for instance Refs.~\cite{art:Mastrolia:2012an, art:Zhang:2012ce, art:Badger:2013gxa}.

Integrand reduction is able to significantly reduce the number of integrals needed to evaluate an amplitude. Yet the number remains uncomfortably large, for the propagator structure corresponding to the first diagram of Fig.~\ref{fig:feynmantwoloop}, $84$ non-zero terms remain after the reduction. To reduce the number further, we will need to go beyond integrand-level relations to relations that hold only at the integral-level - the so-called IBP identities which are described in section~\ref{sec:IBPs} in the main text.

\subsection{The differential equation method}

After reducing the set of Feynman integrals that needs to be solved to a minimal set, we can no longer postpone discussing methods for solving them. Traditional text-book methods such as Feynman parametrization are rarely sufficient beyond one-loop as they themselves yield multi-dimensional integrals that are hard or impossible to solve. The method that will be described here is that of differential equations\cite{art:KOTIKOV1991158, art:Remiddi:1997ny, art:Gehrmann:1999as, art:Gehrmann:2000xj, art:Papadopoulos:2014lla} which is one of the main methods being used for current Feynman integral computations.

In its traditional version, the differential equation method consists of relating the derivative (with respect to a kinematic variable) of the Feynman integral which one wants to solve, with the Feynman integral itself along with Feynman integrals of similar or lower complexity. Doing so allows one to integrate the differential equation using traditional methods for first-order differential equations, usually giving in a result in terms of gamma functions, hyper-geometric functions, or some generalization there-of.

Yet for more involved Feynman integrals, there may be more than one master integral with a given topology. That is the case for the double-box calculated in ref. \cite{art:Gehrmann:2000xj}. For that case the differential equations for the two double-box topologies will couple, giving a second order differential equation for each of the master integrals which can not be solved in general. An approach for solving this problem is to pick the two master integrals in a way such that the derivatives of the two master integrals is proportional to the $\epsilon = (4-d)/2$ from dimensional regularization, as it was done in ref. \cite{art:Gehrmann:2000xj}.

This allows for an solution of the differential equation system at each order in $\eps$, which after all is all that is needed, as only terms up to a fized order in $\eps$ can contain physical information. So writing each of the master integrals $f(\eps,d)$ (with $x$ denoting all the kinematic variables) as 
\begin{align}
f(\eps,x) = \sum_{j=j_{\text{min}}\!\!\!\!\!\!\!\!\!}^{\infty} \eps^j f^{(j)}(x),
\end{align}
allows us to compute $f^{(i)}$ in terms of $f^{(i-1)}$ along with the boundary condition which can be calculated in some convenient point, such as the point where all kinematic variables equal one.

A recently developed systematic approach to this decoupling is denoted the canonical form \cite{art:Henn:2013pwa}. A differential equation system in the canonical form is given as
\begin{align}
d \bar{f}(\eps,x) &= \eps \bigg( \! \sum_k A_k \, d \log \! \big( y_k(x) \big) \! \bigg) \bar{f}(\eps,x).
\label{eq:canonical}
\end{align}
Here $\bar{f}$ is a vector made of all the master integrals, $y_j(x)$ are algebraic functions of the kinematic variables $x$, and the matrices $A_j$ consist of rational numbers only. Specifically this corresponds to a differential equation in each variable
\begin{align}
\frac{\partial}{\partial x_i} f_a^{(j)}(x) &= \sum_k {A_k}_{ab} \, \frac{\partial \log( y_k )}{\partial x_i} f_b^{(j-1)}(x),
\label{eq:canonicaldif}
\end{align}
which can be solved using general methods described in the next sections.

How does one get a differential equation system into canonical form? There is no general method which works in all cases. One method described in \cite{art:Henn:2014qga}, consist of imposing that the leading singularity of the Feynman-integrals are constant. All canonical integrals have this property\cite{art:ArkaniHamed:2010gh}, so reducing the search space to such integrals simplifies the problem of finding the canonical basis significantly. Other approaches include the application of Magnus series\cite{art:Argeri:2014qva} which is the method used in the example in the main text and App. \ref{sec:oneloopappendix}, and many other approaches are available as well\cite{art:Gehrmann:2014bfa, art:Lee:2014ioa, art:Meyer:2016slj}, none of which will be described here. For further recent developments, see \cite{Meyer:2017joq, art:Primo:2016ebd, Frellesvig:2017aai, Primo:2017ipr}.

For the mass-less double box of the example in the previous chapter the canonical form is given in ref. \cite{art:Henn:2013pwa}. For that case the only modifications that are needed are the addition of certain kinematical prefactors and the raising of the power of some of the propagators from one to two. For more involved cases, such as the $gg \rightarrow gH$ integrals computed in \cite{art:Bonciani:2016qxi}, the same is true for the canonical form of many of the master integrals, yet for some the canonical master integrals have to be expressed as sums of integrals.

\subsection{Generalized Polylogarithms \label{sec:GPLs}}

In general the solutions of differential equation systems in canonical form, may be expressed in terms of a function-class denoted generalized polylogarithms (GPLs)\cite{art:Goncharov:2011ar}. Generalized polylogarithms (also known as Goncharov polylogarithms or hyperlogarithms) are defined recursively as
\begin{align}
G(a_1,\ldots,a_n;x) = \int_0^{x} \! \frac{\id z}{z - a_1} G(a_2,\ldots,a_n;z),
\label{eq:GPLdef}
\end{align}
where the integration path is a straight line, and where $G(;x) \define 1$. The exception is the case where all the $a$-indices equal zero, in which case
\begin{align}
G(\bar{0}_n;x) = \frac{1}{n!} \log^n(x).
\end{align}
A piece of terminology that we will need later is the concept of {\it weight} which is defined as the number of $a$-indices or correspondingly as the number of recursive integrals.

Many functions often encountered in the ($\eps$-expanded) results for Feynman integrals, are special cases of generalized polylogarithms
\begin{align}
\log(x) &= G(0,x) & \Li_n(x) = -G(\bar{0}_{n-1},1,x),
\end{align}
where the latter function is the classical (or Euler) polylogarithm, defined recursively as
\begin{align}
\Li_n(x) &= \int_0^x \! \frac{\id z}{z} \Li_{n-1}(z),
\end{align}
with $\Li_1(x) = -\log(1-x)$.

At the one-loop level all Feynman integrals may be expressed (up to the $\eps^0$ part) in terms of the functions $\log$ and $\Li_2$ and thus they are always given in terms of generalized polylogarithms. At higher weights this is true not for all, but for a large class of functions, since whenever the differential equation system for a Feynman integral can be expressed in canonical form, the result may be expressed in terms of GPLs.

For the case where the algebraic functions $y(x)$ of Eqs.~\eqref{eq:canonical} and \eqref{eq:canonicaldif} may be expressed as polynomials, this is easy to see. In that case the logarithms of Eqs.~\eqref{eq:canonical} and \eqref{eq:canonicaldif} factorize, and Eq.~\eqref{eq:canonicaldif} may be expressed as
\begin{align}
\frac{\partial}{\partial x_i} f_a^{(j)}(x) &= \sum_k {A_k}_{ab} \, \frac{1}{x_i - a_k} f_b^{(j-1)}(x),
\label{eq:canGPL}
\end{align}
which we see has the exact form of Eq.~\eqref{eq:GPLdef}, allowing for a result given directly in terms of generalized polylogarithms. Whether or not this property is present for all algebraic forms of $y(x)$ including cases where no variables change can be found that allows for a rationalization, must be considered an open question.

GPLs fulfill a large number of relations between themselves. They go under names such as the re-scaling relation, and the shuffle and stuffle relations. We shall not summarize those here, see for instance refs. \cite{art:Borwein:1999aa, art:Vollinga:2004sn, art:Duhr:2012po, art:Panzer:2014caa}. We will however mention one property - namely the fact that all GPL can be reduced to a certain minimal set of functions, of lower complexity than the general GPL\cite{art:Duhr:2012po}. In general this statement is a conjecture, but for GPLs of weight $\leq 4$ which is all that is needed for Feynman integrals with two loops, this has been shown explicitly\cite{art:Frellesvig:2016ske} and the minimal set for that case is the functions $\log$, $\Li_n$ with $n \leq 4$, and the special function $\Litt$, which in the language of GPLs is given as
\begin{align}
\Litt(x,y) &= G \big( 0,1,0,\tfrac{1}{y}; x \big).
\end{align}

Evaluating a GPL numerically, may be done by reexpressing is as an infinite sum, which may be truncated once the desired numerical precision has been optained. A \texttt{GiNaC} implementation which can be used to evauate any GPL using such techniqes, is presented in ref. \cite{art:Vollinga:2004sn}. Another option is to first reduce the GPLs to a minimal set of functions, and then use specialized tools to evaluate those, such as those presented in refs. \cite{art:Frellesvig:2016ske, art:Kirchner:2016hmb}.

\subsection{Symbols \label{sec:symbols}}

The large number of relations between the GPLs make the task of reducing them to a minimal set quite challenging. One mathematical tool which simplifies the application of these relations very significantly is that of {\em symbols}~\cite{art:Goncharov:2010jf, art:Duhr:2012po}. Symbols are short for {\em Chen symbols}, as they utilize the iterated structure of the polylogarithmic functions, which are of the form of {\em Chen iterated integrals}~\cite{art:chen1977}. Popularly expressed, the algebra of these symbols captures the algebraic parts of the relations between the polylogarithmic functions but leaves out the analytical parts. This is shown by the fact that the rules of symbol calculus explicitly set to zero factors such as $\pi$, which arise from the analyticity, as the branch-cut of the logarithm is given as $2 \pi i$ and those of the poly-logarithms contain similar factors.

Explicitly, the symbol $\kur{S}$ of a GPL is defined recursively, and given as~\cite{art:Dixon:2013eka}
\begin{align}
\kur{S}(G(a_1,\ldots,a_n;x)) &= \sum_{1}^{n} \bigg( \kur{S}\big( G(a_1,\ldots,\hat{a}_i,\ldots,a_n;x) \big) \otimes (a_i - a_{i-1}) \nn \\
& \quad\quad - \kur{S}\big( G(a_1,\ldots,\hat{a}_i,\ldots,a_n;x) \big) \otimes (a_i - a_{i+1}) \bigg),
\end{align}
where $a_{n+1} \define 0$ and $a_{0} \define x$, and where $\hat{a}_i$ denotes that the $a_i$ entry is left out. The recursions ends as $\kur{S}(\log(x))=x$.
In simple cases this reduces to
\begin{align}
\kur{S}\big( \log^n(x) \big) &= n! \, \big( x \otimes \cdots \otimes x \big) \nn \\
\kur{S}\big( \Li_n(x) \big) &= - \; \big( (1-x) \otimes x \otimes \cdots \otimes x \big),
\end{align}
both with $n$ entries in the symbol.

The individual terms in the symbol, are thus tensorial factors of the general form $\alpha_1 \otimes \cdots \otimes \alpha_n$, where $n$ is the weight of the original polylogarithmic function. These terms fulfil algebraic rules such as
\begin{align}
a \otimes bc \otimes d &= a \otimes b \otimes d \; + \; a \otimes c \otimes d,
\end{align}
which reflect the corresponding property of the logarithm. From this we get that $a \otimes 1 \otimes b = 0$ and likewise for any root of unity. From the mathematical properties, it is guaranteed that expressions which have the same symbol, are identical up to factors for which the symbol vanish, such as factors\footnote{The fact that the symbol of $\pi$ vanishes, can be seen from the fact that $\kur{S}(\Li_2(1)) = -(0 \otimes 1) = 0$, and $\Li_2(1) = \pi^2/6$.} containing $\pi$ and other of what is known as transcendental constants.

To show a simple example
\begin{align}
\kur{S} \big( \Li_2(x) \big) &= - \; \big( (1-x) \otimes x \big), \nn \\
\kur{S} \big( \Li_2(1/x) \big) &= \big( (1-x) \otimes x \big) - \big( x \otimes x \big), \nn \\
\kur{S} \big( \log^2(-x) \big) &= 2 \big( x \otimes x \big),
\end{align}
from which we may deduce the relation
\begin{align}
\Li_2\big( \tfrac{1}{x} \big) &= - \Li_2(x) - \tfrac{1}{2} \log^2(-x) + c,
\end{align}
where $c$ is a term containing factors for which the symbol vanishes. Such factors can usually be found numerically, and for this case it tuns out that $c = - \pi^2/6$.

The minimal set of objects which may appear inside the symbol is called the alphabet and the set of possible terms in the symbol which contain the letters of that alphabet are called words. In the example above the alphabet consist of two letters $x$ and $1-x$, but in the general case the set may be much larger, the alphabet of the largest of the integral families of Ref.~\cite{art:Bonciani:2016qxi} contained 49 letters. 

The general strategy~\cite{art:Goncharov:2010jf, art:Duhr:2012po} for using symbols to simplify expressions containing GPLs (often denoted the DGR algorithm after the authors of Ref.~\cite{art:Duhr:2012po}) is as follows:
\begin{enumerate}
\item Find the symbol $S$ of the expressions that needs reducing.
\item Find a ``basis'' of functions containing all the words present in $S$.
\item Invert the system to get an expression with matching symbol.
\item Find remaining factors not captured by the symbol.
\end{enumerate}
Point 2 on that list is the difficult part. For a logarithm of a specific argument, we see that a criterion for that logarithm to be present in the basis is that its argument can be written as a product of alphabet letters. For the classical polylogarithm $\Li_n(x)$ the criterion is that both $x$ and $1-x$ has that property, and for $\Litt$ there is a similar but stricter criterion. However, there is an infinite number of functions fulfilling these criteria, and to find a set that is complete and not over-complete is a significant (computer-)algebraic challenge.

Point 4 on the list is also far from easy, as the set of functions containing factors for which the symbol vanish is rather large in general. There is a development of the symbol framework, utilizing the mathematical concept of coproducts~\cite{art:Duhr:2012fh}, which may be used to capture most of these terms with a similar algorithm, but these developments will not be described here.

Yet the symbol is not only for simplification, but may be used in general to extract the result from a differential equation in canonical form without going through the step of GPLs. Whenever we have a differential equation on the form of Eq.~\eqref{eq:canonical}, the symbol may be found as
\begin{align}
d f_i^{(n)}(x) &= \sum_{j,k} {A_k}_{ij} \, d \log \! \big( y_k(x) \big) f_{j}^{(n-1)}(x) \;\; \Leftrightarrow \nn \\
\kur{S} \big( f_i^{(n)}(x) \big) &= \sum_{j,k} {A_k}_{ij} \kur{S} \big( f_{j}^{(n-1)}(x) \big) \otimes y_k(x)
\label{eq:symfromdif}
\end{align}
This replaces point 1 in the above procedure, which then may be used to solve the differential equation at each order in $\epsilon$. The remaining factors in point 4 are fixed at each order by the boundary conditions of the differential equation, making that step much simpler than it is in the general case.

No assumption was made in Eq.~\eqref{eq:symfromdif} that the letters $y(x)$ had to be polynomial, as it was assumed in for instance Eq.~\eqref{eq:canGPL}. And thus we see that an expression in terms of GPLs may be obtained for any differential equation in canonical form, as long as a suitable basis of functions to which to fit can be found.

It is this procedure which was used to solve the canonical integrals in Ref.~\cite{art:Bonciani:2016qxi}, with one exception: The size of the alphabet made it too impractical to find a basis of genuine polylogarithms to span the result. Only up to weight two was this possible, so at weights three and four a one-fold integral was used to express the result.

\subsection{Elliptic integrals}

\begin{figure}
\begin{center}
  \begin{subfigure}{0.32\textwidth}
    \begin{center}
    \includegraphics[width=1.\textwidth]{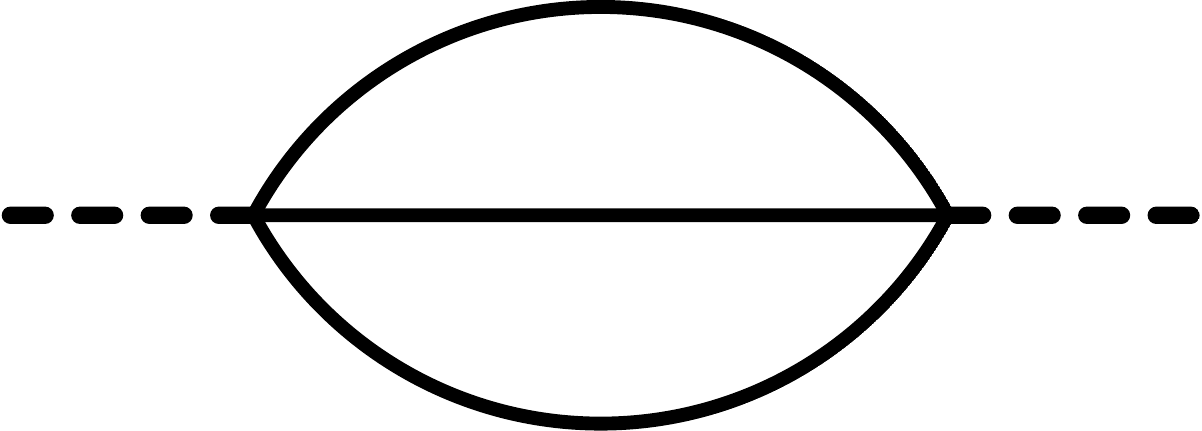}
\caption{}
    \end{center}
  \end{subfigure}
  \begin{subfigure}{0.32\textwidth}
    \begin{center}
    \includegraphics[width=1.\textwidth]{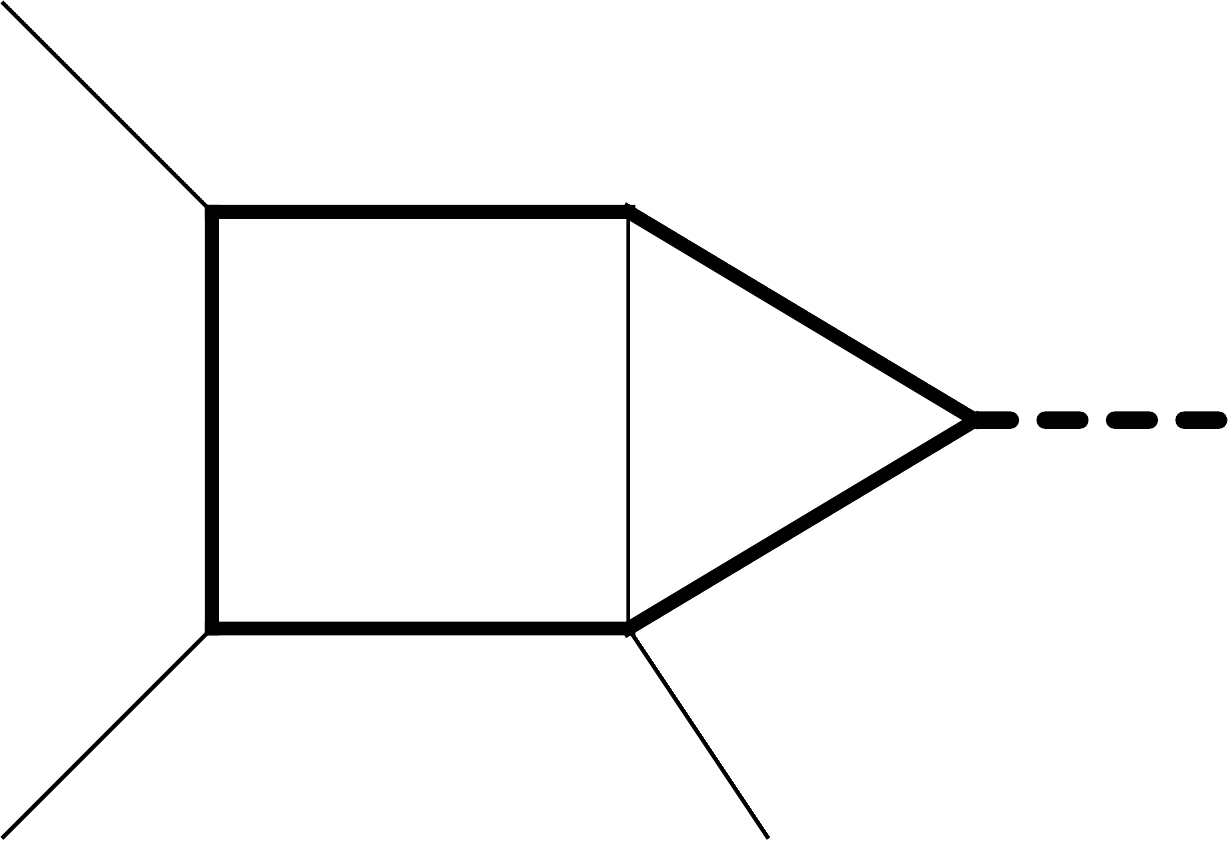}
\caption{}
    \end{center}
  \end{subfigure}
  \begin{subfigure}{0.32\textwidth}
    \begin{center}
    \includegraphics[width=1.\textwidth]{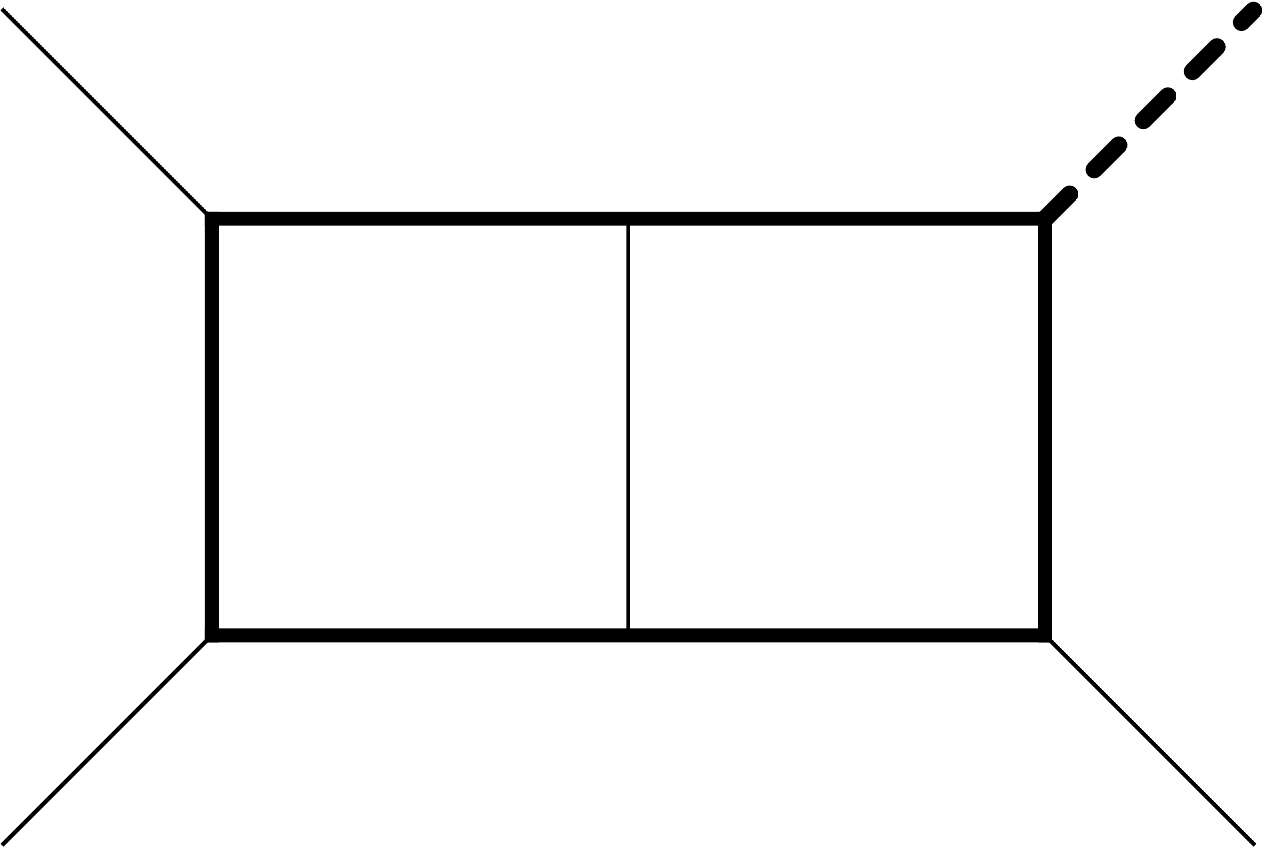}
\caption{}
    \end{center}
  \end{subfigure}
\end{center}
\caption{(a) is the well-known elliptic massive sunrise diagram. (b) and (c) are the elliptic topologies arising in the context of planar $gg \rightarrow gH$ at two-loop: the box-triangle $I_{\text{bt}}^A$ and the double-box $I_{\text{db}}^A$}.
\label{fig:elliptic}
\end{figure}

It is not all Feynman integrals which can be expressed in terms of GPLs, once we move beyond the one-loop case. The best known example is the so-called massive sunrise diagram~\cite{art:Caffo:1998du, art:Laporta:2004rb, art:Adams:2013kgc} shown in Fig.~\ref{fig:elliptic}(a)
\begin{align}
I_{\text{massive sunrise}} &= \int \!\! \int \frac{\id^d k_1 \id^d k_2}{(i \pi^{d/2})^2} \, \frac{1}{\big( k_1^2 - m^2 \big) \big( (k_2+p)^2 - m^2 \big) \big( (k_1-k_2)^2 - m^2 \big)},
\end{align}
where $p^2=s$, for which the result contains the functions
\begin{align}
K(k) &\define \int_0^1 \! \frac{\id x}{\sqrt{(1-x^2) (1- k^2 x^2)}}, \\
E(k) &\define \int_0^1 \! \sqrt{\frac{1 - k^2 x^2}{(1-x^2)}} \; \id x,
\end{align}
which are denoted the complete elliptic integral of the first and second kind respectively, and which are independent functions not expressible in terms of generalized polylogarithms.

There have been a lot of studies in the previous years of this class of functions~\cite{art:CaronHuot:2012ab, art:Bloch:2013tra, art:Sogaard:2014jla, art:Henn:2014qga, art:Adams:2014vja, art:Adams:2015ydq, art:Bloch:2016izu} particularly attempting to generalize polylogarithmic functions to this larger class. But so far no mathematical tool-set exist which is anywhere nearly as powerful as that of canonical basis and symbols for GPLs.

Elliptic integrals also show up in the results for eight of the planar $gg \rightarrow gH$ integrals of Ref.~\cite{art:Bonciani:2016qxi}. This is not in the form of the massive sunset graph, but rather in the form of the two topologies $I^A_{\text{bt}}$ and $I^A_{\text{db}}$ shown on Fig.~\ref{fig:elliptic}, which contain four master integrals each. For the six-propagator box-triangle topology $I^A_{\text{bt}}$, the ellipticity reveals itself when getting the topology to canonical form turns out to be impossible, and the best that can be done is to obtain a differential equation of the form
\begin{align}
\frac{\partial}{\partial x} \bar{f}_{\text{bt}}(\eps,x) &= A(x) \bar{f}_{\text{bt}}(\eps,x) + \eps B(x) \bar{f}_{\text{bt}}(\eps,x) + \eps \sum \bar{C}(x) f_{\text{lower}\,i}(\eps,x).
\end{align}
A basis can be found where the matrix $A$ can be expressed as
\begin{align}
A(x) &= \left[ \begin{array}{cccc}
a_{11}(x) & a_{12}(x) & 0 & 0 \\
a_{21}(x) & a_{22}(x) & 0 & 0 \\
a_{31}(x) & a_{32}(x) & a_{33}(x) & 0 \\
a_{41}(x) & a_{42}(x) & 0 & a_{44}(x) \end{array} \right]
\end{align}
where we see that at the level of the homogeneous term, the first two entries of $\bar{f}_{\text{bt}}$ decouple and form their own two-by-two sub-system.
This property may be used to write a second-order differential equation for the homogeneous part of the equation, which may be solved with the result given in terms of elliptic integrals. Using the ``method of the variation of constants'', the general solution may then be expressed as a one-fold integral over the homogeneous solution multiplied with the lower order terms as well as with various other factors, yielding the full result for the two first entries of $\bar{f}_{\text{bt}}$ as integrals over elliptic integrals. The remaining two entries may then, from the differential equation, be written as integrals over those solutions, though it turns out to be possible to reduce the number of recursive integrals for this case by one, using methods that will not be described here.

For the seven-propagator topology $I^A_{\text{db}}$, the ellipticity enters only through the elliptic sub-topology. It is thus possible to describe the differential equation as
\begin{align}
\frac{\partial}{\partial x} \bar{f}_{\text{db}}(\eps,x) &= \eps D(x) \bar{f}_{\text{db}}(\eps,x) + E(x) \bar{f}_{\text{bt}}(\eps,x) + \eps \sum \bar{F}(x) f_{\text{lower}\,i}(\eps,x),
\end{align}
where the ``lower'' in the last term includes contributions from the $I_{\text{bt}}$ topology. It is possible to remove the $E(x) \bar{f}_{\text{bt}}(\eps,x)$ term and get the equation to an $\epsilon$-factorized form~\cite{art:Gehrmann:2014bfa}, but as the $I_{\text{bt}}$-topology itself involve elliptic integrals, doing so does not remove any elliptic integrals from the result, it merely ensures that no new ones appear. For further considerations of the elliptic functions appearing in these topologies, see Refs.~\cite{art:Primo:2016ebd, Frellesvig:2017aai, Primo:2017ipr, Bosma:2017ens, Harley:2017qut}.

\subsubsection*{Final remarks}

The result of the calculation of the planar Feynman integrals contributing to the process $gg \rightarrow gH$ as described in Ref.~\cite{art:Bonciani:2016qxi} takes up around $500$ MB and to evaluate the expression in one phase-space point with eight digits of accuracy, takes about $20$ minutes on one CPU core. This is primarily due to the fact that the result is expressed as an integral (or for the elliptic cases, as double or triple iterated integrals) which has to be performed numerically. 

Needless to say, such a timing is not suitable for inclusion in for instance a Monte Carlo event generator, and this fact should motivate the search for methods to approximate the result, such as those described in Refs.~\cite{art:Neumann:2016dny, art:Melnikov:2016qoc} as well as in sections~\ref{sec:NLOinHEFT} and \ref{sec:NNLOinHEFT}.


\clearpage



\bibliographystyle{atlasnote}
\bibliography{biblio}


\end{document}